\documentclass{article}
\usepackage{float}

\usepackage{graphicx}  
\usepackage[left=2cm,top=2cm,right=2cm,bottom=2cm,nohead]{geometry}
\usepackage{amsmath,amssymb}  
\usepackage{natbib} 
\usepackage{longtable}
\usepackage{lscape}
\usepackage{enumerate}
\usepackage{bm}
\usepackage{booktabs,caption}
\usepackage{threeparttable}
\usepackage{url}

\usepackage{bbm}
\usepackage{authblk}

\usepackage{mathtools}
\mathtoolsset{showonlyrefs,showmanualtags}

\usepackage[table]{xcolor}
\newif\ifcoauthorview
\coauthorviewfalse

\newcommand{\be}{\begin{enumerate}}
\newcommand{\ee}{\end{enumerate}}
\newcommand{\bi}{\begin{itemize}}
\newcommand{\ei}{\end{itemize}}

\title{Realized Stochastic Volatility Models with Skew-t Distributions for Volatility and Tail Risk Forecasting\thanks{The authors gratefully acknowledge the Ministry of Education, Culture, Sports, Science and Technology of the Japanese Government through Grant-in-Aid for Scientific Research (Nos. 22K13376, 23H00048, and 24H00142), the Hitotsubashi Institute for Advanced Study, and Hosei University’s Innovation Management Research Center.}}
\author[1]{Makoto Takahashi\thanks{Corresponding author. Email: \texttt{m-takahashi@hosei.ac.jp}}}
\author[2]{Yuta Yamauchi}
\author[3]{Toshiaki Watanabe}
\author[4]{Yasuhiro Omori}
\affil[1]{Faculty of Business Administration, Hosei University}
\affil[2]{Graduate School of Economics, Nagoya University}
\affil[3]{Graduate School of Social Data Science, Hitotsubashi University}
\affil[4]{Faculty of Economics, University of Tokyo}
\date{\empty}

\begin{document}
\maketitle
\ifcoauthorview\input{coauthor_note.tex}\fi

\begin{abstract}
Volatility forecasts underlie tail risk measures: value-at-risk (VaR) and expected shortfall (ES).
This paper compares realized stochastic volatility (RSV) models, which add a realized measure to the stochastic volatility (SV) model, with their SV counterparts under normal, Student's $t$, two skew-normal, and three skew-$t$ return distributions, and against EGARCH and REGARCH benchmarks.
Estimated by Markov chain Monte Carlo on the Dow Jones Industrial Average and the Nikkei 225, the models are compared by their one-day-ahead volatility, VaR, and ES forecasts (the predictive mean for volatility).
Adding realized volatility improves volatility forecasts over the main sample for both indices under every proxy and return distribution; the gains are uniformly significant for the Nikkei 225 but, for the Dow Jones, only under the realized variance used in estimation.
During the COVID-19 period the gains persist for the Nikkei 225 but vanish for the Dow Jones, where the REGARCH benchmark is best.
For VaR and ES the gains are confined to the Nikkei 225 and to the $5\%$ level for the Dow Jones, and few matched-pair differences are significant.
Skewed innovations lower the $1\%$ joint loss for the Dow Jones, but no specification dominates across indices, risk levels, and sample periods.
\end{abstract}

\noindent%

\section{Introduction}
\label{sec:intro}
Volatility forecasts are the main input for tail risk measures such as value-at-risk (VaR) and expected shortfall (ES).
Two classes of models dominate the literature on time-varying volatility: the generalized autoregressive conditional heteroskedasticity (GARCH) family \citep{engle_autoregressive_1982, bollerslev_generalized_1986}, including the exponential GARCH (EGARCH) model of \citet{nelson_conditional_1991}, and the stochastic volatility (SV) model \citep{taylor_modelling_1986}.
Realized volatility (RV), computed from intraday returns, provides a more precise daily volatility measure \citep{andersen_realized_2009, mcaleer_realized_2008}, although it is affected by market microstructure noise and non-trading hours, which motivates robust estimators such as the realized kernel (RK) of \citet{barndorff-nielsen_designing_2008, barndorff-nielsen_realized_2009}.
Hybrid models that add a realized measure to a parametric volatility model combine the two approaches: the realized stochastic volatility (RSV) model \citep{takahashi_estimating_2009, dobrev_information_2010, koopman_analysis_2013} and the realized GARCH (RGARCH) and realized EGARCH (REGARCH) models \citep{hansen_realized_2012, hansen_exponential_2016}.
Comprehensive comparisons across these classes remain limited; see \citet{takahashi_stochastic_2023, takahashi_forecasting_2024} for recent attempts.
Tail risk forecasting also requires an adequate conditional return distribution, since returns remain skewed and leptokurtic after accounting for time-varying volatility.
Skewed Student's \(t\)-distributions have therefore been widely used in volatility models \citep{abanto-valle_bayesian_2015, kobayashi_skew_2016}, and the generalized hyperbolic (GH) skew-\(t\) distribution in particular has performed well in SV \citep{nakajima_stochastic_2012, leao_bayesian_2017} and RSV models \citep{trojansebastian_regime_2013, nugroho_realized_2014, nugroho_boxcox_2016, takahashi_volatility_2016}.
Score-driven models \citep{creal_generalized_2013, harvey_dynamic_2013} offer a related route to skewed and heavy-tailed volatility dynamics; \citet{catania_forecasting_2022} apply the GH skew-\(t\) distribution in that framework.\footnote{An extensive and regularly updated list of score-driven models is available at \url{https://www.gasmodel.com/}.}
We focus on SV and RSV models and regard score-driven specifications as a complementary direction.

This paper compares RSV models with three types of skew-\(t\) distributions, the Azzalini-type skew-\(t\) distribution \citep{azzalini_class_1985}, the Fern\'{a}ndez and Steel (FS) skew-\(t\) distribution \citep{fernandez_bayesian_1995}, and the GH skew-\(t\) distribution \citep{aas_generalized_2006}, against their SV counterparts and against EGARCH and REGARCH benchmarks.
All models are estimated by Bayesian Markov chain Monte Carlo (MCMC) methods, so that the comparison incorporates parameter and state uncertainty for every model class, and are applied to daily returns of the Dow Jones Industrial Average (DJIA) and the Nikkei 225 (N225).
Volatility forecasts are evaluated with the quasi-likelihood (QLIKE) loss against four realized proxies, using the predictive mean as the common point forecast for all models, and VaR and ES forecasts with the joint loss of \citet{fissler_higher_2016} as specified in \citet{patton_dynamic_2019}.
Forecast comparisons rely on the model confidence set (MCS) of \citet{hansen_model_2011} and on the pairwise test of \citet{giacomini_tests_2006} for the seven pairs of RSV and SV models that share the same return distribution.
The design separates two questions: what adding a realized measure contributes, holding the return distribution fixed, and what a more flexible return distribution contributes, holding the volatility model fixed.

The empirical findings can be summarized as follows.
(i) Adding realized volatility improves volatility forecasts over the main sample for both indices: every RSV model attains a lower QLIKE than its SV counterpart under all four proxies, and the pairwise test rejects equal accuracy at the $1\%$ level for every pair for the N225, while for the DJIA it does so under RV, the measure used in estimation, with weaker evidence under the RK, bipower, and median RV proxies.
During the COVID-19 period, the gains persist for the N225, with smaller margins, but vanish for the DJIA.
(ii) The Bayesian REGARCH benchmarks are competitive: they attain the lowest QLIKE for both indices during the COVID-19 period and belong to the model confidence set for the DJIA under the RK, bipower, and median RV proxies, but they fall behind the RSV models under RV and for the N225 over the main sample.
(iii) For VaR and ES, the evidence is mixed.
Gains from realized volatility are most apparent for the N225 and at the \(5\%\) level for the DJIA, whereas at the \(1\%\) level for the DJIA every SV model attains a lower joint loss than its RSV counterpart and most models, RSV and SV alike, fail the coverage tests; in most pairs the pairwise test does not reject equal accuracy.
During the COVID-19 period, the tail risk forecasts of the RSV models relative to the SV models remain mixed.
(iv) These tail risk results depend on a restriction that we maintain in the main specification, under which log realized volatility moves one for one with latent log volatility in the measurement equation of the RSV model.
Relaxing it over the COVID-19 sample improves both the volatility and the tail risk forecasts of every RSV specification for the DJIA and reverses the comparison against the matched SV models at both risk levels, whereas for the N225 it improves neither the volatility nor the tail risk forecasts.
(v) The return distribution matters less for volatility forecasting than for tail risk: skew-\(t\) specifications offer limited incremental gains in QLIKE, but skewed distributions lower the \(1\%\) joint loss for the DJIA within both the SV and the RSV class, and no single distribution dominates across indices, risk levels, and sample periods.
Taken together, realized volatility delivers clear improvements in volatility forecasting over the main sample for both indices, most robustly for the N225, whereas under the stress of 2020 its advantage survives only for the N225, and its benefits for tail risk forecasting are conditional.

The remainder of the paper is organized as follows.
Section~\ref{sec:model} presents the SV and RSV models with skewed innovations and the GARCH-type benchmarks.
Section~\ref{sec:bayes_method} outlines the Bayesian estimation and the forecasting procedure.
Section~\ref{sec:application} reports the empirical results.
Section~\ref{sec:conc} concludes.
\section{Model}
\label{sec:model}

\subsection{Stochastic Volatility Model}
\label{sec:SV}

Let \( y_t \) denote the daily log return of an asset:
\[
y_t = \log p_t - \log p_{t-1},
\]
where \( p_t \) is the closing price on day \( t \).
In the SV model the log volatility \( h_t \) is a latent process, and the return and its log volatility are specified as
\begin{align}
y_t &= \epsilon_t \exp \left( \frac{h_t}{2} \right), \quad \epsilon_t \sim \mathcal{N}(0, 1), \quad t = 1, \ldots, n, \label{eq:sv-leverage-1} \\
h_{t+1} &= \mu + \phi (h_t - \mu) + \eta_t, \quad \eta_t \mid \epsilon_t \sim \mathcal{N} \left( \rho \sigma_\eta \epsilon_t, (1 - \rho^2) \sigma_\eta^2 \right), \quad t = 1, \ldots, n-1, \quad |\phi| < 1, \label{eq:sv-leverage-2} \\
h_1 &\sim \mathcal{N} \left( \mu, \frac{\sigma_{\eta}^2}{1 - \phi^2} \right).
\end{align}
Here, \( \phi \) is the persistence parameter, and \( \mu \) is the unconditional mean of log volatility.
Stationarity is ensured by \( |\phi| < 1 \).
The initial state \( h_1 \) is drawn from its unconditional distribution.

The bivariate error term \( (\epsilon_t, \eta_t)' \) follows a jointly normal distribution with correlation \( \rho \).
This structure allows for a leverage effect: a negative return shock (\( \epsilon_t < 0 \)) tends to increase future volatility (\( h_{t+1} \)), reflected by \( \rho < 0 \).
This feature captures the empirically observed asymmetry in financial markets, where volatility tends to rise after negative returns.

\subsection{Realized Volatility}
Realized volatility (RV), the sum of squared intraday returns, is a consistent estimator of the integrated variance of the day under ideal conditions, but two biases arise in practice \citep{andersen_realized_2009, mcaleer_realized_2008}.
First, intraday data do not cover non-trading hours, so that RV underestimates daily volatility; following \citet{hansen_forecast_2005} we scale RV so that its average matches the sample variance of daily returns,
\begin{align}
RV_t^{HL} = c_{HL} RV_t, \quad c_{HL} = \frac{ \sum_{t=1}^n (y_t - \bar{y})^2 }{ \sum_{t=1}^n RV_t }, \quad \bar{y} = \frac{1}{n} \sum_{t=1}^n y_t. \label{eqn:HL}
\end{align}
Second, market microstructure noise biases RV at high sampling frequencies, which motivates robust estimators such as the realized kernel (RK) of \citet{barndorff-nielsen_designing_2008, barndorff-nielsen_realized_2009} and, for evaluation, the use of several proxies.
The RSV model of \citet{takahashi_estimating_2009}, introduced next, treats the remaining bias of the realized measure as a parameter of the measurement equation.
\subsection{Realized Stochastic Volatility Model}
\label{sec:RSV}

\citet{takahashi_estimating_2009} proposed enhancing the SV model by incorporating additional information from RV.
In their framework, the daily log-return \( y_t \) and the logarithm of realized volatility \( x_t = \log RV_t \) are jointly modeled, with the latent log volatility \( h_t = \log IV_t \) serving as a common driver.

The RSV model is specified as:
\begin{align}
x_t &= \xi + h_t + u_t, \quad u_t \sim \mathcal{N}(0, \sigma_u^2), \quad t = 1, \ldots, n, \label{eqn:RSV-x} \\
y_t &= \epsilon_t \exp \left( \frac{h_t}{2} \right), \quad \epsilon_t \sim \mathcal{N}(0, 1), \quad t = 1, \ldots, n, \label{eqn:RSV-R} \\
h_{t+1} &= \mu + \phi (h_t - \mu) + \eta_t, \quad \eta_t \mid \epsilon_t \sim \mathcal{N} \left( \rho \sigma_\eta \epsilon_t, (1 - \rho^2) \sigma_\eta^2 \right), \quad t = 1, \ldots, n-1, \quad |\phi| < 1, \label{eqn:RSV-h}
\end{align}
where \( h_1 \) follows its unconditional distribution as in the SV model of Section~\ref{sec:SV}.
The error term \( u_t \) is assumed to be independent of \( (\epsilon_t, \eta_t) \).

Equation~\eqref{eqn:RSV-x} may be generalized as \( x_t = \xi + \psi h_t + u_t \).
We fix \( \psi = 1 \) for parsimony and comparability across the RSV specifications, following \citet{takahashi_estimating_2009}.
When \( \psi \) is estimated freely on the full sample, its posterior 95\% credible interval excludes unity (Table~S6 of the Supplementary Material).
A sensitivity analysis reported in Sections~S10.3 and S11.3 of the Supplementary Material, comparing free-\( \psi \) and \( \psi = 1 \) forecasts during the COVID-19 period, shows that every RSV specification attains better forecast accuracy under a free \( \psi \) for the DJIA and none does for the N225.

The parameter \( \xi \) adjusts for systematic bias in \( x_t = \log RV_t \).
When \( \xi = 0 \), \( x_t \) is an unbiased estimate of \( h_t \) .
In practice, however, \( RV_t \) may be biased downward due to non-trading hours and upward due to microstructure noise.
\citet{hansen_realized_2006} show that the net bias can be either positive or negative.
Therefore, the sign of \( \xi \) reflects the relative magnitude of these two opposing effects.

The distribution of \( \epsilon_t \) captures residual return behavior after controlling for time-varying volatility.
While the SV structure already accounts for volatility clustering and excess kurtosis, it is well documented that financial returns often exhibit additional heavy tails and asymmetry.
To model these features more effectively, several studies have incorporated non-Gaussian innovations.

For the SV model, \citet{nakajima_stochastic_2012} and \citet{leao_bayesian_2017} employed the GH skew-\(t\) distribution from \citet{aas_generalized_2006}.
Other approaches include the skew-\(t\) distribution by \citet{azzalini_distributions_2003} \citep[used in][]{abanto-valle_bayesian_2015}, the skew exponential power distribution \citep{kobayashi_skew_2016}, and the flexible skew-\(t\) model by \citet{fernandez_bayesian_1995}, as adopted in \citet{steel_bayesian_1998}.
In the context of RSV models, \citet{trojansebastian_regime_2013}, \citet{nugroho_realized_2014, nugroho_boxcox_2016}, and \citet{takahashi_volatility_2016} applied the GH skew-\(t\) distribution.

Building on this literature, we extend the RSV model defined in equations~\eqref{eqn:RSV-x}--\eqref{eqn:RSV-h}, hereafter referred to as the RSV-N model, by incorporating skewed and heavy-tailed distributions for \( \epsilon_t \).
This extension allows for more accurate modeling of return asymmetry and tail risk, which is essential for reliable risk-forecasting applications.

\subsection{RSV Model with Student's \( t \) Distribution}\label{subsec:rsv-t}

We extend the RSV-N model by allowing the return innovation \( \epsilon_t \) to follow a standardized Student's \( t \) distribution.
The resulting model, referred to as the RSV-T model, replaces the Gaussian specification of \( \epsilon_t \) and \( \eta_t \) in the RSV-N model as follows:
\begin{align}
\epsilon_t &= z_t \sqrt{ \frac{\lambda_t}{\mu_{\lambda}} }, \quad z_t \sim \mathcal{N}(0, 1), \quad \eta_t \mid z_t \sim \mathcal{N} \left( \rho \sigma_{\eta} z_t, (1 - \rho^2) \sigma_{\eta}^2 \right), \label{eqn:rsv-t-eps}
\end{align}
where
\begin{align}
\lambda_t &\overset{\text{i.i.d.}}{\sim} \mathcal{IG} \left( \frac{\nu}{2}, \frac{\nu}{2} \right), \quad \mu_{\lambda} = \mathrm{E}[\lambda_t] = \frac{\nu}{\nu - 2}, \quad \nu > 2. \label{eqn:mu-lambda}
\end{align}
Here, \( \lambda_t \) is an inverse-gamma mixing variable, and \( \nu \) is the degrees of freedom parameter controlling tail thickness.
As \( \nu \to \infty \), the model converges to the Gaussian RSV-N specification.

The use of Student’s \( t \) distribution allows the model to accommodate excess kurtosis beyond what is captured by the stochastic volatility component \( h_t \).
For finite \( \nu \), the distribution of \( \epsilon_t \) exhibits heavier tails, improving the model’s ability to reflect extreme return events.

Importantly, the contemporaneous correlation between \( \epsilon_t \) and \( \eta_t \) becomes a function of both \( \rho \) and \( \nu \), given by:
\begin{align}
\mathrm{Corr}[\epsilon_t, \eta_t] = \frac{ \mathrm{E} \left[ \sqrt{\lambda_t} \right] }{ \sqrt{\mu_{\lambda}} } \rho,
\end{align}
where
\begin{align}
\mathrm{E} \left[ \sqrt{\lambda_t} \right] = \sqrt{ \frac{\nu}{2} } \cdot \frac{ \Gamma \left( \frac{\nu - 1}{2} \right) }{ \Gamma \left( \frac{\nu}{2} \right) }. \label{eqn:inv-gam-moment}
\end{align}
Note that the variance of \( \epsilon_t \) exists only when \( \nu > 2 \), and its fourth moment exists when \( \nu > 4 \).
In empirical applications, \( \nu \) is typically estimated from the data.

\subsection{RSV Model with GH Skew-\( t \) Distribution}

\citet{takahashi_volatility_2016} extended the RSV-N model in equations~\eqref{eqn:RSV-x}--\eqref{eqn:RSV-h} by incorporating the GH skew-\( t \) distribution, resulting in the RSV-GH-ST model.
This is achieved by replacing \( \epsilon_t \) with the following mixture representation:
\begin{align}
\epsilon_t &= \frac{ \beta (\lambda_t - \mu_\lambda) + \sqrt{\lambda_t} z_t }{ \sqrt{ \beta^2 \sigma_\lambda^2 + \mu_\lambda } }, \label{eqn:rsv-gh-st-eps}
\end{align}
where \( z_t \), \( \eta_t \mid z_t \), \( \lambda_t \), and \( \mu_\lambda \) are as in the RSV-T model, and
\begin{align}
\sigma_\lambda^2 = \mathrm{Var}[\lambda_t] = \frac{2\nu^2}{(\nu - 2)^2 (\nu - 4)}, \quad \nu > 4. \label{eqn:sigma-lambda}
\end{align}
The denominator in \eqref{eqn:rsv-gh-st-eps} standardizes \( \epsilon_t \) to have unit variance, so that the variance of \( y_t \) given \( h_t \) equals \( \exp(h_t) \).

A distinctive feature of the GH skew-\( t \) distribution, originally emphasized by \citet{aas_generalized_2006}, is its asymmetric tail behavior: one tail exhibits polynomial (power-law) decay, while the opposite tail decays exponentially, depending on the sign of the skewness parameter \( \beta \).
This feature makes the GH skew-\( t \) distribution particularly flexible in capturing return asymmetry and tail risk in financial applications.

In the present formulation, the skewness and tail behavior of \( \epsilon_t \) are jointly governed by \( \beta \) and \( \nu \).
The parameter \( \beta \) introduces asymmetry and determines which tail becomes heavier, while \( \nu \) controls the degree of polynomial tail thickness.
When \( \beta = 0 \), the model reduces to the RSV-T specification in Section~\ref{subsec:rsv-t}.
As \( \nu \to \infty \), or if \( \lambda_t = 1 \) for all \( t \), the distribution converges to the Gaussian case regardless of \( \beta \).
Panel (a) of Figure~\ref{fig:density-row} shows the standardized density for \( \nu = 10 \) and \( \beta = 0, -1, -2 \): a more negative \( \beta \) produces a heavier left tail, and \( \beta = 0 \) recovers the symmetric Student's \( t \) distribution.

\subsection{RSV Model with Azzalini Skew-\(t\) Distribution}

We extend the RSV-N model defined in equations~\eqref{eqn:RSV-x}--\eqref{eqn:RSV-h} by adopting the skew-\(t\) distribution proposed by \citet{azzalini_class_1985}.\footnote{See also \citet{azzalini_distributions_2003} and \citet{sahu_new_2003}.} The resulting model, referred to as RSV-AZ-ST, modifies the return innovation \( \epsilon_t \) as follows:
\begin{align}
\epsilon_t &= \frac{ \delta(z_{0t} - c) + \sqrt{1 - \delta^2} \, z_t }{ \sqrt{1 - c^2 \delta^2} } \cdot \sqrt{ \frac{ \lambda_t }{ \mu_\lambda } }, \quad z_{0t} \overset{\text{i.i.d.}}{\sim} \mathcal{TN}_{(0,\infty)}(0, 1), \quad |\delta| < 1, \label{eqn:rsv-az-st-eps}
\end{align}
where \( z_t \), \( \eta_t \mid z_t \), \( \lambda_t \), and \( \mu_\lambda \) are as in the RSV-T and RSV-GH-ST models, and \( \mathcal{TN}_{(0,\infty)} \) denotes a standard normal distribution truncated to the interval \( (0, \infty) \).
The constants
\[
c = \mathrm{E}[z_{0t}] = \sqrt{ \frac{2}{\pi} }, \quad \mathrm{Var}[z_{0t}] = 1 - c^2
\]
are used to ensure that \( \epsilon_t \) has unit variance, so that the variance of \( y_t \) given \( h_t \) remains \( \exp(h_t) \).

Setting \( \delta = 0 \) recovers the RSV-T model.
Moreover, as \( \nu \to \infty \) or \( \lambda_t = 1 \), the model simplifies to the RSV-AZ-SN model with a skew-normal distribution.
The shape of the distribution of \( \epsilon_t \) is governed jointly by \( \delta \) and \( \nu \), which respectively control skewness and tail thickness.
Panel (b) of Figure~\ref{fig:density-row} shows the standardized density for \( \nu = 10 \) and \( \delta = 0, -0.8, -0.95 \): a more negative \( \delta \) produces a heavier left tail, and \( \delta = 0 \) recovers the symmetric Student's \( t \) distribution.

\subsection{RSV Model with Fern\'{a}ndez–Steel Skew-\(t\) Distribution}

Following the approach of \citet{fernandez_bayesian_1995}, we define the Fern\'{a}ndez–Steel (FS) skew-\(t\) density as
\begin{align}
p_T(w \mid \gamma, \nu) = \frac{2}{\gamma + \gamma^{-1}} \left\{
f_T\left( \frac{w}{\gamma} \mid \nu \right) \mathbbm{1}\{ w \geq 0 \}
+ f_T\left( \gamma w \mid \nu \right) \mathbbm{1}\{ w < 0 \}
\right\}, \quad \gamma > 0,
\end{align}
where \( f_T(\cdot \mid \nu) \) denotes the probability density function of the standard Student’s \(t\) distribution:
\begin{align}
f_T(\tilde{w} \mid \nu) = c_\nu \left(1 + \frac{\tilde{w}^2}{\nu} \right)^{ -\frac{\nu + 1}{2} }, \quad
c_\nu = \frac{ \Gamma\left( \frac{\nu + 1}{2} \right) }{ \Gamma\left( \frac{\nu}{2} \right) \Gamma\left( \frac{1}{2} \right) \sqrt{\nu} }, \quad \nu > 2.
\label{eq:t_mean0}
\end{align}
Here, \( \mathbbm{1}\{ \cdot \} \) denotes the indicator function.

The mean and variance of \( w \sim p_T(w \mid \gamma, \nu) \) are:
\begin{align}
\mu_* &= \mathrm{E}[w] = M_1(\gamma - \gamma^{-1}), \label{eq:fsst_mean} \\
\sigma_*^2 &= \mathrm{Var}[w] = M_2 \cdot \frac{ \gamma^3 + \gamma^{-3} }{ \gamma + \gamma^{-1} } - \left\{ M_1(\gamma - \gamma^{-1}) \right\}^2, \label{eq:fsst_variance}
\end{align}
where
\begin{align}
M_1 &= 2 \int_0^\infty \tilde{w} f_T(\tilde{w} \mid \nu) d\tilde{w} = \frac{2 c_\nu \nu}{\nu - 1}, \label{eq:m1_gfsst} \\
M_2 &= 2 \int_0^\infty \tilde{w}^2 f_T(\tilde{w} \mid \nu) d\tilde{w} = \frac{\nu}{\nu - 2}. \label{eq:m2_gfsst}
\end{align}

To obtain a standardized version of the FS skew-\(t\) density, we define:
\[
w_* = \frac{w - \mu_*}{\sigma_*},
\]
so that \( \mathrm{E}[w_*] = 0 \) and \( \mathrm{Var}[w_*] = 1 \).
The corresponding standardized density becomes:
\begin{align}
q_T(w_* \mid \gamma, \nu) &= \frac{2\sigma_*}{\gamma + \gamma^{-1}} \left[ 
f_T\left( \frac{ \sigma_* w_* + \mu_* }{ \gamma } \mid \nu \right) \mathbbm{1} \left\{ w_* \geq -\frac{\mu_*}{\sigma_*} \right\} \right. \nonumber \\
&\quad \left. + f_T\left( \gamma( \sigma_* w_* + \mu_* ) \mid \nu \right) \mathbbm{1} \left\{ w_* < -\frac{\mu_*}{\sigma_*} \right\}
\right]. \label{eqn:dens-fsst}
\end{align}
Replacing \( f_T \) with the standard normal density \( f_N \) in the above expression yields the standardized FS skew-normal density \( q_N(\cdot \mid \gamma) \).

We now define the RSV-FS-ST model by assuming:
\begin{align}
\epsilon_t &\sim q_T(\epsilon_t \mid \gamma, \nu), \label{eqn:rsv-fs-st-eps} \\
\eta_t \mid \epsilon_t &\sim \mathcal{N}(\rho \sigma_\eta \epsilon_t, (1 - \rho^2) \sigma_\eta^2), \label{eqn:rsv-fs-st-eta}
\end{align}
where \( q_T(\cdot \mid \gamma, \nu) \) is the standardized FS skew-\(t\) density.
Replacing \( q_T \) with \( q_N \) yields the RSV-FS-SN model.

\begin{figure}[tbp]
\centering
\includegraphics[width = .32\textwidth]{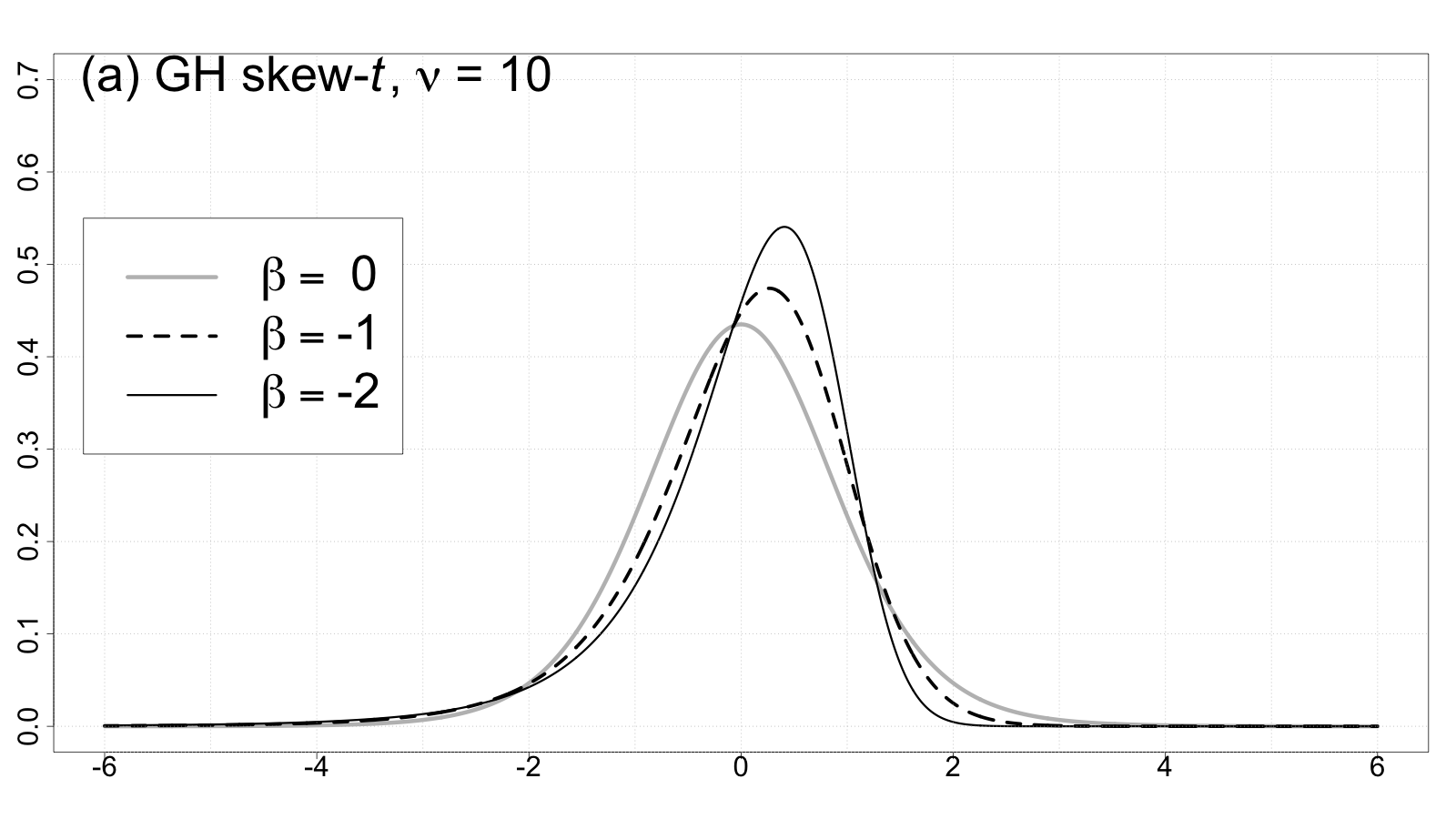}\hfill
\includegraphics[width = .32\textwidth]{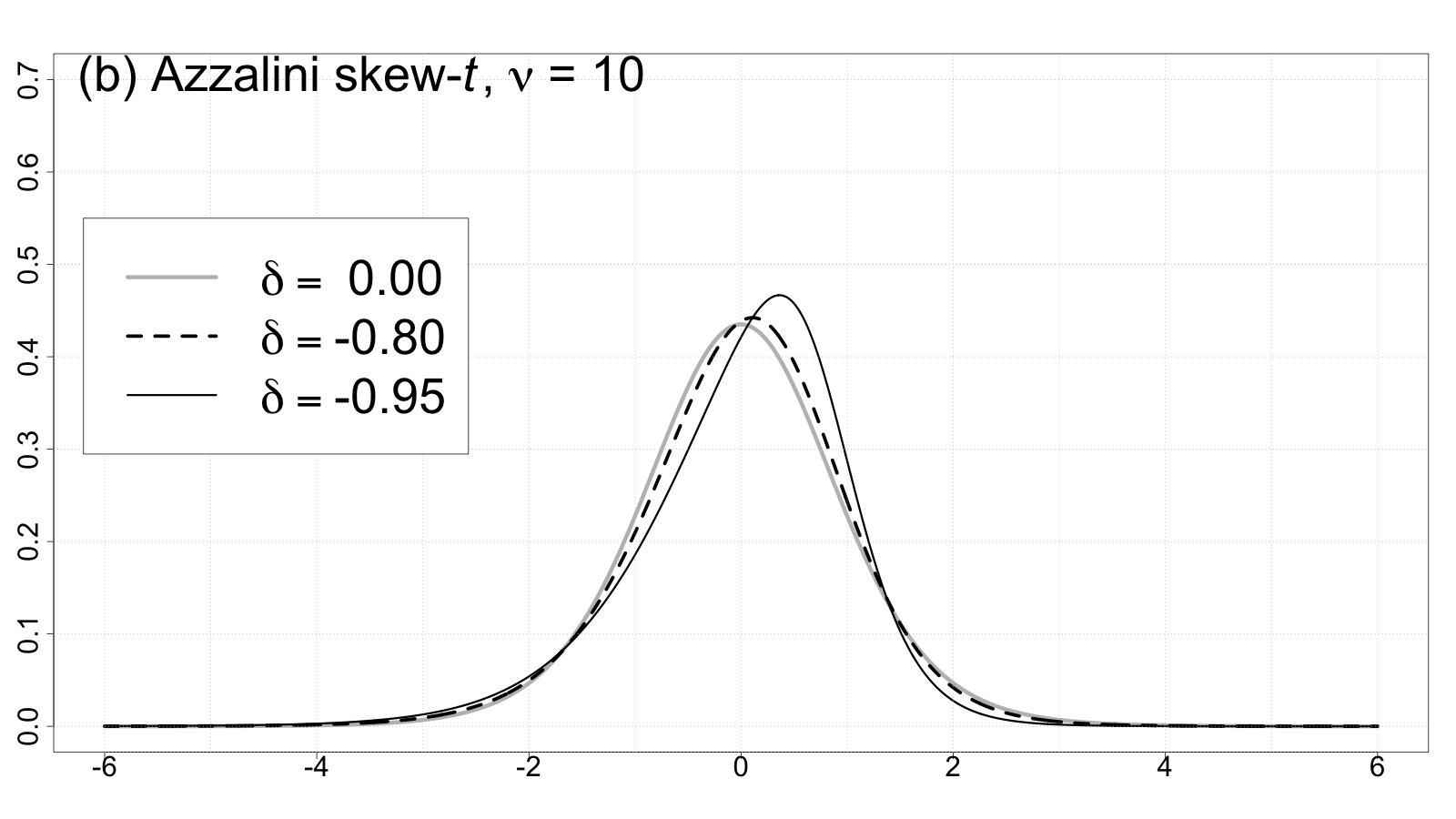}\hfill
\includegraphics[width = .32\textwidth]{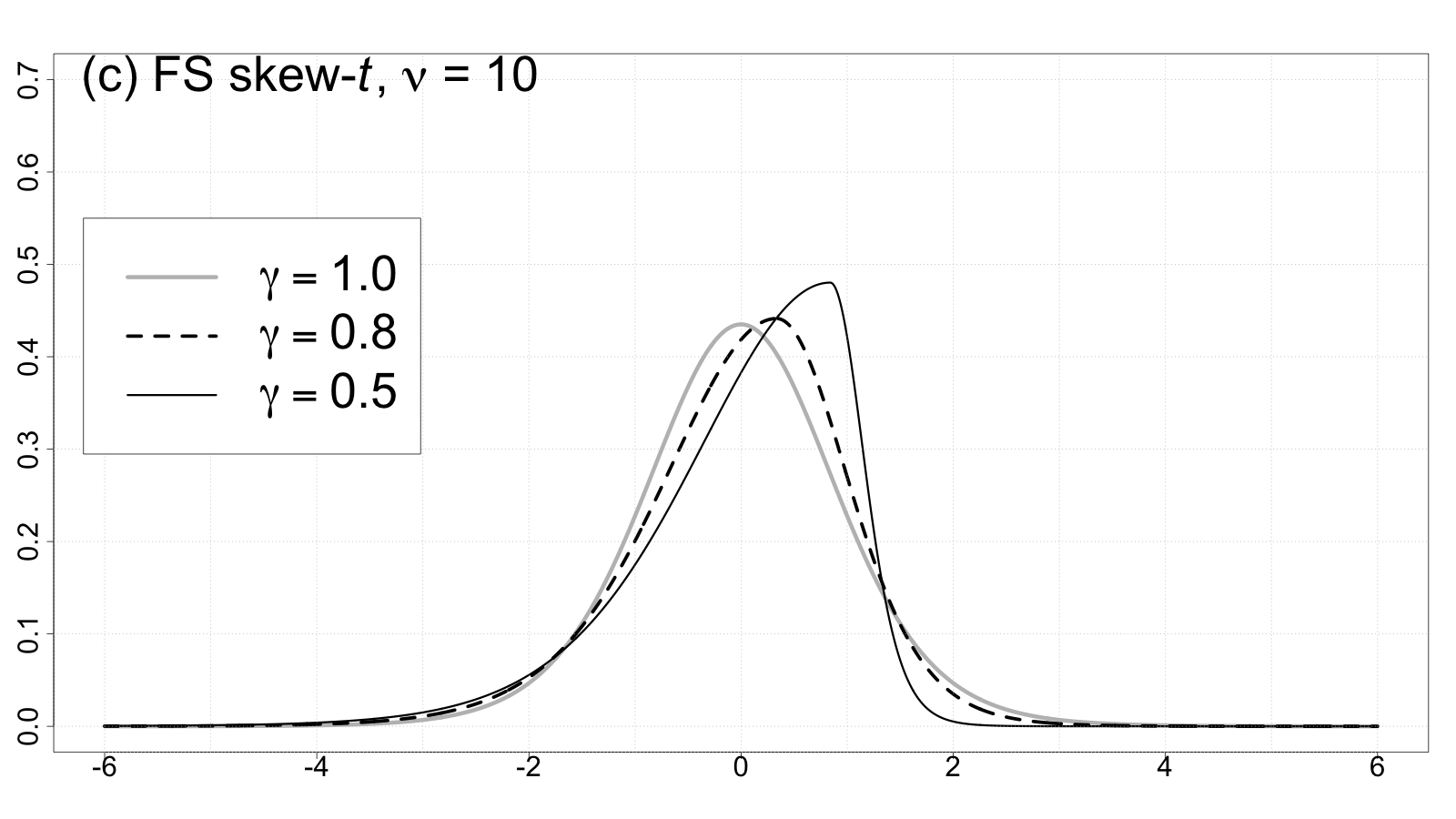}
\caption{Standardized densities with $\nu = 10$: GH skew-$t$ in equation~\eqref{eqn:rsv-gh-st-eps} with $\beta = 0, -1, -2$ (a), Azzalini skew-$t$ in equation~\eqref{eqn:rsv-az-st-eps} with $\delta = 0, -0.8, -0.95$ (b), and FS skew-$t$ in equation~\eqref{eqn:dens-fsst} with $\gamma = 1, 0.8, 0.5$ (c).}
\label{fig:density-row}
\end{figure}

The skewness of the distribution is governed by the parameter \( \gamma \), where \( \gamma = 1 \) corresponds to a symmetric \(t\)-distribution.
If \( \gamma < 1 \), the distribution becomes left-skewed, while \( \gamma > 1 \) results in right-skewness.
The parameter \( \nu \) controls tail heaviness.
Panel (c) of Figure~\ref{fig:density-row} shows the standardized density for \( \nu = 10 \) and \( \gamma = 1, 0.8, 0.5 \): a smaller \( \gamma \) produces a heavier left tail, and \( \gamma = 1 \) recovers the symmetric Student's \( t \) distribution.
\subsection{EGARCH Model}
\label{subsec:egarch}

GARCH models are widely used to describe time-varying volatility.
Among numerous variants, we consider an EGARCH-type model.
Following \citet{hansen_exponential_2016}, we replace the term $\gamma(|\epsilon_t| - \mathrm{E}|\epsilon_t|)$ in the original specification of \citet{nelson_conditional_1991} with the quadratic term $\gamma(\epsilon_t^2 - 1)$, which is differentiable at $\epsilon_t = 0$ and yields
\begin{align}
y_t &= \epsilon_t \exp\!\left(\frac{h_t}{2}\right),
\quad \epsilon_t \sim \mathcal{N}(0,1),
\quad t = 1,\ldots,n,
\label{eq:egarch-1}\\
h_{t+1} &= \omega + \varphi (h_t - \omega)
          + \tau \epsilon_t
          + \gamma (\epsilon_t^2 - 1),
\quad |\varphi| < 1,
\quad t = 1,\ldots,n-1,
\label{eq:egarch-2}
\end{align}
with the initial condition \(h_1 = \omega\).
The parameter~\(\tau\) captures the asymmetric response of conditional volatility to return shocks, and when~\(\tau < 0\), negative shocks increase volatility more than positive ones, reflecting the leverage effect. 
For expositional clarity, we present the model above under the Gaussian assumption for $\epsilon_t$. 
In Section~\ref{sec:application}, we relax this assumption and consider a standardized Student-\textit{t} distribution with degrees-of-freedom parameter $\nu$, allowing for heavy-tailed return behavior.

\paragraph{Remark.}
\citet{nelson_conditional_1991} adopted the generalized error distribution (GED) rather than the Student-\textit{t} distribution because, under Student-\textit{t} innovations, the conditional variance $\exp(h_t)$ and the return $y_t$ have no finite unconditional moments \citep[p.~352 and Theorem~A1.2]{nelson_conditional_1991}.
This also applies to the quadratic specification \eqref{eq:egarch-2}, but it does not affect the conditional likelihood or the one-day-ahead forecasts used in this paper.
Since \citet{bollerslev_arch_1994} and \citet{watanabe_excess_2000} find that the Student-\textit{t} distribution fits daily U.S.\ and Japanese stock index returns better than the GED, we use the Student-\textit{t} distribution for $\epsilon_t$.

\subsection{REGARCH Model}
\label{subsec:regarch}

\citet{hansen_realized_2012} and \citet{hansen_exponential_2016} extend the conventional GARCH and EGARCH models to the RGARCH and REGARCH frameworks, respectively, by incorporating realized measures into the volatility dynamics. 
Because the REGARCH model provides a more flexible characterization of the joint dynamics between returns and volatility than the RGARCH model, we adopt the REGARCH specification, which can be written as
\begin{align}
y_t &= \epsilon_t \exp\!\left(\frac{h_t}{2}\right), 
\quad \epsilon_t \sim \mathcal{N}(0,1), 
\quad t = 1,\ldots,n, 
\label{eq:regarch-1}\\
h_{t+1} &= \omega + \varphi (h_t - \omega)
          + \tau_1 \epsilon_t
          + \tau_2 (\epsilon_t^2 - 1)
          + \gamma \upsilon_t,
\quad |\varphi|<1, \quad t=1,\ldots,n-1,
\label{eq:regarch-2}\\
x_t &= \zeta + h_t
       + \delta_1 \epsilon_t
       + \delta_2 (\epsilon_t^2 - 1)
       + \upsilon_t,
\quad \upsilon_t \sim \mathcal{N}(0,\sigma_\upsilon^2),
\quad t = 1,\ldots,n,
\label{eq:regarch-3}
\end{align}
with the initial condition \(h_1 = \omega\). 
As in the EGARCH model, in Section~\ref{sec:application} we also consider a standardized Student-\textit{t} distribution for $\epsilon_t$ in the REGARCH model, with degrees-of-freedom parameter $\nu$, allowing for heavy-tailed return behavior.

The latent log-volatility process \(h_t\) and the realized measure \(x_t\) each include a quadratic form of the leverage function, given by \(\tau_1 \epsilon_t + \tau_2 (\epsilon_t^2 - 1)\) and \(\delta_1 \epsilon_t + \delta_2 (\epsilon_t^2 - 1)\), respectively.
Following \citet{hansen_exponential_2016}, one can introduce an additional scaling parameter~\(\psi\) as
\begin{equation*}
x_t = \zeta + \psi h_t + \delta_1 \epsilon_t + \delta_2 (\epsilon_t^2 - 1) + \upsilon_t,
\end{equation*}
The parameter~\(\zeta\) captures the bias in~\(\log RV_t\); \(\zeta=0\) implies an unbiased realized measure.
As with the RSV specification, we fix \(\psi=1\) for parsimony and comparability, rather than treating it as an additional free parameter.
See Sections~S10.3 and S11.3 of the Supplementary Material for a sensitivity analysis in which \(\psi\) is estimated freely.

\section{Bayesian Estimation Methodology}
\label{sec:bayes_method}
Each RSV model introduced in Section~\ref{sec:model} is a nonlinear state space model with a large number of latent variables, whose likelihood is intractable in closed form.
We therefore estimate all models by Bayesian MCMC simulation.
The prior distributions of the common parameters are
\begin{align}
\mu &\sim \mathcal{N}(m_\mu, s_\mu^2), \quad
\frac{\phi + 1}{2} \sim \mathcal{B}(a_{\phi 0}, b_{\phi 0}), \quad
\frac{\rho + 1}{2} \sim \mathcal{B}(a_{\rho 0}, b_{\rho 0}), \quad
\sigma_\eta^2 \sim \mathcal{IG}\left( \frac{n_\eta}{2}, \frac{S_\eta}{2} \right), \\
\xi &\sim \mathcal{N}(m_\xi, s_\xi^2), \quad
\sigma_u^2 \sim \mathcal{IG} \left( \frac{n_u}{2}, \frac{S_u}{2} \right),
\end{align}
where \( \mathcal{B}(\cdot, \cdot) \) and \( \mathcal{IG}(\cdot, \cdot) \) denote the beta and inverse-gamma distributions; for the model-specific parameters we use \( (\delta + 1)/2 \sim \mathcal{B}(a_{\delta 0}, b_{\delta 0}) \), \( \gamma \sim \mathcal{G}(n_{\gamma 0}, S_{\gamma 0}) \), and \( \nu \sim \mathcal{G}(n_{\nu 0}, S_{\nu 0}) \) truncated to the region in which the moment conditions of Section~\ref{sec:model} hold, where \( \mathcal{G}(\cdot, \cdot) \) denotes the gamma distribution.
The hyperparameter values are reported in Section~\ref{sec:application}.

Let \( \bm{y} = (y_1, \ldots, y_n)' \), \( \bm{x} = (x_1, \ldots, x_n)' \), and \( \bm{h} = (h_1, \ldots, h_n)' \), and let \( \bm{\theta} \) denote the parameter vector.
The MCMC algorithm cycles through the full conditional distributions of the parameters, of the latent log-volatilities \( \bm{h} \), which are sampled with the multi-move sampler, and, where present, of the mixing variables \( \bm{\lambda} \) and \( \bm{z}_0 \); parameters without a closed-form full conditional are updated by Metropolis--Hastings (MH) steps.

The sampling schemes for the RSV-AZ-ST and RSV-FS-ST models, including all full conditional distributions, are given in Sections~S2.1 and S2.2 of the Supplementary Material; those for the RSV-N and RSV-GH-ST models are in \citet{takahashi_estimating_2009} and \citet{takahashi_volatility_2016}, and the SV counterparts omit the measurement equation.
Section~S5 of the Supplementary Material reports the convergence diagnostics and the checks applied to the rolling window estimates.
\subsection{One-day-ahead Forecast}
\label{subsec:method-1day-forecast}
To obtain one-day-ahead forecasts of returns and volatilities, we use the predictive distribution within each state-space model.
Let $\bm{\theta}^{(i)}$ and $\bm{h}^{(i)}$ denote the $i$th posterior draw of the parameters and latent log-volatilities in the MCMC simulation, respectively, together with the draws of the mixing variables $\lambda_n^{(i)}$ (RSV-T, RSV-GH-ST, and RSV-AZ-ST) and $z_{0n}^{(i)}$ (RSV-AZ-ST) for the last observation.
The one-step-ahead predictive samples are then generated as follows:
\begin{enumerate}
\item Generate $h_{n+1}^{(i)} \mid \bm{x}, \bm{y}, \bm{h}^{(i)}, \bm{\theta}^{(i)} \sim N(\mu_{n+1}^{(i)}, (\sigma_{n+1}^{(i)})^{2})$, where
\begin{align}
\mu_{n+1}^{(i)} &= \mu^{(i)} + \phi^{(i)} \left( h_n^{(i)} - \mu^{(i)} \right) + \rho^{(i)} \sigma_\eta^{(i)} w_{n}^{(i)}, \label{eq:forecast-h} \\
(\sigma_{n+1}^{(i)})^{2} &= (1 - (\rho^{(i)})^2) (\sigma_\eta^{(i)})^2,
\end{align}
and $w_n^{(i)}$ is the innovation on which $\eta_n$ is conditioned in the leverage specification of the model, recovered from the last return through $\epsilon_n^{(i)} = y_n \exp(-h_n^{(i)}/2)$.
For the RSV-N and RSV-FS models, in which $\eta_t$ is conditioned on $\epsilon_t$, $w_n^{(i)} = \epsilon_n^{(i)}$.
For the RSV-T, RSV-GH-ST, and RSV-AZ-ST models, in which $\eta_t$ is conditioned on the standard normal component $z_t$, $w_n^{(i)} = z_n^{(i)}$ is obtained by inverting equations \eqref{eqn:rsv-t-eps}, \eqref{eqn:rsv-gh-st-eps}, and \eqref{eqn:rsv-az-st-eps}:
\begin{align}
z_n^{(i)} =
\begin{cases}
\epsilon_n^{(i)} \sqrt{ \mu_\lambda^{(i)} / \lambda_n^{(i)} }, & \text{RSV-T}, \\[4pt]
\left( \sqrt{ (\beta^{(i)})^2 (\sigma_\lambda^{(i)})^2 + \mu_\lambda^{(i)} } \, \epsilon_n^{(i)} - \beta^{(i)} \big( \lambda_n^{(i)} - \mu_\lambda^{(i)} \big) \right) \Big/ \sqrt{ \lambda_n^{(i)} }, & \text{RSV-GH-ST}, \\[4pt]
\left( \sqrt{ 1 - c^2 (\delta^{(i)})^2 } \, \epsilon_n^{(i)} \sqrt{ \mu_\lambda^{(i)} / \lambda_n^{(i)} } - \delta^{(i)} \big( z_{0n}^{(i)} - c \big) \right) \Big/ \sqrt{ 1 - (\delta^{(i)})^2 }, & \text{RSV-AZ-ST}.
\end{cases}
\label{eq:forecast-z}
\end{align}
The same construction applies to the SV counterparts, with $\bm{x}$ omitted.
\item Generate $\epsilon_{n+1}^{(i)}$ from the return distribution of the model given $\bm{\theta}^{(i)}$, drawing the mixing variables for $n+1$ from their priors.
\item Compute $y_{n+1}^{(i)} = \epsilon_{n+1}^{(i)} \exp( h_{n+1}^{(i)} / 2 )$.
\end{enumerate}
Repeating the procedure for $i = 1, \ldots, M$ yields the predictive samples $\{h_{n+1}^{(i)}\}_{i=1}^{M}$ and $\{y_{n+1}^{(i)}\}_{i=1}^{M}$.
The volatility forecast is the predictive mean $M^{-1} \sum_{i=1}^{M} \exp(h_{n+1}^{(i)})$, which is the point forecast used for every model in Section~\ref{sec:application}.
The one-day-ahead VaR forecast at level $\alpha$, $\mbox{VaR}_{n+1}(\alpha)$, is defined by $\mbox{Pr}(y_{n+1} < \mbox{VaR}_{n+1}(\alpha) \mid \mathcal{I}_{n}) = \alpha$, where $\mathcal{I}_{n}$ denotes the information set available at time $n$, and the corresponding ES forecast is $\mbox{ES}_{n+1}(\alpha) = E[y_{n+1} \mid y_{n+1} < \mbox{VaR}_{n+1}(\alpha), \mathcal{I}_{n}]$.
They are obtained, respectively, as the $\alpha$th quantile of $\{y_{n+1}^{(i)}\}_{i=1}^{M}$ and the average of the draws falling below this quantile.

We evaluate the volatility forecasts with the Gaussian quasi-likelihood (QLIKE) loss
\begin{align*}
L_{QLIKE}(x, f) = \frac{x}{f} - \log \frac{x}{f} - 1,
\end{align*}
where $x$ and $f$ denote a volatility proxy and a volatility forecast.
QLIKE is robust in the sense of \cite{patton_volatility_2011}, so that rankings are consistent whether based on the true volatility or on a conditionally unbiased proxy, and \cite{patton_evaluating_2009} document its greater power relative to the mean squared error in predictive accuracy tests.
The VaR and ES forecasts are evaluated jointly with the FZ0 loss of \cite{fissler_higher_2016} in the form used by \cite{patton_dynamic_2019},
\begin{align}
L_{FZ0}(y, v, e; \alpha) = -\frac{1}{\alpha e} \mathbbm{1}\{y \leq v\} (v - y) + \frac{v}{e} + \log(-e) - 1,
\label{eq:fz0}
\end{align}
where $y$, $v$, and $e$ denote a return, VaR, and ES; its loss differences are homogeneous of degree zero.
To assess whether differences in average losses are statistically significant, we use the unconditional predictive ability test of \cite{giacomini_tests_2006} (GW test) on the loss differences, which accommodates the rolling window scheme of Section~\ref{sec:application} and applies to nested and non-nested models alike; its statistic is that of \cite{diebold_comparing_1995}, computed with the Newey--West variance, and is asymptotically standard normal under the null.
The conditional version of the test and the associated win proportions are described in Section~S3 of the Supplementary Material.
\subsection{EGARCH and REGARCH Models}

Following \citet{mitsui_bayesian_2003}, we use the MH algorithm to sample the parameters of the EGARCH and REGARCH models from the posterior distribution.\footnote{For other Bayesian approaches to RGARCH-type models and their applications to volatility and tail risk forecasting, see \citet{chen_bayesian_2023} and \citet{wang_bayesian_2024}.} 
First, we obtain the mode of the posterior density, and then set the proposal density to a multivariate normal distribution centered at the posterior mode, with covariance matrix proportional to the inverse Hessian of the log posterior density evaluated at the mode, scaled by a factor of 1.2. 
The scaling factor is introduced to ensure sufficient exploration of the tails of the posterior distribution. 
\citet{asai_comparison_2006} shows that this method is efficient for sampling the parameters of GARCH-type models from the posterior distribution.

The prior distributions for the EGARCH model are specified as follows:
\begin{align}
&\omega \sim \mathcal{N}(0,100), \quad 
\frac{\varphi+1}{2} \sim \mathcal{B}(1,1), \quad 
-\tau \sim \mathcal{G}(0.1,0.1), \quad 
\gamma \sim \mathcal{G}(0.1,0.1), \quad 
\nu \sim \mathcal{G}(5,0.5)\mathbb{I}(\nu>4),    
\end{align}
For the additional parameters of the REGARCH model, the prior distributions are given by
\[
-\tau_1 \sim \mathcal{G}(0.1,0.1), \quad
\tau_2 \sim \mathcal{G}(0.1,0.1), \quad
\zeta \sim \mathcal{N}(0,10), \quad 
-\delta_1 \sim \mathcal{G}(0.1,0.1), \quad 
\delta_2 \sim \mathcal{G}(0.1,0.1), \quad 
\sigma_\upsilon^2 \sim \mathcal{IG}(2.5,0.5).
\]
The parameters $\tau$, $\tau_1$, $\tau_2$, $\gamma$, $\delta_1$, and $\delta_2$ are, in principle, unrestricted in sign.
However, empirical evidence typically suggests $\tau<0$ for EGARCH, and $\tau_1<0$, $\tau_2>0$, $\gamma>0$, $\delta_1<0$, and $\delta_2>0$ for REGARCH models.
Accordingly, we employ Gamma priors with appropriate sign restrictions to confine these parameters to economically meaningful regions.

Based on the posterior draws, one-day-ahead forecasts of volatility, VaR, and ES are obtained. 
This approach allows us to jointly sample both parameters and the implied conditional volatilities from the posterior distribution, thereby explicitly incorporating parameter uncertainty into volatility and return forecasts.

Importantly, this feature addresses concerns raised in the literature that ignoring parameter uncertainty in GARCH-type models may lead to unfair comparisons with stochastic volatility models; see, for example, \citet{ardia_forecasting_2018}, who document that accounting for parameter uncertainty improves volatility and risk forecasts in GARCH frameworks. 
By adopting a fully Bayesian estimation strategy, our framework ensures that parameter uncertainty is treated consistently across models, providing a fair basis for comparison and yielding coherent predictive distributions for one-step-ahead volatility and return forecasts.

\section{Empirical Study}
\label{sec:application}

\subsection{Data and Descriptive Statistics}

We estimate the SV and RSV models described in Section~\ref{sec:model} using daily (close-to-close) returns and RVs for two major stock indices: the DJIA and the N225. Following \citet{liu_does_2015}, among the available realized measures we adopt the 5-minute RV estimator, computed during trading hours only.
The DJIA data are obtained from the Oxford-Man Institute’s Realized Library,\footnote{The website \url{https://realized.oxford-man.ox.ac.uk/} is no longer accessible.} while the N225 data are constructed from the Nikkei NEEDS-TICK dataset.\footnote{See \citet{ubukata_pricing_2014} for details on the construction of the N225 dataset.}

The sample period spans from June 2009 to September 2019 for both indices.
Specifically, the DJIA sample covers 2,596 trading days from June 1, 2009, to September 27, 2019, and the N225 sample includes 2,532 trading days from June 1, 2009, to September 30, 2019.
Table~\ref{tab:stats} reports descriptive statistics of the daily returns and of the logarithms of the 5-minute RVs; the time series plots and histograms are shown in Figures~S4 and S5 of the Supplementary Material.
The mean return is significantly different from zero for the DJIA but negligible in magnitude, so we do not demean the series, and the Ljung--Box statistic \citep{ljung_measure_1978}, adjusted for heteroskedasticity as in \citet{diebold_empirical_1988}, does not reject the absence of autocorrelation up to 10 lags in either return series.
Both return series are significantly negatively skewed and leptokurtic, and the Jarque--Bera (JB) test rejects normality, which motivates the skewed-$t$ innovations of Section~\ref{sec:model}.
The log-RVs are strongly autocorrelated, positively skewed, and leptokurtic, so that the normality of \(u_t\) in equation~\eqref{eqn:RSV-x}, which we retain for tractability, is rejected; a skewed measurement error would be a natural extension that we leave for future work.
\begin{table}[t]
\centering
\begin{threeparttable}
\caption{Descriptive statistics of daily returns and log-RVs for the DJIA and N225.}
\label{tab:stats}
\begin{tabular}{llcccccccc}
\toprule
& & Mean & SD & Skew & Kurt & Min & Max & JB & LB \\
\midrule
Return & DJIA & \(0.044\) & \(0.895\) & \(-0.448\) & \(6.675\) & \(-5.562\) & \(4.857\) & \(0.00\) & \(0.52\) \\
& & \( (0.018) \) & & \( (0.048) \) & \( (0.096) \) \\
& N225 & \(0.033\) & \(1.321\) & \(-0.544\) & \(8.101\) & \(-11.153\) & \(7.426\) & \(0.00\) & \(0.55\) \\
& & \( (0.026) \) & & \( (0.049) \) & \( (0.097) \) \\
\midrule
Log-RV & DJIA & \(-0.987\) & \(1.048\) & \(0.317\) & \(3.177\) & \(-3.953\) & \(4.087\) & \(0.00\) & \(0.00\) \\
& & \( (0.021) \) & & \( (0.048) \) & \( (0.096) \) \\
& N225 & \(-0.858\) & \(0.844\) & \(0.589\) & \(4.219\) & \(-3.321\) & \(3.258\) & \(0.00\) & \(0.00\) \\
& & \( (0.017) \) & & \( (0.049) \) & \( (0.097) \) \\
\bottomrule
\end{tabular}
\begin{tablenotes}
\footnotesize
\item \textit{Notes:} Returns are in percentage points and RVs are 5-minute realized variances.
Standard errors are shown in parentheses.
JB refers to the $p$-value of the Jarque--Bera test.
LB denotes the $p$-value of the \citet{ljung_measure_1978} statistic up to 10 lags, adjusted as in \citet{diebold_empirical_1988}.
\end{tablenotes}
\end{threeparttable}
\end{table}
\subsection{Competing Models}

We evaluate the forecasting performance of the following RSV models, which differ in the distributional assumption for $\epsilon_t$:
\begin{enumerate}
    \item \textbf{RSV-N}: $\epsilon_t$ follows a normal distribution.
    \item \textbf{RSV-T}: $\epsilon_t$ follows a Student's $t$ distribution.
    \item \textbf{RSV-GH-ST}: $\epsilon_t$ follows the GH skew-$t$ distribution.
    \item \textbf{RSV-AZ-SN}: $\epsilon_t$ follows the Azzalini skew normal distribution.
    \item \textbf{RSV-AZ-ST}: $\epsilon_t$ follows the Azzalini skew-$t$ distribution.
    \item \textbf{RSV-FS-SN}: $\epsilon_t$ follows the Fern\'{a}ndez–Steel skew normal distribution.
    \item \textbf{RSV-FS-ST}: $\epsilon_t$ follows the Fern\'{a}ndez–Steel skew-$t$ distribution.
\end{enumerate}
The prior distributions for the parameters common to the SV and RSV specifications are specified as follows:
\begin{align}
\mu &\sim \mathcal{N}(0, 100), \quad \frac{\phi+1}{2} \sim \mathcal{B}(20, 1.5), \quad \frac{\rho+1}{2} \sim \mathcal{B}(1, 2), \quad \sigma_{\eta}^{2} \sim \mathcal{IG}(2.5, 0.025), \\
\xi &\sim \mathcal{N}(0, 1), \quad \sigma_{u}^{2} \sim \mathcal{IG}(2.5, 0.1).
\end{align}
The prior for $(\phi+1)/2$ places most of its mass on strongly persistent volatility, with a prior mean of $0.86$ for $\phi$, and the prior for $(\rho+1)/2$ favours a negative leverage effect, with a prior mean of $-1/3$ for $\rho$.
Both are standard in the stochastic volatility literature.
The remaining priors are weakly informative and allow the data to dominate posterior inference.

For the additional distributional parameters, we set:
\begin{align}
\nu \sim \mathcal{G}(5, 0.5)\mathbbm{1}\{ \nu > 4 \}, \quad \beta \sim \mathcal{N}(0, 1), \quad \frac{\delta+1}{2} \sim \mathcal{B}(1, 2).
\end{align}
The truncation $\nu>4$ ensures the existence of the fourth moment, which is required for volatility forecasting and risk evaluation.
Where $\psi$ is estimated freely rather than fixed at unity, we use $\psi \sim \mathcal{N}(1, 1)$.

The Fern\'{a}ndez--Steel specifications are estimated with a separate sampler and use their own, more diffuse, priors:
\begin{align}
\mu, \xi &\sim \mathcal{N}(0, 10^{8}), \quad \frac{\phi+1}{2}, \frac{\rho+1}{2} \sim \mathcal{B}(1, 1), \quad \sigma_{\eta}^{2}, \sigma_{u}^{2} \sim \mathcal{IG}(0.05, 0.05), \\
\gamma &\sim \mathcal{G}(10^{-5}, 10^{-5}), \quad \nu \sim \mathcal{G}(0.01, 0.2)\mathbbm{1}\{ \nu > 2 \}.
\end{align}
Here the truncation is the condition for the variance of the innovation to exist rather than the fourth moment.
This is immaterial in practice: across all estimation windows and both indices the posterior of $\nu$ for these models concentrates between $7$ and $31$, so the fourth moment exists throughout.

In addition, to isolate the contribution of RV, we estimate standard SV models with the same innovation distributions as in the RSV specifications.
These include SV-N, SV-T, SV-GH-ST, SV-AZ-SN, SV-AZ-ST, SV-FS-SN, and SV-FS-ST, mirroring the distributional assumptions used in the RSV models.
This setup allows us to directly assess the added value of incorporating RV in terms of predictive performance.

To further benchmark the RSV models, we also consider the EGARCH and REGARCH models described in Sections~\ref{subsec:egarch} and~\ref{subsec:regarch}, respectively.
These models are estimated within a Bayesian framework using the MH algorithm under standard normal and Student’s $t$ innovations.
One-day-ahead forecasts of volatility, VaR, and ES are then obtained based on the posterior draws.

\subsection{Out-of-sample Forecasting}

Following the approach of \cite{takahashi_volatility_2016}, we implement a rolling window estimation procedure to evaluate the out-of-sample performance of each model.
The window size is kept fixed throughout the forecasting period.
For the DJIA, we use a window of 1,993 observations, generating forecasts from May 1, 2017, to September 27, 2019. For the N225, the window consists of 1,942 observations, with forecast dates ranging from May 1, 2017, to September 30, 2019. After each estimation step, we produce one-day-ahead forecasts of volatility, VaR, and ES.

At each forecast point, we draw 15,000 predictive samples from the posterior predictive distribution.
For every model the volatility forecast is the mean of the predictive draws of $\exp(h_{n+1})$, so that all models are compared with the same point forecast; VaR and ES are obtained from the predictive quantiles of the return distribution.
This procedure yields 603 forecasts for the DJIA and 590 for the N225, covering the period from early May 2017 to late September 2019.

\subsubsection{Volatility Forecasts}
\label{sec:forecast-vol}

To evaluate the accuracy of volatility forecasts, we compute the average loss using the QLIKE loss function, as described in Section~\ref{subsec:method-1day-forecast}.
The QLIKE loss is known for its robustness, yielding model rankings that align closely with those based on latent volatility, provided the proxy is conditionally unbiased.
Simulation evidence from \cite{patton_evaluating_2009} and empirical findings by \cite{hansen_forecast_2005} and \cite{patton_volatility_2011} suggest that QLIKE has greater statistical power than MSE to discriminate between models.

To mitigate market microstructure noise, we employ multiple volatility proxies: realized kernel (RK) with a flat-top Tukey-Hanning${}_2$ kernel \citep{barndorff-nielsen_designing_2008}, bipower variation (BV) \citep{barndorff-nielsen_power_2004}, and median realized volatility (Med) \citep{andersen_jump_2012}, in addition to the standard 5-minute realized volatility (RV5).

To correct for biases due to non-trading hours, we apply the adjustment of \cite{hansen_forecast_2005} in equation~\eqref{eqn:HL} to each proxy with a rolling correction factor computed from the estimation window.
For the forecast made at origin $t$ for the target day $t+1$, the factor uses the $n$ days of the estimation window, $s = t-n+1, \ldots, t$, and is applied to the raw proxy of the target day:
\begin{align}
\tilde{x}_{t+1} = c_{HL,t} \, x_{t+1}, \quad c_{HL,t} = \frac{ \sum_{s=t-n+1}^{t} (y_{s} - \bar{y}_{t})^{2} }{ \sum_{s=t-n+1}^{t} x_{s} }, \quad \bar{y}_{t} = \frac{1}{n} \sum_{s=t-n+1}^{t} y_{s},
\end{align}
where $x_s$ denotes the raw volatility proxy on day $s$ and $\tilde{x}_{t+1}$ is the adjusted proxy against which the forecast of $\exp(h_{t+1})$ is evaluated.
The factor therefore uses only information available at the forecast origin, and the adjusted proxy and the forecast refer to the same day.


\begin{table}[tbp]
\centering
\begin{threeparttable}
\setlength{\tabcolsep}{4pt}
\caption{QLIKE scores of volatility forecasts for the DJIA and N225.}
\label{tab:forecast-vol-qlike}
\begin{tabular}{lllllllllllrr}
\toprule
& \multicolumn{4}{l}{DJIA} & & \multicolumn{4}{l}{N225} & & \multicolumn{2}{l}{Average} \\
\cmidrule(lr){2-5} \cmidrule(lr){7-10} \cmidrule(lr){12-13}
& RV5 & RK & BV & Med & & RV5 & RK & BV & Med & & Score & Rank \\
\midrule
SV-N & $0.287$ & $0.203^{*}$ & $0.252^{**}$ & $0.286^{**}$ &  & $0.233$ & $0.242$ & $0.226$ & $0.187$ & & $0.240$ & $14.2$ \\ 
SV-T & $0.281$ & $0.198^{*}$ & $0.251^{**}$ & $0.285^{*}$ &  & $0.243$ & $0.254$ & $0.235$ & $0.197$ & & $0.243$ & $15.5$ \\ 
SV-AZ-SN & $0.267$ & $0.189^{**}$ & $0.242^{**}$ & $0.278^{**}$ &  & $0.229$ & $0.237$ & $0.222$ & $0.184$ & & $0.231$ & $10.5$ \\ 
SV-AZ-ST & $0.278$ & $0.197^{*}$ & $0.252^{**}$ & $0.287^{*}$ &  & $0.238$ & $0.247$ & $0.231$ & $0.190$ & & $0.240$ & $14.5$ \\ 
SV-FS-SN & $0.268$ & $0.193^{**}$ & $0.245^{**}$ & $0.280^{**}$ &  & $0.227$ & $0.234$ & $0.220$ & $0.182$ & & $0.231$ & $10.6$ \\ 
SV-FS-ST & $0.278$ & $0.198^{*}$ & $0.252^{**}$ & $0.287^{*}$ &  & $0.232$ & $0.240$ & $0.225$ & $0.186$ & & $0.237$ & $13.2$ \\ 
SV-GH-ST & $0.275$ & $0.192^{**}$ & $0.247^{**}$ & $0.282^{**}$ &  & $0.241$ & $0.250$ & $0.235$ & $0.192$ & & $0.239$ & $13.9$ \\ 
\cmidrule(lr){1-13}
RSV-N & $0.192^{**}$ & $0.166^{**}$ & $0.226^{**}$ & $0.257^{**}$ &  & $0.162^{*}$ & $0.166$ & $0.156$ & $0.111^{**}$ & & $0.179$ & $4.0$ \\ 
RSV-T & $0.190^{**}$ & $0.162^{**}$ & $0.222^{**}$ & $0.250^{**}$ &  & $0.163$ & $0.168$ & $0.158$ & $0.112^{**}$ & & $0.178$ & $3.4$ \\ 
RSV-AZ-SN & $0.190^{**}$ & $0.166^{**}$ & $0.227^{**}$ & $0.256^{**}$ &  & $0.162^{*}$ & $0.166$ & $0.156$ & $0.112^{**}$ & & $0.179$ & $4.5$ \\ 
RSV-AZ-ST & $0.193^{**}$ & $0.165^{**}$ & $0.225^{**}$ & $0.253^{**}$ &  & $0.163$ & $0.168$ & $0.158$ & $0.112^{**}$ & & $0.180$ & $5.2$ \\ 
RSV-FS-SN & $0.186^{**}$ & $0.165^{**}$ & $0.225^{**}$ & $0.256^{**}$ &  & $0.161^{**}$ & $0.164^{**}$ & $0.155^{**}$ & $0.112^{**}$ & & $0.178$ & $2.9$ \\ 
RSV-FS-ST & $0.189^{**}$ & $0.164^{**}$ & $0.224^{**}$ & $0.254^{**}$ &  & $0.162^{*}$ & $0.166$ & $0.156$ & $0.112^{**}$ & & $0.178$ & $3.1$ \\ 
RSV-GH-ST & $0.194^{*}$ & $0.163^{**}$ & $0.223^{**}$ & $0.251^{**}$ &  & $0.163$ & $0.168$ & $0.158$ & $0.112^{**}$ & & $0.179$ & $5.1$ \\ 
\cmidrule(lr){1-13}
EGARCH-N & $0.300$ & $0.221$ & $0.279$ & $0.317$ &  & $0.236$ & $0.248$ & $0.231$ & $0.184$ & & $0.252$ & $15.5$ \\ 
EGARCH-T & $0.311$ & $0.229$ & $0.285$ & $0.319$ &  & $0.259$ & $0.272$ & $0.253$ & $0.207$ & & $0.267$ & $18.0$ \\ 
\cmidrule(lr){1-13}
REGARCH-N & $0.212^{*}$ & $0.181^{**}$ & $0.239^{**}$ & $0.260^{**}$ &  & $0.172$ & $0.179$ & $0.168$ & $0.118$ & & $0.191$ & $8.4$ \\ 
REGARCH-T & $0.218$ & $0.181^{**}$ & $0.237^{**}$ & $0.256^{**}$ &  & $0.175$ & $0.182$ & $0.171$ & $0.121$ & & $0.193$ & $8.4$ \\ 
\bottomrule
\end{tabular}
\begin{tablenotes}
\footnotesize
\item \textit{Notes:} A double asterisk ($^{**}$) and a single asterisk ($^{*}$) indicate that the model belongs to the 75\% and the 90\% model confidence set (MCS), respectively; the 75\% MCS is nested within the 90\% MCS.
The average columns report the mean QLIKE scores and ranks across the two indices.
\end{tablenotes}
\end{threeparttable}
\end{table}

Table~\ref{tab:forecast-vol-qlike} reports QLIKE scores across all models and volatility proxies.
Following \cite{hansen_model_2011}, throughout the empirical analysis we report both the 75\% and the 90\% MCS for every loss function: a double asterisk ($^{**}$) denotes inclusion in the 75\% MCS and a single asterisk ($^{*}$) inclusion in the 90\% MCS.\footnote{The rolling window scheme satisfies the stationarity assumption required by the MCS bootstrap; we use 1,000 bootstrap replications with a block size of 10.
Since the MCS $p$-value of a model does not depend on the confidence level, a model belongs to the 75\% (90\%) MCS whenever its MCS $p$-value exceeds $0.25$ ($0.10$).
For the QLIKE loss the two sets are far smaller than the full set of models, whereas for the FZ0 loss, which is driven by infrequent tail events, they are much larger and can diverge substantially; reporting both levels makes this difference in resolution visible.}
For both indices, every RSV model attains a lower QLIKE score than its SV counterpart under all four proxies.
For the N225, the RSV models attain scores of $0.155$ to $0.168$ under RV5, RK, and BV and of $0.111$ to $0.112$ under Med, against $0.18$ to $0.25$ for the SV models, $0.12$ to $0.18$ for the REGARCH models, and $0.18$ to $0.27$ for the EGARCH models, so that the RSV models also dominate the REGARCH benchmarks under every proxy.
For the DJIA, the margins over the SV models are of the same size under RV5, the measure used in estimation, where the RSV scores range from $0.186$ (RSV-FS-SN) to $0.194$ against $0.267$ to $0.287$ for the SV models and $0.212$ and $0.218$ for the REGARCH models, but they shrink to about a third under RK, BV, and Med (from $0.077$--$0.095$ under RV5 to $0.015$--$0.037$); the RSV models remain first under these proxies (RSV-T, with $0.162$, $0.222$, and $0.250$), while the REGARCH models enter the $75\%$ MCS under all three proxies, as do all seven SV models under BV, three under RK, and four under Med.
On average across the two indices, the Fern\'{a}ndez--Steel and Student's $t$ RSV specifications deliver the best performance, with small margins between the seven RSV specifications.
\begin{table}[tbp]
\centering
\begin{threeparttable}
\caption{Giacomini--White tests for the seven matched pairs of RSV and SV models.}
\label{tab:gw-matched}
\footnotesize
\begin{tabular}{lllllllllllll}
\toprule
& \multicolumn{4}{l}{QLIKE, DJIA} & \multicolumn{4}{l}{QLIKE, N225} & \multicolumn{2}{l}{FZ0, DJIA} & \multicolumn{2}{l}{FZ0, N225} \\
\cmidrule(lr){2-5} \cmidrule(lr){6-9} \cmidrule(lr){10-11} \cmidrule(lr){12-13}
Pair & RV5 & RK & BV & Med & RV5 & RK & BV & Med & 1\% & 5\% & 1\% & 5\% \\
\midrule
RSV-N vs SV-N & $<$0.01 & 0.01 & 0.15 & 0.10 & $<$0.01 & $<$0.01 & $<$0.01 & $<$0.01 & 0.26$^{\dagger}$ & 0.31 & 0.34 & 0.25 \\
RSV-T vs SV-T & $<$0.01 & 0.02 & 0.07 & 0.03 & $<$0.01 & $<$0.01 & $<$0.01 & $<$0.01 & 0.28$^{\dagger}$ & 0.35 & 0.12 & 0.61 \\
RSV-AZ-SN vs SV-AZ-SN & $<$0.01 & 0.14 & 0.40 & 0.17 & $<$0.01 & $<$0.01 & $<$0.01 & $<$0.01 & 0.34$^{\dagger}$ & 0.07 & 0.36 & 0.39 \\
RSV-AZ-ST vs SV-AZ-ST & $<$0.01 & 0.02 & 0.11 & 0.03 & $<$0.01 & $<$0.01 & $<$0.01 & $<$0.01 & 0.76$^{\dagger}$ & 0.01 & 0.22 & 0.31 \\
RSV-FS-SN vs SV-FS-SN & $<$0.01 & 0.04 & 0.21 & 0.12 & $<$0.01 & $<$0.01 & $<$0.01 & $<$0.01 & 0.49$^{\dagger}$ & 0.14 & 0.45 & 0.47 \\
RSV-FS-ST vs SV-FS-ST & $<$0.01 & 0.02 & 0.08 & 0.03 & $<$0.01 & $<$0.01 & $<$0.01 & $<$0.01 & 0.40$^{\dagger}$ & 0.11 & 0.39 & 0.47 \\
RSV-GH-ST vs SV-GH-ST & $<$0.01 & 0.02 & 0.11 & 0.03 & $<$0.01 & $<$0.01 & $<$0.01 & $<$0.01 & 0.72$^{\dagger}$ & 0.03 & 0.05 & 0.28 \\
\bottomrule
\end{tabular}
\begin{tablenotes}
\footnotesize
\item \textit{Notes:} Entries are $p$-values of the unconditional Giacomini--White test of equal predictive ability for the RSV model against the SV model with the same return distribution, computed with the Newey--West variance.
A dagger ($^{\dagger}$) marks pairs in which the SV model attains the lower average loss; in all other pairs the RSV model does.
\end{tablenotes}
\end{threeparttable}
\end{table}

Table~\ref{tab:gw-matched} reports the unconditional test of \cite{giacomini_tests_2006} for the seven pairs of RSV and SV models that share the same return distribution, with the Newey--West variance and a two-sided normal $p$-value; the conditional version of the test and the pairwise comparisons across all models are reported in Section~S3 and Figures~S16--S23 of the Supplementary Material.
For both indices, the RSV model attains the lower average loss in every pair and under every proxy.
For the N225, equal predictive ability is rejected at the $1\%$ level throughout.
For the DJIA, it is rejected at the $1\%$ level for every pair under RV5, at the $5\%$ level for six pairs under RK and for four pairs under Med, and for no pair under BV, so that the evidence weakens as the proxy departs from the measure used in estimation.
The cumulative loss differences reported in Figures~S8--S15 of the Supplementary Material show that the advantage accrues steadily over the three years for both indices: under RV5 the RSV-N model has a lower QLIKE than the SV-N model on $69\%$ of the forecast days for the DJIA and on $68\%$ for the N225.

In summary, every RSV model attains a lower average QLIKE than its SV counterpart for both indices under all four proxies, in line with \citet{takahashi_stochastic_2023, takahashi_forecasting_2024}, but a lower average loss and a significant improvement must be distinguished: for the N225 the improvement is significant for every pair under every proxy, whereas for the DJIA it is significant for every pair only under RV5, the realized measure that also enters the measurement equation, and for fewer pairs or none under the other proxies.
The DJIA evidence therefore depends on the realized measure used in estimation, which is a limitation of the comparison: a proxy that is also the model input is not an independent benchmark, and the gains measured against proxies that do not enter the model are smaller and mostly not significant.
Within each class, the choice of return distribution matters less than the choice between SV and RSV: under RV5, the spread of QLIKE across the seven RSV specifications is $0.008$ for the DJIA and $0.003$ for the N225, and the $90\%$ MCS retains all seven of them for the DJIA and four for the N225.
The next section examines whether these results carry over to tail risk measures.

\subsubsection{VaR and ES forecasts}
Table \ref{tab:forecast-var-es-loss} presents the empirical violation rates ($\hat{\alpha}$), the $p$-values of the unconditional coverage (UC) test of \cite{kupiec_techniques_1995} and of the dynamic quantile (DQ) test of \cite{engle_caviar_2004},\footnote{The UC test is the likelihood ratio test of the violation frequency against the nominal level $\alpha$. The DQ test regresses the centred violation indicator, $\mathbbm{1}\{y_t \leq \mathrm{VaR}_t\} - \alpha$, on a constant, its first four lags, and the VaR forecast for day $t$ over the forecast days; its statistic, $\hat{b}' X' X \hat{b} / (\alpha(1-\alpha))$ with $\hat{b}$ the least squares coefficients and $X$ the regressor matrix, is asymptotically $\chi^2$ with six degrees of freedom under the null of correct conditional coverage.} the average FZ0 loss values (as defined in Equation \eqref{eq:fz0}), and the MCS $p$-values for the DJIA at the $1\%$ and $5\%$ levels.
At the $1\%$ level, all models underestimate tail risk: the violation rates, from $1.49\%$ to $3.15\%$, exceed the nominal level, and the deviation is particularly pronounced for the RSV models, from $1.99\%$ to $2.82\%$.
The coverage tests confirm this: the UC test rejects correct coverage at the $5\%$ significance level for 14 of the 18 models, including all seven RSV models, the exceptions being SV-FS-SN, SV-FS-ST, SV-GH-ST, and EGARCH-T with nine violations each ($1.49\%$) over the 603 forecast days, and the DQ test rejects at the $1\%$ significance level for every model except SV-FS-SN and EGARCH-T.
For $\alpha = 5\%$, the violation rates are well aligned with the target and neither test rejects for any model.
In terms of the FZ0 loss, the SV-AZ-SN and RSV-AZ-ST models achieve the lowest losses at $\alpha = 1\%$ and $5\%$, respectively.
The matched pairs point in opposite directions at the two levels: at the $1\%$ level every RSV specification has a higher FZ0 loss than its SV counterpart, whereas at the $5\%$ level every RSV specification has a lower loss.
The MCS does not separate the models sharply: every model belongs to the $90\%$ MCS at both levels except SV-AZ-ST at $\alpha = 5\%$, and all models except EGARCH-N and SV-AZ-ST at $\alpha = 5\%$ also belong to the $75\%$ MCS.
\begin{table}[t]
\centering
\begin{threeparttable}
\caption{Violation rates and FZ0 losses of VaR and ES forecasts for the DJIA.}
\label{tab:forecast-var-es-loss}
\begin{tabular}{lrrrrlrrrrrl}
\toprule
& \multicolumn{5}{l}{$\alpha = 1\%$} & & \multicolumn{5}{l}{$\alpha = 5\%$} \\
\cline{2-6} \cline{8-12}
& \multicolumn{1}{c}{$\hat{\alpha}$} & \multicolumn{1}{c}{$p_{UC}$} & \multicolumn{1}{c}{$p_{DQ}$} & \multicolumn{1}{c}{FZ0} & \multicolumn{1}{c}{$p_{MCS}$} & & \multicolumn{1}{c}{$\hat{\alpha}$} & \multicolumn{1}{c}{$p_{UC}$} & \multicolumn{1}{c}{$p_{DQ}$} & \multicolumn{1}{c}{FZ0} & \multicolumn{1}{c}{$p_{MCS}$} \\
\midrule
SV-N & $2.32$ & $0.01$ & $0.00$ & $1.0714$ & $0.84^{**}$ &  & $4.64$ & $0.68$ & $0.79$ & $0.5965$ & $0.29^{**}$ \\
SV-T & $1.99$ & $0.03$ & $0.00$ & $1.0483$ & $0.88^{**}$ &  & $4.31$ & $0.43$ & $0.59$ & $0.5809$ & $0.53^{**}$ \\
SV-AZ-SN & $1.99$ & $0.03$ & $0.00$ & $0.9752$ & $1.00^{**}$ &  & $4.15$ & $0.32$ & $0.78$ & $0.5899$ & $0.38^{**}$ \\
SV-AZ-ST & $2.16$ & $0.01$ & $0.00$ & $1.0109$ & $0.93^{**}$ &  & $4.31$ & $0.43$ & $0.59$ & $0.6030$ & $0.06^{\hspace{0.8em}}$ \\
SV-FS-SN & $1.49$ & $0.26$ & $0.07$ & $0.9786$ & $0.96^{**}$ &  & $4.15$ & $0.32$ & $0.78$ & $0.5947$ & $0.41^{**}$ \\
SV-FS-ST & $1.49$ & $0.26$ & $0.00$ & $0.9899$ & $0.94^{**}$ &  & $4.15$ & $0.32$ & $0.78$ & $0.5954$ & $0.40^{**}$ \\
SV-GH-ST & $1.49$ & $0.26$ & $0.00$ & $0.9870$ & $0.96^{**}$ &  & $4.31$ & $0.43$ & $0.56$ & $0.5916$ & $0.33^{**}$ \\
\cmidrule(lr){1-12}
RSV-N & $2.82$ & $0.00$ & $0.00$ & $1.2345$ & $0.30^{**}$ &  & $5.80$ & $0.38$ & $0.08$ & $0.5567$ & $0.75^{**}$ \\
RSV-T & $2.65$ & $0.00$ & $0.00$ & $1.1709$ & $0.44^{**}$ &  & $5.80$ & $0.38$ & $0.09$ & $0.5480$ & $0.75^{**}$ \\
RSV-AZ-SN & $2.49$ & $0.00$ & $0.00$ & $1.0957$ & $0.84^{**}$ &  & $5.31$ & $0.73$ & $0.24$ & $0.5301$ & $0.97^{**}$ \\
RSV-AZ-ST & $2.16$ & $0.01$ & $0.00$ & $1.0388$ & $0.93^{**}$ &  & $5.31$ & $0.73$ & $0.26$ & $0.5291$ & $1.00^{**}$ \\
RSV-FS-SN & $2.32$ & $0.01$ & $0.00$ & $1.0490$ & $0.93^{**}$ &  & $5.64$ & $0.48$ & $0.24$ & $0.5411$ & $0.77^{**}$ \\
RSV-FS-ST & $2.49$ & $0.00$ & $0.00$ & $1.0958$ & $0.82^{**}$ &  & $5.31$ & $0.73$ & $0.25$ & $0.5365$ & $0.77^{**}$ \\
RSV-GH-ST & $1.99$ & $0.03$ & $0.00$ & $1.0186$ & $0.94^{**}$ &  & $4.98$ & $0.98$ & $0.27$ & $0.5307$ & $0.97^{**}$ \\
\cmidrule(lr){1-12}
EGARCH-N & $2.49$ & $0.00$ & $0.00$ & $1.2822$ & $0.30^{**}$ &  & $4.64$ & $0.68$ & $0.68$ & $0.6206$ & $0.16^{*\hspace{0.4em}}$ \\
EGARCH-T & $1.49$ & $0.26$ & $0.85$ & $1.0605$ & $0.93^{**}$ &  & $4.98$ & $0.98$ & $0.47$ & $0.5826$ & $0.75^{**}$ \\
\cmidrule(lr){1-12}
REGARCH-N & $3.15$ & $0.00$ & $0.00$ & $1.3048$ & $0.30^{**}$ &  & $5.31$ & $0.73$ & $0.53$ & $0.5580$ & $0.75^{**}$ \\
REGARCH-T & $2.16$ & $0.01$ & $0.00$ & $1.0476$ & $0.93^{**}$ &  & $5.47$ & $0.60$ & $0.56$ & $0.5413$ & $0.77^{**}$ \\
\bottomrule
\end{tabular}
\begin{tablenotes}
\footnotesize
\item \textit{Notes:} $\hat{\alpha}$ represents the empirical violation rate (\%).
$p_{UC}$ indicates the $p$-value of the unconditional coverage test of \cite{kupiec_techniques_1995}.
$p_{DQ}$ indicates the $p$-value of the dynamic quantile test of \cite{engle_caviar_2004}, with four lags of the violation indicator and the VaR forecast as regressors.
FZ0 denotes the average FZ0 loss.
$p_{MCS}$ indicates the MCS $p$-value.
A double asterisk ($^{**}$) and a single asterisk ($^{*}$) on $p_{MCS}$ indicate that the model belongs to the 75\% and the 90\% MCS, respectively (equivalently, $p_{MCS}>0.25$ and $p_{MCS}>0.10$); the 75\% MCS is nested within the 90\% MCS.
\end{tablenotes}
\end{threeparttable}
\end{table}

Table \ref{tab:forecast-var-es-loss-n225} summarizes the results for the N225.
Across both levels, most models yield violation rates close to the nominal levels; the UC test does not reject for any model, and the DQ test rejects at the $5\%$ significance level only for RSV-AZ-SN at $\alpha = 1\%$ ($p = 0.047$, shown as $0.05$ in the table).
The lowest FZ0 loss is attained by RSV-GH-ST at the $1\%$ level and by RSV-AZ-ST at the $5\%$ level, and at both levels every RSV specification has a lower loss than its SV counterpart (by $0.045$--$0.098$ at the $1\%$ level and $0.014$--$0.031$ at the $5\%$ level), whereas for the DJIA this holds at the $5\%$ level only.
Only the EGARCH models fall outside the $75\%$ MCS, and EGARCH-N falls outside the $90\%$ MCS at the $5\%$ level.
\begin{table}[t]%
\centering
\begin{threeparttable}
\caption{Violation rates and FZ0 losses of VaR and ES forecasts for the N225.}
\label{tab:forecast-var-es-loss-n225}
\begin{tabular}{lrrrrlrrrrrl}
\toprule
& \multicolumn{5}{l}{$\alpha = 1\%$} & & \multicolumn{5}{l}{$\alpha = 5\%$} \\
\cline{2-6} \cline{8-12}
& \multicolumn{1}{c}{$\hat{\alpha}$} & \multicolumn{1}{c}{$p_{UC}$} & \multicolumn{1}{c}{$p_{DQ}$} & \multicolumn{1}{c}{FZ0} & \multicolumn{1}{c}{$p_{MCS}$} & & \multicolumn{1}{c}{$\hat{\alpha}$} & \multicolumn{1}{c}{$p_{UC}$} & \multicolumn{1}{c}{$p_{DQ}$} & \multicolumn{1}{c}{FZ0} & \multicolumn{1}{c}{$p_{MCS}$} \\
\midrule
SV-N & $1.02$ & $0.97$ & $1.00$ & $1.2034$ & $0.44^{**}$ &  & $3.90$ & $0.20$ & $0.53$ & $0.8163$ & $0.37^{**}$ \\
SV-T & $0.85$ & $0.70$ & $1.00$ & $1.2158$ & $0.40^{**}$ &  & $3.39$ & $0.06$ & $0.55$ & $0.7973$ & $0.95^{**}$ \\
SV-AZ-SN & $0.85$ & $0.70$ & $1.00$ & $1.1948$ & $0.69^{**}$ &  & $3.73$ & $0.14$ & $0.49$ & $0.8095$ & $0.66^{**}$ \\
SV-AZ-ST & $0.85$ & $0.70$ & $1.00$ & $1.2026$ & $0.48^{**}$ &  & $3.39$ & $0.06$ & $0.59$ & $0.8097$ & $0.66^{**}$ \\
SV-FS-SN & $0.85$ & $0.70$ & $0.99$ & $1.1979$ & $0.74^{**}$ &  & $3.39$ & $0.06$ & $0.58$ & $0.8105$ & $0.64^{**}$ \\
SV-FS-ST & $0.85$ & $0.70$ & $0.99$ & $1.1870$ & $0.77^{**}$ &  & $3.39$ & $0.06$ & $0.56$ & $0.8093$ & $0.72^{**}$ \\
SV-GH-ST & $0.68$ & $0.40$ & $0.98$ & $1.2267$ & $0.44^{**}$ &  & $3.39$ & $0.06$ & $0.59$ & $0.8181$ & $0.39^{**}$ \\
\cmidrule(lr){1-12}
RSV-N & $1.53$ & $0.23$ & $0.41$ & $1.1468$ & $0.92^{**}$ &  & $5.08$ & $0.92$ & $0.84$ & $0.7854$ & $0.95^{**}$ \\
RSV-T & $1.36$ & $0.41$ & $0.62$ & $1.1357$ & $1.00^{**}$ &  & $4.92$ & $0.92$ & $0.85$ & $0.7833$ & $0.95^{**}$ \\
RSV-AZ-SN & $1.53$ & $0.23$ & $0.05$ & $1.1343$ & $1.00^{**}$ &  & $4.92$ & $0.92$ & $0.86$ & $0.7846$ & $0.95^{**}$ \\
RSV-AZ-ST & $1.19$ & $0.66$ & $0.60$ & $1.1355$ & $1.00^{**}$ &  & $5.25$ & $0.78$ & $0.85$ & $0.7816$ & $1.00^{**}$ \\
RSV-FS-SN & $1.19$ & $0.66$ & $0.55$ & $1.1533$ & $0.77^{**}$ &  & $5.25$ & $0.78$ & $0.38$ & $0.7887$ & $0.95^{**}$ \\
RSV-FS-ST & $1.19$ & $0.66$ & $0.58$ & $1.1301$ & $1.00^{**}$ &  & $5.25$ & $0.78$ & $0.87$ & $0.7865$ & $0.95^{**}$ \\
RSV-GH-ST & $1.19$ & $0.66$ & $0.59$ & $1.1283$ & $1.00^{**}$ &  & $4.75$ & $0.78$ & $0.83$ & $0.7871$ & $0.95^{**}$ \\
\cmidrule(lr){1-12}
EGARCH-N & $1.02$ & $0.97$ & $0.89$ & $1.2944$ & $0.44^{**}$ &  & $3.56$ & $0.09$ & $0.42$ & $0.8480$ & $0.07^{\hspace{0.8em}}$ \\
EGARCH-T & $1.02$ & $0.97$ & $0.77$ & $1.2448$ & $0.15^{*\hspace{0.4em}}$ &  & $3.56$ & $0.09$ & $0.36$ & $0.8319$ & $0.19^{*\hspace{0.4em}}$ \\
\cmidrule(lr){1-12}
REGARCH-N & $1.69$ & $0.12$ & $0.08$ & $1.1744$ & $0.77^{**}$ &  & $4.41$ & $0.50$ & $0.79$ & $0.7988$ & $0.66^{**}$ \\
REGARCH-T & $1.02$ & $0.97$ & $0.54$ & $1.1343$ & $1.00^{**}$ &  & $4.75$ & $0.78$ & $0.80$ & $0.7922$ & $0.72^{**}$ \\
\bottomrule
\end{tabular}
\begin{tablenotes}
\footnotesize
\item \textit{Notes:} See Table \ref{tab:forecast-var-es-loss} for additional details.
\end{tablenotes}
\end{threeparttable}
\end{table}
The last four columns of Table~\ref{tab:gw-matched} report the unconditional GW test for the matched pairs under the FZ0 loss.
The test rejects equal predictive ability at the $5\%$ level only for the Azzalini and GH skew-$t$ pairs for the DJIA at $\alpha = 5\%$, so the differences between the RSV and SV specifications, although systematic in sign, are in general not statistically significant; the tail-risk losses are driven by infrequent violations and are estimated far less precisely than the QLIKE losses.
The cumulative loss differences relative to SV-N, reported in Figures~S25--S28 of the Supplementary Material together with the conditional tests and the pairwise comparisons across all models (Figures~S29--S32), show that the $5\%$ advantage of the RSV and REGARCH specifications over SV-N is present on virtually every day of the forecast period for both indices, and that at the $1\%$ level every RSV specification stays ahead of SV-N for the N225, whereas for the DJIA four of the seven RSV specifications fall behind SV-N, the shortfall being incurred in the volatility episodes of February 2018 and the second half of 2019, when their VaR forecasts are violated.

In summary, the tail risk results are more nuanced than the volatility results.
Incorporating realized volatility lowers the FZ0 loss for the N225 at both levels and for the DJIA at the $5\%$ level, but raises it for the DJIA at the $1\%$ level, and only two of these differences, both for the DJIA at the $5\%$ level, are significant in the pairwise test.
At the $1\%$ level for the DJIA, moreover, the coverage tests reject for most models of both classes, so the ranking by FZ0 loss at this level compares forecasts whose violation rates all exceed the nominal level.
The return distribution matters more here than for volatility forecasting: at the $1\%$ level for the DJIA, the skewed specifications reduce the loss within both classes, from $1.071$ for SV-N to $0.975$ for SV-AZ-SN and from $1.235$ for RSV-N to $1.019$ for RSV-GH-ST.
Distributional characteristics of the one-day-ahead return forecasts under the RSV models are reported in Table~S5 of the Supplementary Material.
\subsection{Performance during the COVID-19 Stress Period}
\label{sec:covid}
A natural concern is whether the patterns documented above carry over to periods of extreme market stress.
We use independent samples spanning January 2012 to December 2020, which cover the COVID-19 volatility spike, and re-run the full rolling forecast exercise over the 2020 calendar year for both indices.
Table~\ref{tab:covid-summary} summarizes the QLIKE loss under RV5 and the FZ0 losses for the matched pairs, for the RSV specifications with $\psi$ estimated freely, and for the Student's $t$ benchmarks; the descriptive statistics, parameter estimates, full loss tables with all proxies, violation rates, coverage tests, MCS $p$-values, and pairwise tests are collected in Sections~S10 and S11 of the Supplementary Material.
RSV-GH-ST is excluded for the DJIA because its posterior degenerates over this sample, under a fixed and a free $\psi$ and under a fresh seed, as documented in Section~S10 of the Supplementary Material; SV-GH-ST is retained for both indices.
\begin{table}[tbp]
\centering
\begin{threeparttable}
\caption{QLIKE and FZ0 losses over the COVID-19 period (2020).}
\label{tab:covid-summary}
\footnotesize
\begin{tabular}{llllllllll}
\toprule
& \multicolumn{3}{l}{QLIKE (RV5)} & \multicolumn{3}{l}{FZ0, 1\%} & \multicolumn{3}{l}{FZ0, 5\%} \\
\cmidrule(lr){2-4} \cmidrule(lr){5-7} \cmidrule(lr){8-10}
Distribution & SV & RSV & RSV, free $\psi$ & SV & RSV & RSV, free $\psi$ & SV & RSV & RSV, free $\psi$ \\
\midrule
\multicolumn{10}{l}{\textit{DJIA}} \\
N & 0.258 & 0.274 & 0.243 & 2.107 & 2.407 & 1.937 & 1.395 & 1.427 & 1.326 \\
T & 0.264 & 0.266 & 0.240 & 2.014 & 2.243 & 1.838 & 1.408 & 1.403 & 1.308 \\
AZ-SN & 0.260 & 0.272 & 0.235 & 2.018 & 2.170 & 1.813 & 1.378 & 1.387 & 1.304 \\
AZ-ST & 0.264 & 0.264 & 0.234 & 2.024 & 2.039 & 1.777 & 1.392 & 1.379 & 1.302 \\
FS-SN & 0.274 & 0.289 & 0.253 & 1.958 & 2.060 & 1.843 & 1.360 & 1.400 & 1.330 \\
FS-ST & 0.276 & 0.281 & 0.246 & 2.079 & 2.135 & 1.932 & 1.393 & 1.400 & 1.330 \\
GH-ST & 0.259 & -- & -- & 1.922 & -- & -- & 1.394 & -- & -- \\
EGARCH-T & \multicolumn{2}{l}{0.362} & -- & \multicolumn{2}{l}{2.264} & -- & \multicolumn{2}{l}{1.466} & -- \\
REGARCH-T & \multicolumn{2}{l}{0.246} & 0.248 & \multicolumn{2}{l}{2.137} & 2.097 & \multicolumn{2}{l}{1.405} & 1.403 \\
\midrule
\multicolumn{10}{l}{\textit{N225}} \\
N & 0.345 & 0.293 & 0.312 & 1.400 & 1.531 & 1.507 & 1.065 & 1.108 & 1.137 \\
T & 0.350 & 0.290 & 0.314 & 1.407 & 1.424 & 1.442 & 1.066 & 1.099 & 1.133 \\
AZ-SN & 0.335 & 0.293 & 0.310 & 1.398 & 1.478 & 1.495 & 1.052 & 1.102 & 1.129 \\
AZ-ST & 0.341 & 0.289 & 0.311 & 1.398 & 1.402 & 1.463 & 1.060 & 1.097 & 1.129 \\
FS-SN & 0.343 & 0.300 & 0.303 & 1.402 & 1.443 & 1.456 & 1.048 & 1.110 & 1.119 \\
FS-ST & 0.336 & 0.296 & 0.301 & 1.404 & 1.443 & 1.479 & 1.051 & 1.101 & 1.108 \\
GH-ST & 0.326 & 0.289 & -- & 1.429 & 1.415 & -- & 1.060 & 1.089 & -- \\
EGARCH-T & \multicolumn{2}{l}{0.340} & -- & \multicolumn{2}{l}{1.387} & -- & \multicolumn{2}{l}{1.082} & -- \\
REGARCH-T & \multicolumn{2}{l}{0.284} & 0.282 & \multicolumn{2}{l}{1.374} & 1.386 & \multicolumn{2}{l}{1.093} & 1.097 \\
\bottomrule
\end{tabular}
\begin{tablenotes}
\footnotesize
\item \textit{Notes:} Average losses over the 253 (DJIA) and 242 (N225) forecast days of 2020. The RSV columns use $\psi = 1$ (the SV models have no measurement equation); the free-$\psi$ columns re-estimate the RSV (and REGARCH-T) specifications with $\psi$ estimated freely. RSV-GH-ST is excluded for the DJIA, and its free-$\psi$ variant for both indices, because the posterior degenerates (see text). Full tables with all proxies, violation rates, coverage tests, and MCS $p$-values are in Tables~S8--S11 (DJIA) and S13--S16 (N225) of the Supplementary Material.
\end{tablenotes}
\end{threeparttable}
\end{table}

For volatility forecasting, the two indices part company.
For the N225, every RSV specification attains a lower QLIKE than its SV counterpart under all four proxies, with scores of $0.289$ to $0.300$ against $0.326$ to $0.350$ under RV5, but the margins are smaller than over the main sample, the pairwise test rejects equal accuracy at the $5\%$ level only for the Student's $t$ pair under RV5, RK, and Med and for four pairs under BV, and the REGARCH benchmarks attain the lowest scores ($0.284$).
For the DJIA, the advantage of realized volatility disappears: the SV specifications attain scores of $0.258$ to $0.276$ against $0.264$ to $0.289$ for the RSV specifications under RV5, only for the Azzalini skew-$t$ pair does the RSV model not fall behind, and there the two scores coincide to three decimals, no pairwise difference is significant, and REGARCH-T attains the lowest score ($0.246$), followed by REGARCH-N and SV-N.

For VaR and ES, the evidence runs the other way, and more sharply than in the main sample.
At the $1\%$ level, every matched pair favours the SV specification for the DJIA and all but the GH skew-$t$ pair for the N225, and the RSV violation rates, between $3.16\%$ and $4.35\%$ for the DJIA and between $1.65\%$ and $2.48\%$ for the N225, are far above the nominal level, against $1.98\%$ to $3.16\%$ and $0.41\%$ to $0.83\%$ for the SV models.
The lowest $1\%$ loss is attained by SV-GH-ST for the DJIA and by REGARCH-T for the N225.
At the $5\%$ level, four of the six DJIA pairs and all seven N225 pairs favour the SV specification, with SV-FS-SN attaining the lowest loss for both indices; for the DJIA, SV-AZ-SN and RSV-AZ-ST follow within $0.02$.
The pairwise test rejects equal accuracy only for the Student's $t$ pair at the $1\%$ level for the DJIA and is borderline for the N225 at the $5\%$ level, so most of these differences are not statistically significant.

The two indices therefore qualify the message of the main sample in different ways: the volatility forecasting gains from incorporating realized volatility survive the stress of 2020 for the N225 but not for the DJIA, and the tail risk forecasts of the RSV models with $\psi = 1$ fall behind their SV counterparts for both indices, most clearly at the $1\%$ level.
A plausible interpretation is that under abrupt and persistent shocks, the conditional asymmetry captured by skew-$t$ innovations matters less than the rapid revision of the volatility level itself, which is delivered by all RV-based specifications, including REGARCH-T.

Because the measurement equations fix $\psi = 1$, we also re-run both COVID-19 exercises with $\psi$ estimated freely; the free-$\psi$ columns of Table~\ref{tab:covid-summary} report the RV5 QLIKE and FZ0 losses, and the full results are in Sections~S10.3 and S11.3 of the Supplementary Material.
For the DJIA, the free-$\psi$ variants are the more accurate ones on both dimensions: all six RSV specifications attain a lower QLIKE under all four proxies, by between $14\%$ and $17\%$ on average, and a lower FZ0 loss at both risk levels, with the $1\%$ violation rates falling from between $3.16\%$ and $4.35\%$ to between $1.98\%$ and $3.56\%$.
This is enough to overturn the matched-pair comparison reported above.
Under a freely estimated $\psi$ every RSV specification attains a lower FZ0 loss than its SV counterpart at both risk levels, and the best free-$\psi$ specification reaches $1.78$ at the $1\%$ level and $1.30$ at the $5\%$ level, against $1.92$ and $1.36$ for the best SV specification.
For the DJIA over this sample, therefore, the tail-risk shortfall of the RSV models disappears once $\psi$ is estimated freely; this is evidence from one index and one year, and whether it generalizes is not established here.

For the N225, a free $\psi$ helps on neither dimension: it raises the average QLIKE of every RSV specification, by up to $8\%$, lowering it only for RSV-FS-SN under the BV and Med proxies, and raises the FZ0 loss for every RSV specification at the $5\%$ level and for all but RSV-N at the $1\%$ level.
For REGARCH, a free $\psi$ buys nothing on either index: it raises QLIKE under all four proxies for REGARCH-N on both indices and for REGARCH-T on the DJIA, and lowers it only marginally for REGARCH-T on the N225, where the average falls from $0.268$ to $0.267$.
The FZ0 losses move only slightly for REGARCH and not systematically in either direction, for both the DJIA and the N225.

These departures from $\psi = 1$ are consistent in direction with the in-sample estimates over the main sample reported in Table~S6 of the Supplementary Material: when $\psi$ is estimated freely in a single full-sample estimation, its $95\%$ credible interval lies entirely below unity for every RSV specification and at or above unity for REGARCH.
Those estimates only indicate where $\psi$ lies, however; they do not measure forecast accuracy, which we do not evaluate under a free $\psi$ over the main sample.
RSV-GH-ST with a free $\psi$ diverges to a degenerate posterior mode, as it does over the main sample, and is excluded from this comparison.
\section{Conclusion}
\label{sec:conc}
This study has compared RSV models with three types of skew-$t$ innovations against their SV counterparts and against Bayesian EGARCH and REGARCH benchmarks for one-day-ahead volatility, VaR, and ES forecasting of the DJIA and the N225, with the volatility forecasts of all models evaluated under a common point forecast, the predictive mean.
Incorporating realized volatility improves volatility forecasts over the main sample for both indices, under every proxy and for every return distribution, with uniformly significant gains for the N225 and, for the DJIA, gains that are significant only under the realized measure that also enters the model, a dependence that limits the DJIA evidence; under the stress of 2020 the advantage survives for the N225 but not for the DJIA, where the Bayesian REGARCH-T model attains the lowest loss, and REGARCH-T is a competitive benchmark throughout.
For tail risk, the benefits are conditional: realized volatility lowers the FZ0 loss for the N225 and at the $5\%$ level for the DJIA, but not at the $1\%$ level for the DJIA, where most models, RSV and SV alike, fail the coverage tests, nor during the stress period with $\psi = 1$, and only two of the matched-pair differences over the main sample are statistically significant, while relaxing the restriction $\psi = 1$ restores the advantage of the RSV models for the DJIA over the COVID-19 sample.
The return distribution matters little for volatility forecasting but skewed innovations reduce the $1\%$ tail risk loss for the DJIA over the main sample within both classes, and no single distribution dominates across indices, risk levels, and periods.
Practitioners interested in tail risk should therefore complement RSV-based forecasts with Bayesian REGARCH-type benchmarks and assess the sensitivity of conclusions to the realized proxy, the quantile level, and stress periods; extending the measurement equation to alternative or multiple realized measures, and comparing with score-driven specifications \citep{catania_forecasting_2022}, are natural next steps.
\section*{Declaration of generative AI and AI-assisted technologies\\ in the manuscript preparation process}

During the preparation of this work, the authors used Claude (Anthropic) in order to draft and revise the wording of the manuscript, to cross-check the numbers reported in the text against the estimation and evaluation output, and to assist in writing the analysis and typesetting scripts.
The authors also used ChatGPT (OpenAI) to assist with proofreading the manuscript.
The authors reviewed and edited the output as needed and take full responsibility for the content of the published article.

\bibliography{rsvst202608}
\bibliographystyle{agsm}

\end{document}


\maketitle

\section{Additional Details of the Models}
\label{sec:suppl-models}

This section collects the material on realized volatility and on the higher-order moments of the skewed distributions that Section~2 of the main text summarizes; the innovation of each model is restated at the start of its subsection so that the section is self-contained.
Equation numbers in this section refer to this supplement; equations of the main text are referred to in words.

\subsection{Realized Volatility}

In the SV model, volatility \( \exp(h_t) \) is treated as a latent variable, since the log-volatility \( h_t \) is unobserved.
With the increasing availability of high-frequency data, RV has emerged as a prominent nonparametric estimator of true volatility.

Let \( p(s) \) denote the logarithm of the asset price at time \( s \), and suppose it follows the continuous-time diffusion process:
\begin{equation}
dp(s) = \mu(s) ds + \sigma(s) dW(s), \label{eqn:ito}
\end{equation}
where \( \mu(s) \) and \( \sigma(s)^2 \) are the drift and instantaneous variance, respectively, and \( W(s) \) is a standard Brownian motion.
The integrated volatility (IV) over a single day is defined as
\begin{equation}
IV_t = \int_{t-1}^{t} \sigma^2(s) ds. \label{eqn:IV}
\end{equation}

Assume that \( m \) intraday returns are observed on day \( t \), denoted by \( \{r_{t-1+1/m}, r_{t-1+2/m}, \ldots, r_{t}\} \).
Then, the realized volatility is defined as
\begin{equation}
RV_t = \sum_{i=1}^{m} r_{t-1+i/m}^2. \label{eqn:RV}
\end{equation}
Under ideal conditions (e.g., no microstructure noise or missing observations), \( RV_t \) converges to \( IV_t \) as \( m \to \infty \).

However, these ideal conditions are often violated in practice.
One source of bias arises from non-trading hours, which are not captured in intraday data.
Ignoring these periods leads to an underestimation of daily volatility.
To address this, \citet{hansen_forecast_2005} proposed a scaling adjustment:
\begin{align}
RV_t^{HL} = c_{HL} RV_t, \quad c_{HL} = \frac{ \sum_{t=1}^n (y_t - \bar{y})^2 }{ \sum_{t=1}^n RV_t }, \quad \bar{y} = \frac{1}{n} \sum_{t=1}^n y_t. \label{eqn:HL}
\end{align}
This adjustment ensures that the average of the scaled RV matches the sample variance of daily returns.

Another significant bias stems from microstructure noise.
Observed prices deviate from the efficient price due to bid–ask bounce, price discreteness, and asynchronous trading, introducing autocorrelation and distortions in high-frequency returns.
The bias tends to worsen as the sampling interval becomes shorter.
Optimal sampling frequency and alternative estimators have been studied extensively \citep{ait-sahalia_how_2005, bandi_separating_2006, bandi_microstructure_2008, liu_does_2015}.

Among various alternatives, the RK estimator \citep{barndorff-nielsen_designing_2008} has been widely adopted due to its robustness.
It is defined as:
\begin{align}
RK_t = \sum_{h=-H}^{H} k \left( \frac{h}{H+1} \right) \gamma_h, \label{eq:rk}
\end{align}
where \( H \) is the bandwidth, \( k(\cdot) \) is a weight function, and
\begin{align}
\gamma_h = \sum_{i=|h|+1}^{m} r_{t-1+i/m} \cdot r_{t-1+(i - |h|)/m}.
\end{align}
\citet{barndorff-nielsen_realized_2009} provide guidelines for choosing the bandwidth \( H \).

To account for these biases in modeling, hybrid frameworks such as the RSV and RGARCH models have been developed.
In particular, the RSV model of \citet{takahashi_estimating_2009} addresses RV bias by jointly estimating model parameters, as detailed in Section~2.3 of the main text.

\subsection{GH Skew-\( t \) Distribution}

In the RSV-GH-ST model of Section~2.5 of the main text, the return innovation is
\begin{align}
\epsilon_t = \frac{ \beta (\lambda_t - \mu_\lambda) + \sqrt{\lambda_t} \, z_t }{ \sqrt{ \beta^2 \sigma_\lambda^2 + \mu_\lambda } }, \quad z_t \sim \mathcal{N}(0, 1), \quad \lambda_t \overset{\text{i.i.d.}}{\sim} \mathcal{IG} \left( \frac{\nu}{2}, \frac{\nu}{2} \right), \label{eqn:suppl-gh-eps}
\end{align}
where \( \mu_\lambda = \mathrm{E}[\lambda_t] = \nu / (\nu - 2) \), \( \sigma_\lambda^2 = \mathrm{Var}[\lambda_t] = 2 \nu^2 / \{ (\nu - 2)^2 (\nu - 4) \} \) with \( \nu > 4 \), and the log-volatility innovation is \( \eta_t \mid z_t \sim \mathcal{N}( \rho \sigma_\eta z_t, (1 - \rho^2) \sigma_\eta^2 ) \).
By construction \( \mathrm{E}[\epsilon_t] = 0 \) and \( \mathrm{Var}[\epsilon_t] = 1 \); \( \beta < 0 \) produces a left-skewed distribution, and \( \beta = 0 \) gives the standardized Student's \( t \) distribution of the RSV-T model.

The correlation between \( \epsilon_t \) and \( \eta_t \) depends on \( \rho \), \( \nu \), and \( \beta \), and is given by:
\begin{align}
\mathrm{Corr}[\epsilon_t, \eta_t] = \frac{ \mathrm{E}[ \sqrt{\lambda_t} ] }{ \sqrt{ \beta^2 \sigma_\lambda^2 + \mu_\lambda } } \rho,
\end{align}
where \( \mathrm{E}[ \sqrt{\lambda_t} ] \) is obtained from equation~\eqref{eqn:inv-gam-moment-general} below with \( m = 1 \).

\begin{figure}[tbp]
\centering
\begin{tabular}{c}
\includegraphics[width = .9\textwidth]{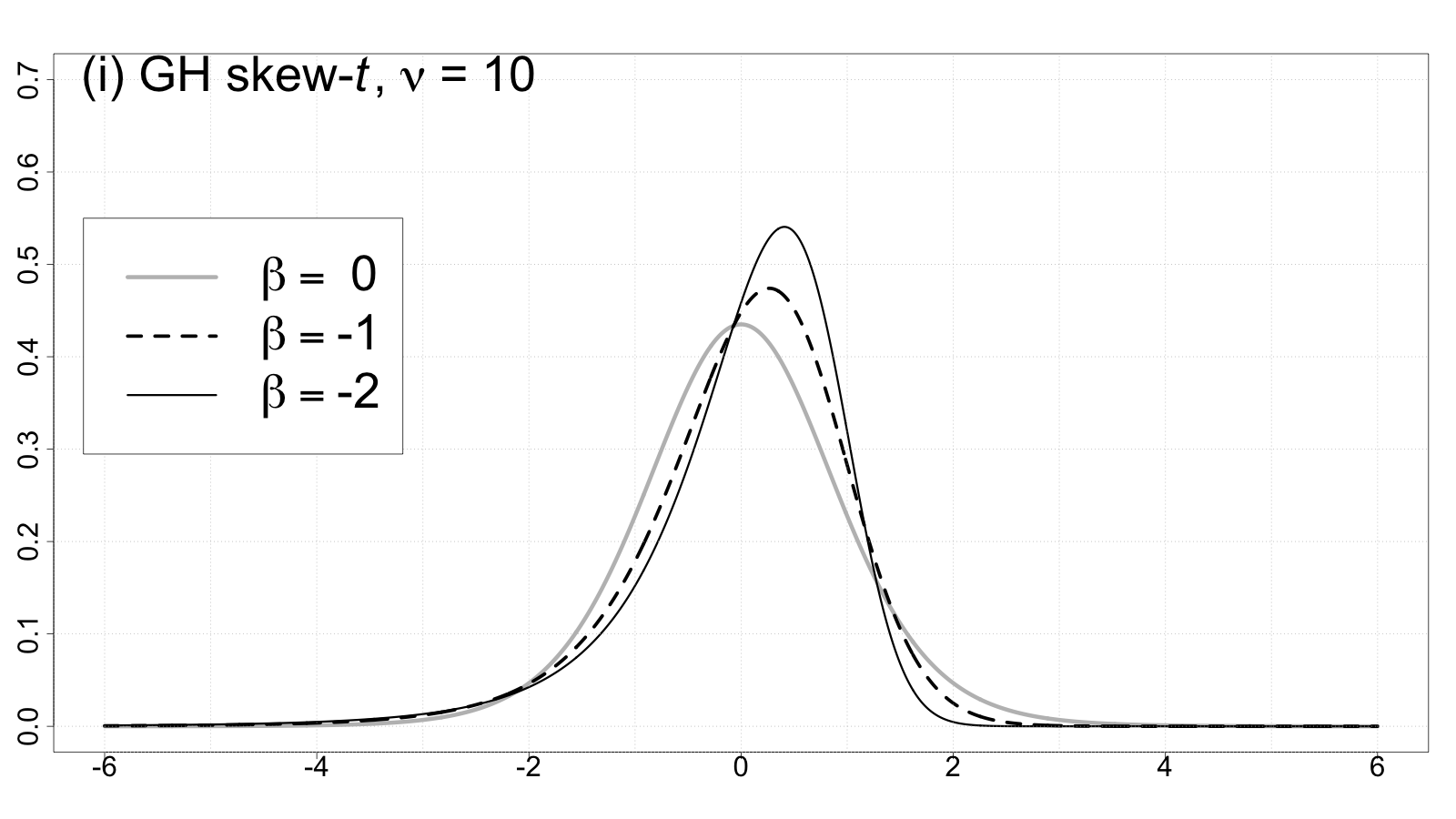} \\
\includegraphics[width = .9\textwidth]{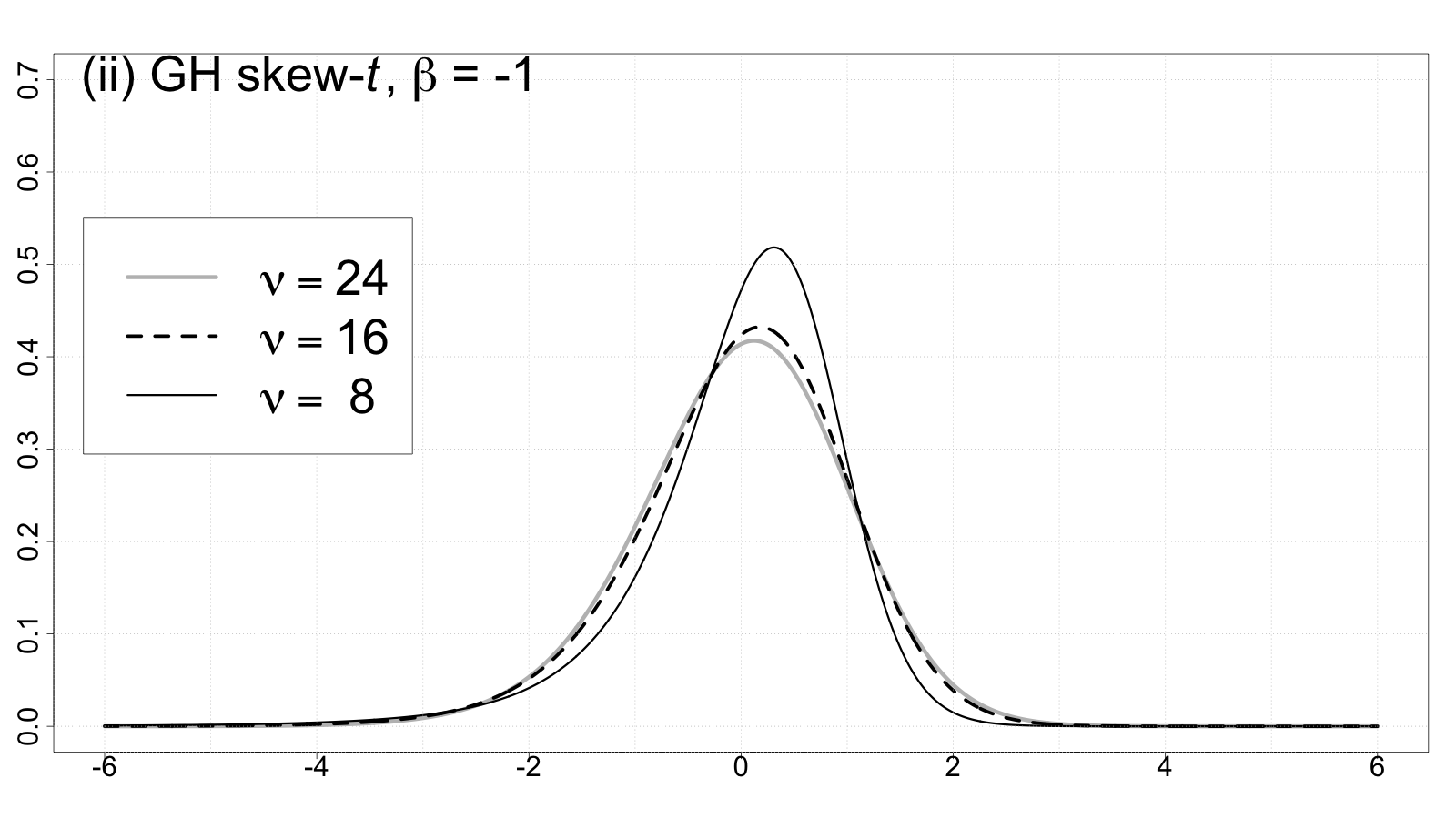}
\end{tabular}
\caption{Density of the standardized GH skew-\( t \) distribution in Section~2.5 of the main text.
(i) Varying \( \beta = 0, -1, -2 \) with fixed \( \nu = 10 \).
(ii) Varying \( \nu = 24, 16, 8 \) with fixed \( \beta = -1 \).}
\label{fig:density-ghst}
\end{figure}

To illustrate the influence of \( \beta \) and \( \nu \), Figure~\ref{fig:density-ghst} plots the densities of \( \epsilon_t \) under various parameter values.
Panel (i) shows that increasing negative values of \( \beta \) yield more pronounced left-skewness and heavier polynomial tails.
In contrast, Panel (ii) indicates that as \( \nu \) increases, the distribution becomes more symmetric and approaches normality, exhibiting thinner tails.

\subsection{Azzalini Skew-\( t \) Distribution}

In the RSV-AZ-ST model of Section~2.6 of the main text, the return innovation is
\begin{align}
\epsilon_t = \frac{ \delta (z_{0t} - c) + \sqrt{1 - \delta^2} \, z_t }{ \sqrt{1 - c^2 \delta^2} } \cdot \sqrt{ \frac{ \lambda_t }{ \mu_\lambda } }, \quad z_{0t} \overset{\text{i.i.d.}}{\sim} \mathcal{TN}_{(0,\infty)}(0, 1), \quad |\delta| < 1, \label{eqn:suppl-az-eps}
\end{align}
where \( z_t \), \( \lambda_t \), \( \mu_\lambda \), and \( \eta_t \mid z_t \) are as in equation~\eqref{eqn:suppl-gh-eps}, \( \mathcal{TN}_{(0,\infty)}(0, 1) \) denotes the standard normal distribution truncated to \( (0, \infty) \), and \( c = \mathrm{E}[z_{0t}] = \sqrt{2 / \pi} \).
The centering by \( c \) and the scaling by \( \sqrt{1 - c^2 \delta^2} \) make \( \mathrm{E}[\epsilon_t] = 0 \) and \( \mathrm{Var}[\epsilon_t] = 1 \); \( \delta = 0 \) gives the RSV-T model.

The contemporaneous correlation between \( \epsilon_t \) and \( \eta_t \) is now a function of \( \rho \), \( \nu \), and \( \delta \), and is given by
\begin{align}
\mathrm{Corr}[\epsilon_t, \eta_t] = \sqrt{ \frac{ 1 - \delta^2 }{ (1 - c^2 \delta^2) \mu_\lambda } } \cdot \mathrm{E}[ \sqrt{ \lambda_t } ] \cdot \rho,
\end{align}
where \( \mathrm{E}[ \sqrt{ \lambda_t } ] \) is again given by equation~\eqref{eqn:inv-gam-moment-general} with \( m = 1 \).

From the general moment expression
\begin{align}
\mathrm{E}[\lambda_t^{m/2}] = \left( \frac{\nu}{2} \right)^{m/2} \cdot \frac{ \Gamma\left( \frac{\nu - m}{2} \right) }{ \Gamma\left( \frac{\nu}{2} \right) }, \quad \nu > m, \label{eqn:inv-gam-moment-general}
\end{align}
we derive the third and fourth moments of \( \epsilon_t \) as:
\begin{align}
\mathrm{E}[\epsilon_t^3] &= \frac{ 4 - \pi }{ 2 } \cdot \frac{ c^3 \delta^3 }{ (1 - c^2 \delta^2)^{3/2} } \cdot \left( \frac{ \nu - 2 }{ 2 } \right)^{3/2} \cdot \frac{ \Gamma\left( \frac{ \nu - 3 }{ 2 } \right) }{ \Gamma\left( \frac{ \nu }{ 2 } \right) }, \\
\mathrm{E}[\epsilon_t^4] &= \left( 3 + 2 (\pi - 3) \cdot \frac{ c^4 \delta^4 }{ (1 - c^2 \delta^2)^2 } \right) \cdot \frac{ \nu - 2 }{ \nu - 4 }.
\end{align}

\begin{figure}[tbp]
\centering
\begin{tabular}{c}
\includegraphics[width = .9\linewidth]{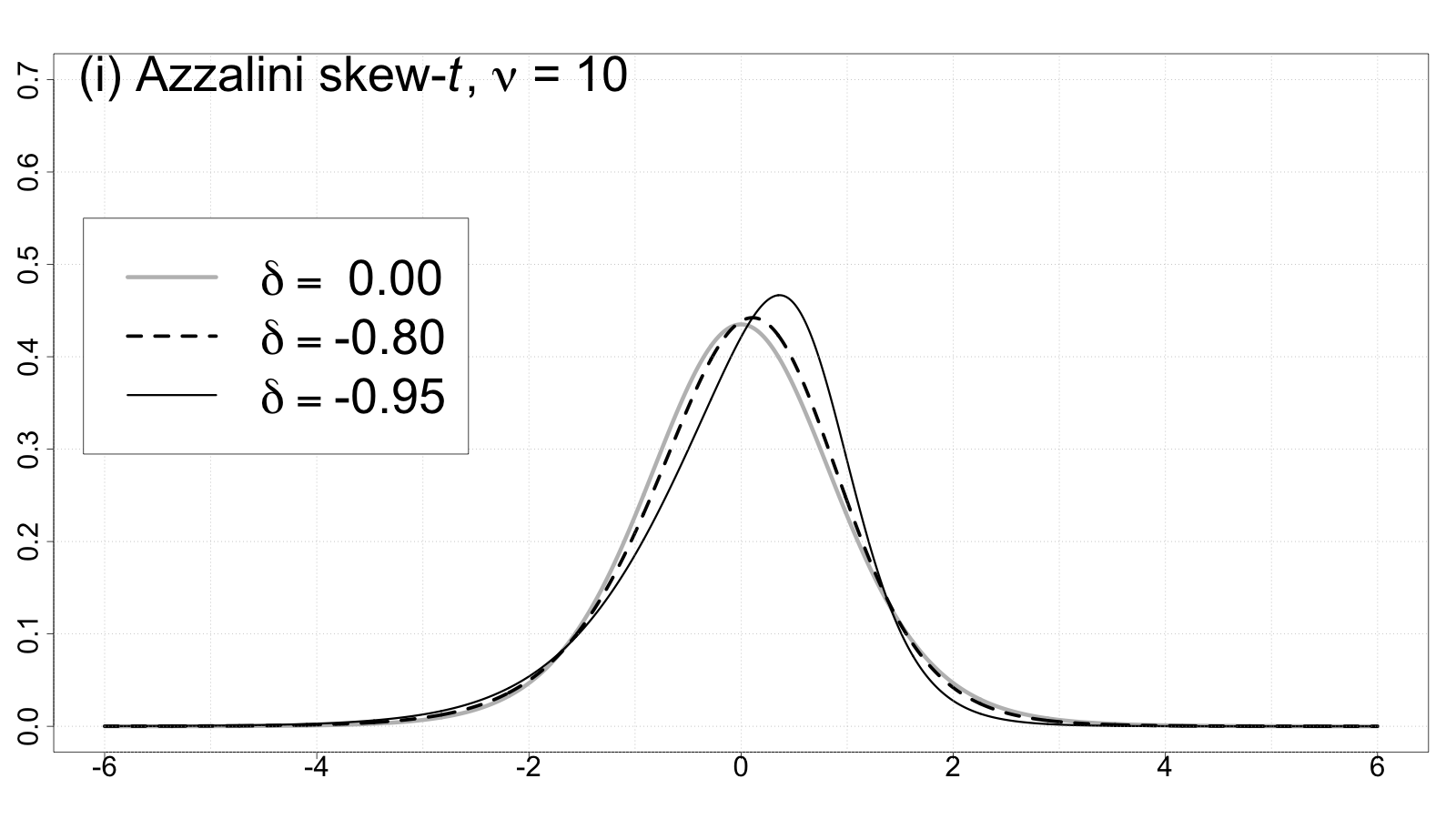} \\
\includegraphics[width = .9\linewidth]{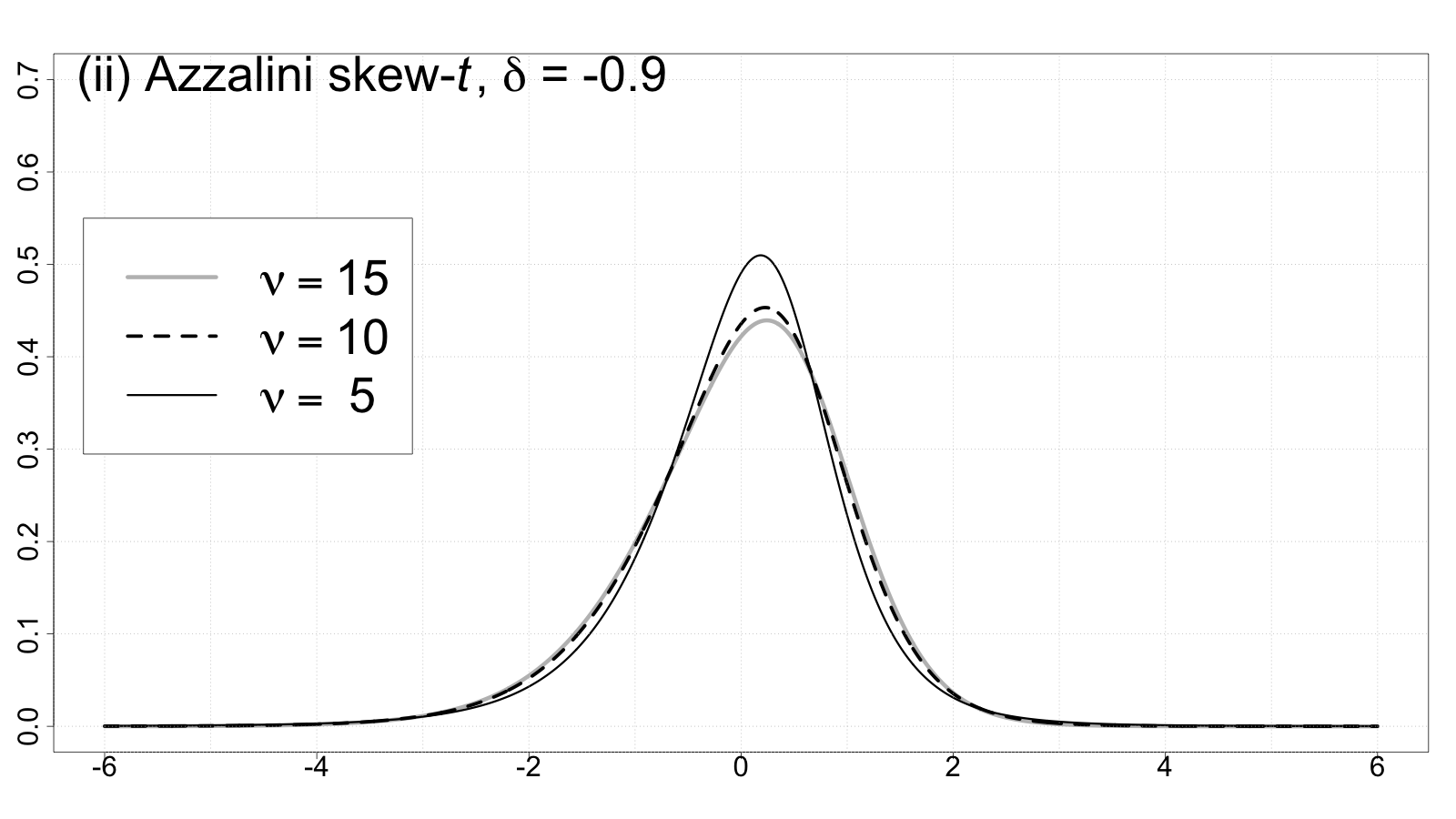}
\end{tabular}
\caption{Density of the standardized Azzalini skew-\( t \) distribution in the innovation equation of the RSV-AZ-ST model in Section~2.6 of the main text.
(i) \( \delta = 0, -0.8, -0.95 \) with \( \nu = 10 \) fixed.
(ii) \( \nu = 15, 10, 5 \) with \( \delta = -0.9 \) fixed.}
\label{fig:density-azst}
\end{figure}

As with the GH skew-\( t \) distribution, the skewness and kurtosis of the return distribution are jointly influenced by the parameters.
While \( \nu \) governs tail behavior, \( \delta \) introduces asymmetry.
Figure~\ref{fig:density-azst} illustrates the impact of these parameters on the shape of \( \epsilon_t \).
Panel (i) shows that more negative values of \( \delta \) lead to stronger left skewness, with \( \delta = 0 \) yielding a symmetric \( t \)-distribution.
Panel (ii) demonstrates that larger values of \( \nu \) reduce tail thickness and approximate the normal distribution.
Notably, even when \( \nu \to \infty \), the distribution remains skewed unless \( \delta = 0 \), in which case the skew-normal distribution is recovered.

\subsection{Fern\'{a}ndez--Steel Skew-\( t \) Distribution}

In the RSV-FS-ST model of Section~2.7 of the main text, the return innovation is the standardized variable \( \epsilon_t = w_* = (w - \mu_*) / \sigma_* \), where \( w \) follows the FS skew-\( t \) density
\begin{align}
p_T(w \mid \gamma, \nu) = \frac{2}{\gamma + \gamma^{-1}} \left\{
f_T\left( \frac{w}{\gamma} \,\middle|\, \nu \right) \mathbbm{1}\{ w \geq 0 \}
+ f_T\left( \gamma w \,\middle|\, \nu \right) \mathbbm{1}\{ w < 0 \}
\right\}, \quad \gamma > 0, \label{eqn:suppl-fs-dens}
\end{align}
\( f_T(\cdot \mid \nu) \) is the density of the standard Student's \( t \) distribution with \( \nu \) degrees of freedom, and \( \mu_* = \mathrm{E}[w] \) and \( \sigma_*^2 = \mathrm{Var}[w] \) are given in Section~2.7 of the main text; the log-volatility innovation is \( \eta_t \mid \epsilon_t \sim \mathcal{N}( \rho \sigma_\eta \epsilon_t, (1 - \rho^2) \sigma_\eta^2 ) \).
Values \( \gamma < 1 \) produce a left-skewed distribution, and \( \gamma = 1 \) recovers the symmetric Student's \( t \) distribution.

Following \citet{trottier_higher_2016}, higher-order moments for the FS skew-$t$ distribution and their standardized forms are derived using the absolute moments of the underlying Student's $t$ distribution.
For the absolute moments, we refer to the unified results provided by \citet{kirkby_moments_2025}.
The $k$-th absolute moment, denoted by $M_k = 2 \int_0^\infty \tilde{w}^k f_T(\tilde{w} \mid \nu) d\tilde{w}$, is generally given by:
\begin{align}
M_k = \nu^{k/2} \frac{\Gamma\left( \frac{k+1}{2} \right) \Gamma\left( \frac{\nu-k}{2} \right)}{\sqrt{\pi} \, \Gamma\left( \frac{\nu}{2} \right)}, \quad k < \nu.
\end{align}
Specifically, for $k=3$ and $k=4$, we obtain:
\begin{align}
M_3 &= \frac{\nu^{3/2} \, \Gamma\left(\frac{\nu-3}{2}\right)}{\sqrt{\pi}\,\Gamma\left(\frac{\nu}{2}\right)} \quad (\nu > 3), \\
M_4 &= \frac{3\nu^2}{(\nu-2)(\nu-4)} \quad (\nu > 4).
\end{align}

Using the general raw-moment formula provided in \citet{trottier_higher_2016}:
\[
\mu_r' = E[w^r] = M_r \frac{\gamma^{r+1} + (-1)^r \gamma^{-(r+1)}}{\gamma + \gamma^{-1}}, \quad r=1,2,3,4,
\]
the centered third and fourth moments are defined as:
\[
\mu_3 = \mu_3' - 3\mu_* \mu_2' + 2\mu_*^3, \qquad
\mu_4 = \mu_4' - 4\mu_* \mu_3' + 6\mu_*^2 \mu_2' - 3\mu_*^4.
\]
Consequently, the skewness and excess kurtosis of the standardized variable $w_* = (w - \mu_*)/\sigma_*$ are given by:
\begin{align}
\mathrm{Skew}(w_*) &= \frac{\mu_3}{\sigma_*^3}, \\[6pt]
\mathrm{Kurt}(w_*) &= \frac{\mu_4}{\sigma_*^4} - 3.
\end{align}
These expressions depend on $(\gamma, \nu)$, with $\nu$ entering through the moments $M_k$: $\gamma$ governs asymmetry, whereas $\nu$ determines tail thickness.

\begin{figure}[tbp]
\centering
\begin{tabular}{c}
\includegraphics[width = .9\linewidth]{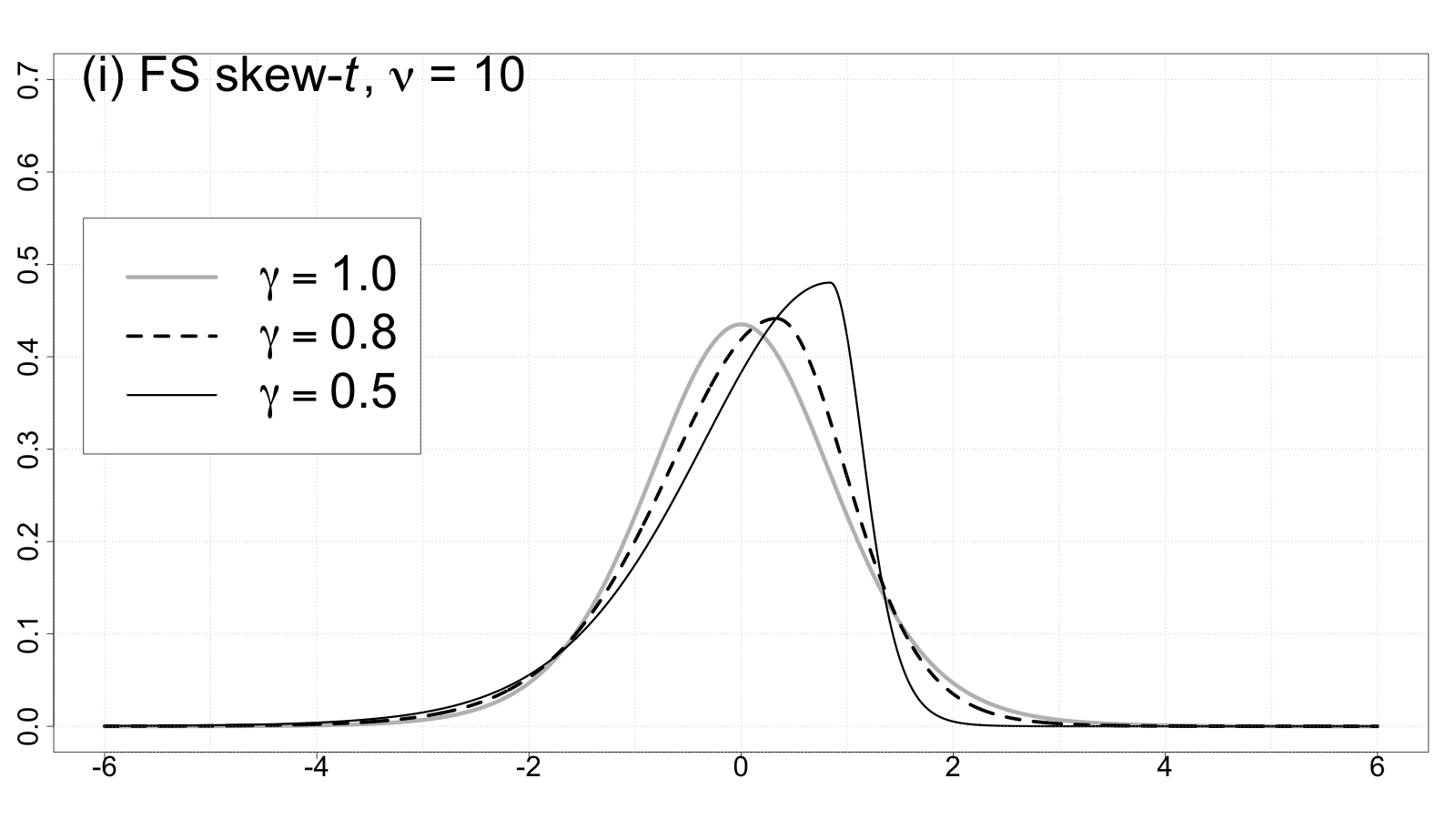} \\
\includegraphics[width = .9\linewidth]{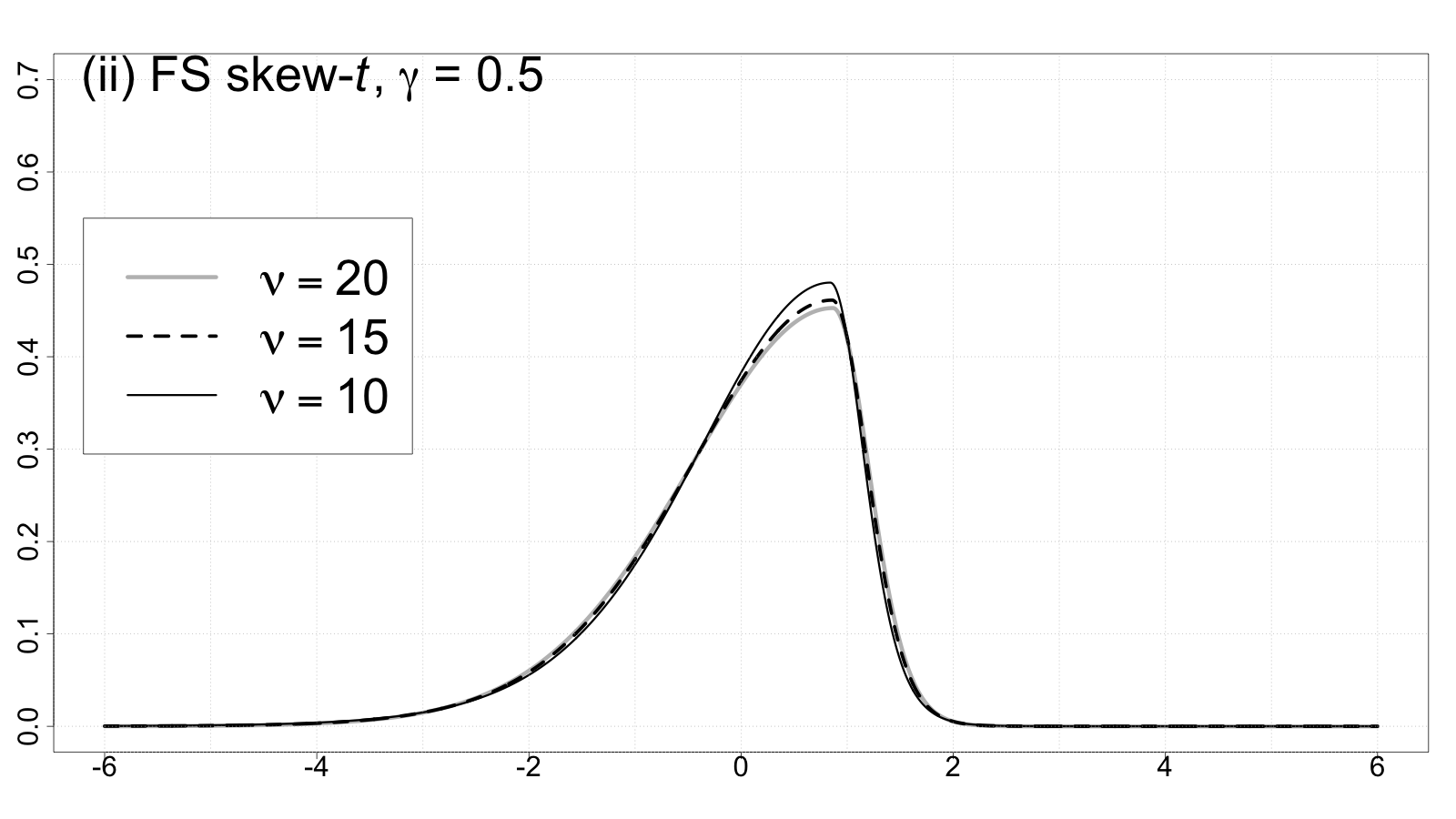}
\end{tabular}
\caption{Density of the standardized FS skew-\(t\) distribution in Section~2.7 of the main text.
(i) \( \gamma = 1, 0.8, 0.5 \) with \( \nu = 10 \).
(ii) \( \nu = 20, 15, 10 \) with \( \gamma = 0.5 \).}
\label{fig:FSST}
\end{figure}

Figure~\ref{fig:FSST} illustrates how the shape of the distribution varies with \( \gamma \) and \( \nu \).
Panel (i) shows that decreasing \( \gamma \) induces stronger negative skewness, with \( \gamma = 1 \) yielding symmetry.
Panel (ii) demonstrates that increasing \( \nu \) reduces tail thickness.
Notably, as \( \nu \to \infty \), the distribution converges to a skew-normal form when \( \gamma \ne 1 \), and to a standard normal when \( \gamma = 1 \).

\clearpage
\section{MCMC Sampling Scheme}
\label{sec:suppl-mcmc}

This section gives the sampling schemes for the RSV-AZ-ST and RSV-FS-ST models outlined in Section~3 of the main text, with the prior distributions of Section~4.2 of the main text.
The schemes for the RSV-N and RSV-GH-ST models are given in \citet{takahashi_estimating_2009} and \citet{takahashi_volatility_2016}, and the SV counterparts omit the measurement equation.

\subsection{RSV-AZ-ST Model}
\label{sec:appendix-azst}

Let \( \bm{\theta} = (\mu, \phi, \rho, \sigma_\eta^2, \delta, \xi, \sigma_u^2, \nu)' \) denote the full vector of model parameters, and define the latent variable vectors \( \bm{\lambda} = (\lambda_1, \ldots, \lambda_n)' \) and \( \bm{z}_0 = (z_{01}, \ldots, z_{0n})' \).
For the prior distributions of the model-specific parameters \( \delta \) and \( \nu \), we assume:
\begin{align}
\frac{\delta + 1}{2} &\sim \mathcal{B}(a_{\delta 0}, b_{\delta 0}), \quad
\nu \sim \mathcal{G}(n_{\nu 0}, S_{\nu 0})\mathbbm{1}\{ \nu > 4 \},
\end{align}
where \( \mathcal{G}(\cdot, \cdot) \) denotes the gamma distribution.

The MCMC algorithm proceeds as follows:
\begin{enumerate}
    \item Initialize \( \bm{h} \), \( \bm{\theta} \), \( \bm{z}_0 \), and \( \bm{\lambda} \).
    \item Sample \( \mu \mid \bm{\theta}_{-\mu}, \bm{h}, \bm{z}_0, \bm{\lambda}, \bm{x}, \bm{y} \).
    \item Sample \( \phi \mid \bm{\theta}_{-\phi}, \bm{h}, \bm{z}_0, \bm{\lambda}, \bm{x}, \bm{y} \).
    \item Jointly sample \( (\rho, \sigma_\eta^2) \mid \bm{\theta}_{-(\rho,\sigma_\eta^2)}, \bm{h}, \bm{z}_0, \bm{\lambda}, \bm{x}, \bm{y} \).
    \item Sample \( \delta \mid \bm{\theta}_{-\delta}, \bm{h}, \bm{z}_0, \bm{\lambda}, \bm{x}, \bm{y} \).
    \item Sample \( \xi \mid \bm{\theta}_{-\xi}, \bm{h}, \bm{z}_0, \bm{\lambda}, \bm{x}, \bm{y} \).
    \item Sample \( \sigma_u^2 \mid \bm{\theta}_{-\sigma_u^2}, \bm{h}, \bm{z}_0, \bm{\lambda}, \bm{x}, \bm{y} \).
    \item Sample \( \nu \mid \bm{\theta}_{-\nu}, \bm{h}, \bm{z}_0, \bm{\lambda}, \bm{x}, \bm{y} \).
    \item Sample \( \bm{z}_0 \mid \bm{\theta}, \bm{h}, \bm{\lambda}, \bm{x}, \bm{y} \).
    \item Sample \( \bm{h} \mid \bm{\theta}, \bm{z}_0, \bm{\lambda}, \bm{x}, \bm{y} \).
    \item Sample \( \bm{\lambda} \mid \bm{\theta}, \bm{h}, \bm{z}_0, \bm{x}, \bm{y} \).
    \item Repeat steps 2–11 for a sufficiently large number of iterations.
\end{enumerate}
The full conditional distribution used in each step is given below.


The joint probability density of \((y_{t}, h_{t+1})\) conditional on \(z_{0t}\), \(h_t\), and parameters \(\bm{\theta}\) is given by
\begin{align*}
f(y_t, h_{t+1}| z_{0t}, h_t,\bm{\theta}) = 
& \frac{1}{2\pi k(\delta)\sqrt{{(1-\rho^2)\sigma_{\eta}^2}}}
  \sqrt{\frac{\mu_{\lambda}}{\lambda_t}}
  \exp\left(-\frac{1}{2}h_t \right) \\
& \times \exp\left\{-\frac{\mu_{\lambda}\exp(-h_t)}{2\lambda_tk(\delta)^2(1-\rho^2)}
  \left[y_{t} - \sqrt{\frac{\lambda_t}{\mu_{\lambda}}}
    \frac{\exp(h_t/2)\delta(z_{0t}-c)}{\sqrt{1-c^2\delta^2}} \right. \right. \\
& \hspace{4.5cm}
  \left. \left. - \sqrt{\frac{\lambda_t}{\mu_{\lambda}}} \rho \exp(h_t/2) \frac{k(\delta)}{\sigma_{\eta}}(h_{t+1}-\mu - \phi(h_t-\mu)) \right]^2 \right\} \\
& \times \exp\left\{-\frac{1}{2\sigma_{\eta}^2}(h_{t+1} - \mu - \phi(h_t-\mu))^2\right\},
\end{align*}
where
\[
k(\delta) = \sqrt{\frac{1-\delta^2}{1-c^2\delta^2}}.
\]
The full conditional posterior density is then given by
\begin{align*}
\pi(\bm{h},\bm{\theta}, \bm{z}_0, \bm{\lambda}|\bm{x}, \bm{y}) \propto 
& k(\delta)^{-T}(1-\rho^2)^{-\frac{T-1}{2}} \sigma_{\eta}^{-T} (1-\phi^2)^{\frac{1}{2}} \\
& \times \exp\left\{
  -\frac{1}{2}\sum_{t=1}^T h_t 
  -\frac{\mu_{\lambda} \exp(-h_T)}{2\lambda_T k(\delta)^2}
  \left[y_T - \sqrt{\frac{\lambda_T}{\mu_{\lambda}}} \frac{\exp(h_T/2)\delta(z_{0T}-c)}{\sqrt{1-c^2\delta^2}} \right]^2 \right\} \\
& \times \exp\left\{
  -\sum_{t=1}^{T-1} \frac{\mu_{\lambda} \exp(-h_t)}{2\lambda_t k(\delta)^2(1-\rho^2)}
  \left[y_t - \sqrt{\frac{\lambda_t}{\mu_{\lambda}}} \frac{\exp(h_t/2)\delta(z_{0t}-c)}{\sqrt{1-c^2\delta^2}} \right. \right. \\
& \hspace{4.8cm}
  \left. \left. - \sqrt{\frac{\lambda_t}{\mu_{\lambda}}} \rho \exp(h_t/2) \frac{k(\delta)}{\sigma_{\eta}} (h_{t+1} - \mu - \phi(h_t - \mu)) \right]^2 \right\} \\
& \times \exp\left\{ -\frac{1}{2\sigma_{\eta}^2} \sum_{t=1}^{T-1} (h_{t+1} - \mu - \phi(h_t - \mu))^2 - \frac{(1 - \phi^2)}{2\sigma_{\eta}^2} (h_1 - \mu)^2 \right\} \\
& \times \exp\left\{ -\frac{1}{2} \sum_{t=1}^T z_{0t}^2 I(z_{0t} > 0) \right\}
  \times \prod_{t=1}^T \frac{\mu_{\lambda}^{1/2} (\nu/2)^{\nu/2}}{\Gamma(\nu/2)} \lambda_t^{-(\nu+1)/2 - 1} \exp\left( -\frac{\nu}{2\lambda_t} \right) \\
& \times \sigma_u^{-T} \times \exp\left\{ -\frac{1}{2\sigma_u^2} \sum_{t=1}^T (x_t - \xi - h_t)^2 \right\}
  \times \pi(\bm{\theta}),
\end{align*}
where \(\pi(\bm{\theta})\) denotes the prior distribution of the model parameters.

\subsubsection{Generation of $\mu$}

The conditional posterior distribution of \(\mu\) follows a normal distribution:
\[
\mu \mid \cdot \sim \mathcal{N}(\mu_1, \sigma_{\mu 1}^2),
\]
where
\begin{align*}
\sigma_{\mu 1}^{-2} &= 
\sigma_{\mu 0}^{-2} + \sigma_{\eta}^{-2} \left[ 1 - \phi^2 + (T - 1)(1 - \phi)^2 (1 - \rho^2)^{-1} \right], \\
\mu_1 &= \sigma_{\mu 1}^2 \left\{
\sigma_{\mu 0}^{-2} \mu_0 
+ \sigma_{\eta}^{-2} \left[
(1 - \phi^2) h_1 +
\frac{1 - \phi}{1 - \rho^2} \sum_{t = 1}^{T - 1}
\left( h_{t + 1} - \phi h_t
\right. \right. \right. \\
& \hspace{4.3cm}
\left. \left. \left.
- \sqrt{\frac{\mu_{\lambda}}{\lambda_t}} \cdot \frac{\rho \sigma_{\eta} \exp(-h_t / 2)}{k(\delta)} \cdot
\left( y_t - \sqrt{\frac{\lambda_t}{\mu_{\lambda}}} \cdot \frac{\exp(h_t / 2)\delta(z_{0t} - c)}{\sqrt{1 - c^2 \delta^2}} \right)
\right) \right]
\right\}.
\end{align*}

\subsubsection{Generation of $\phi$}

The conditional posterior density of \(\phi\) is given by
\begin{align*}
\pi(\phi \mid \cdot) \propto \ &
(1 + \phi)^{a_{\phi 0} - 1}(1 - \phi)^{b_{\phi 0} - 1}(1 - \phi^2)^{1/2} \\
& \times \exp\left\{ -\sum_{t = 1}^{T - 1} \frac{\mu_{\lambda} \exp(-h_t)}{2 \lambda_t k(\delta)^2 (1 - \rho^2)} 
\left[ y_t - \sqrt{\frac{\lambda_t}{\mu_{\lambda}}} \cdot \frac{\exp(h_t / 2) \delta (z_{0t} - c)}{\sqrt{1 - c^2 \delta^2}} \right. \right. \\
& \hspace{5.0cm} 
\left. \left. - \sqrt{\frac{\lambda_t}{\mu_{\lambda}}} \cdot \rho \cdot \exp(h_t / 2) \cdot \frac{k(\delta)}{\sigma_{\eta}} (h_{t + 1} - \mu - \phi(h_t - \mu)) \right]^2 \right\} \\
& \times \exp\left\{ -\frac{1}{2\sigma_{\eta}^2} \sum_{t = 1}^{T - 1} (h_{t + 1} - \mu - \phi(h_t - \mu))^2 - \frac{1 - \phi^2}{2\sigma_{\eta}^2}(h_1 - \mu)^2 \right\} \\
\propto \ & g(\phi) \cdot \exp\left\{ -\frac{1}{2 \sigma_{\phi}^2} (\phi - \mu_{\phi})^2 \right\},
\end{align*}
where
\begin{align*}
g(\phi) &= (1 + \phi)^{a_{\phi 0} - 1/2} (1 - \phi)^{b_{\phi 0} - 1/2}, \\
\sigma_{\phi}^2 &= \frac{\sigma_{\eta}^2 (1 - \rho^2)}{\rho^2 (h_1 - \mu)^2 + \sum_{t = 2}^{T - 1} (h_t - \mu)^2}, \\
\mu_{\phi} &= \frac{\sigma_{\phi}^2}{\sigma_{\eta}^2 (1 - \rho^2)} 
\sum_{t = 1}^{T - 1} \left[
(h_{t + 1} - \mu) -
\sqrt{\frac{\mu_{\lambda}}{\lambda_t}} \cdot \frac{\rho \sigma_{\eta} \exp(-h_t / 2)}{k(\delta)} 
\left( y_t - \sqrt{\frac{\lambda_t}{\mu_{\lambda}}} \cdot \frac{\exp(h_t / 2) \delta (z_{0t} - c)}{\sqrt{1 - c^2 \delta^2}} \right)
\right] (h_t - \mu).
\end{align*}
We then propose a candidate \(\phi^{\dagger} \sim \mathcal{N}(\mu_{\phi}, \sigma_{\phi}^2)\) and accept it with probability \(\min\left\{ 1, \frac{g(\phi^{\dagger})}{g(\phi)} \right\}\).

\subsubsection*{Generation of $\sigma_{\eta}^{2}$ and $\rho$}

Instead of sampling \(\sigma_{\eta}^{2}\) and \(\rho\) separately, we sample them jointly.
Let \(\vartheta = (\sigma_{\eta}, \rho)'\).
The conditional posterior density of \(\vartheta\) is given by
\begin{align*}
\pi(\vartheta \mid \cdot)
&\propto (\sigma_{\eta}^2)^{-(n_0 + T)/2 - 1} \exp\left( -\frac{S_0}{2\sigma_{\eta}^2} \right)
(1 + \rho)^{a_{\rho 0} - 1}(1 - \rho)^{b_{\rho 0} - 1} (1 - \rho^2)^{-(T - 1)/2} \\
& \quad \times \exp\left\{ -\sum_{t = 1}^{T - 1} \frac{\mu_{\lambda} \exp(-h_t)}{2 \lambda_t k(\delta)^2 (1 - \rho^2)}
\left[ y_t - \sqrt{\frac{\lambda_t}{\mu_{\lambda}}} \cdot \frac{\exp(h_t/2) \delta (z_{0t} - c)}{\sqrt{1 - c^2 \delta^2}} \right. \right. \\
& \hspace{5.4cm} \left. \left.
- \sqrt{\frac{\lambda_t}{\mu_{\lambda}}} \cdot \rho \cdot \exp(h_t/2) \cdot \frac{k(\delta)}{\sigma_{\eta}} (h_{t+1} - \mu - \phi(h_t - \mu)) \right]^2 \right\} \\
& \quad \times \exp\left\{ -\frac{1}{2\sigma_{\eta}^2} \sum_{t = 1}^{T - 1} (h_{t + 1} - \mu - \phi(h_t - \mu))^2
- \frac{1 - \phi^2}{2\sigma_{\eta}^2}(h_1 - \mu)^2 \right\}.
\end{align*}
To handle the parameter constraints \(\sigma_{\eta} > 0\) and \(|\rho| < 1\), we perform a transformation:
\[
\omega_1 = \log \sigma_{\eta}, \quad \omega_2 = \log(1 + \rho) - \log(1 - \rho).
\]
This transformation allows unconstrained sampling of \(\omega = (\omega_1, \omega_2)\).
Let \(\hat{\vartheta}\) be the mode of the conditional posterior, and \(\hat{\omega}\) its corresponding transformed value.
We approximate the conditional posterior of \(\omega\) by a normal distribution \(\mathcal{N}(\omega_{*}, \Sigma_{\eta*})\), where
\begin{align*}
\omega_{*} &= \hat{\omega} + \Sigma_{\eta*} \cdot \left. \frac{\partial \log \pi(\omega \mid \cdot)}{\partial \omega} \right|_{\omega = \hat{\omega}}, \\
\Sigma_{\eta*}^{-1} &= - \left. \frac{\partial^2 \log \pi(\omega \mid \cdot)}{\partial \omega \partial \omega'} \right|_{\omega = \hat{\omega}}.
\end{align*}
We then propose a candidate \(\omega^{\dagger} \sim \mathcal{N}(\omega_{*}, \Sigma_{\eta*})\), and accept it with probability
\[
\min \left\{
1,
\frac{ \pi(\vartheta^{\dagger} \mid \cdot) \, f_N(\omega \mid \omega_{*}, \Sigma_{\eta*}) \, |J(\vartheta)| }
     { \pi(\vartheta \mid \cdot) \, f_N(\omega^{\dagger} \mid \omega_{*}, \Sigma_{\eta*}) \, |J(\vartheta^{\dagger})| }
\right\},
\]
where \(f_N(\cdot \mid \mu, \Sigma)\) denotes the density of a multivariate normal distribution, \(J(\cdot)\) is the Jacobian of the inverse transformation, and \(\vartheta^{\dagger}\) is the back-transformed candidate.

\subsubsection{Generation of $\delta$}

The conditional posterior density of \(\delta\) is given by
\begin{align*}
\pi(\delta \mid \cdot) \propto \ &
(1 + \delta)^{a_{\delta 0} - 1} (1 - \delta)^{b_{\delta 0} - 1} \cdot k(\delta)^{-T} \\
& \times \exp\left\{
-\frac{\mu_{\lambda} \exp(-h_T)}{2 \lambda_T k(\delta)^2}
\left[ y_T - \sqrt{\frac{\lambda_T}{\mu_{\lambda}}} \cdot \frac{\exp(h_T/2) \delta (z_{0T} - c)}{\sqrt{1 - c^2 \delta^2}} \right]^2
\right\} \\
& \times \exp\left\{
-\sum_{t = 1}^{T - 1} \frac{\mu_{\lambda} \exp(-h_t)}{2 \lambda_t k(\delta)^2 (1 - \rho^2)}
\left[ y_t - \sqrt{\frac{\lambda_t}{\mu_{\lambda}}} \cdot \frac{\exp(h_t/2) \delta (z_{0t} - c)}{\sqrt{1 - c^2 \delta^2}} \right. \right. \\
& \hspace{5.3cm}
\left. \left. - \sqrt{\frac{\lambda_t}{\mu_{\lambda}}} \cdot \rho \cdot \exp(h_t/2) \cdot \frac{k(\delta)}{\sigma_{\eta}} (h_{t+1} - \mu - \phi(h_t - \mu)) \right]^2
\right\}.
\end{align*}
We approximate the conditional posterior of \(\delta\) by a normal distribution \(\mathcal{N}(\hat{\delta}, \sigma_{\hat{\delta}}^2)\), where \(\hat{\delta}\) is the mode of the conditional density and
\[
\sigma_{\hat{\delta}}^{-2} = - \left.
\frac{\partial^2 \log \pi(\delta \mid \cdot)}{\partial \delta^2} \right|_{\delta = \hat{\delta}}.
\]
We then propose a candidate value \(\delta^{\dagger} \sim \mathcal{TN}_{(-1,1)}(\hat{\delta}, \sigma_{\hat{\delta}}^2)\) and accept it with probability
\[
\min\left\{
1,
\frac{ \pi(\delta^{\dagger} \mid \cdot) \, f_N(\delta \mid \hat{\delta}, \sigma_{\hat{\delta}}^2) }
     { \pi(\delta \mid \cdot) \, f_N(\delta^{\dagger} \mid \hat{\delta}, \sigma_{\hat{\delta}}^2) }
\right\},
\]
where \(\mathcal{TN}_{(a,b)}(\mu, \sigma^2)\) denotes a normal distribution truncated over the interval \((a, b)\), and \(f_N(\cdot \mid \mu, \sigma^2)\) is the density function of a normal distribution with mean \(\mu\) and variance \(\sigma^2\).

\subsubsection{Generation of $\xi$}

The conditional posterior distribution of \(\xi\) is normal:
\[
\xi \mid \cdot \sim \mathcal{N}(\mu_{\xi}, \sigma_{\xi}^2),
\]
where
\begin{align*}
\sigma_{\xi}^{-2} &= \sigma_{\xi 0}^{-2} + T \sigma_u^{-2}, \\
\mu_{\xi} &= \sigma_{\xi}^2 \left[ \sigma_{\xi 0}^{-2} \mu_{\xi 0} + \sigma_u^{-2} \sum_{t = 1}^T (x_t - h_t) \right].
\end{align*}

\subsubsection{Generation of $\sigma_u^2$}

The conditional posterior distribution of \(\sigma_u^2\) follows an inverse gamma distribution:
\[
\sigma_u^2 \mid \cdot \sim \mathcal{IG}\left( \frac{n_{u1}}{2}, \frac{S_{u1}}{2} \right),
\]
where
\begin{align*}
n_{u1} &= n_{u0} + T, \\
S_{u1} &= S_{u0} + \sum_{t = 1}^{T} (x_t - \xi - h_t)^2.
\end{align*}

\subsubsection{Generation of $\nu$}

The conditional posterior density of \(\nu\) is given by
\begin{align*}
\pi(\nu \mid \cdot) \propto \ &
\exp\left\{
\frac{T}{2} \log \mu_{\lambda}
+ \frac{\nu T}{2} \log\left( \frac{\nu}{2} \right)
- T \log \Gamma\left( \frac{\nu}{2} \right)
- \frac{\nu}{2} \sum_{t = 1}^{T} \left( \lambda_t^{-1} + \log \lambda_t \right)
+ (n_{\nu 0} - 1) \log \nu - S_{\nu 0} \nu
\right\} \\
& \times \exp\left\{
- \sum_{t = 1}^{T - 1} \frac{\mu_{\lambda} \exp(-h_t)}{2 \lambda_t k(\delta)^2 (1 - \rho^2)}
\left[ y_t - \sqrt{\frac{\lambda_t}{\mu_{\lambda}}} \cdot \frac{\exp(h_t/2) \delta (z_{0t} - c)}{\sqrt{1 - c^2 \delta^2}} \right. \right. \\
& \hspace{5.2cm} \left. \left.
- \sqrt{\frac{\lambda_t}{\mu_{\lambda}}} \cdot \rho \cdot \exp(h_t/2) \cdot \frac{k(\delta)}{\sigma_{\eta}} (h_{t+1} - \mu - \phi(h_t - \mu)) \right]^2
\right\} \\
& \times \exp\left\{
- \frac{\mu_{\lambda} \exp(-h_T)}{2 \lambda_T k(\delta)^2}
\left[ y_T - \sqrt{\frac{\lambda_T}{\mu_{\lambda}}} \cdot \frac{\exp(h_T/2) \delta (z_{0T} - c)}{\sqrt{1 - c^2 \delta^2}} \right]^2
\right\}.
\end{align*}
We approximate the conditional posterior of \(\nu\) by a normal distribution centered at its mode and apply a Metropolis-Hastings (MH) step, as done for the generation of \(\delta\).

\subsubsection{Generation of $\bm{z}_{0}$}

The conditional posterior distribution of \(z_{0t}\) follows a truncated normal distribution:
\[
z_{0t} \mid \cdot \sim \mathcal{TN}_{(0, \infty)}(\mu_{zt}, \sigma_{zt}^2),
\]
where
\[
\sigma_{zt}^{-2} = 
\begin{cases}
1 + \dfrac{\delta^2}{(1 - \delta^2)(1 - \rho^2)}, & t = 1, \ldots, T - 1, \\
\dfrac{1}{1 - \delta^2}, & t = T,
\end{cases}
\]
and
\[
\mu_{zt} = 
\begin{cases}
\sigma_{zt}^2 \cdot \sqrt{\dfrac{\mu_{\lambda}}{\lambda_t}} \cdot \dfrac{\exp(-h_t / 2) \delta}{k(\delta)^2 (1 - \rho^2) \sqrt{1 - c^2 \delta^2}} \cdot
\left[
y_t + \sqrt{\dfrac{\lambda_t}{\mu_{\lambda}}} \cdot \dfrac{\exp(h_t / 2) c \delta}{\sqrt{1 - c^2 \delta^2}} \right.
\\[0.8em]
\hspace{4.5cm} \left.
- \sqrt{\dfrac{\lambda_t}{\mu_{\lambda}}} \cdot \rho \cdot \exp(h_t / 2) \cdot \dfrac{k(\delta)}{\sigma} (h_{t+1} - \mu - \phi(h_t - \mu))
\right], & t = 1, \ldots, T - 1, \\[1.2em]
\sigma_{zT}^2 \cdot \sqrt{\dfrac{\mu_{\lambda}}{\lambda_T}} \cdot \dfrac{\exp(-h_T / 2) \delta}{k(\delta)^2 \sqrt{1 - c^2 \delta^2}} \cdot
\left[
y_T + \sqrt{\dfrac{\lambda_T}{\mu_{\lambda}}} \cdot \dfrac{\exp(h_T / 2) c \delta}{\sqrt{1 - c^2 \delta^2}}
\right], & t = T.
\end{cases}
\]

\subsubsection{Generation of $\bm{h}$}

We first rewrite the RSV-AZ-ST model, that is, the measurement, return, and log-volatility equations of Section~2.3 of the main text together with the innovation equation of the RSV-AZ-ST model in Section~2.6 of the main text, as
\begin{align}
x_{t} &= \mu_{x} + \alpha_{t} + u_{t}, \quad t = 1, \ldots, n, \\
y_{t} &= (\delta \bar{z}_{0t} + \sqrt{1 - \delta^2} z_t) \sqrt{\lambda_{t}} \exp(\alpha_{t}/2) \gamma, \quad t = 1, \ldots, n, \\
\alpha_{t+1} &= \phi \alpha_{t} + \eta_{t}, \quad t = 0, \ldots, n - 1,
\end{align}
where
\begin{align}
\mu_{x} &= \xi + \mu, \quad
\alpha_t = h_t - \mu, \quad
\bar{z}_{0t} = z_{0t} - c, \quad
\gamma = \frac{\exp(\mu / 2)}{\sqrt{(1 - c^2 \delta^2)\mu_{\lambda}}}.
\end{align}
The log-likelihood of \((y_t, x_t)\) given \(\alpha_t\), \(\alpha_{t+1}\), and other parameters (excluding constants) is
\[
l_{t} = -\frac{\alpha_t}{2} - \frac{(y_t - \mu_t)^2}{2\sigma_t^2} - \frac{(x_t - \mu_x - \alpha_t)^2}{2\sigma_u^2},
\]
where
\begin{align}
\mu_t &= 
\begin{cases}
\left[ \delta \bar{z}_{0t} + \sqrt{1 - \delta^2} \dfrac{\rho}{\sigma_{\eta}} (\alpha_{t+1} - \phi \alpha_t) \right] \sqrt{\lambda_t} \exp(\alpha_t / 2) \gamma, & t = 1, \ldots, n-1, \\
\delta \bar{z}_{0t} \sqrt{\lambda_t} \exp(\alpha_t / 2) \gamma, & t = n,
\end{cases} \\[0.5em]
\sigma_t^2 &= 
\begin{cases}
(1 - \rho^2)(1 - \delta^2) \lambda_t \exp(\alpha_t) \gamma^2, & t = 1, \ldots, n-1, \\
(1 - \delta^2) \lambda_t \exp(\alpha_t) \gamma^2, & t = n.
\end{cases}
\end{align}
Using this log-likelihood, the latent states \((\alpha_1, \ldots, \alpha_n)\) can be efficiently sampled via a multi-move sampler based on \cite{omori_stochastic_2007, omori_block_2008, omori_stochastic_2015}.
For further details, see also \cite{takahashi_volatility_2016, takahashi_stochastic_2023}.

\subsubsection{Generation of $\bm{\lambda}$}

The conditional probability density function of $\lambda_t$ is
\begin{align*}
\pi(\lambda_t|\cdot) \propto \ &
	\lambda_t^{-(\frac{\nu+1}{2} + 1)} \\
& \times \exp\left\{
	- \frac{\nu}{2\lambda_t}
	- \frac{\mu_{\lambda} \exp(-h_t)}{2\lambda_t k(\delta)^2 (1 - \rho^2)} 
	\left[
		y_t - \sqrt{\frac{\lambda_t}{\mu_{\lambda}}} \cdot \frac{\exp(h_t/2)\delta (z_{0t} - c)}{\sqrt{1 - c^2\delta^2}}
	\right.\right. \\
& \left.\left.\hspace{4cm}
		- \sqrt{\frac{\lambda_t}{\mu_{\lambda}}} \cdot \rho \cdot \exp(h_t/2) \cdot \frac{k(\delta)}{\sigma_{\eta}}(h_{t+1} - \mu - \phi(h_t - \mu)) \cdot I(t < T)
	\right]^2
\right\}.
\end{align*}
We rewrite this density as a product of an inverse gamma kernel and a correction term:
\begin{align*}
\pi(\lambda_t|\cdot) \propto
	\lambda_t^{-(\frac{a}{2} + 1)}
	\exp\left( -\frac{b_t}{2\lambda_t} \right)
	\times g(\lambda_t),
\end{align*}
where
\begin{align*}
a &= \nu + 1, \\
b_t &=
\begin{cases}
	\nu + \dfrac{\mu_{\lambda} \exp(-h_t) y_t^2}{k(\delta)^2 (1 - \rho^2)}, & t = 1, \ldots, n-1, \\[0.8em]
	\nu + \dfrac{\mu_{\lambda} \exp(-h_t) y_t^2}{k(\delta)^2}, & t = n,
\end{cases} \\[1.2em]
g(\lambda_t) &=
\begin{cases}
	\exp\left\{
		\sqrt{\dfrac{\mu_{\lambda}}{\lambda_t}} \cdot
		\dfrac{\exp(-h_t/2) y_t}{k(\delta)^2 (1 - \rho^2)} 
		\left[
			\dfrac{\delta (z_{0t} - c)}{\sqrt{1 - c^2 \delta^2}} 
			+ \rho \cdot \dfrac{k(\delta)}{\sigma_{\eta}}(h_{t+1} - \mu - \phi(h_t - \mu))
		\right]
	\right\}, & t = 1, \ldots, n-1, \\[1em]
	\exp\left\{
		\sqrt{\dfrac{\mu_{\lambda}}{\lambda_t}} \cdot
		\dfrac{\exp(-h_t/2) y_t}{k(\delta)^2} \cdot
		\dfrac{\delta (z_{0t} - c)}{\sqrt{1 - c^2 \delta^2}}
	\right\}, & t = n.
\end{cases}
\end{align*}
We propose a candidate draw from the inverse gamma distribution, 
\[
\lambda_t^{\dagger} \sim \mathcal{IG}\left( \frac{a}{2}, \frac{b_t}{2} \right),
\]
and accept it with probability 
\[
\min\left\{ 1, \frac{g(\lambda_t^{\dagger})}{g(\lambda_t)} \right\}.
\]


\subsection{RSV-FS-ST Model}
\label{sec:appendix-fsst}


Let \( \bm{\theta} = (\mu, \phi, \rho, \sigma_\eta^2, \gamma, \xi, \sigma_u^2, \nu)' \) denote the full set of model parameters.
For the model-specific parameters \( \gamma \) and \( \nu \), we assign the following prior distributions:
\begin{align}
\gamma \sim \mathcal{G}(n_{\gamma 0}, S_{\gamma 0}), \quad
\nu \sim \mathcal{G}(n_{\nu 0}, S_{\nu 0})\mathbbm{1}\{ \nu > 2 \}.
\end{align}
The MCMC algorithm proceeds as follows:
\begin{enumerate}
    \item Initialize \( \bm{h} \) and \( \bm{\theta} \).
    \item Sample \( \mu \mid \bm{\theta}_{-\mu}, \bm{h}, \bm{x}, \bm{y} \).
    \item Sample \( \phi \mid \bm{\theta}_{-\phi}, \bm{h}, \bm{x}, \bm{y} \).
    \item Sample \( \rho \mid \bm{\theta}_{-\rho}, \bm{h}, \bm{x}, \bm{y} \).
    \item Sample \( \sigma_\eta^2 \mid \bm{\theta}_{-\sigma_\eta^2}, \bm{h}, \bm{x}, \bm{y} \).
    \item Sample \( \xi \mid \bm{\theta}_{-\xi}, \bm{h}, \bm{x}, \bm{y} \).
    \item Sample \( \sigma_u^2 \mid \bm{\theta}_{-\sigma_u^2}, \bm{h}, \bm{x}, \bm{y} \).
    \item Sample \( \gamma \mid \bm{\theta}_{-\gamma}, \bm{h}, \bm{x}, \bm{y} \).
    \item Sample \( \nu \mid \bm{\theta}_{-\nu}, \bm{h}, \bm{x}, \bm{y} \).
    \item Sample \( \bm{h} \mid \bm{\theta}, \bm{x}, \bm{y} \).
    \item Repeat steps 2–10 for a sufficiently large number of iterations.
\end{enumerate}
The full conditional distribution used in each step is given below.


The joint probability density of $(y_t, h_{t+1})$ is given by
\begin{align*}
f(y_t, h_{t+1} \mid \bm{\theta}, \lambda_t) = &
\ q_T\left( y_t \exp(-h_t / 2) \mid \gamma, \nu \right) 
\exp\left( -\frac{1}{2} h_t \right) \\
& \times \frac{1}{\sqrt{2\pi(1 - \rho^2)\sigma_{\eta}^2}} 
\exp\left\{ -\frac{1}{2(1 - \rho^2)\sigma_{\eta}^2}
\left[ h_{t+1} - \mu - \phi(h_t - \mu) - \rho \sigma_{\eta} \exp(-h_t / 2)y_t \right]^2 \right\},
\end{align*}
where $q_T$ is defined in Section~2.7 of the main text.
Then, the joint posterior density is expressed as
\begin{align*}
\pi(\bm{h}, \bm{\theta} \mid \bm{x}, \bm{y}) \propto &
\ (1 - \rho^2)^{-\frac{T - 1}{2}} \sigma_{\eta}^{-T}(1 - \phi^2)^{-\frac{1}{2}}
\prod_{t = 1}^T q_T\left( y_t \exp(-h_t / 2) \mid \gamma, \nu \right)
\exp\left( -\frac{1}{2}\sum_{t = 1}^T h_t \right) \\
& \times \exp\left\{ -\frac{1}{2(1 - \rho^2)\sigma_{\eta}^2}\sum_{t = 1}^{T - 1}
\left[ h_{t+1} - \mu - \phi(h_t - \mu) - \rho \sigma_{\eta} \exp(-h_t / 2)y_t \right]^2 \right\} \\
& \times \exp\left\{ -\frac{(1 - \phi^2)}{2\sigma_{\eta}^2}(h_1 - \mu)^2 \right\}
\sigma_u^{-T} 
\exp\left\{ -\frac{1}{2\sigma_u^2}\sum_{t = 1}^T (x_t - \xi - h_t)^2 \right\}
\pi(\bm{\theta}).
\end{align*}

\subsubsection{Generation of $\mu$}

The conditional posterior distribution of $\mu$ follows a normal distribution:
\[
\mu \mid \cdot \sim \mathcal{N}(\mu_1, \sigma_{\mu 1}^2),
\]
where
\begin{align*}
\sigma_{\mu 1}^{-2} &= 
\sigma_{\mu 0}^{-2} + \sigma_{\eta}^{-2}\left[ 1 - \phi^2 + (T - 1)(1 - \phi)^2 (1 - \rho^2)^{-1} \right], \\
\mu_1 &= \sigma_{\mu 1}^2 \left\{
\sigma_{\mu 0}^{-2} \mu_0 
+ \sigma_{\eta}^{-2} \left[
(1 - \phi^2) h_1 +
\frac{1 - \phi}{1 - \rho^2} \sum_{t = 1}^{T - 1}
\left( h_{t + 1} - \phi h_t
- \rho \sigma_{\eta} \exp(-h_t / 2)y_t
\right)
\right]
\right\}.
\end{align*}

\subsubsection{Generation of $\phi$}

The conditional posterior density of $\phi$ is given by
\begin{align*}
\pi(\phi \mid \cdot) \propto & \
(1 - \phi)^{a_{\phi 0} - 1} (1 + \phi)^{b_{\phi 0} - 1} (1 - \phi^2)^{1/2} \\
& \times \exp\left\{ -\frac{1}{2(1 - \rho^2)\sigma_{\eta}^2} \sum_{t = 1}^{T - 1}
\left[ h_{t + 1} - \mu - \phi(h_t - \mu) - \rho \sigma_{\eta} \exp(-h_t / 2)y_t \right]^2 \right\} \\
& \times \exp\left\{ -\frac{1 - \phi^2}{2\sigma_{\eta}^2}(h_1 - \mu)^2 \right\} \\
\propto & \
k(\phi) \times \exp\left\{ -\frac{1}{2\sigma_{\phi}^2}(\phi - \mu_{\phi})^2 \right\},
\end{align*}
where
\begin{align*}
k(\phi) &= (1 - \phi)^{a_{\phi 0} - 1/2}(1 + \phi)^{b_{\phi 0} - 1/2}, \\
\sigma_{\phi}^{2} &= \frac{\sigma_{\eta}^2 (1 - \rho^2)}{\rho^2 (h_1 - \mu)^2 + \sum_{t = 2}^{T - 1} (h_t - \mu)^2}, \\
\mu_{\phi} &= \frac{\sigma_{\phi}^{2}}{\sigma_{\eta}^2 (1 - \rho^2)}
\sum_{t = 1}^{T - 1}\left[
h_{t + 1} - \mu - \rho \sigma_{\eta} \exp(-h_t / 2)y_t
\right](h_t - \mu).
\end{align*}
We then employ the MH algorithm, proposing a candidate \(\phi^{\dagger} \sim \mathcal{N}(\mu_{\phi}, \sigma_{\phi}^2)\), and accept the candidate with probability \(\min\left\{ 1, k(\phi^{\dagger})/k(\phi) \right\}\).

\subsubsection{Generation of $\rho$}

The conditional posterior density of $\rho$ is given by
\begin{align*}
\pi(\rho \mid \cdot) \propto & \
(1 + \rho)^{a_{\rho 0} - 1}(1 - \rho)^{b_{\rho 0} - 1}(1 - \rho^2)^{-\frac{T - 1}{2}} \\
& \times \exp\left\{ -\frac{1}{2(1 - \rho^2)\sigma_{\eta}^2}\sum_{t = 1}^{T - 1}
\left[ h_{t + 1} - \mu - \phi(h_t - \mu) - \rho \sigma_{\eta} \exp(-h_t / 2)y_t \right]^2 \right\}.
\end{align*}
We approximate the conditional posterior distribution by \( \mathcal{N}(\hat{\rho}, \sigma_{\hat{\rho}}^2) \), where \( \hat{\rho} \) is the mode of the conditional density and
\[
\sigma_{\hat{\rho}}^{-2} = -\left.\frac{\partial^2 \log \pi(\rho \mid \cdot)}{\partial \rho^2}\right|_{\rho = \hat{\rho}}.
\]
We then propose a candidate \( \rho^{\dagger} \sim \mathcal{TN}_{(-1, 1)}(\hat{\rho}, \sigma_{\hat{\rho}}^2) \), and accept the candidate with probability
\[
\min\left\{ 1, \frac{\pi(\rho^{\dagger} \mid \cdot) f_N(\rho \mid \hat{\rho}, \sigma_{\hat{\rho}}^2)}{\pi(\rho \mid \cdot) f_N(\rho^{\dagger} \mid \hat{\rho}, \sigma_{\hat{\rho}}^2)} \right\}.
\]

\subsubsection{Generation of $\sigma_{\eta}^2$}

The conditional posterior density of $\sigma_{\eta}^2$ is given by
\begin{align*}
\pi(\sigma_{\eta}^2 \mid \cdot) \propto & \ 
(\sigma_{\eta}^2)^{-\frac{n_0 + T}{2} - 1} \exp\left\{ -\frac{S_0}{2\sigma_{\eta}^2} \right\}
\times \exp\left\{ -\frac{1 - \phi^2}{2\sigma_{\eta}^2}(h_1 - \mu)^2 \right\} \\
& \times \exp\left\{ -\frac{1}{2(1 - \rho^2)\sigma_{\eta}^2}\sum_{t = 1}^{T - 1}
\left[ h_{t + 1} - \mu - \phi(h_t - \mu) - \rho \sigma_{\eta} \exp(-h_t / 2)y_t \right]^2 \right\} \\
\propto & \ 
k(\sigma_{\eta}^2) \times (\sigma_{\eta}^2)^{-\frac{n_1}{2} - 1} \exp\left\{ -\frac{S_1}{2\sigma_{\eta}^2} \right\},
\end{align*}
where
\begin{align*}
n_1 & = n_0 + T, \\
S_1 & = S_0 + (1 - \phi^2)(h_1 - \mu)^2
+ \frac{1}{1 - \rho^2}\sum_{t = 1}^{T - 1}\left[ h_{t + 1} - \mu - \phi(h_t - \mu) \right]^2, \\
k(\sigma_{\eta}^2) & = 
\exp\left\{ \frac{\rho}{(1 - \rho^2)\sigma_{\eta}}\sum_{t = 1}^{T - 1} y_t \exp(-h_t / 2)
\left[ h_{t + 1} - \mu - \phi(h_t - \mu) \right] \right\}.
\end{align*}
We then propose a candidate \( \sigma_{\eta}^{2\dagger} \sim \mathcal{IG}(n_1 / 2, S_1 / 2) \), and accept the candidate with probability \( \min\{1, k(\sigma_{\eta}^{2\dagger}) / k(\sigma_{\eta}^2)\} \).

\subsubsection{Generation of $\xi$}

The conditional posterior distribution of $\xi$ follows a normal distribution:
\[
\xi \mid \cdot \sim \mathcal{N}(\mu_{\xi}, \sigma_{\xi}^2),
\]
where
\begin{align*}
\sigma_{\xi}^{-2} &= \sigma_{\xi 0}^{-2} + T \sigma_u^{-2}, \\
\mu_{\xi} &= \sigma_{\xi}^{2}\left[ \sigma_{\xi 0}^{-2}\mu_{\xi 0} + \sigma_u^{-2}\sum_{t = 1}^T (x_t - h_t) \right].
\end{align*}

\subsubsection{Generation of $\sigma_u^2$}

The conditional posterior distribution of $\sigma_u^2$ follows an inverse-gamma distribution:
\[
\sigma_u^2 \mid \cdot \sim \mathcal{IG}\left( \frac{n_{u1}}{2}, \frac{S_{u1}}{2} \right),
\]
where
\begin{align*}
n_{u1} &= n_{u0} + T, \\
S_{u1} &= S_{u0} + \sum_{t = 1}^{T} (x_{t} - \xi - h_t)^2.
\end{align*}

\subsubsection{Generation of $\gamma$}

The conditional posterior density of $\gamma$ is given by
\begin{align*}
\pi(\gamma \mid \cdot) \propto &
\exp\left\{ (n_{\gamma 0} - 1)\log \gamma - S_{\gamma 0}\gamma 
+ \sum_{t = 1}^T \log q_T\left( y_t \exp(-h_t / 2) \mid \gamma, \nu \right) \right\}.
\end{align*}
We consider the transformation \( \tilde{\gamma} = \log \gamma \), and employ a random walk MH algorithm for \( \tilde{\gamma} \), where we include a Jacobian adjustment term \( \tilde{\gamma} \) in the log conditional posterior density.

\subsubsection{Generation of $\nu$}

The conditional posterior density of $\nu$ is given by
\begin{align*}
\pi(\nu \mid \cdot) \propto 
\exp\left\{
(n_{\nu 0} - 1)\log \nu - S_{\nu 0} \nu
+ \sum_{t = 1}^T \log q_T\left( y_t \exp(-h_t / 2) \mid \gamma, \nu \right)
\right\}.
\end{align*}
We consider the transformation \( \tilde{\nu} = \log(\nu - 2) \), and employ a random walk MH algorithm for \( \tilde{\nu} \), where we include a Jacobian adjustment term \( \tilde{\nu} \) in the log conditional posterior density.

\subsubsection{Generation of $\bm{h}$}

First, we define 
\begin{align*}
g(h_t \mid \gamma, \nu) 
&= -\frac{\nu + 1}{2} \log\left\{ 1 +
\frac{\left( \sigma_{*} y_t \exp(-h_t / 2) + \mu_{*} \right)^2}{\nu}
\left[
\gamma^{-2} I\left( y_t \exp(-h_t / 2) \geq -\frac{\mu_{*}}{\sigma_{*}} \right)
\right.\right. \\
& \hspace{6.0cm} \left.\left.
+ \gamma^{2} I\left( y_t \exp(-h_t / 2) < -\frac{\mu_{*}}{\sigma_{*}} \right)
\right]
\right\}.
\end{align*}
\begin{enumerate}[(1)]
    \item For \( t = 1 \), the conditional posterior density of \( h_1 \) is given by
        \begin{align*}
        \pi(h_1 \mid \cdot)
        \propto & \ 
        \exp\left\{ g(h_1 \mid \gamma, \nu) 
        - \frac{1}{2(1 - \rho^2)\sigma_{\eta}^2}\left[h_2 - \mu - \phi(h_1 - \mu) - \rho \sigma_{\eta} \exp(-h_1 / 2) y_1 \right]^2 \right\} \\
        & \times \exp\left\{ -\frac{1}{2}h_1 
        - \frac{1 - \phi^2}{2\sigma_{\eta}^2}(h_1 - \mu)^2 
        - \frac{1}{2\sigma_u^2}(x_1 - \xi - h_1)^2 \right\}.
        \end{align*}
        Define
        \begin{align*}
        \sigma_{h1}^{-2} 
        &= \sigma_{\eta}^{-2}\left(1 - \phi^2 + \frac{\phi^2}{1 - \rho^2}\right) + \sigma_u^{-2}, \\
        \mu_{h1} 
        &= \sigma_{h1}^2 \left\{
        -\frac{1}{2}
        + \frac{\phi\left[ h_2 - (1 - \phi)\mu \right]}{\sigma_{\eta}^2 (1 - \rho^2)}
        + \frac{(1 - \phi^2)\mu}{\sigma_{\eta}^2}
        + \frac{x_1 - \xi}{\sigma_u^2}
        \right\}, \\
        k(h_1) 
        &= \exp\left\{ g(h_1 \mid \gamma, \nu)
        + \frac{\rho y_1 \exp(-h_1 / 2)}{(1 - \rho^2)\sigma_{\eta}}
        \left[ h_2 - \mu - \phi(h_1 - \mu) \right]
        - \frac{\rho^2 y_1^2 \exp(-h_1)}{2(1 - \rho^2)}
        \right\}.
        \end{align*}
        We then propose a candidate \( h_1^{\dagger} \sim \mathcal{N}(\mu_{h1}, \sigma_{h1}^2) \), and accept it with probability \( \min\{1, k(h_1^{\dagger}) / k(h_1)\} \).
    \item For \( t = 2, \ldots, T - 1 \), the conditional posterior density of \( h_t \) is given by
        \begin{align*}
        \pi(h_t \mid \cdot)
        \propto & \ 
        \exp\left\{ g(h_t \mid \gamma, \nu)
        - \frac{1}{2(1 - \rho^2)\sigma_{\eta}^2}\left[h_{t + 1} - \mu - \phi(h_t - \mu) - \rho \sigma_{\eta} \exp(-h_t / 2) y_t \right]^2 \right\} \\
        & \times \exp\left\{ -\frac{1}{2(1 - \rho^2)\sigma_{\eta}^2}\left[h_t - \mu - \phi(h_{t - 1} - \mu) - \rho \sigma_{\eta} \exp(-h_{t - 1} / 2) y_{t - 1} \right]^2 \right\} \\
        & \times \exp\left\{ -\frac{1}{2}h_t - \frac{1}{2\sigma_u^2}(x_t - \xi - h_t)^2 \right\}.
        \end{align*}
        Define
        \begin{align*}
        \sigma_{ht}^{-2} 
        &= \sigma_{\eta}^{-2}\left( \frac{1 + \phi^2}{1 - \rho^2} \right) + \sigma_u^{-2}, \\
        \mu_{ht} 
        &= \sigma_{ht}^2 \left\{
        -\frac{1}{2}
        + \frac{\phi\left[ h_{t + 1} - (1 - \phi)\mu \right]}{\sigma_{\eta}^2 (1 - \rho^2)}
        + \frac{(1 - \phi)\mu + \phi h_{t - 1}}{\sigma_{\eta}^2 (1 - \rho^2)}
        + \frac{\rho y_{t - 1} \exp(-h_{t - 1} / 2)}{(1 - \rho^2)\sigma_{\eta}}
        + \frac{x_t - \xi}{\sigma_u^2}
        \right\}, \\
        k(h_t) 
        &= \exp\left\{ g(h_t \mid \gamma, \nu)
        - \frac{\rho^2 y_t^2 \exp(-h_t)}{2(1 - \rho^2)}
        + \frac{\rho y_t \exp(-h_t / 2)}{(1 - \rho^2)\sigma_{\eta}}
        \left[ h_{t + 1} - \mu - \phi(h_t - \mu) \right]
        \right\}.
        \end{align*}
        We then propose a candidate \( h_t^{\dagger} \sim N(\mu_{ht}, \sigma_{ht}^2) \), and accept it with probability \( \min\{1, k(h_t^{\dagger}) / k(h_t)\} \).
    \item For \( t = T \), the conditional posterior density of \( h_T \) is given by
        \begin{align*}
        \pi(h_T \mid \cdot)
        \propto & \ 
        \exp\left\{ g(h_T \mid \gamma, \nu)
        - \frac{1}{2(1 - \rho^2)\sigma_{\eta}^2}\left[h_T - \mu - \phi(h_{T - 1} - \mu) - \rho \sigma_{\eta} \exp(-h_{T - 1} / 2) y_{T - 1} \right]^2 \right\} \\
        & \times \exp\left\{ -\frac{1}{2}h_T - \frac{1}{2\sigma_u^2}(x_T - \xi - h_T)^2 \right\}.
        \end{align*}
        Define
        \begin{align*}
        \sigma_{hT}^{-2} 
        &= \sigma_{\eta}^{-2}\left( \frac{1}{1 - \rho^2} \right) + \sigma_u^{-2}, \\
        \mu_{hT} 
        &= \sigma_{hT}^2 \left[
        -\frac{1}{2}
        + \frac{(1 - \phi)\mu + \phi h_{T - 1}}{\sigma_{\eta}^2 (1 - \rho^2)}
        + \frac{\rho y_{T - 1} \exp(-h_{T - 1} / 2)}{(1 - \rho^2)\sigma_{\eta}}
        + \frac{x_T - \xi}{\sigma_u^2}
        \right], \\
        k(h_T) 
        &= \exp\left\{ g(h_T \mid \gamma, \nu) \right\}.
        \end{align*}
        We then propose a candidate \( h_T^{\dagger} \sim N(\mu_{hT}, \sigma_{hT}^2) \), and accept it with probability \( \min\{1, k(h_T^{\dagger}) / k(h_T)\} \).
\end{enumerate}

\clearpage
\section{Conditional Giacomini--White Test}
\label{sec:suppl-gw}

The main text reports the unconditional test of \citet{giacomini_tests_2006} for the seven matched pairs of RSV and SV models.
The pairwise heatmaps in this supplement report, in addition, the conditional version of the test and the associated win proportions defined here.

For the \textit{conditional} predictive ability, the GW test defines the null hypothesis as
\begin{align}
H_{0,h}: \mathrm{E}[\Delta L_{t+1} \mid \mathcal{I}_{t}] = 0, \quad t = n, n+1, \ldots, n+n_{f}-1,
\end{align}
where $\Delta L_{t}$ is the loss difference between two models at forecast date $t$, and $\mathcal{I}_{t}$ is the information set at time $t$.

Given a test function $\mathbbm{h}_{t}$, a $q \times 1$ vector measurable with respect to $\mathcal{I}_{t}$ \citep{stinchcombe_consistent_1998}, the null hypothesis translates to
\begin{align}
H_{0,h}: \mathrm{E}[\mathbbm{h}_{t} \Delta L_{t+1}] = 0, \quad t = n, n+1, \ldots, n+n_{f}-1.
\end{align}
Defining $Z_{t+1} = \mathbbm{h}_{t} \Delta L_{t+1}$ and $\bar{Z}_{n_{f}} = n_{f}^{-1} \sum_{t=n}^{n+n_{f}-1} Z_{t+1}$, the alternative hypothesis becomes
\begin{align}
H_{A,h}: \mathrm{E}[\bar{Z}_{n_{f}}'] \mathrm{E}[\bar{Z}_{n_{f}}] \geq d > 0, \quad \mbox{for sufficiently large $n_{f}$}.
\end{align}
We employ the Wald-type test statistic
\begin{align}
T_{n_{f}} = n_{f} \bar{Z}_{n_{f}}' \hat{\Omega}_{n_{f}}^{-1} \bar{Z}_{n_{f}},
\label{eq:gw-Tstat-cond}
\end{align}
where $\hat{\Omega}_{n_{f}} = n_{f}^{-1} \sum_{t=n}^{n+n_{f}-1} Z_{t+1} Z_{t+1}'$ consistently estimates the variance of $Z_{t+1}$.
Under the null, $T_{n_{f}}$ follows a chi-square distribution with $q$ degrees of freedom, $\chi_{q}^{2}$.
At the significance level $p$, the null hypothesis is rejected if $T_{n_{f}} > \chi_{q,1-p}^{2}$, where $\chi_{q,1-p}^{2}$ is the $(1-p)$ quantile of $\chi_{q}^{2}$.

In practice, the selection of $\mathbbm{h}_{t}$ should distinguish the forecasting performance of the models.
Focusing on one-day-ahead forecasts, we follow the specifications of \cite{takahashi_stochastic_2023} and define $\mathbbm{h}_{t} = (1, \Delta L_{t})'$ for $t = n, n+1, \ldots, n+n_{f}-1$.
If the null is rejected, it implies that the lagged loss differences, $\Delta L_{t}$, can help predict the subsequent loss differences, $\Delta L_{t+1}$.
Using $\hat{b}$ to represent the regression coefficient of $\Delta L_{t+1}$ on $\mathbbm{h}_{t}$, the predicted loss differences $\{\hat{b}' \mathbbm{h}_{t}\}_{t=n}^{n+n_{f}-1}$ serve as indicators for assessing model performances across different time points.

In the pairwise heatmaps of Sections~\ref{sec:suppl-vol}, \ref{sec:suppl-vares}, \ref{sec:covid-djia}, and \ref{sec:covid-n225}, we measure relative performance by the frequency with which one model forecasts larger losses than the other, as expressed by
\begin{align}
I_{n_{f},0} = \frac{1}{n_{f}} \sum_{t=n}^{n+n_{f}-1} I(\hat{b}' \mathbbm{h}_{t} > 0).
\label{eq:gw-prop}
\end{align}
In the heatmaps, the lower triangular part reports the unconditional test statistics and the upper triangular part the win proportions $I_{n_f,0}$; lower values (green) favour the row model and larger values (orange) favour the column model, and double and single asterisks denote significance at the $1\%$ and $5\%$ levels of the unconditional (lower) and conditional (upper) tests.

\clearpage
\section{Simulation Study}

This section reports a simulation study designed to assess the performance of the Bayesian MCMC estimation procedure for the proposed RSV models.
The simulation focuses on parameter recovery and convergence properties under controlled settings.

The sample size in the simulation study is set to $T = 2{,}000$.
For each model, data are generated from the corresponding data-generating process, and the parameters are estimated using the same MCMC specification as in the empirical analysis.
The estimation is based on 50{,}000 MCMC draws after discarding the first 10{,}000 as burn-in.

Tables~\ref{tab:sim-study-1} and \ref{tab:sim-study-2} report the true parameter values, posterior means, posterior standard deviations, and 95\% credible intervals.
The table also reports the $p$-values of the \citet{geweke_evaluating_1992}'s convergence diagnostic test and the inefficiency factors to assess MCMC convergence and sampling efficiency.

Overall, the results indicate that the proposed estimation procedure accurately recovers the true parameters across all models, with credible intervals covering the true values and no evidence of convergence problems.
These findings support the reliability of the Bayesian MCMC approach employed in the empirical analysis.

\clearpage
\section{MCMC Diagnostics}
\label{sec:mcmc-diagnostics}

This section reports convergence diagnostics and prior--posterior comparisons for the Bayesian estimation of the RSV-AZ-ST model in the empirical analysis.
To assess the stability of the Markov chain Monte Carlo (MCMC) procedure, Table~\ref{tab:mcmc-rsvaz-rep} presents results for three representative rolling estimation windows
applied to the DJIA.

For each window, the table reports posterior means, standard deviations, and 95\% credible intervals, together with the $p$-values of the convergence diagnostic proposed by \citet{geweke_evaluating_1992} (CD) and inefficiency factors (IF).
The Geweke statistic does not reject convergence at the $5\%$ level for most parameters; the exceptions are $\nu$ in the first two windows, $\xi$ in the second, and $\rho$ and $\delta$ in the first, and the inefficiency factors of $\nu$, $\rho$, and $\delta$ lie between $77$ and $218$ against at most $76$ for the other parameters.
These are the single-chain diagnostics of the runs reported in the article, with $15{,}000$ retained draws: the mixing of $\nu$, $\rho$, and $\delta$ is slow, while the posterior summaries are stable across the three windows, and Section~\ref{sec:suppl-az-recomputation} reports the multi-chain checks applied to the windows in which such diagnostics are worst.

The table also facilitates a comparison between prior moments and posterior summaries.
While relatively diffuse priors are imposed on the common parameters, the posterior distributions are substantially more concentrated and often shifted away from the prior means, suggesting that posterior inference is driven primarily by the information contained in the data rather than by prior assumptions.
For variance-related parameters, priors are specified on $\sigma_\eta^2$ and $\sigma_u^2$, and the reported prior moments for $\sigma_\eta$ and $\sigma_u$ are computed from the implied distributions.

Overall, the posterior of the RSV-AZ-ST model is well identified across estimation windows and prior dominance is not a concern in the empirical analysis; the slow mixing of the shape and leverage parameters is the reason for the window-by-window checks of Sections~\ref{sec:checks} and \ref{sec:suppl-az-recomputation}.
Similar convergence diagnostics and prior--posterior patterns were observed for other model specifications and estimation windows considered in the empirical analysis.


\subsection{Checks applied to the rolling-window estimates}
\label{sec:checks}

Three checks were applied to the rolling-window estimates before their forecasts entered the evaluation.
First, completeness: one estimate for every forecast date.
Second, for the Student's $t$ and skew-$t$ specifications, whether the posterior mean of $\nu$ sits at the lower bound of its support ($\nu < 4.01$), where the standardization of the distribution breaks down.
Third, whether the log of the one-day-ahead variance forecast deviates by more than $1.0$ from the median of the forecasts of the other specifications of the same class (SV or RSV) on the same date.
A window flagged by the second or the third check was re-estimated from a different random seed under the same prior and the same initial values, and the re-estimate was accepted if it passed all three checks.
A window flagged only by the third check, whose parameters also jumped relative to the adjacent windows, was re-estimated from two further seeds and replaced only if both re-estimates agreed with the adjacent windows; agreement of this kind is what we take as an indication that the original chain had failed to mix, although it does not prove it.

These rules were applied mechanically to every window of the COVID-19 samples of both indices and to every window of the SV-GH-ST model in all four samples, which were re-estimated for this revision; in the final set no window fails any of the three checks.
The second check was binding for the SV-GH-ST model, whose chain fell into a degenerate mode with $\nu$ at the bound and $\mu$ far above the data in nine of roughly forty cold starts and, on three occasions, from the middle of a rolling sequence; in every such case a different seed produced a chain that passed.
Three windows, one in each of the DJIA main sample, the N225 main sample, and the N225 COVID-19 sample, showed a one-window jump in the parameters and in the predicted variance, were re-estimated from two seeds, and were replaced on that basis because both re-estimates agreed with the adjacent windows.

\subsection{Multi-chain confirmation of flagged windows}
\label{sec:suppl-az-recomputation}

The forecasts of the four Azzalini models (SV-AZ-SN, SV-AZ-ST, RSV-AZ-SN, and RSV-AZ-ST) in the 20 series of the article (the four models for the two indices over the main sample and over the 2012--2020 sample, and the free-$\psi$ variants of the two RSV models over the 2012--2020 sample), $7{,}742$ rolling windows in all, were generated from single chains with $15{,}000$ retained draws per window and recorded seeds.
Because the shape and leverage parameters of these models mix slowly (Table~\ref{tab:mcmc-rsvaz-rep}), every window was checked with pre-specified diagnostics and the flagged windows were re-estimated with several chains, as follows.

\paragraph{Diagnostics and acceptance criteria.}
Two diagnostics were applied to every window with the same thresholds for every series: an inefficiency factor of the last log-volatility $h_n$ above $300$, or a ratio of the mean of $\exp(h_{n+1})$ over the first $3{,}000$ retained draws to the mean over the remaining $12{,}000$ outside $[0.5, 2]$.
The 20 windows meeting either criterion, in nine series, together with the 10 windows of SV-AZ-ST over the DJIA main sample that had been flagged by the same diagnostics in a preliminary run, were re-estimated with three independent chains started from the original initial values, from the posterior means of the preceding window, and from dispersed values, each with a burn-in of $15{,}000$ and $15{,}000$ retained draws and with seeds fixed in advance.
A window was accepted when every chain met the single-chain criteria (inefficiency factor of $h_n$ below $300$, Geweke $p$-value above $0.01$, block ratio within $[0.5, 2]$), the posterior means of $h_n$ across chains differed by less than half a posterior standard deviation, the rank-normalized split-$\hat{R}$ of $h_n$, of the parameters, and of the one-step-ahead draws of $\exp(h_{n+1})$ and $y_{n+1}$ was below $1.01$ with bulk and tail effective sample sizes of at least $400$ \citep{vehtari_rank-normalization_2021}, and the Monte Carlo standard errors (batch means, ten batches per chain) of the predictive mean of volatility and of the VaR and ES forecasts were below $0.5\%$ and $1\%$ of their values.
Windows that did not meet the criteria after this first stage were re-run from the same seeds and initial values with the number of retained draws extended, by extrapolation from the shortfall, to between $45{,}000$ and $406{,}000$ per chain, and were judged on the extended chains alone.
These criteria were fixed before the confirmation runs were examined; no window was selected, re-seeded, or discarded on the basis of its loss or of its proximity to other models.

\paragraph{Results.}
Table~\ref{tab:az-confirm} lists the 30 windows with the stage, the number of retained draws per chain, the outcome, and the worst diagnostic values.
Seven windows meet all criteria, among them the two free-$\psi$ RSV windows of the DJIA 2012--2020 sample; for them the draws of the three chains were pooled and the forecasts recomputed from the pooled draws, which replace the single-chain values in every table.
Eighteen windows do not meet the criteria after the extension.
In seven of them, all SV-AZ-ST, the degrees-of-freedom parameter $\nu$ still mixes slowly, with $\hat{R}$ between $1.010$ and $1.026$ or a bulk effective sample size between $137$ and $344$ after up to $194{,}000$ draws per chain (in one of them the Monte Carlo standard error of the predictive mean is $9.1\%$).
In the other eleven, $\hat{R}$ and the effective sample sizes pass but the Monte Carlo standard error of the predictive mean of volatility remains between $0.5\%$ and $3.2\%$ of its value, because the mean of $\exp(h_{n+1})$ is dominated by the right tail of the draws; reaching $0.5\%$ would require up to about forty times as many draws.
For five further windows, all SV specifications over the DJIA main sample, the third chain, started from dispersed values, had not completed its extended run after four days and was stopped; these windows are treated like the ones that do not meet the criteria.
For the windows that are not accepted we keep the single-chain values: they are $23$ of the $7{,}742$ windows ($0.3\%$), all of them SV specifications, and the Monte Carlo error they carry is confined to the SV-AZ rows of the tables.
As a sensitivity check, the forecast dates of these windows ($11$ for the DJIA and $8$ for the N225 over the main samples and $3$ for the N225 over the 2012--2020 sample) were removed for all models and every table was recomputed.
The best model of every table and the sign of every matched-pair comparison are unchanged; the $75\%$ MCS of the N225 main sample under RV5 then contains four RSV specifications instead of one, because three further specifications come within $0.003$ of RSV-FS-SN.

\clearpage
\section{Data: Main Sample}
\label{sec:suppl-data}

\begin{figure}[tbp]
\centering
\begin{tabular}{cc}
DJIA & N225 \\
\includegraphics[width=.45\textwidth]{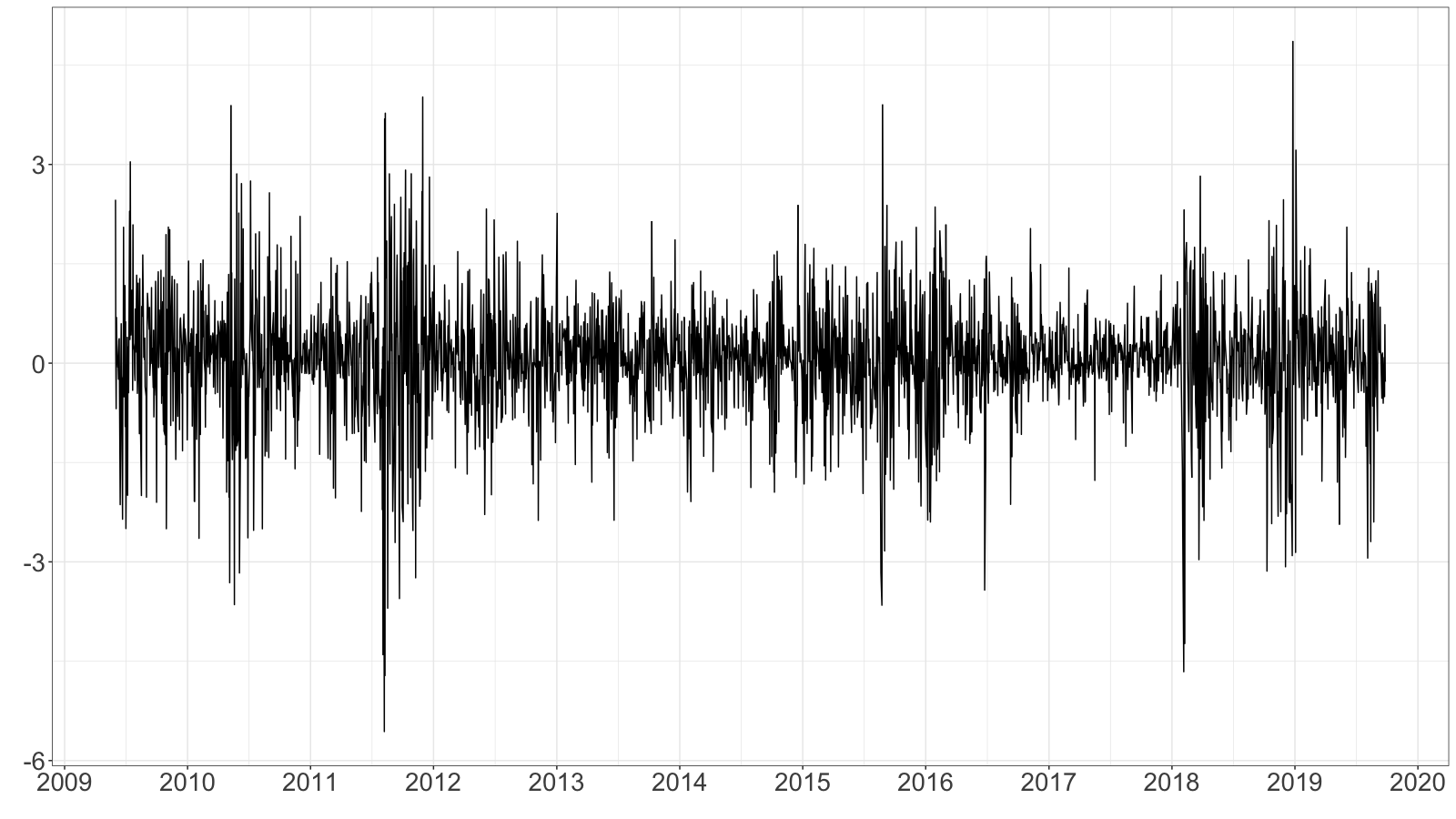} & \includegraphics[width=.45\textwidth]{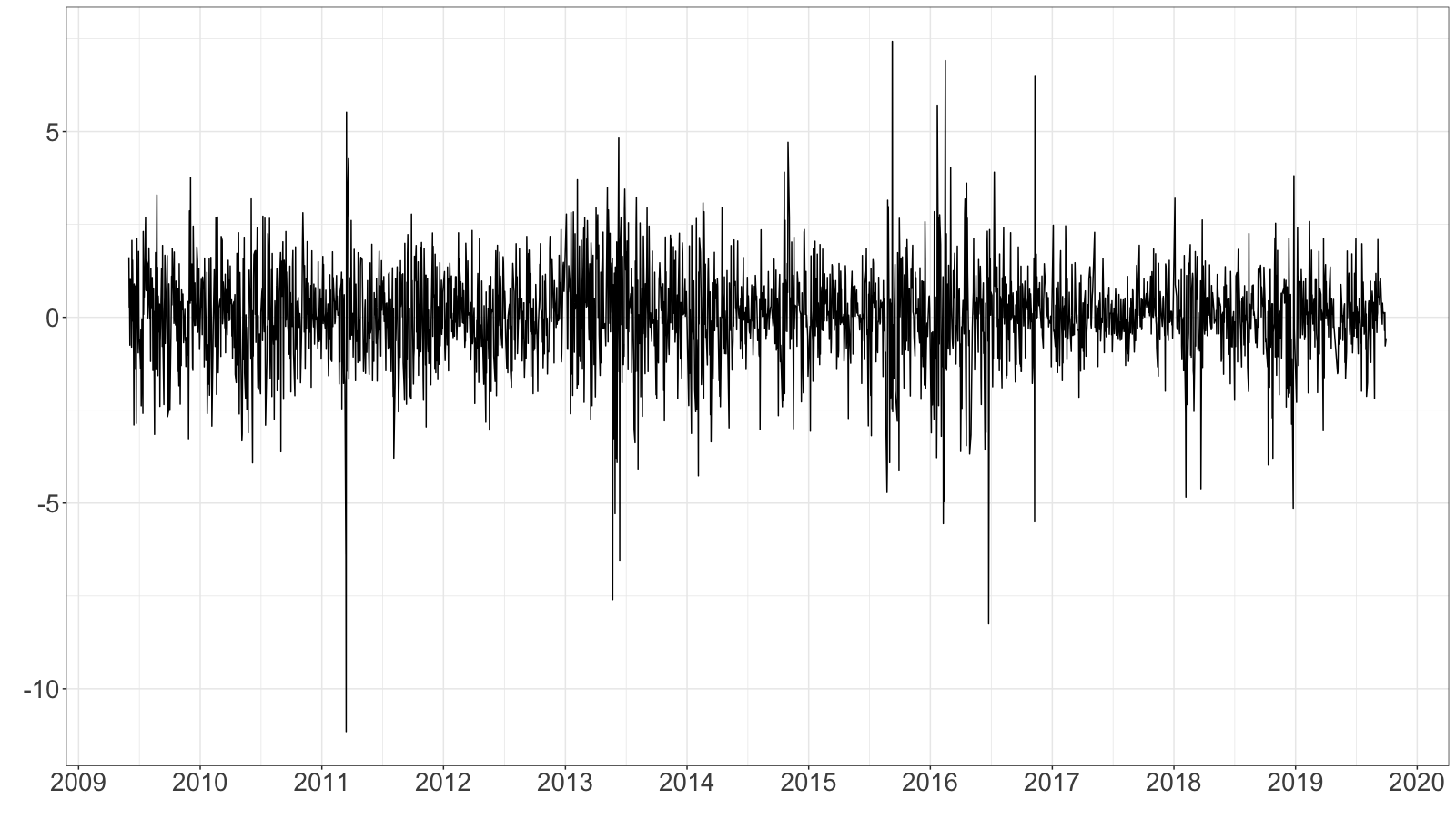} \\
\includegraphics[width=.45\textwidth]{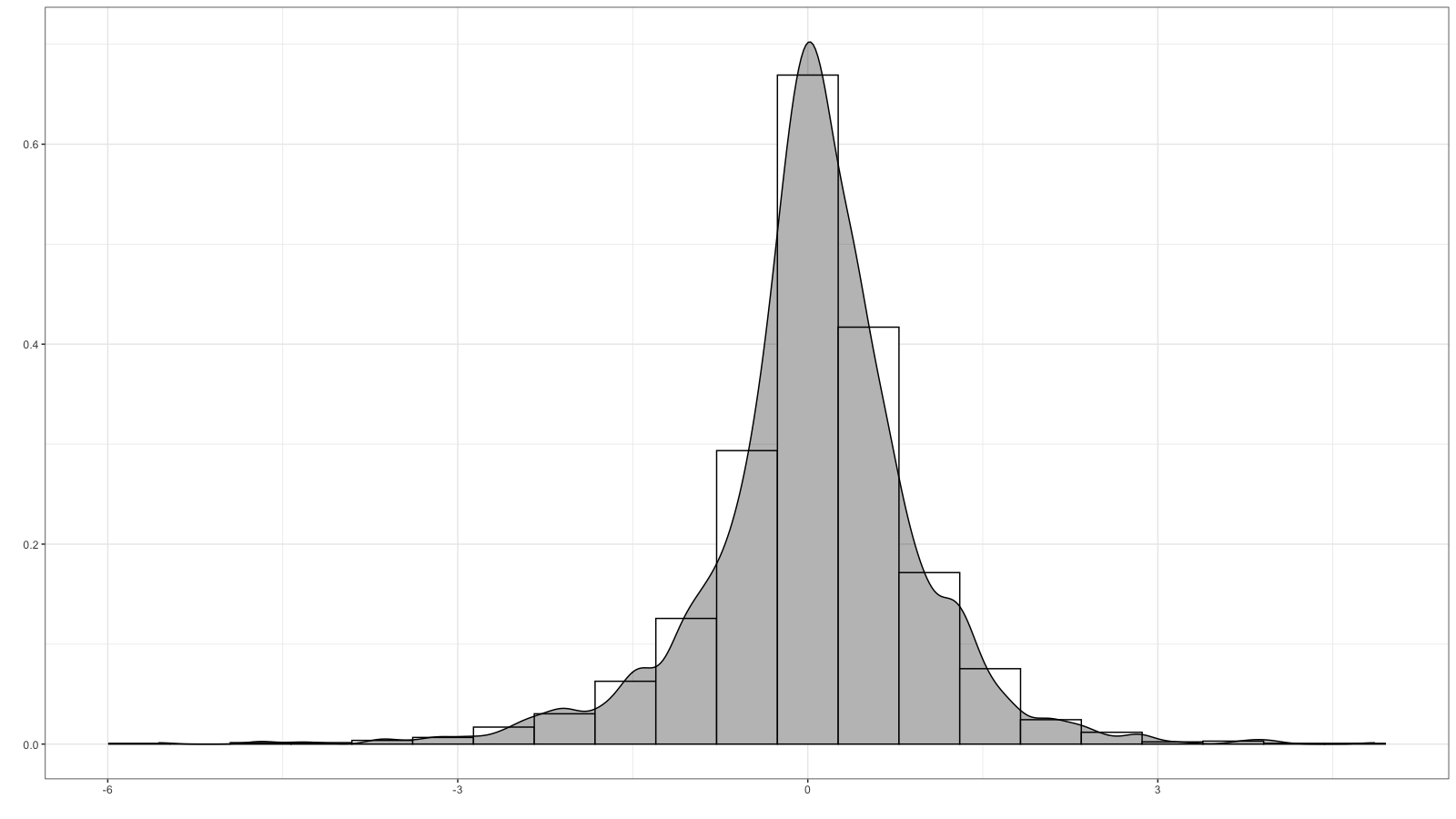} & \includegraphics[width=.45\textwidth]{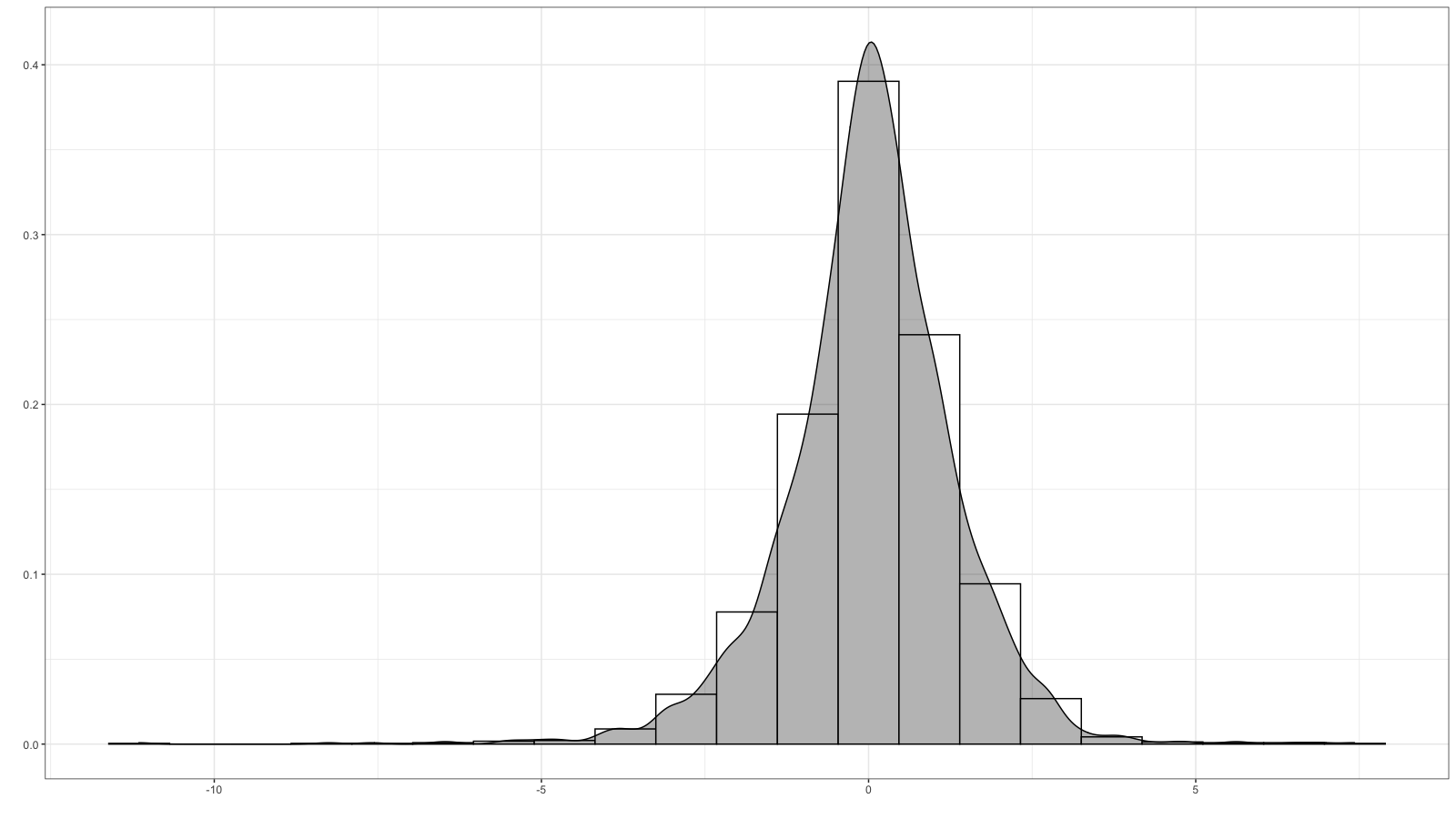}
\end{tabular}
\caption{Time series plots (top) and histograms (bottom) of daily returns (in percentage points) for the DJIA and N225.}
\label{fig:return}
\end{figure}

\begin{figure}[tbp]
\centering
\begin{tabular}{cc}
DJIA & N225 \\
\includegraphics[width=.45\textwidth]{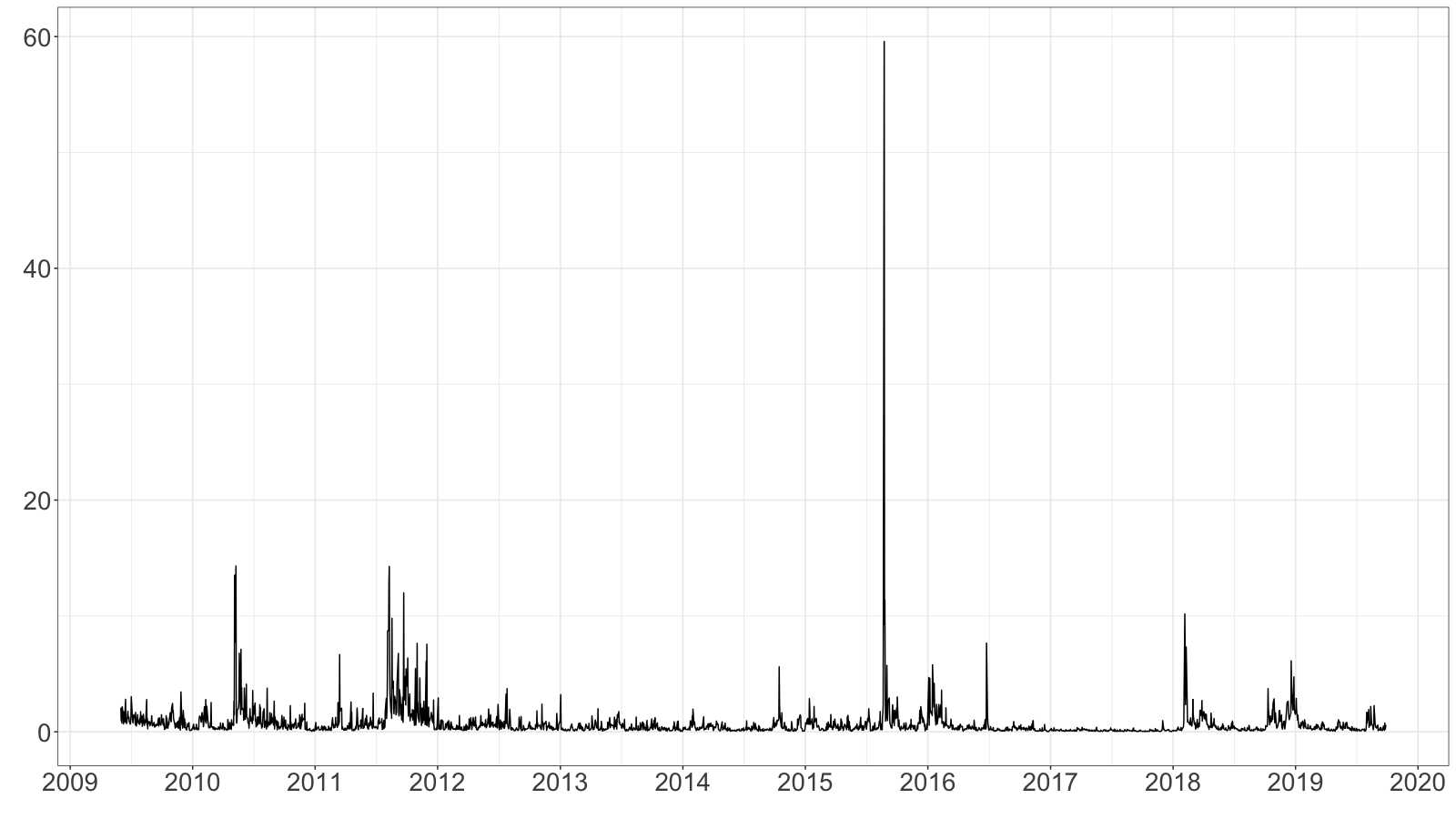} & \includegraphics[width=.45\textwidth]{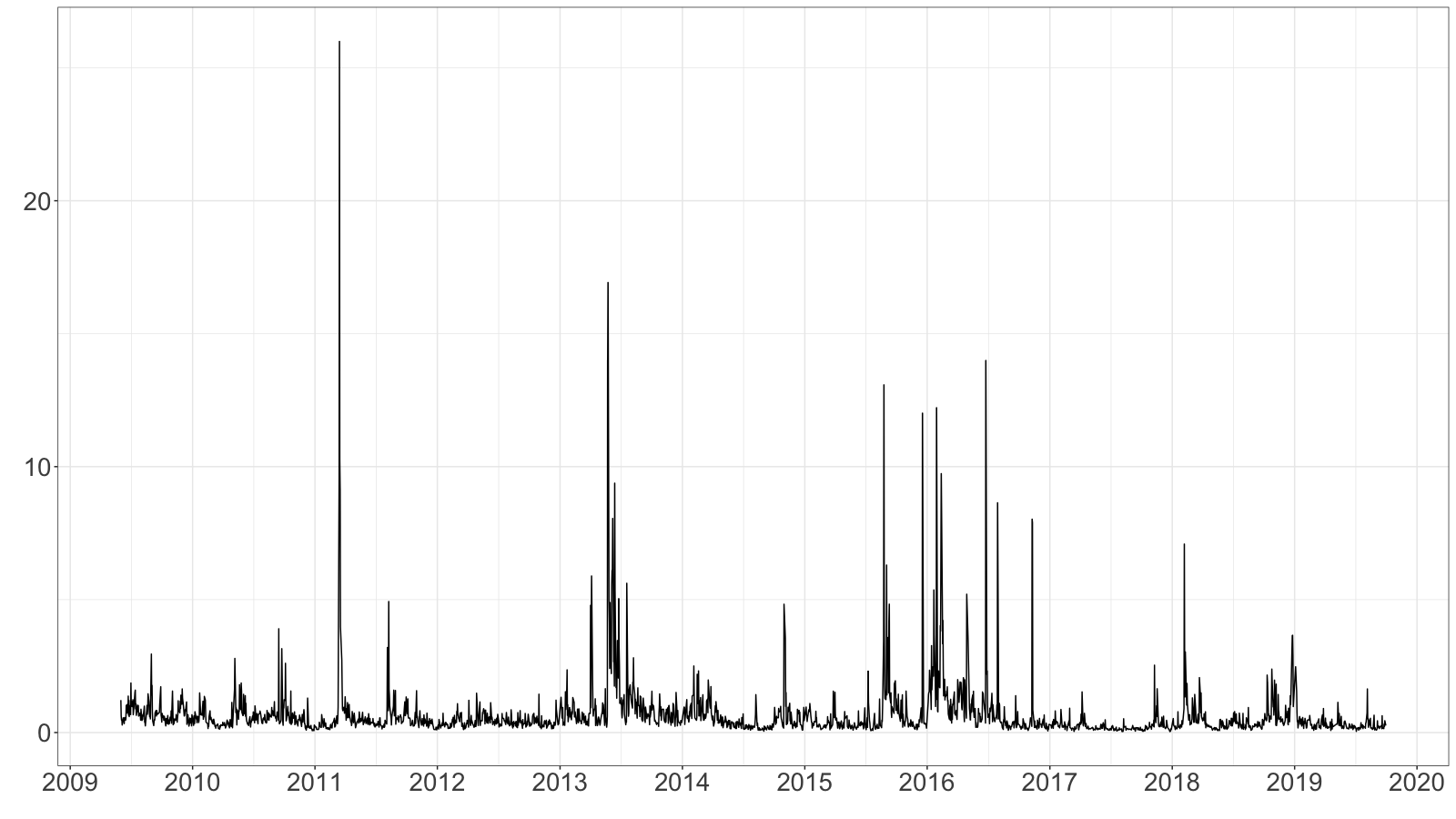} \\
\includegraphics[width=.45\textwidth]{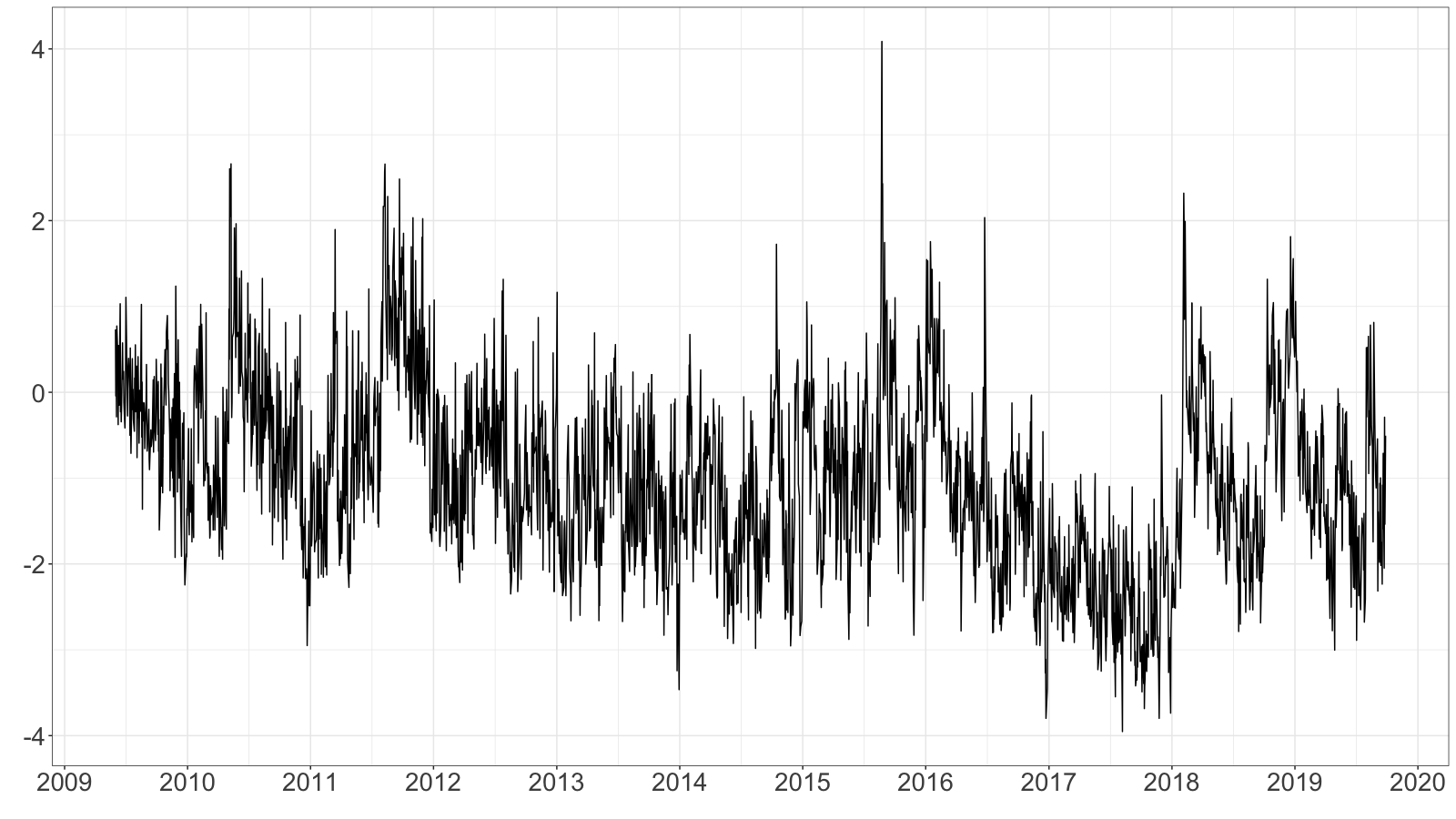} & \includegraphics[width=.45\textwidth]{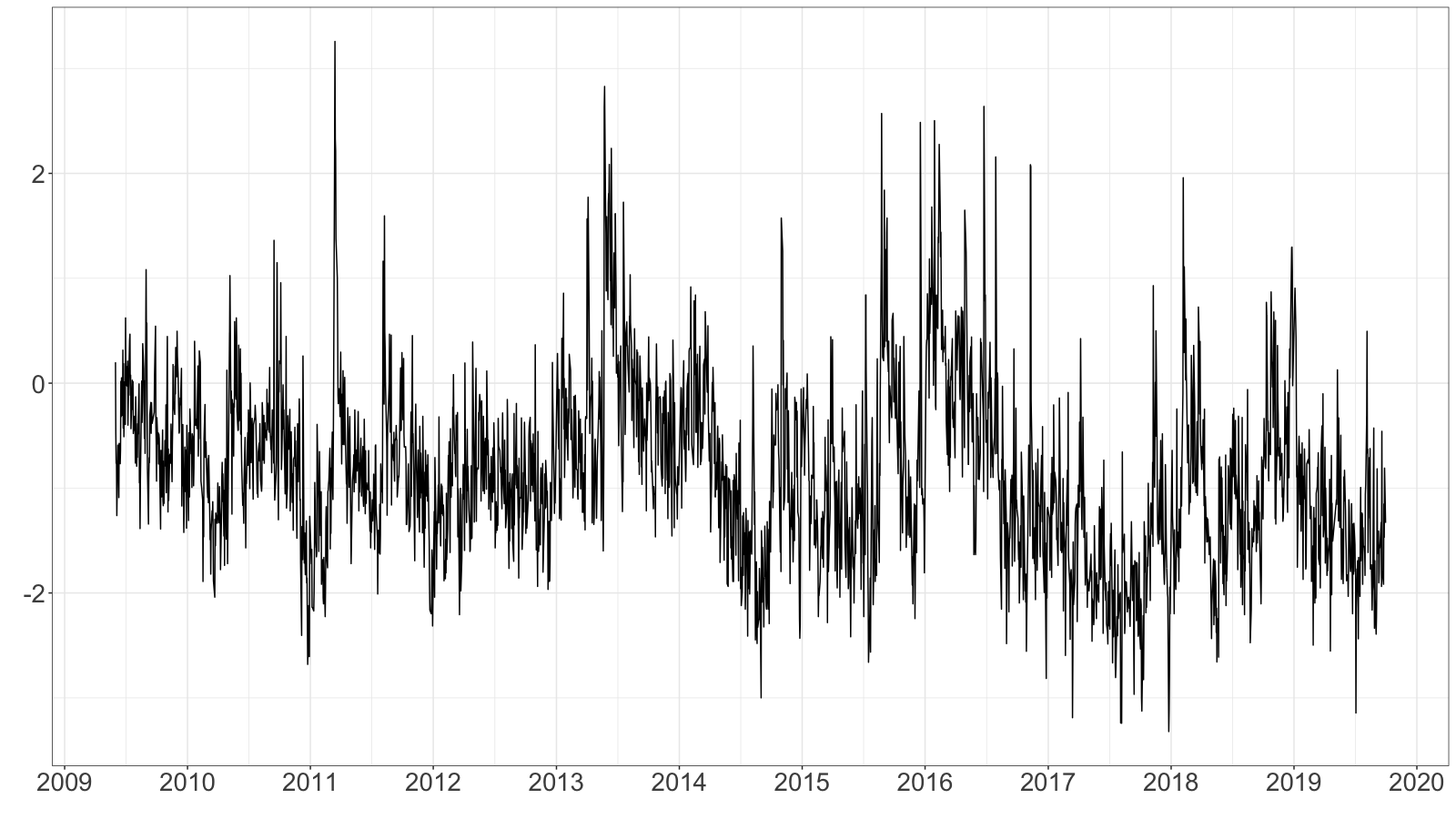} \\
\includegraphics[width=.45\textwidth]{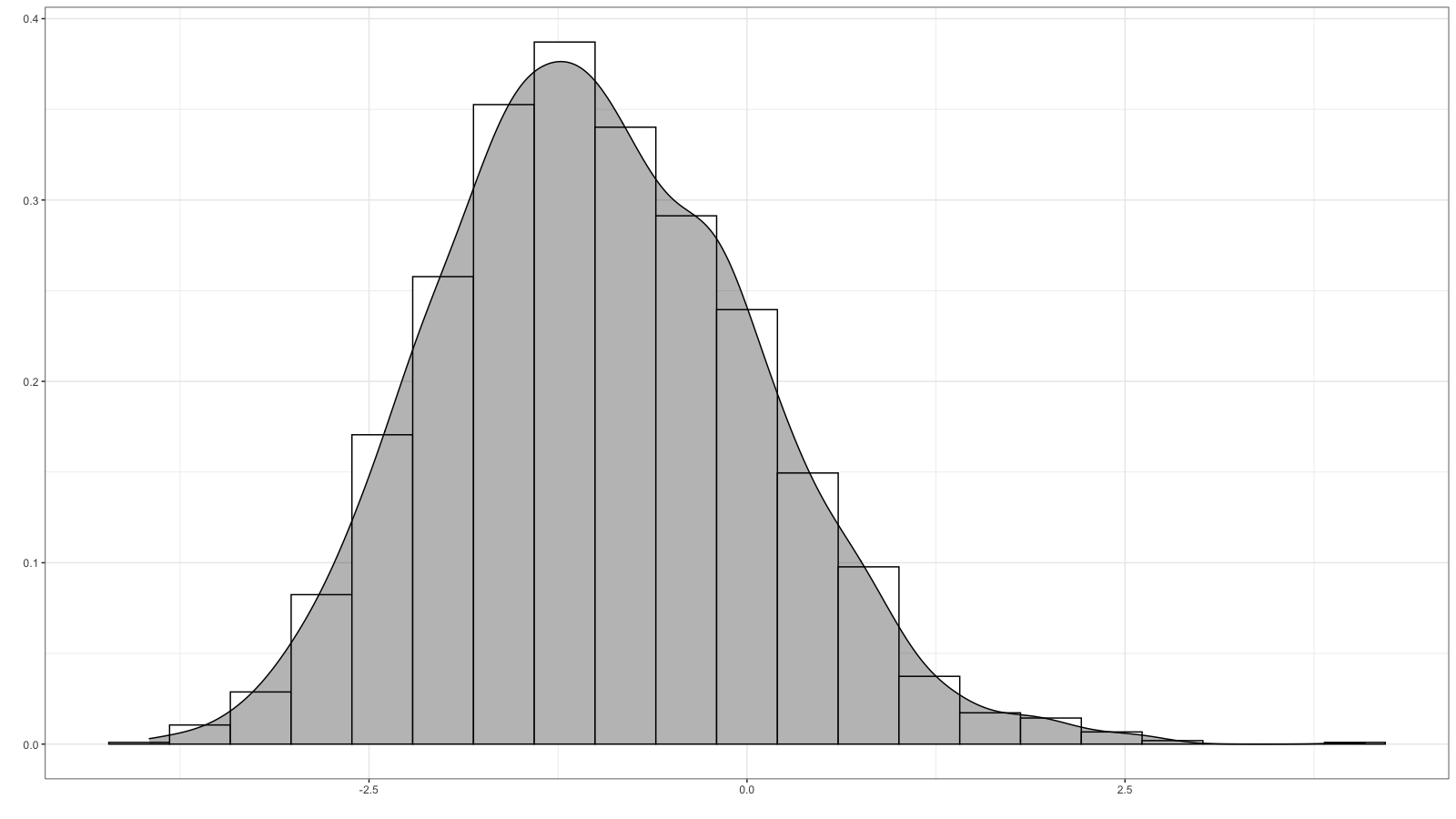} & \includegraphics[width=.45\textwidth]{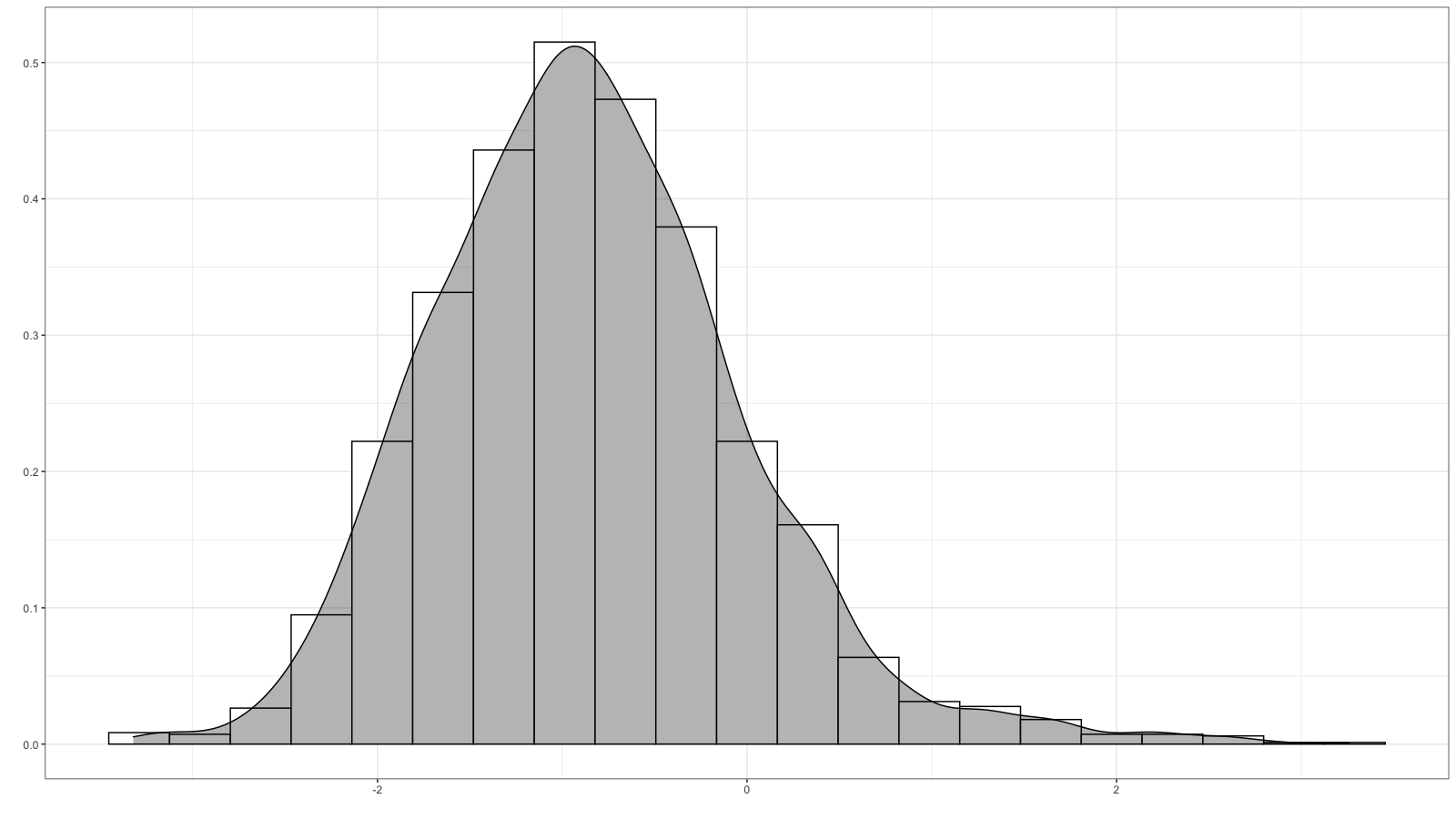}
\end{tabular}
\caption{Time series plots of 5-minute RVs (top), their logarithms (middle), and corresponding histograms (bottom) for the DJIA and N225.}
\label{fig:rv}
\end{figure}

Figures~\ref{fig:return} and \ref{fig:rv} plot the daily returns and the 5-minute RVs of the DJIA and the N225 over the main sample (June 2009 to September 2019), together with their histograms; the descriptive statistics are reported in Table~1 of the main text.
Both return series show the volatility clustering and the negative skewness that the return distributions of Section~2 of the main text are designed to capture, and the log-RV series are strongly persistent and closer to normality than the RVs themselves.

\section{Volatility Forecasts: Main Sample}
\label{sec:suppl-vol}

\begin{figure}[tbp]
\centering
\includegraphics[width = .9\textwidth]{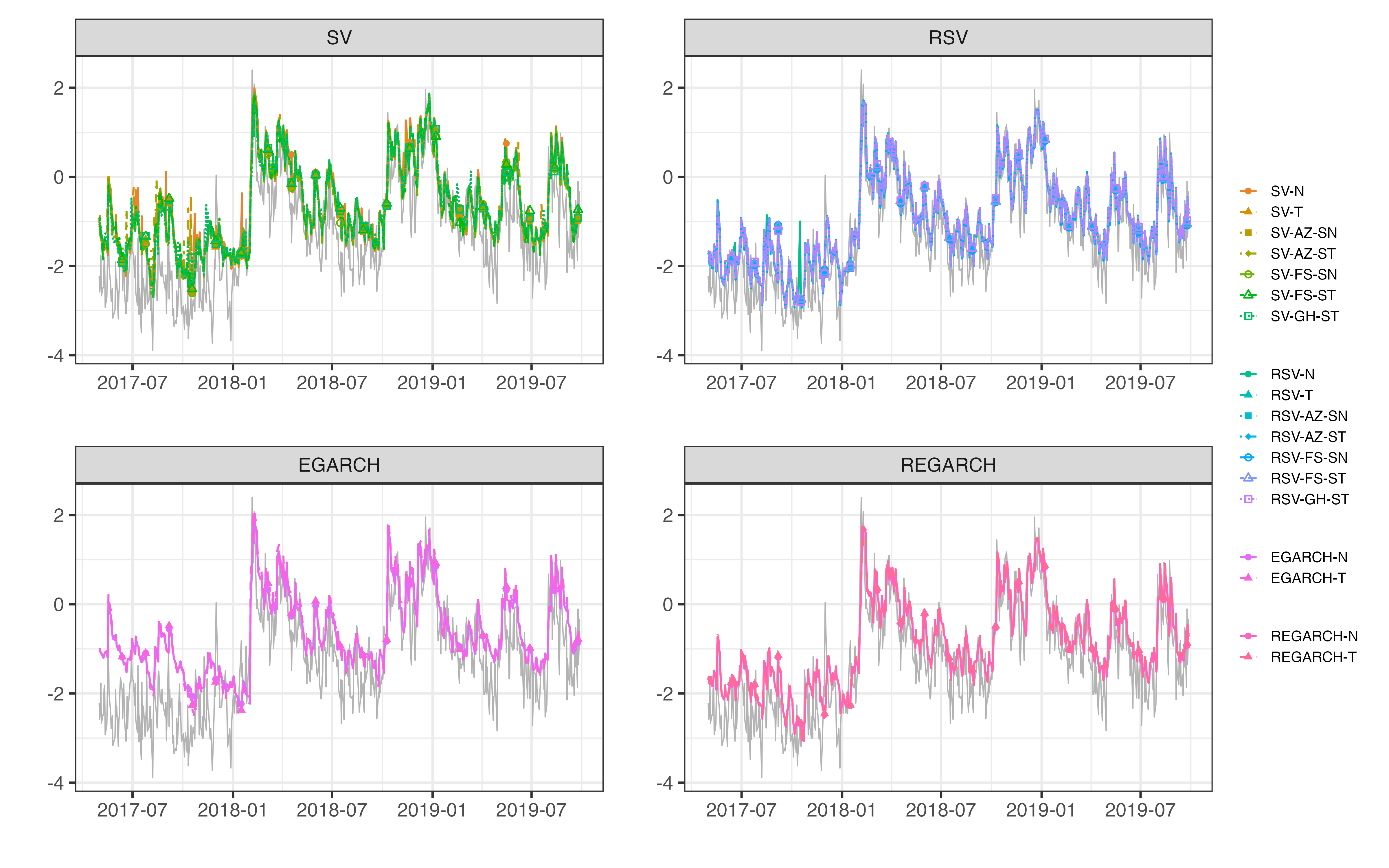}
\caption{Log-scale volatility forecasts together with the adjusted RV5 for the DJIA.}
\label{fig:pred-vol-djia}
\end{figure}

\begin{figure}[tbp]
\centering
\includegraphics[width = .9\textwidth]{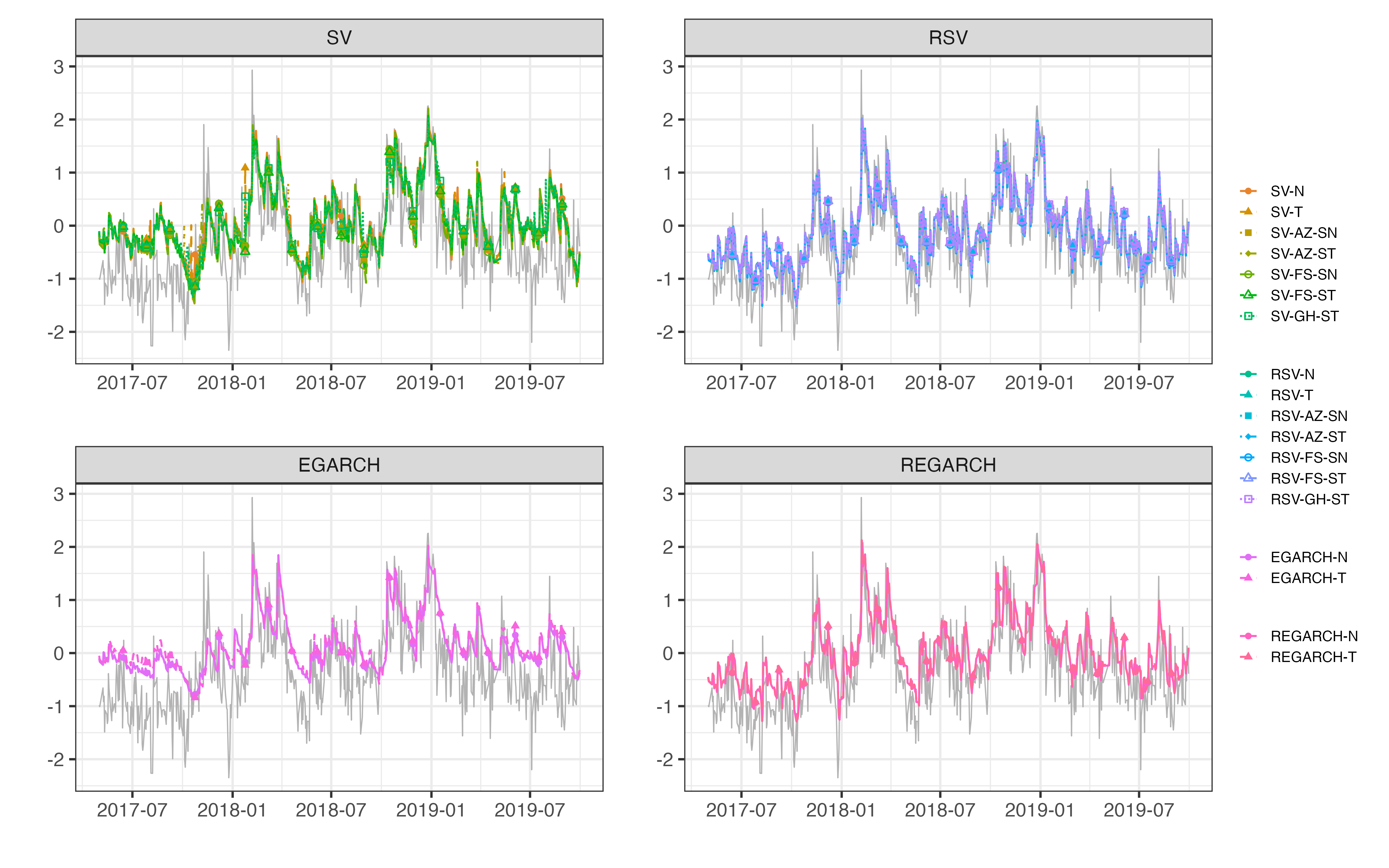}
\caption{Log-scale volatility forecasts together with the adjusted RV5 for the N225.}
\label{fig:pred-vol-n225}
\end{figure}

This section collects the figures on the volatility forecasts of the main sample that complement Tables~2 and 3 of the main text.
Figures~\ref{fig:pred-vol-djia} and \ref{fig:pred-vol-n225} display the log-scale volatility forecasts alongside the adjusted RV5 series.
Forecast patterns are more homogeneous within model classes (SV, RSV, EGARCH, REGARCH) than across return distributions, and the RSV and REGARCH forecasts follow the level of the realized measure more closely than the SV and EGARCH forecasts.

\begin{figure}[tbp]
\centering
\includegraphics[width = \textwidth]{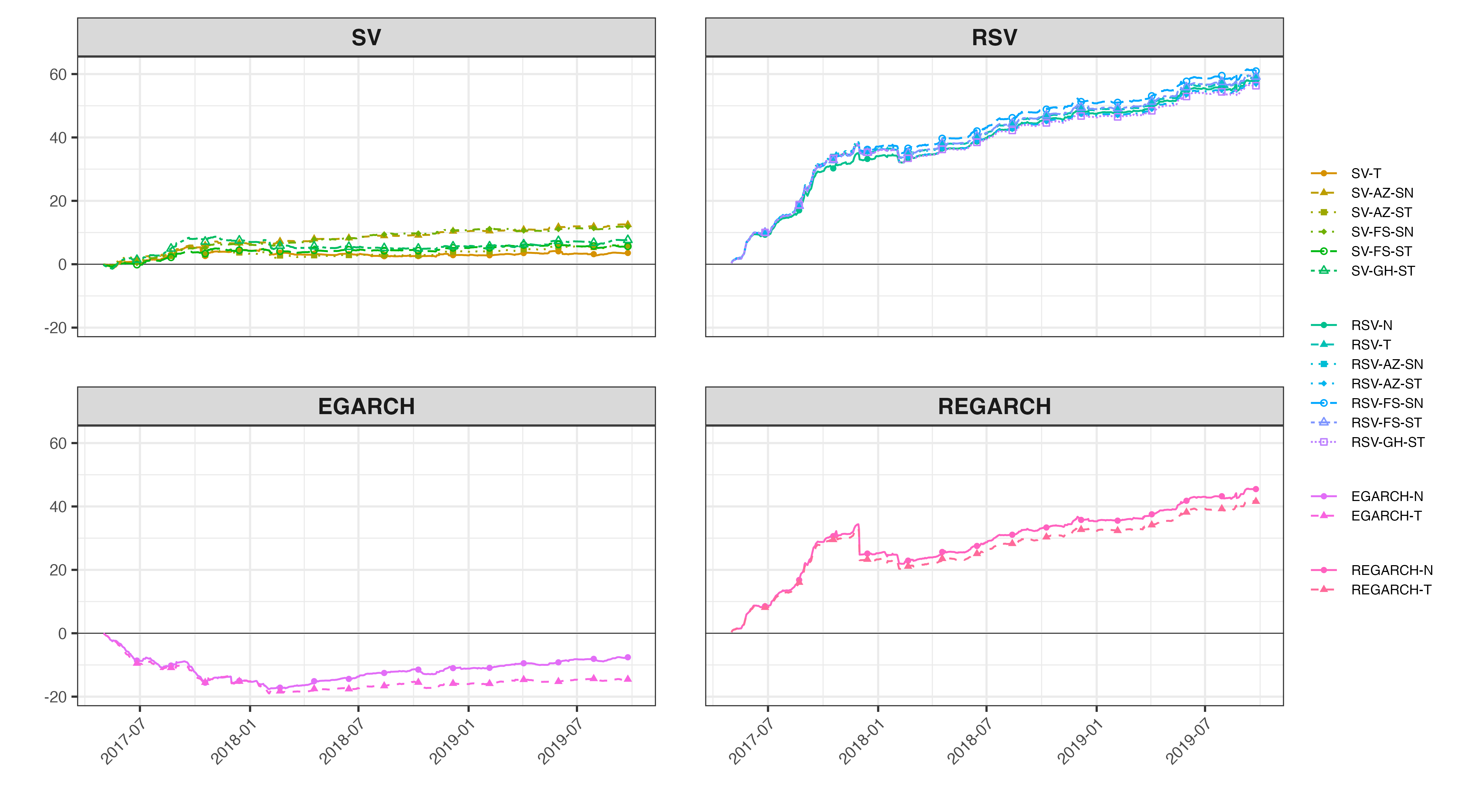}
\caption{Cumulative QLIKE loss differences relative to the SV-N model, based on RV5 as the volatility proxy, for the DJIA. Positive values favour the model over the benchmark.}
\label{fig:pred-vol-cld-djia}
\end{figure}

\begin{figure}[tbp]
\centering
\includegraphics[width = \textwidth]{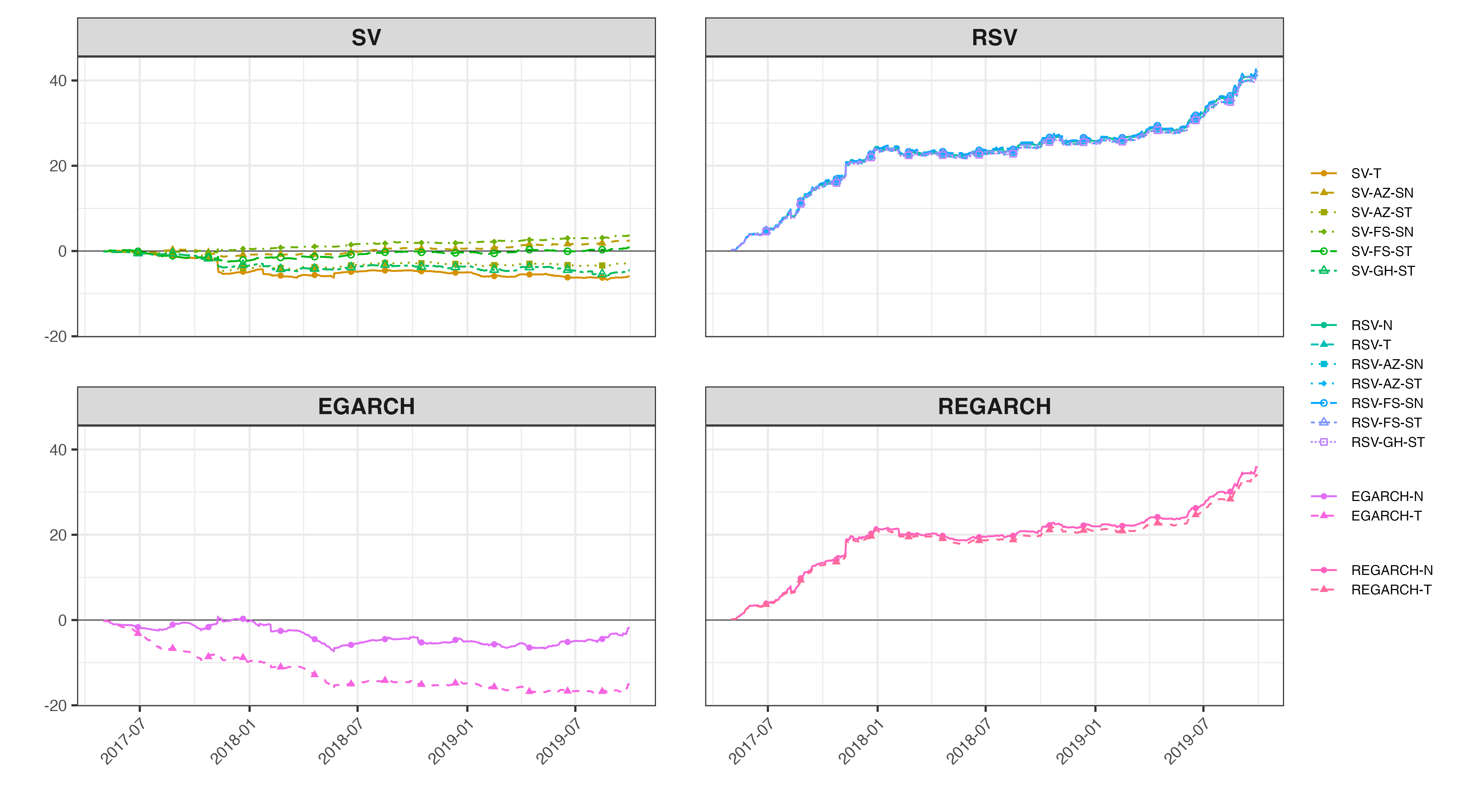}
\caption{Cumulative QLIKE loss differences relative to the SV-N model, based on RV5 as the volatility proxy, for the N225. Positive values favour the model over the benchmark.}
\label{fig:pred-vol-cld-n225}
\end{figure}

\begin{figure}[tbp]
\centering
\includegraphics[width = \textwidth]{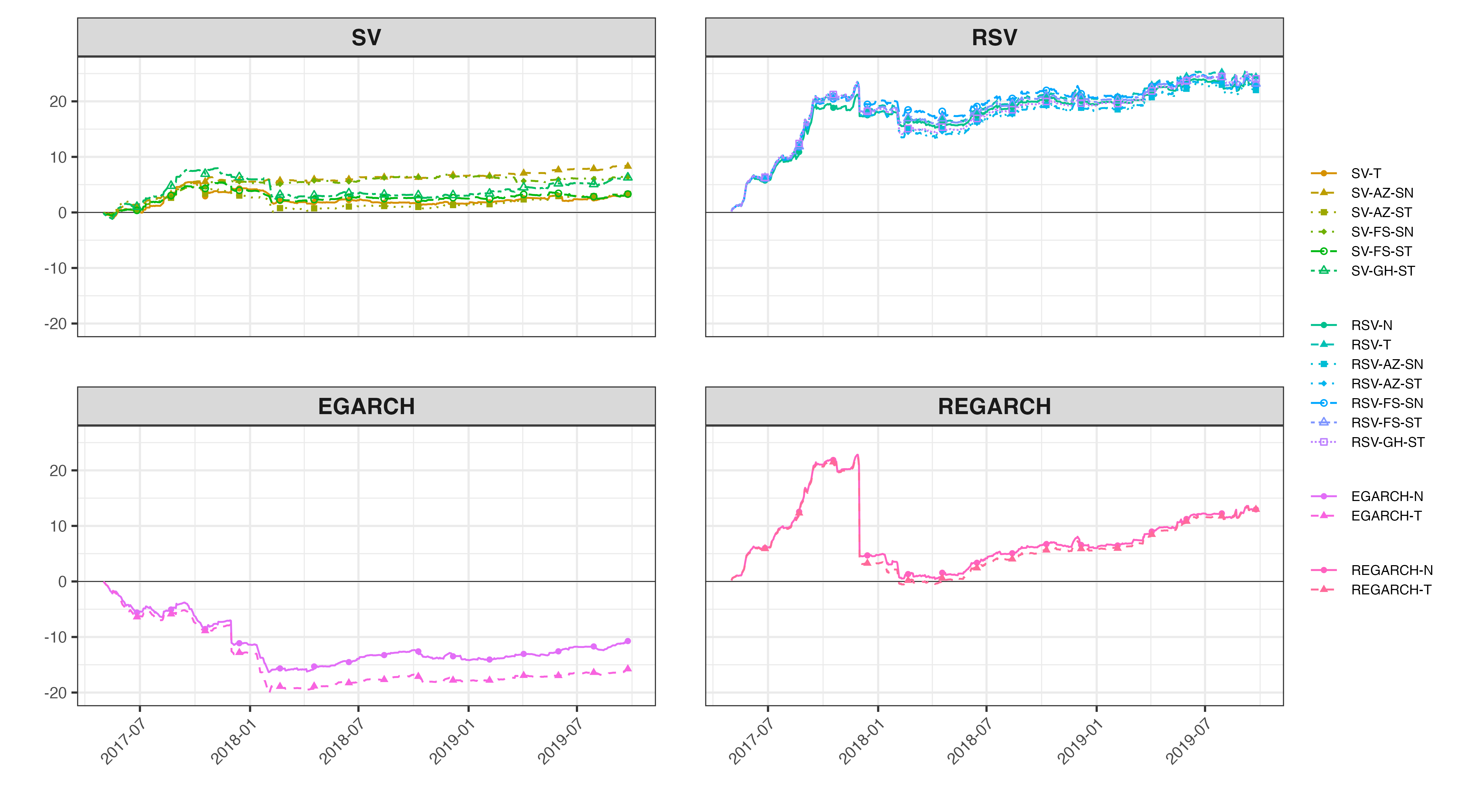}
\caption{Cumulative QLIKE loss differences relative to the SV-N model, based on RK as the volatility proxy, for the DJIA. Main sample (2009--2019).}
\label{fig:suppl-djia-rkth2-vol-qlike-cld}
\end{figure}

\begin{figure}[tbp]
\centering
\includegraphics[width = \textwidth]{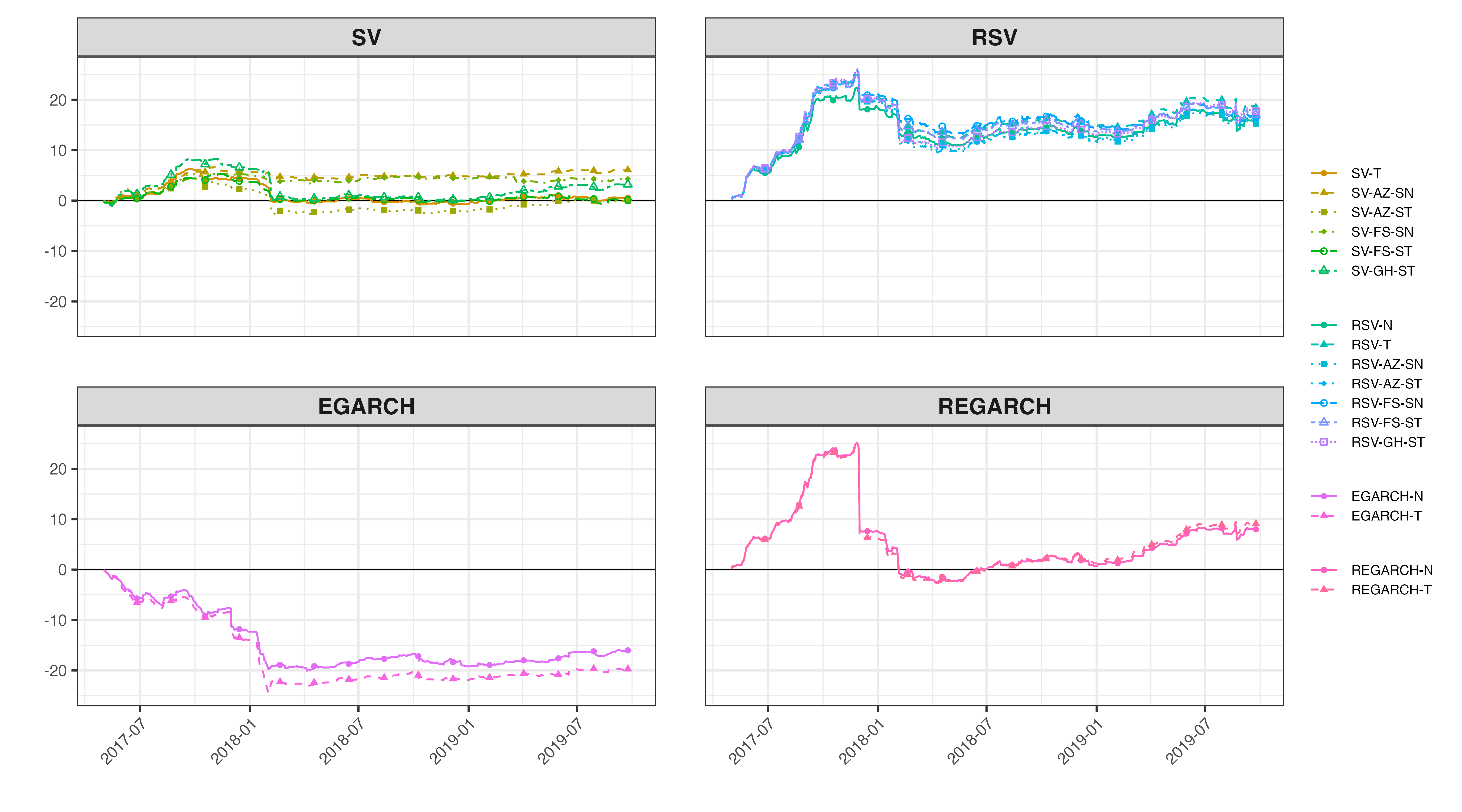}
\caption{Cumulative QLIKE loss differences relative to the SV-N model, based on BV as the volatility proxy, for the DJIA. Main sample (2009--2019).}
\label{fig:suppl-djia-bv-vol-qlike-cld}
\end{figure}

\begin{figure}[tbp]
\centering
\includegraphics[width = \textwidth]{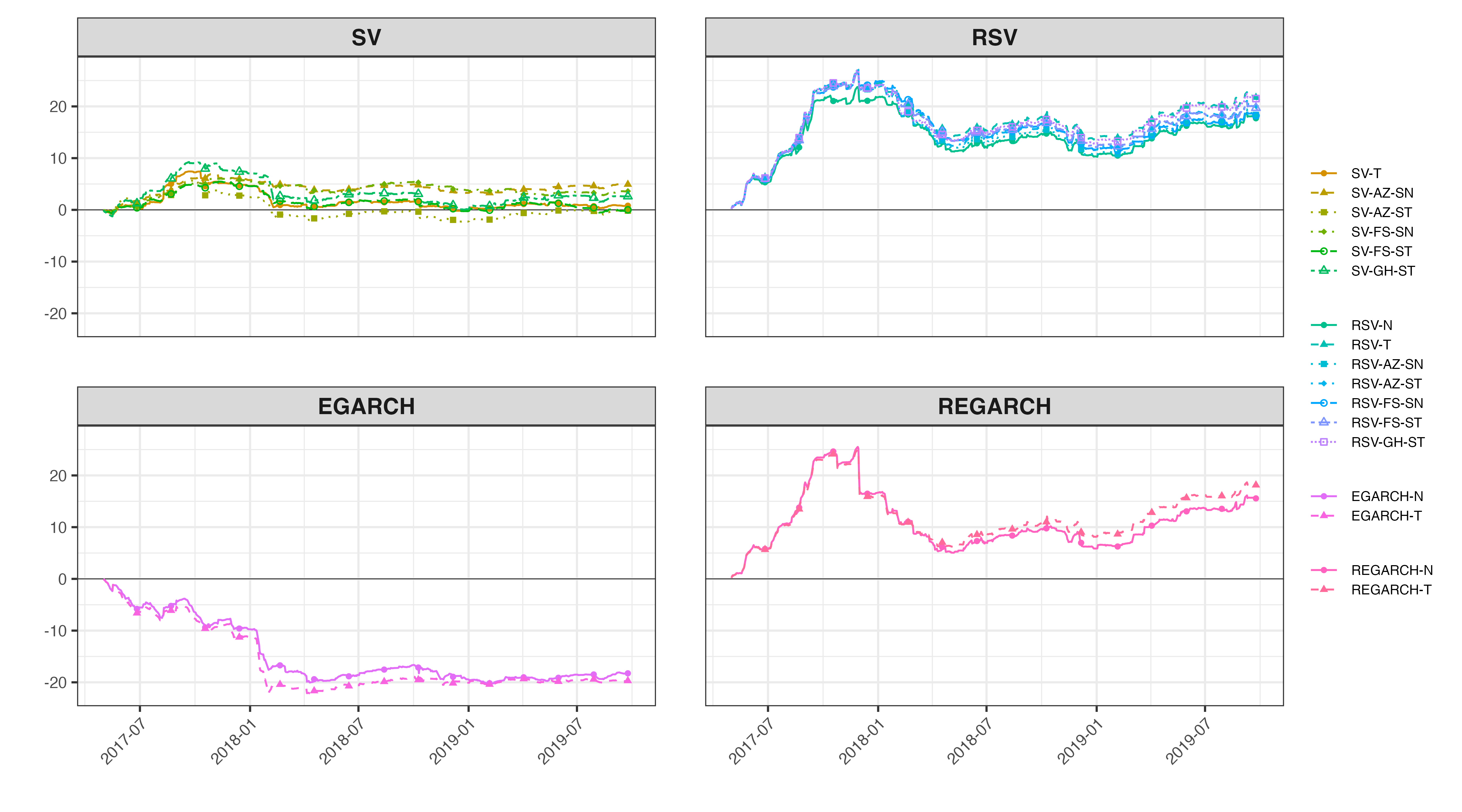}
\caption{Cumulative QLIKE loss differences relative to the SV-N model, based on Med as the volatility proxy, for the DJIA. Main sample (2009--2019).}
\label{fig:suppl-djia-medrv-vol-qlike-cld}
\end{figure}

\begin{figure}[tbp]
\centering
\includegraphics[width = \textwidth]{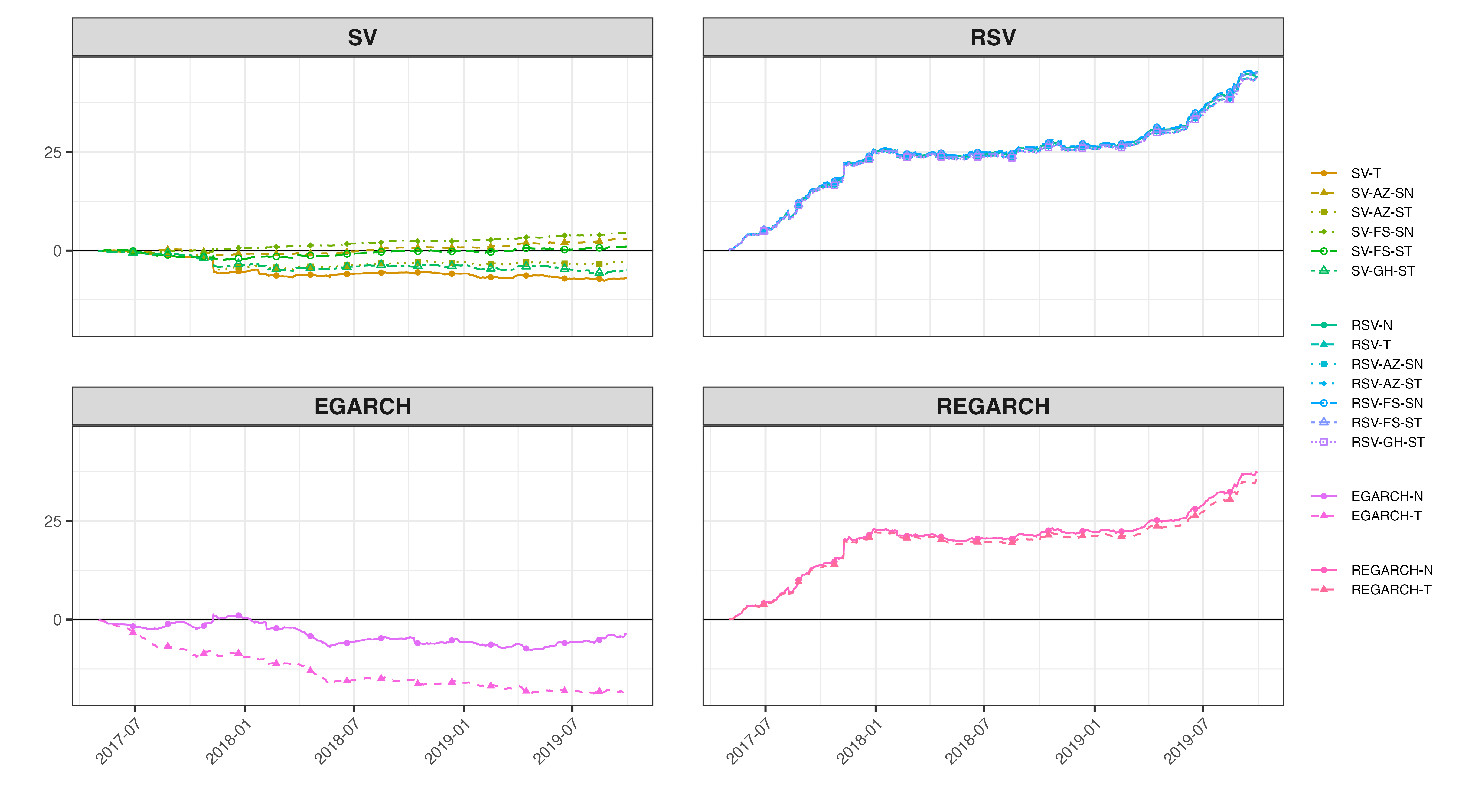}
\caption{Cumulative QLIKE loss differences relative to the SV-N model, based on RK as the volatility proxy, for the N225. Main sample (2009--2019).}
\label{fig:suppl-n225-rkth2-vol-qlike-cld}
\end{figure}

\begin{figure}[tbp]
\centering
\includegraphics[width = \textwidth]{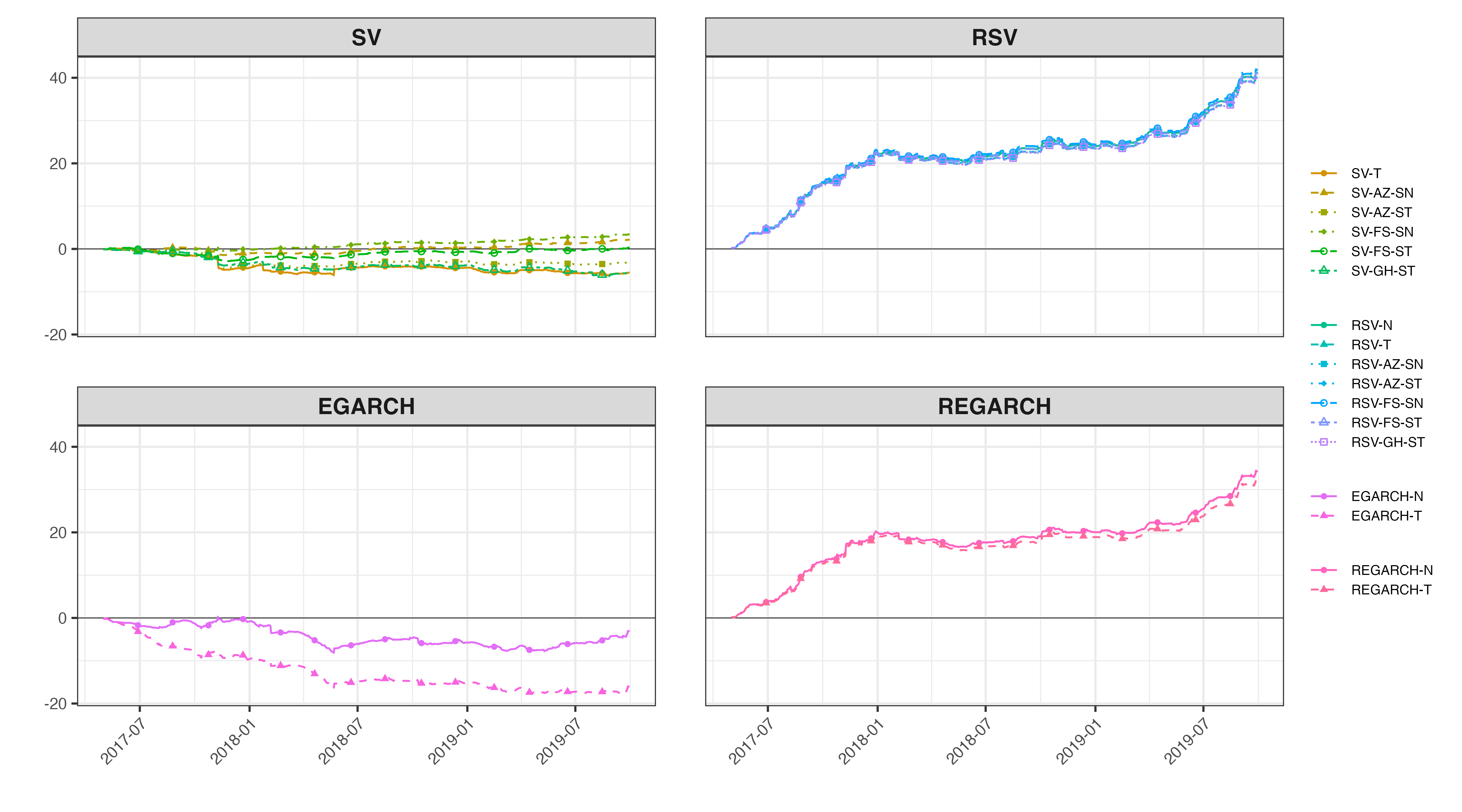}
\caption{Cumulative QLIKE loss differences relative to the SV-N model, based on BV as the volatility proxy, for the N225. Main sample (2009--2019).}
\label{fig:suppl-n225-bv-vol-qlike-cld}
\end{figure}

\begin{figure}[tbp]
\centering
\includegraphics[width = \textwidth]{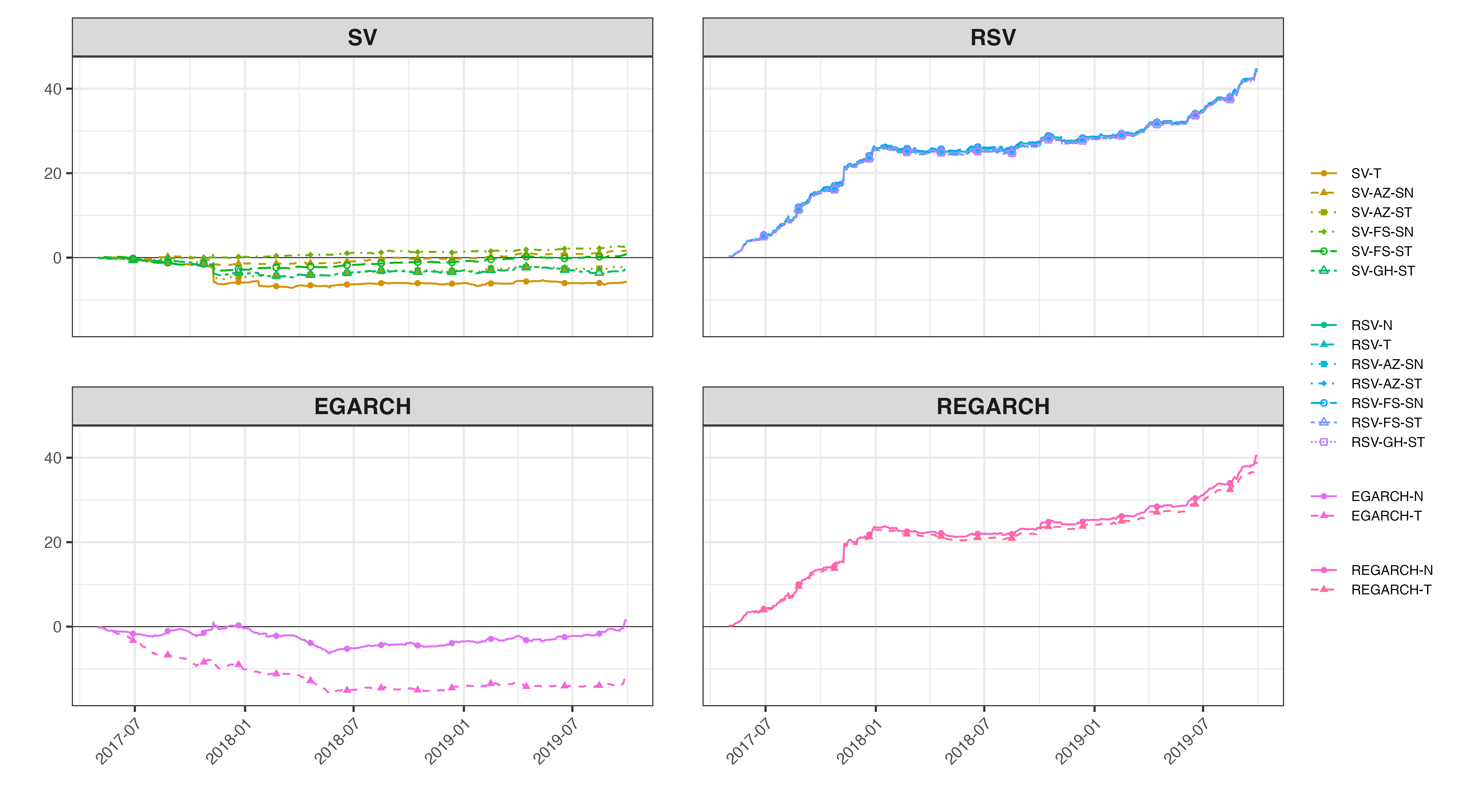}
\caption{Cumulative QLIKE loss differences relative to the SV-N model, based on Med as the volatility proxy, for the N225. Main sample (2009--2019).}
\label{fig:suppl-n225-medrv-vol-qlike-cld}
\end{figure}

Figures~\ref{fig:pred-vol-cld-djia} and \ref{fig:pred-vol-cld-n225} depict the cumulative QLIKE loss differences (CLDs) relative to the SV-N model under RV5, and Figures~\ref{fig:suppl-djia-rkth2-vol-qlike-cld} to \ref{fig:suppl-n225-medrv-vol-qlike-cld} under the RK, BV, and Med proxies; positive values favour the model over the benchmark, and the end point of each trajectory is the average loss difference of Table~2 of the main text up to scaling.
For both indices every RSV and REGARCH specification stays above the benchmark under every proxy, and the advantage accrues over the whole forecast period rather than in a few episodes: for the DJIA the RSV trajectories rise most steeply in the second half of 2017 and keep rising thereafter, while the REGARCH trajectories give back part of their advantage at the beginning of December 2017 and again in the volatility episode of February 2018 before recovering; for the N225 the RSV and REGARCH trajectories rise steadily, most strongly in the second half of 2019.
The EGARCH specifications end below the benchmark for both indices, except EGARCH-N for the N225 under Med.
The other SV specifications end close to the benchmark: for the DJIA all of them end above it under RV5 and RK, all but SV-AZ-ST under BV, and all but SV-AZ-ST and SV-FS-ST under Med; for the N225 only SV-AZ-SN, SV-FS-SN, and SV-FS-ST end above it, under every proxy, and SV-T, SV-AZ-ST, and SV-GH-ST end below it.

\begin{figure}[tbp]
\centering
\includegraphics[width = .9\textwidth]{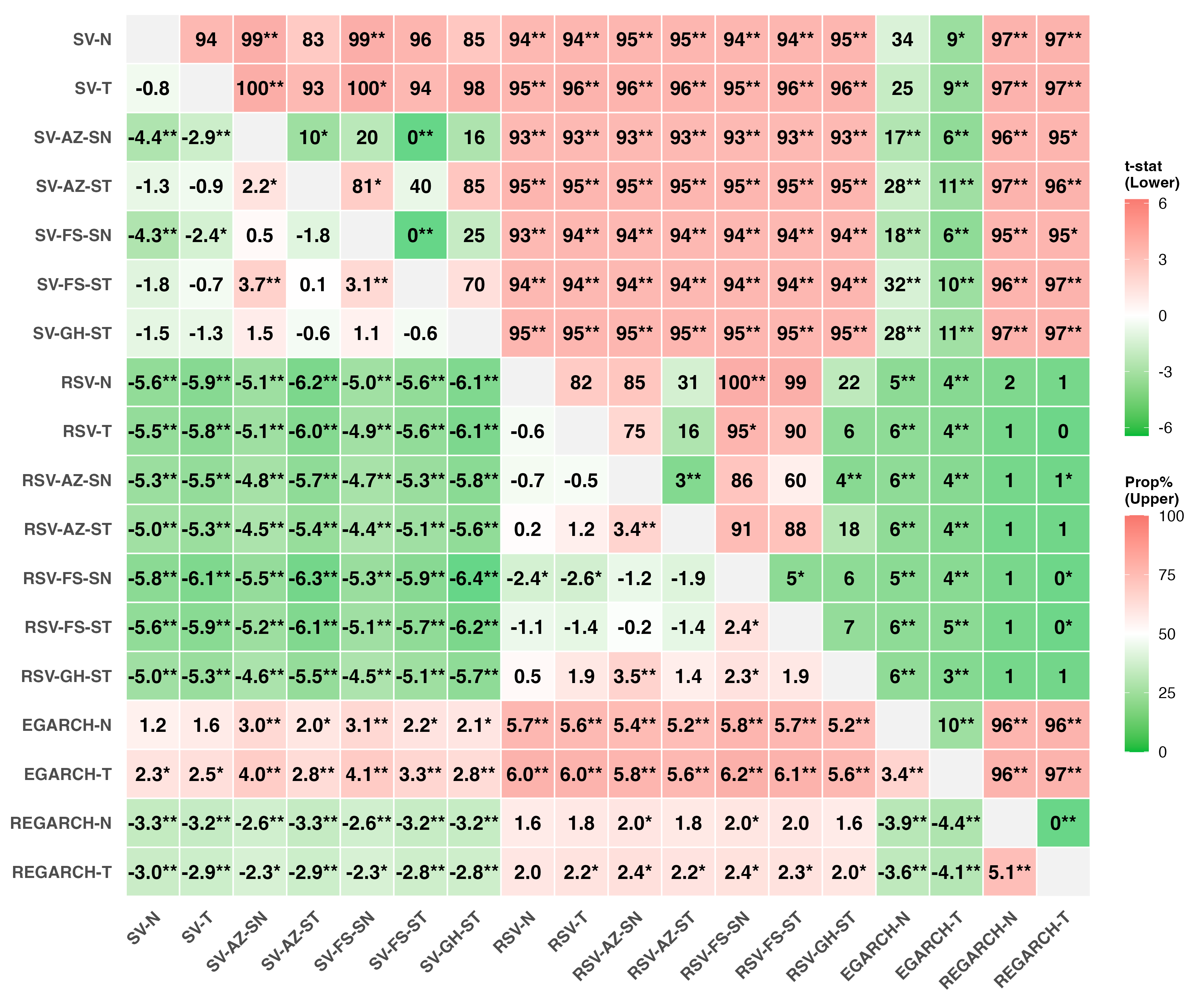}
\caption{Pairwise comparison of GW tests on the QLIKE loss of volatility forecasts, based on RV5 as the volatility proxy, for the DJIA. The lower triangular part reports the unconditional test statistics and the upper triangular part the win proportions of Section~\ref{sec:suppl-gw}; lower values (green) favour the row model and larger values (orange) the column model, and double and single asterisks denote significance at the $1\%$ and $5\%$ levels.}
\label{fig:forecast-djia-vol-gwtest}
\end{figure}

\begin{figure}[tbp]
\centering
\includegraphics[width = .9\textwidth]{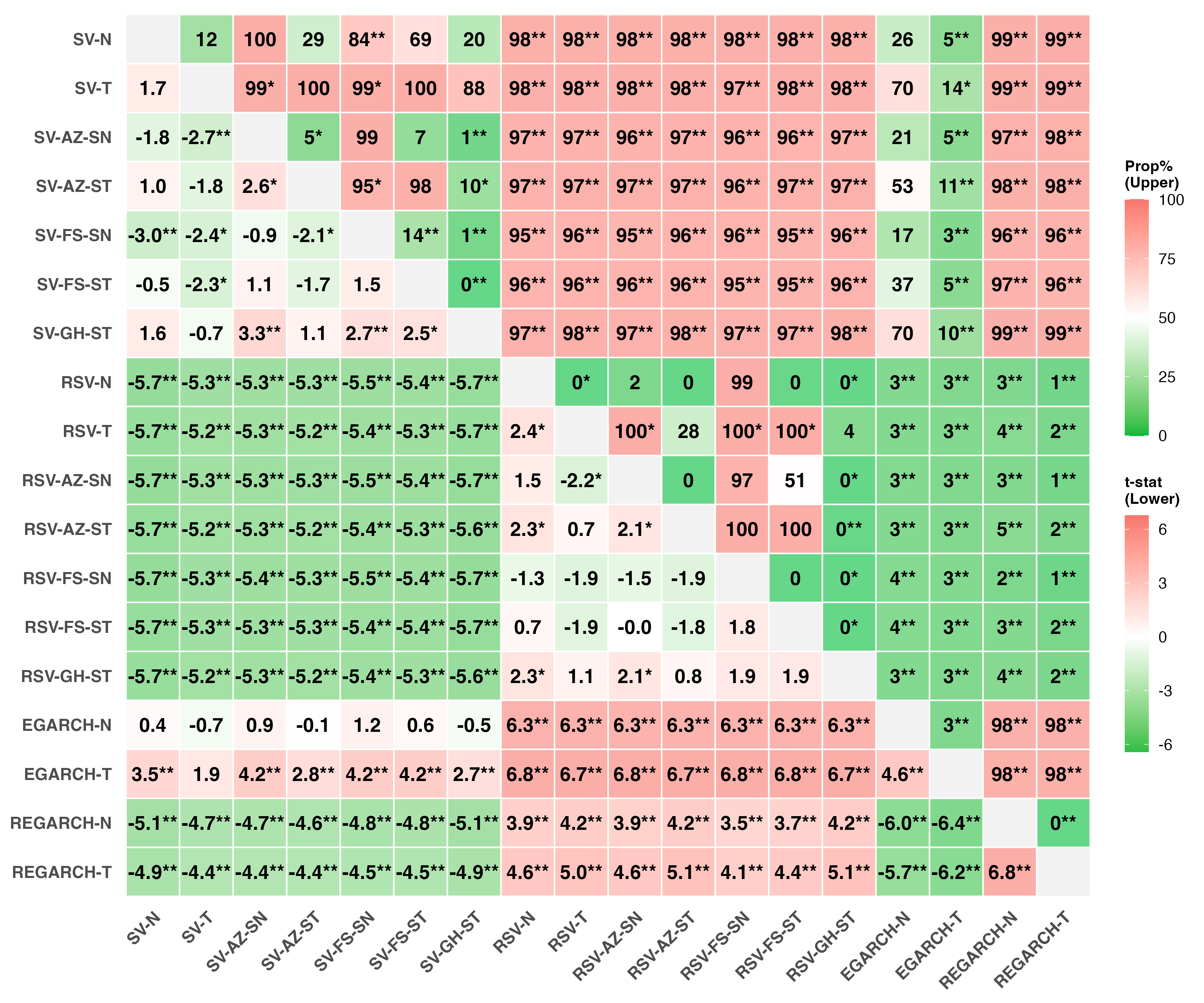}
\caption{Pairwise comparison of GW tests on the QLIKE loss of volatility forecasts, based on RV5 as the volatility proxy, for the N225. The lower triangular part reports the unconditional test statistics and the upper triangular part the win proportions of Section~\ref{sec:suppl-gw}; lower values (green) favour the row model and larger values (orange) the column model, and double and single asterisks denote significance at the $1\%$ and $5\%$ levels.}
\label{fig:forecast-n225-vol-gwtest}
\end{figure}

\begin{figure}[tbp]
\centering
\includegraphics[width = .9\textwidth]{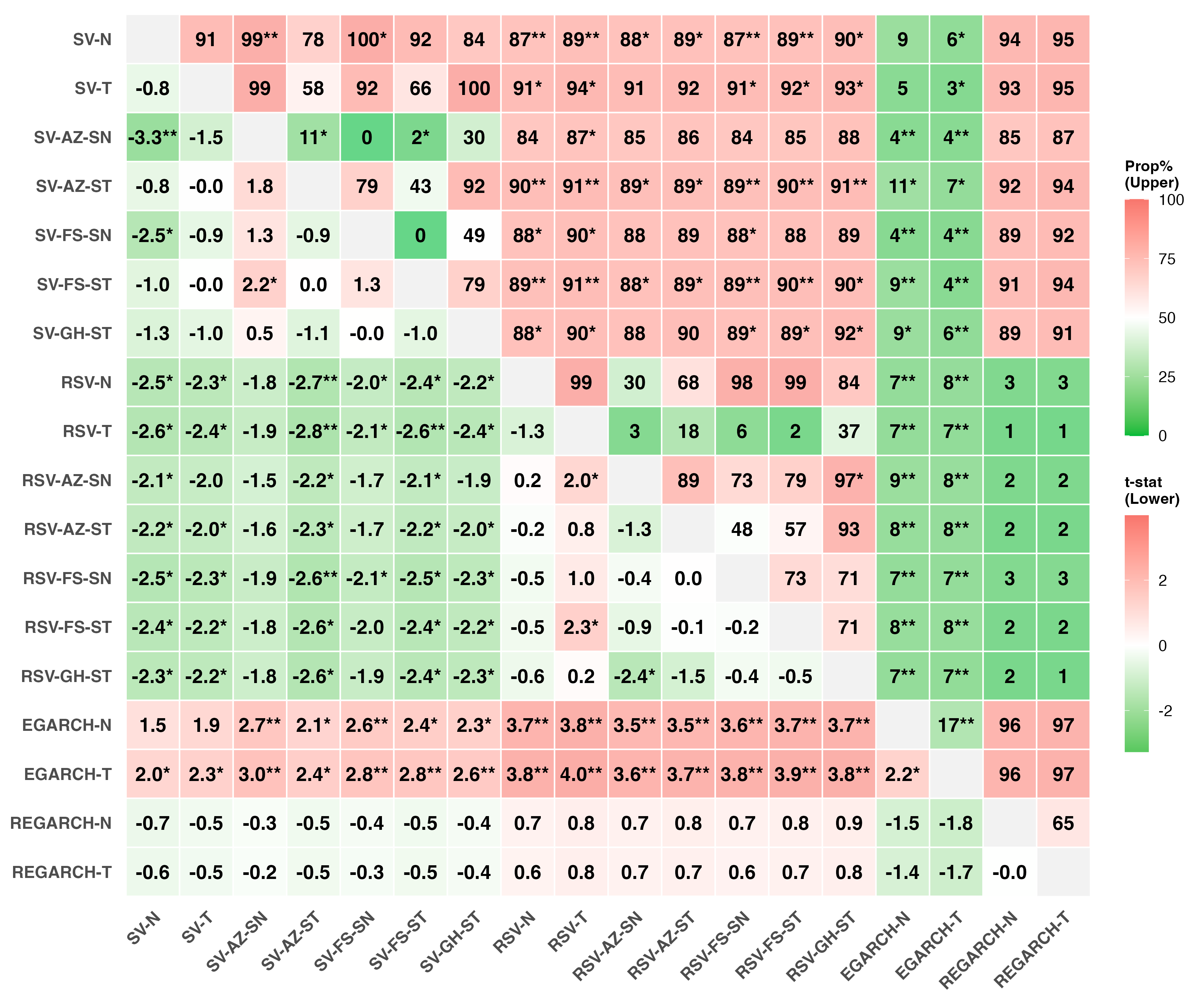}
\caption{Pairwise comparison of GW tests on QLIKE of volatility forecasts, based on RK as the volatility proxy, for the DJIA.
The lower triangular part reports unconditional GW test statistics, while the upper triangular part shows win proportions.
Lower values (green) indicate superior performance of the row model, whereas higher values (orange) favor the column model.
Double and single asterisks ($^{**}$ and $^{*}$) denote statistical significance at the 1\% and 5\% levels, respectively, based on unconditional GW test p-values for the lower triangular part and conditional GW test p-values for the upper triangular part. Main sample (2009--2019).}
\label{fig:forecast-djia-vol-gwtest-suppl}
\end{figure}

\begin{figure}[tbp]
\centering
\includegraphics[width = .9\textwidth]{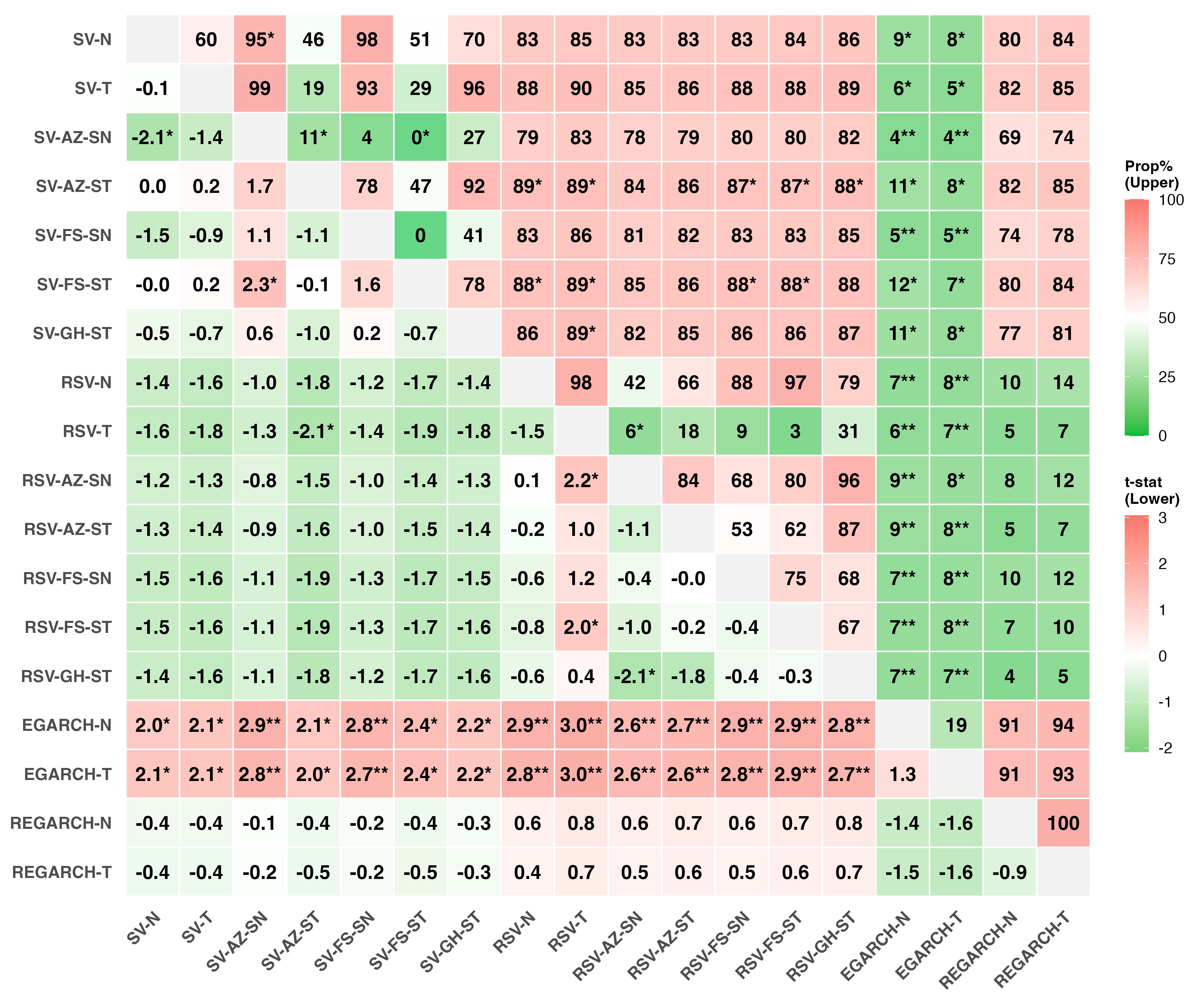}
\caption{Pairwise comparison of GW tests on QLIKE of volatility forecasts, based on BV as the volatility proxy, for the DJIA.
See Figure~\ref{fig:forecast-djia-vol-gwtest-suppl} for additional details. Main sample (2009--2019).}
\label{fig:suppl-djia-gw-vol-bv}
\end{figure}

\begin{figure}[tbp]
\centering
\includegraphics[width = .9\textwidth]{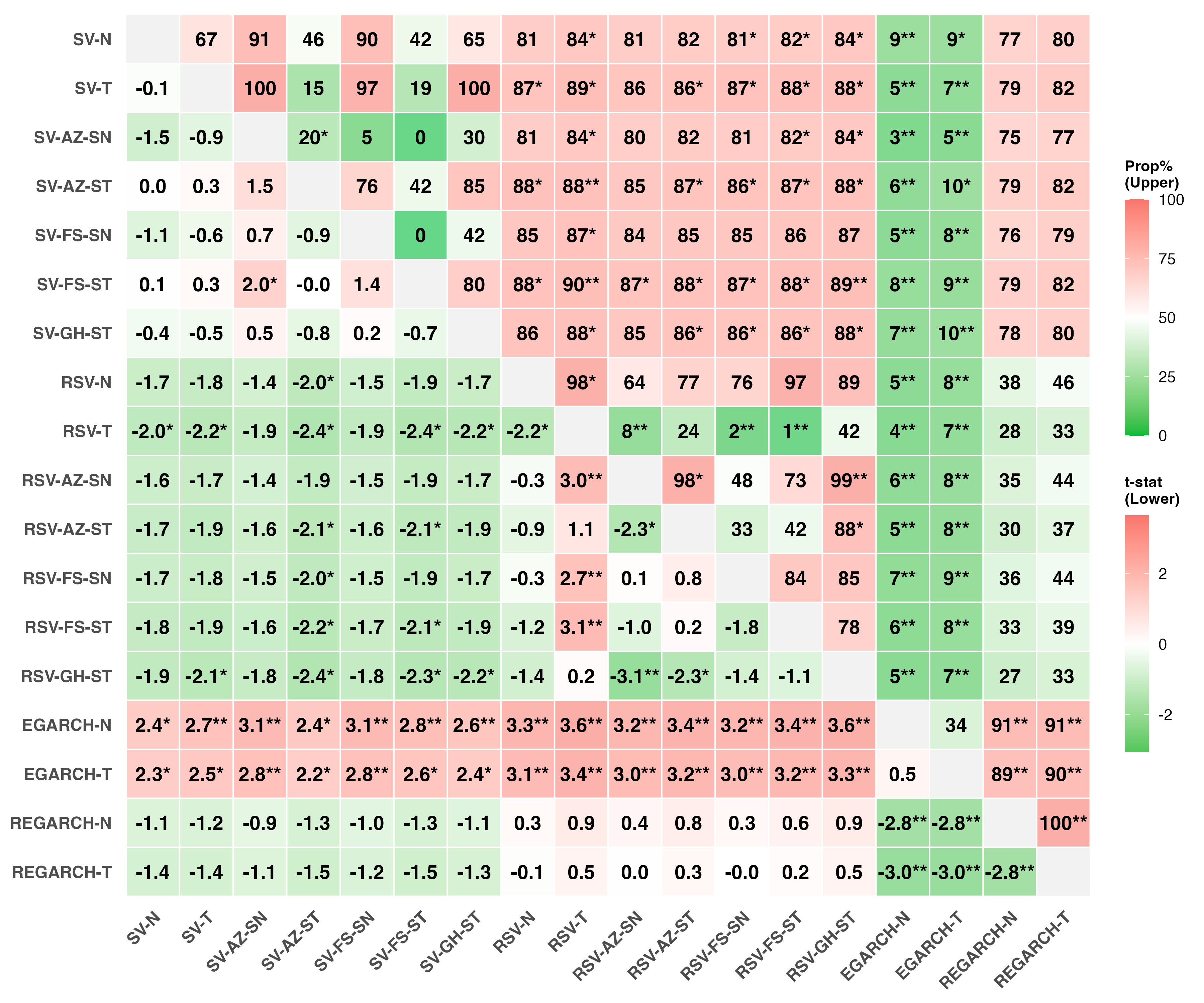}
\caption{Pairwise comparison of GW tests on QLIKE of volatility forecasts, based on Med as the volatility proxy, for the DJIA.
See Figure~\ref{fig:forecast-djia-vol-gwtest-suppl} for additional details. Main sample (2009--2019).}
\label{fig:suppl-djia-gw-vol-medrv}
\end{figure}

\begin{figure}[tbp]
\centering
\includegraphics[width = .9\textwidth]{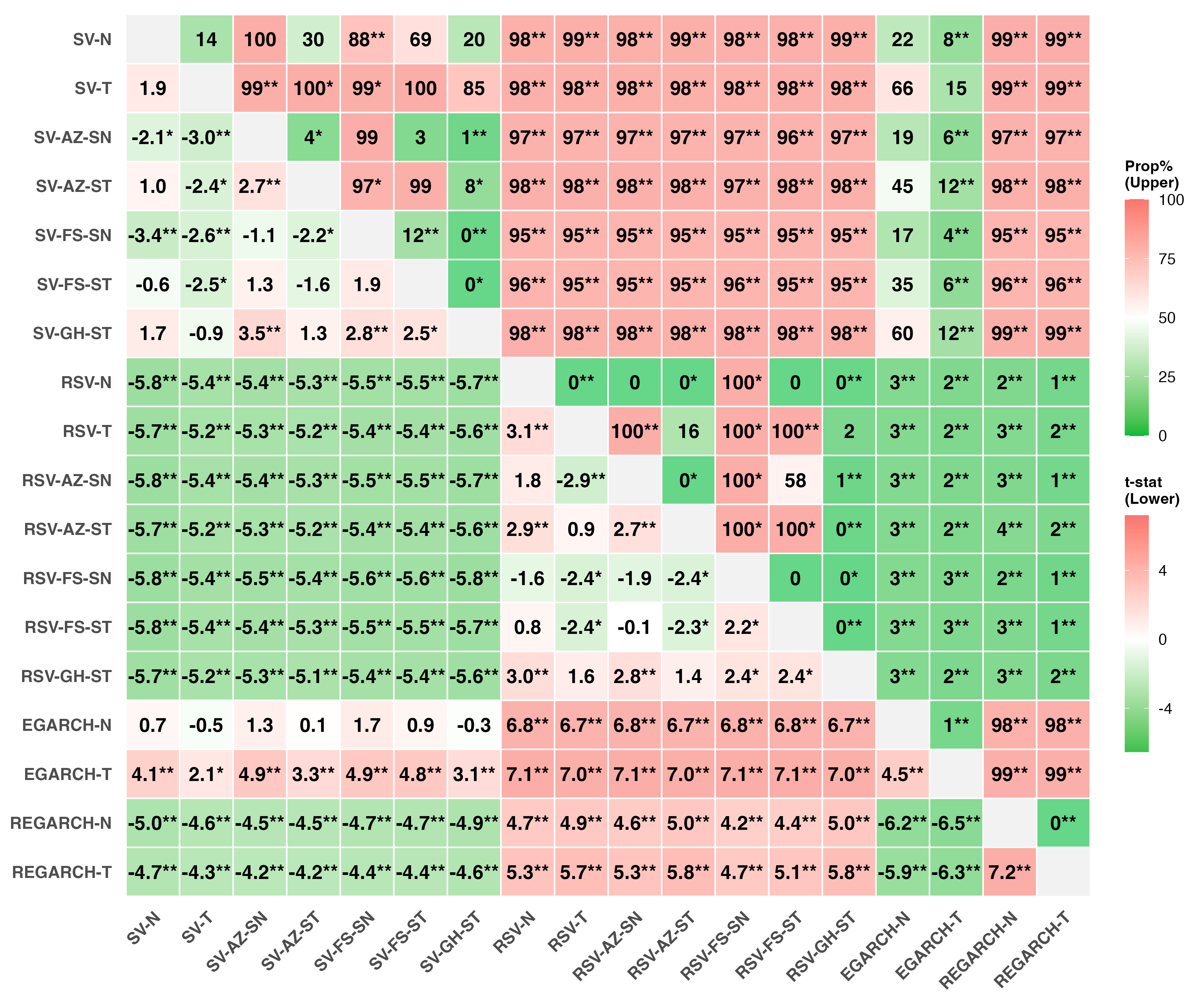}
\caption{Pairwise comparison of GW tests on QLIKE of volatility forecasts, based on RK as the volatility proxy, for the N225. See Figure~\ref{fig:forecast-djia-vol-gwtest-suppl} for additional details. Main sample (2009--2019).}
\label{fig:suppl-n225-gw-vol-rkth2}
\end{figure}

\begin{figure}[tbp]
\centering
\includegraphics[width = .9\textwidth]{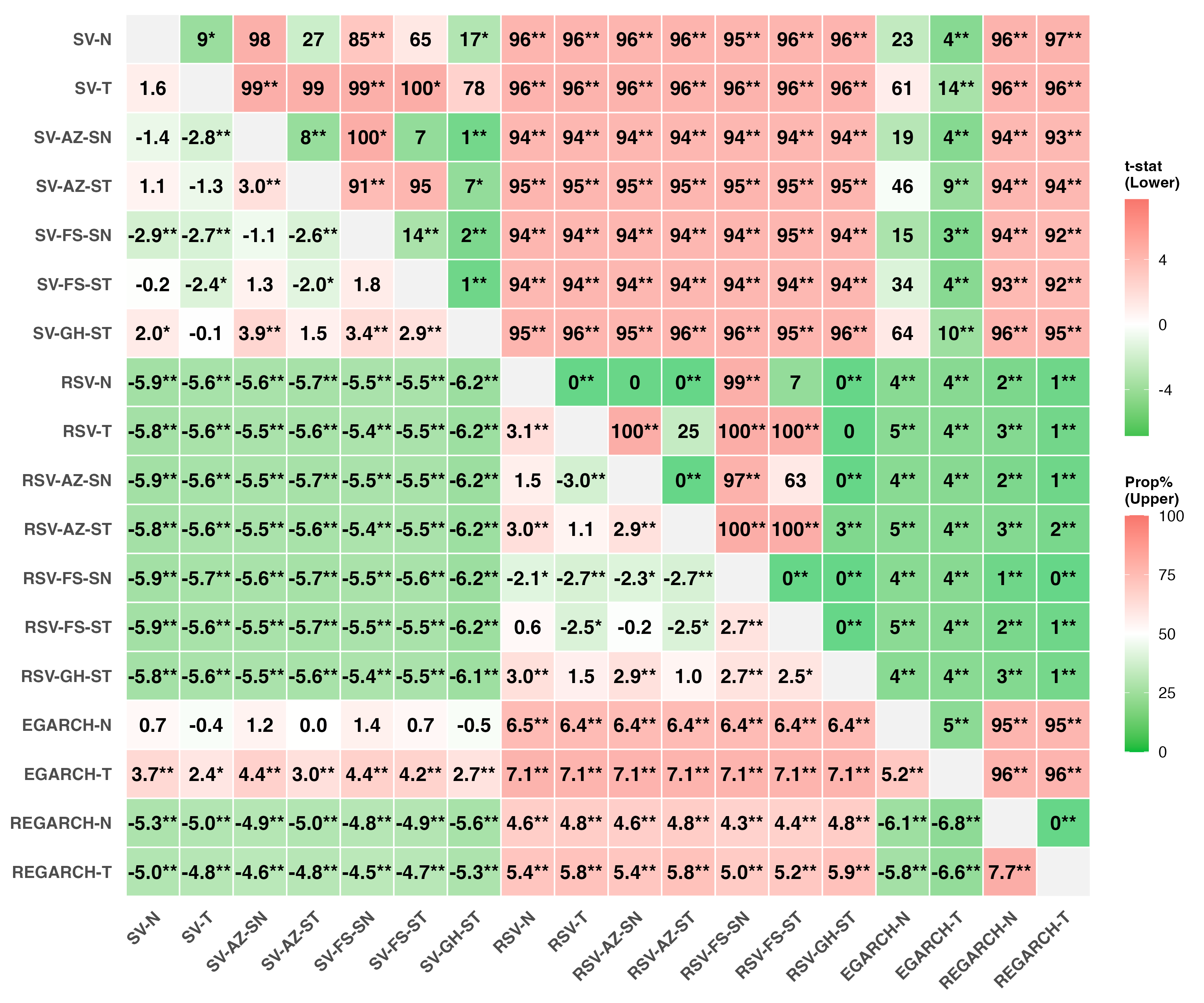}
\caption{Pairwise comparison of GW tests on QLIKE of volatility forecasts, based on BV as the volatility proxy, for the N225. See Figure~\ref{fig:forecast-djia-vol-gwtest-suppl} for additional details. Main sample (2009--2019).}
\label{fig:suppl-n225-gw-vol-bv}
\end{figure}

\begin{figure}[tbp]
\centering
\includegraphics[width = .9\textwidth]{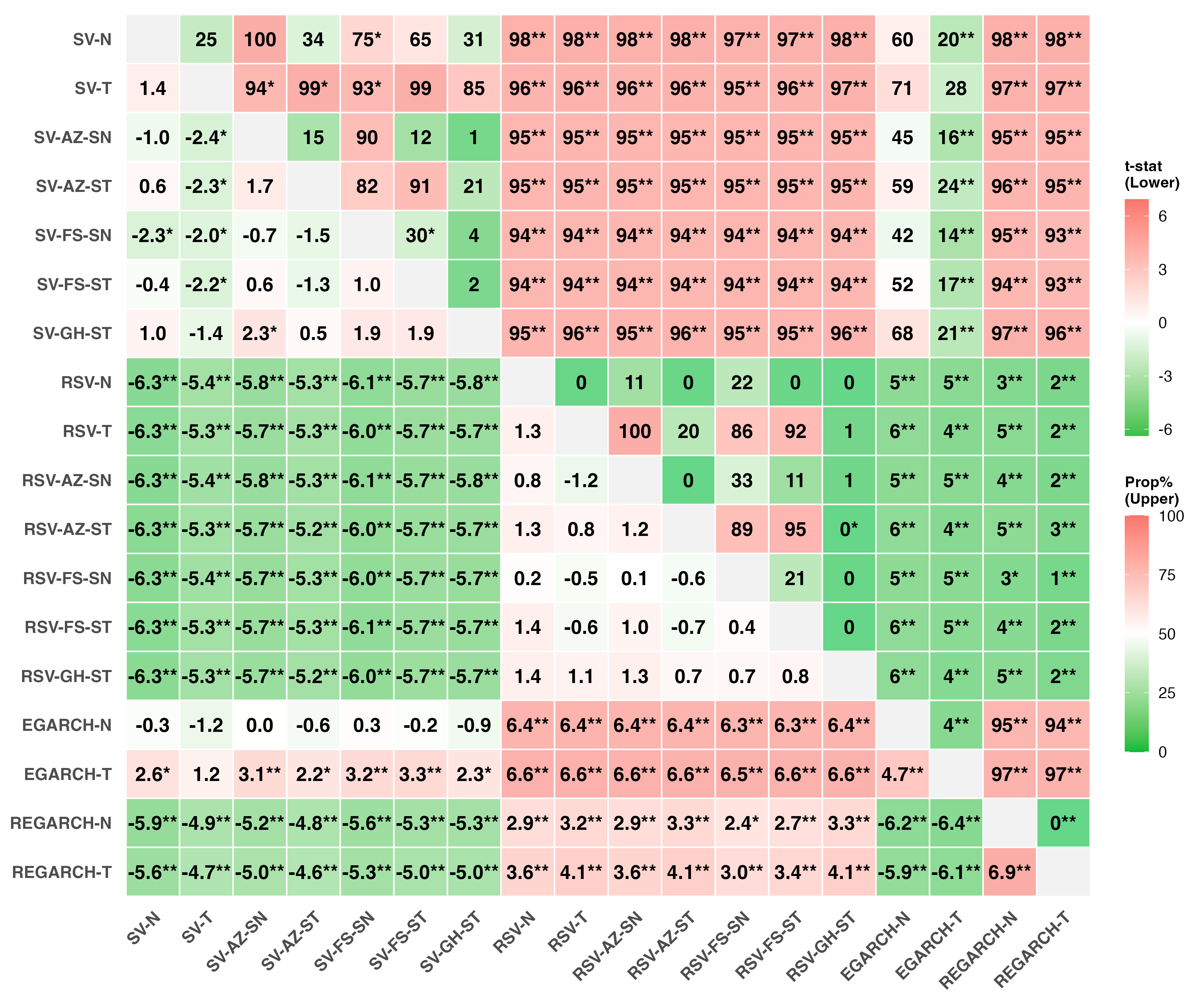}
\caption{Pairwise comparison of GW tests on QLIKE of volatility forecasts, based on Med as the volatility proxy, for the N225. See Figure~\ref{fig:forecast-djia-vol-gwtest-suppl} for additional details. Main sample (2009--2019).}
\label{fig:suppl-n225-gw-vol-medrv}
\end{figure}

Figures~\ref{fig:forecast-djia-vol-gwtest} and \ref{fig:forecast-n225-vol-gwtest} report the pairwise Giacomini--White tests on the QLIKE loss under RV5, and Figures~\ref{fig:forecast-djia-vol-gwtest-suppl} to \ref{fig:suppl-n225-gw-vol-medrv} under the other three proxies, in the heatmap format described in Section~\ref{sec:suppl-gw}.
The entries for the seven matched pairs of RSV and SV models are the unconditional statistics whose $p$-values Table~3 of the main text reports; the remaining entries compare models across classes and across distributions.
For the seven matched pairs the RSV model attains the lower loss in every pair and under every proxy; the test rejects equal accuracy at the $1\%$ level for every pair for the N225 under every proxy and for the DJIA under RV5, and, for the DJIA, at the $5\%$ level for six pairs under RK, for four pairs under Med, and for no pair under BV (Table~3 of the main text).

\section{VaR and ES Forecasts: Main Sample}
\label{sec:suppl-vares}

\begin{figure}[tbp]
\centering
\begin{tabular}{cc}
RSV-AZ-ST for DJIA & RSV-AZ-ST for N225 \\
\includegraphics[width = .45\textwidth]{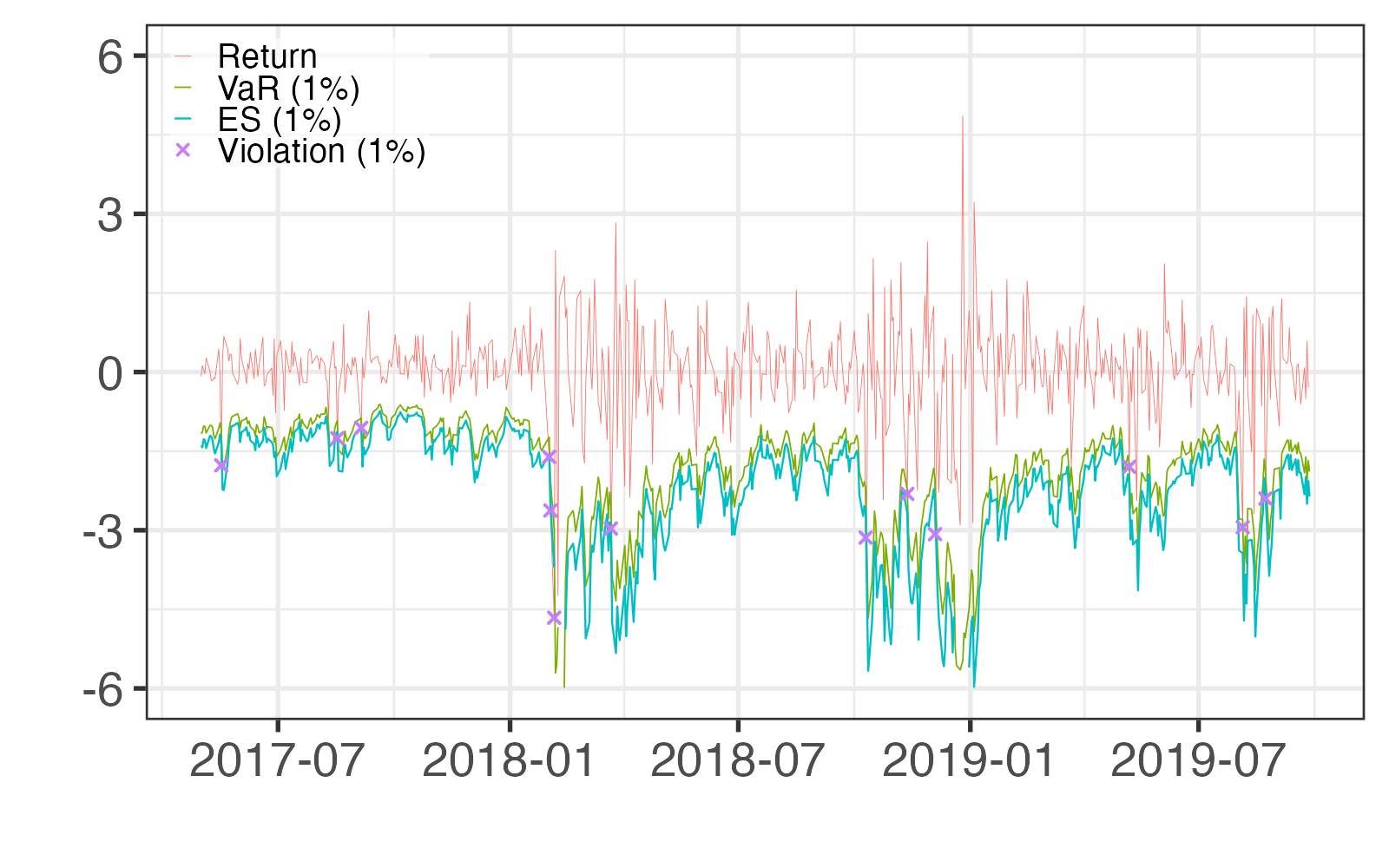} &
\includegraphics[width = .45\textwidth]{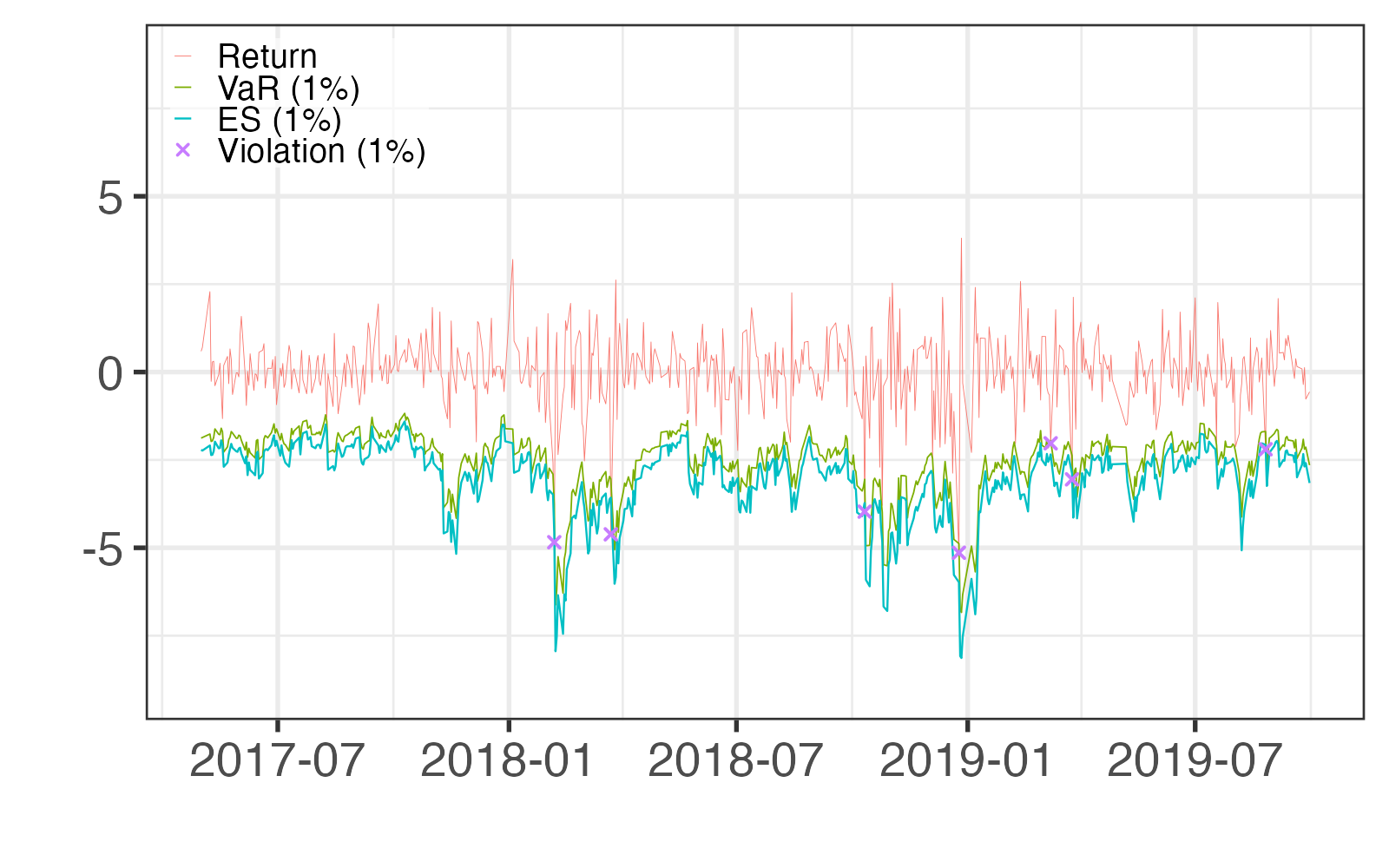} \\
REGARCH-T for DJIA & REGARCH-T for N225 \\
\includegraphics[width = .45\textwidth]{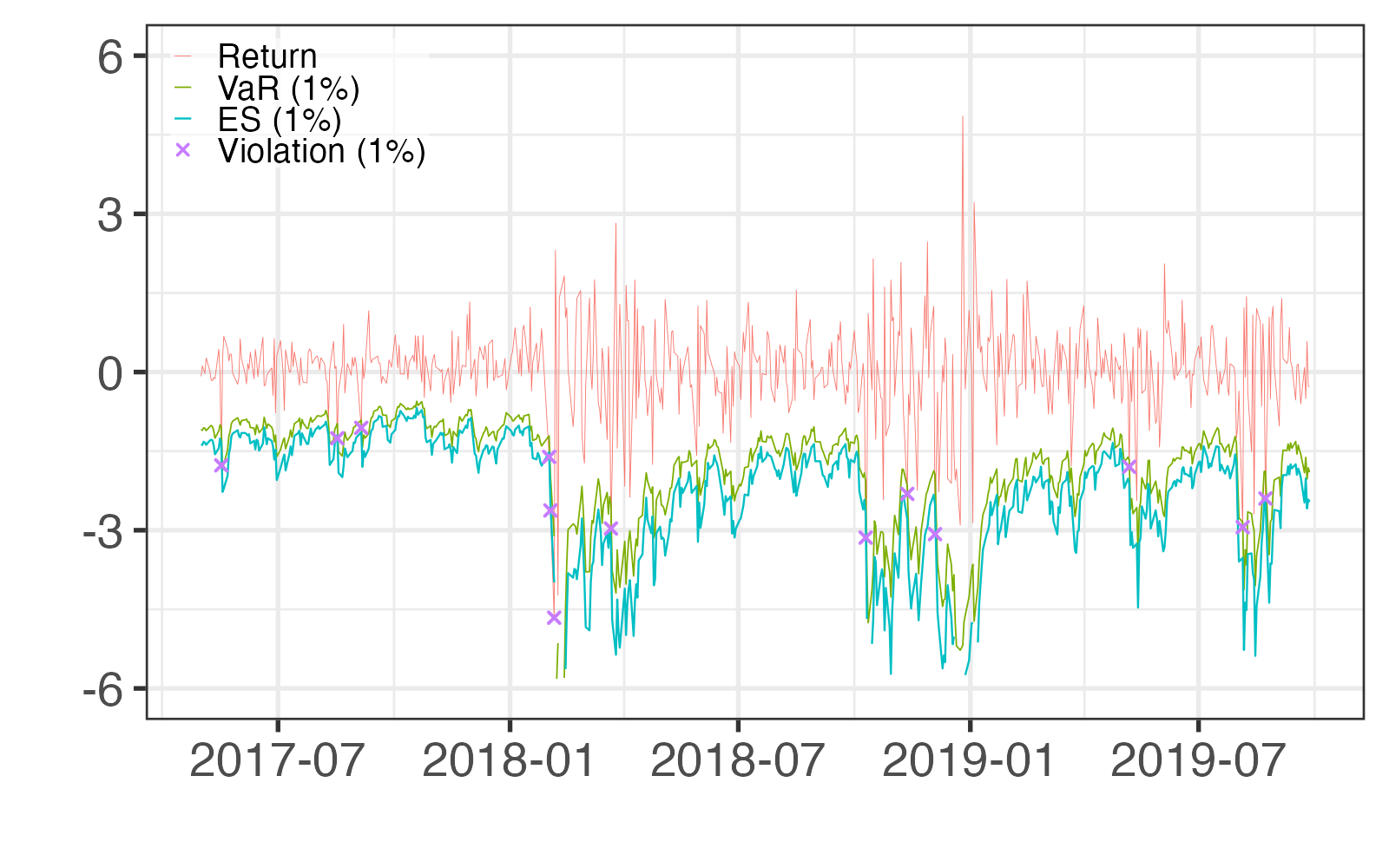} &
\includegraphics[width = .45\textwidth]{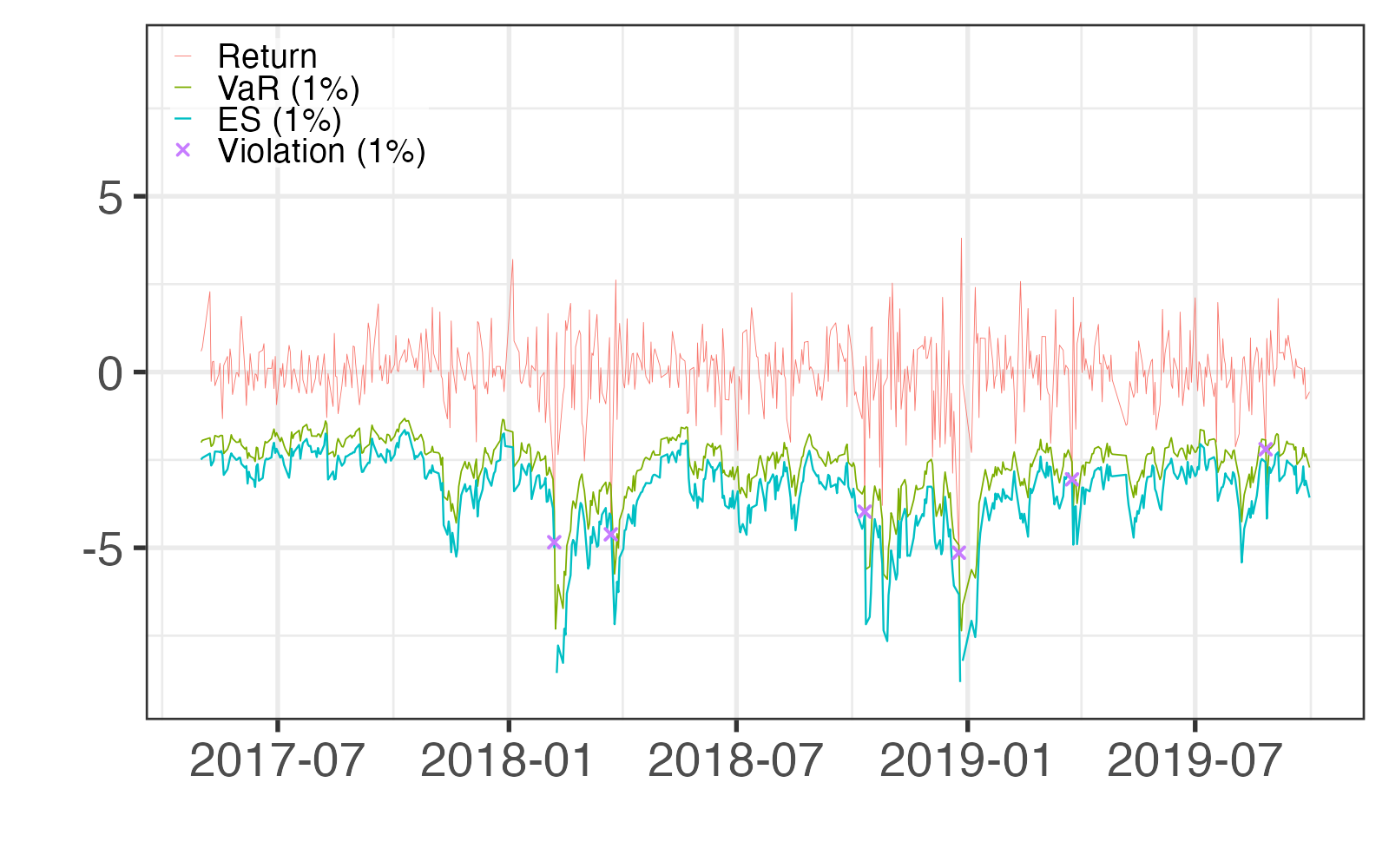}
\end{tabular}
\caption{$1\%$ VaR and ES forecasts of the RSV-AZ-ST and REGARCH-T models for the DJIA and N225.}
\label{fig:pred-var-es}
\end{figure}

This section collects the figures and the additional table on the VaR and ES forecasts of the main sample that complement Tables~3 to 5 of the main text.
Figure~\ref{fig:pred-var-es} illustrates the $1\%$ VaR and ES forecasts of the RSV-AZ-ST and REGARCH-T models for both indices.
VaR violations, that is, days on which the realized return falls below the forecast VaR, occur under both models during the volatility episodes of early 2018 and late 2019 as well as in calmer phases; the forecasts of the other models follow similar patterns with differences in scale and are omitted.

\begin{figure}[tbp]
\centering
\includegraphics[width = \textwidth]{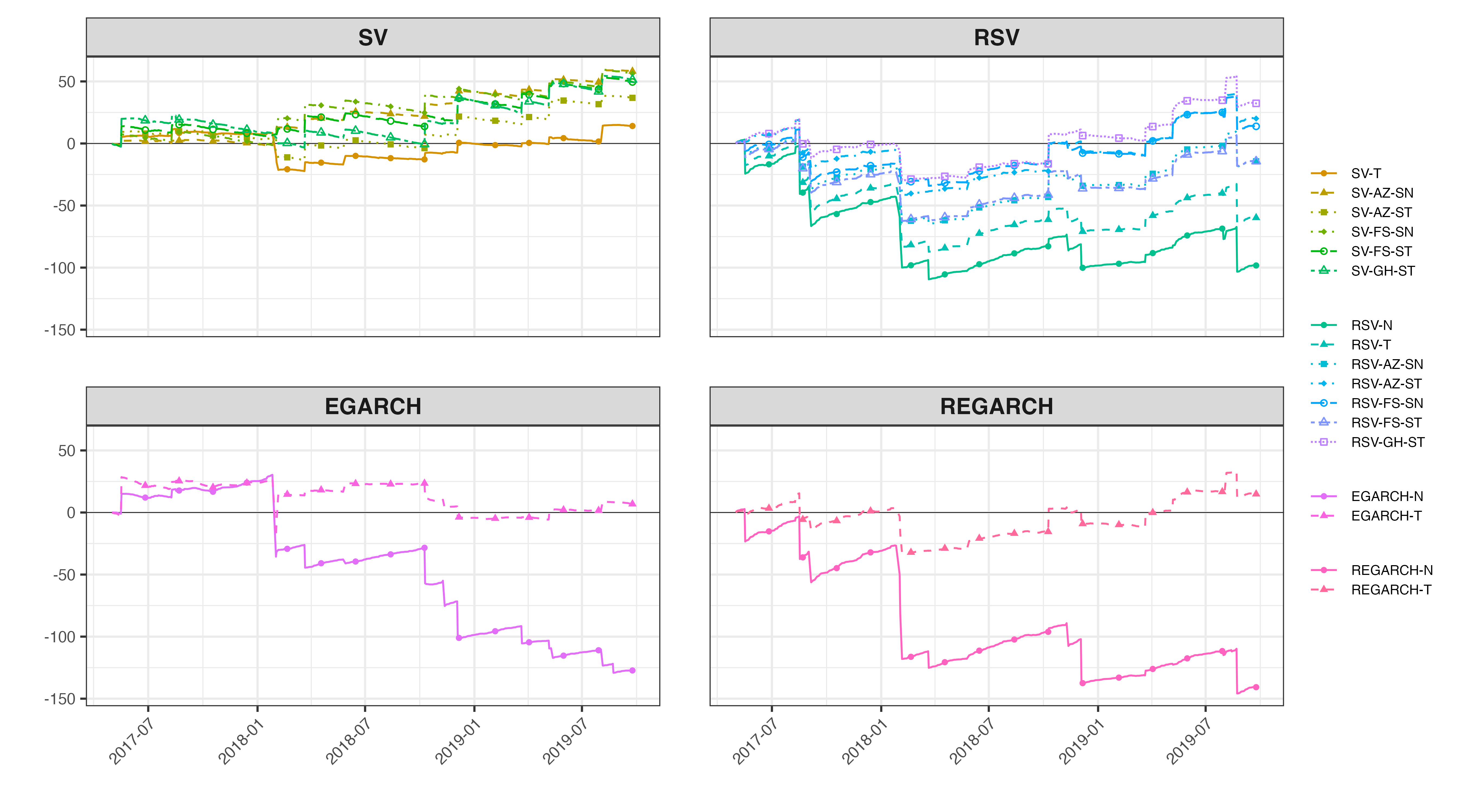}
\caption{Cumulative FZ0 loss differences relative to the SV-N model ($\alpha = 1\%$) for the DJIA. Positive values favour the model over the benchmark.}
\label{fig:pred-fz0-1p-cld-djia}
\end{figure}

\begin{figure}[tbp]
\centering
\includegraphics[width = \textwidth]{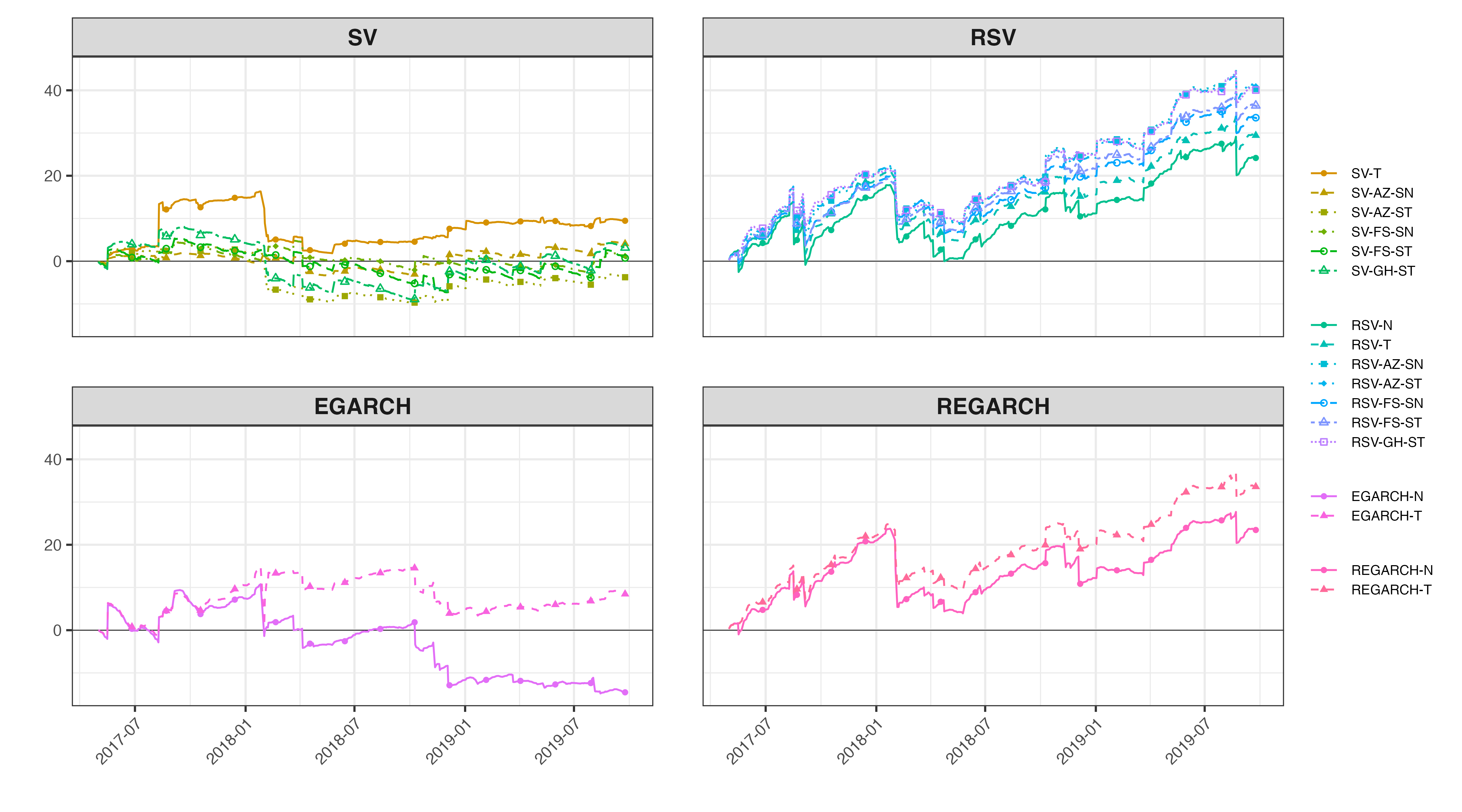}
\caption{Cumulative FZ0 loss differences relative to the SV-N model ($\alpha = 5\%$) for the DJIA. Positive values favour the model over the benchmark.}
\label{fig:pred-fz0-5p-cld-djia}
\end{figure}

\begin{figure}[tbp]
\centering
\includegraphics[width = \textwidth]{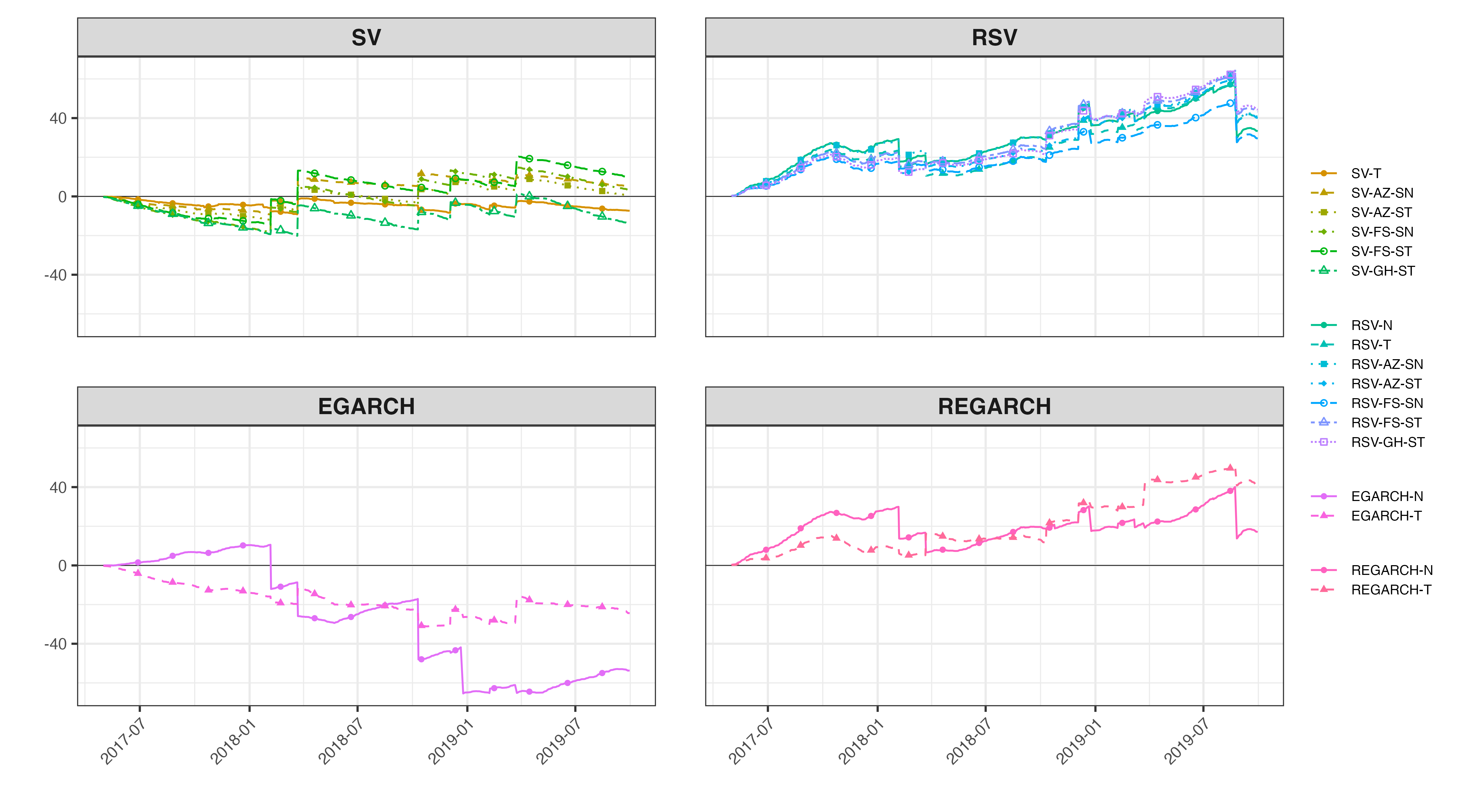}
\caption{Cumulative loss differences (FZ0) relative to the SV-N model ($\alpha = 1\%$) for the N225. Main sample (2009--2019).}
\label{fig:pred-fz0-1p-cld-n225}
\end{figure}

\begin{figure}[tbp]
\centering
\includegraphics[width = \textwidth]{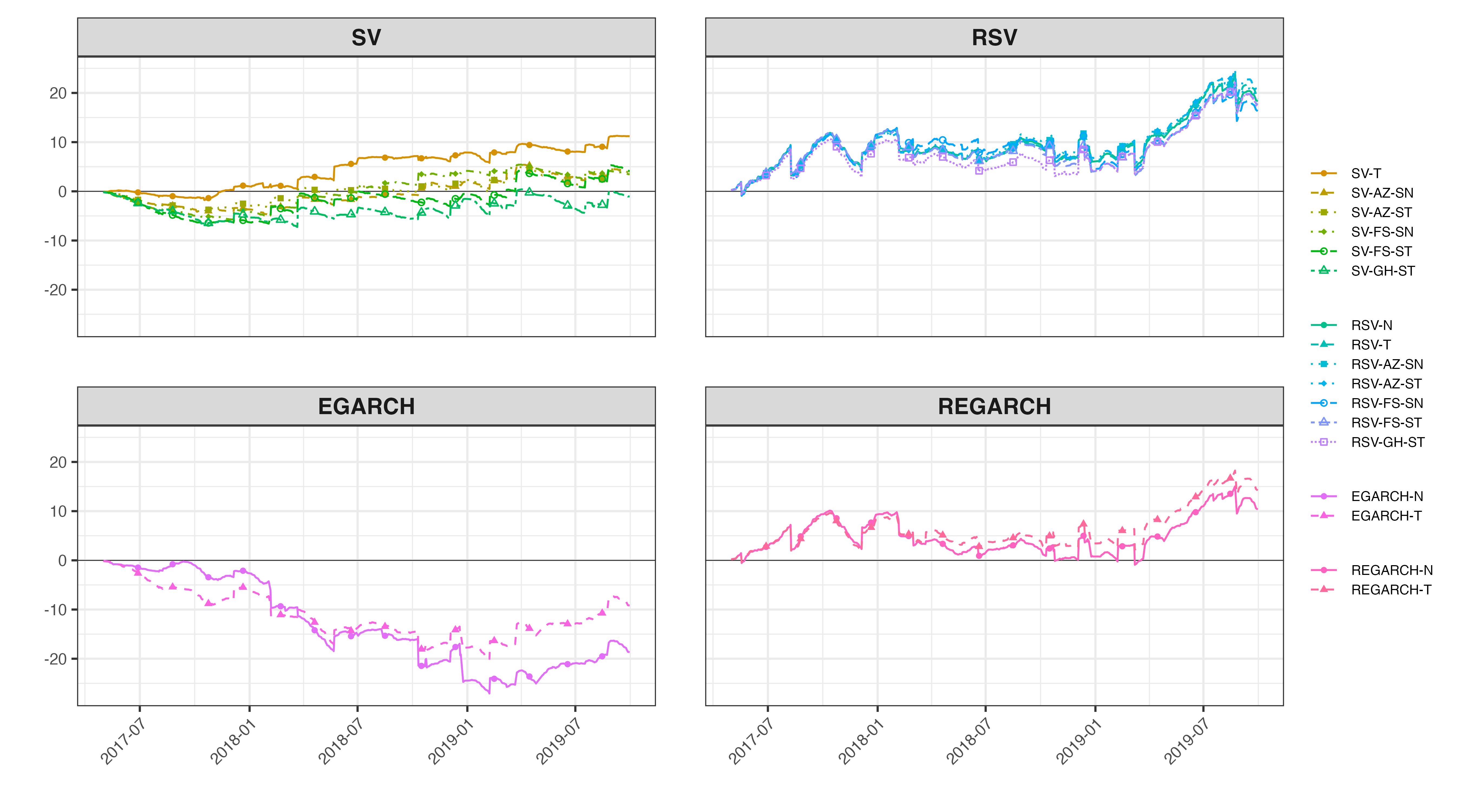}
\caption{Cumulative loss differences (FZ0) relative to the SV-N model ($\alpha = 5\%$) for the N225. Main sample (2009--2019).}
\label{fig:pred-fz0-5p-cld-n225}
\end{figure}

Figures~\ref{fig:pred-fz0-1p-cld-djia} to \ref{fig:pred-fz0-5p-cld-n225} show the cumulative FZ0 loss differences relative to the SV-N model at the $1\%$ and $5\%$ levels, for the DJIA and then for the N225; the end points correspond to the average loss differences implied by Tables~4 and 5 of the main text.
For the DJIA at the $1\%$ level every other SV specification ends above the benchmark, whereas among the RSV specifications only RSV-AZ-ST, RSV-FS-SN, and RSV-GH-ST do; the shortfall of the RSV specifications is incurred in the volatility episodes of February 2018 and the second half of 2019, when their $1\%$ VaR forecasts are violated, and the largest shortfalls are those of EGARCH-N and REGARCH-N, whose Student's $t$ counterparts end above the benchmark.
At the $5\%$ level the ordering reverses: every RSV and REGARCH specification stays above the benchmark on all but a few days around February 2018, and the gap widens steadily, whereas EGARCH-N ends below it.
For the N225 every RSV and REGARCH specification ends above the benchmark at both levels, together with the two Azzalini and the two Fern\'{a}ndez--Steel SV specifications at the $1\%$ level and with the same four SV specifications and SV-T at the $5\%$ level.

\begin{figure}[tbp]
\centering
\includegraphics[width = .9\textwidth]{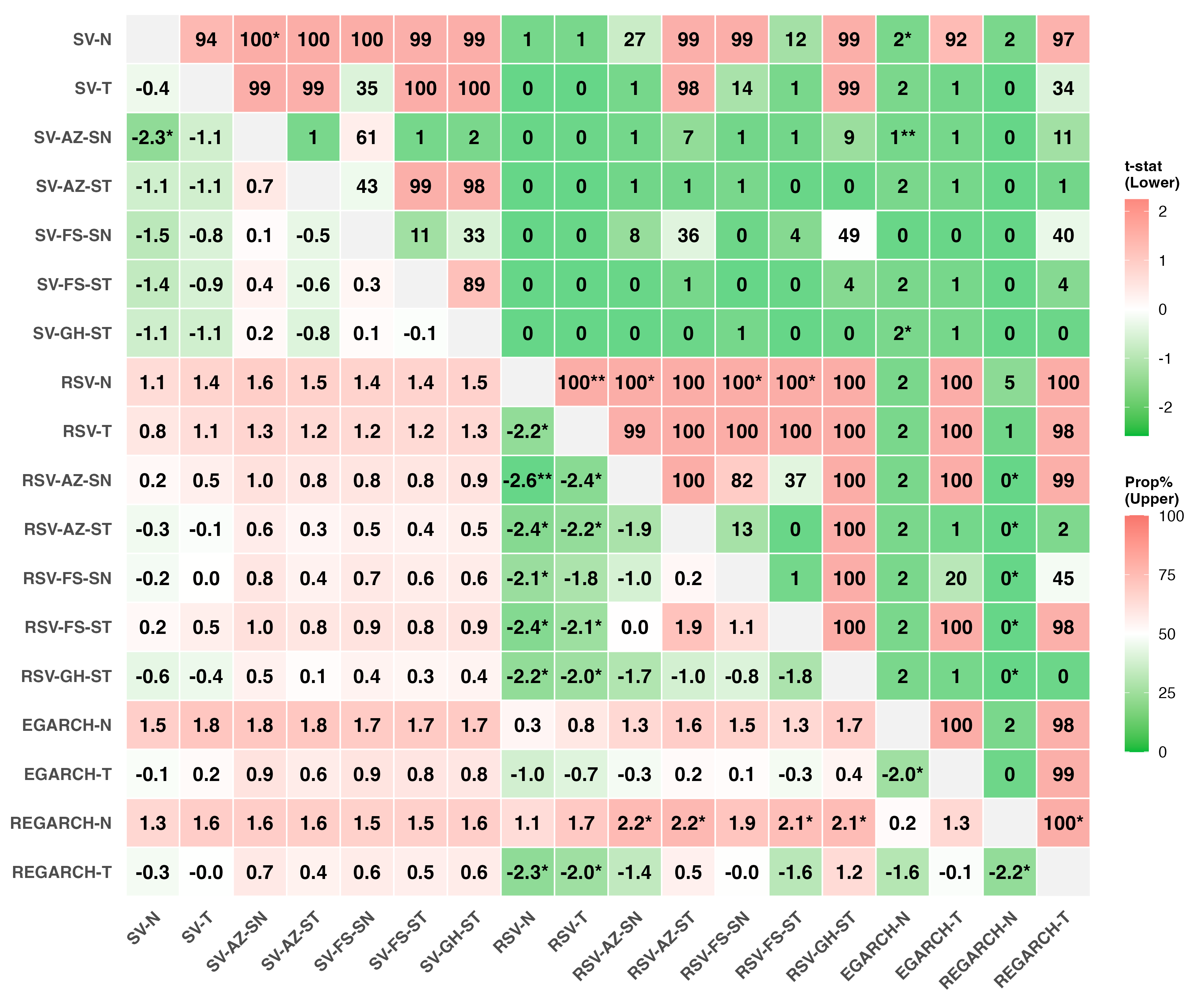}
\caption{Pairwise comparison of GW tests on the FZ0 loss of VaR and ES forecasts ($\alpha = 1\%$) for the DJIA. See Figure~\ref{fig:forecast-djia-vol-gwtest} for the reading of the heatmap.}
\label{fig:forecast-djia-var-es-gwtest-1p}
\end{figure}

\begin{figure}[tbp]
\centering
\includegraphics[width = .9\textwidth]{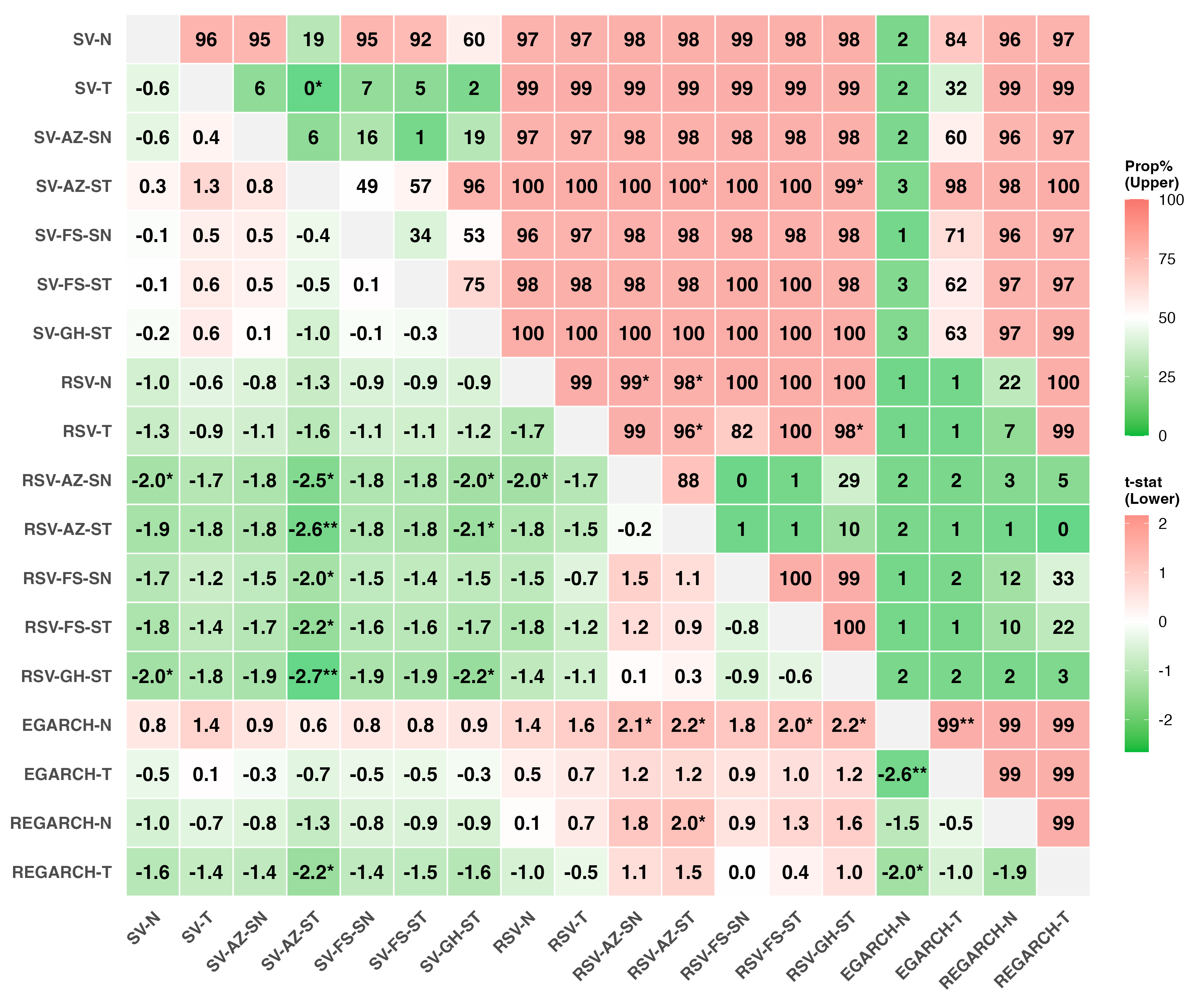}
\caption{Pairwise comparison of GW tests on the FZ0 loss of VaR and ES forecasts ($\alpha = 5\%$) for the DJIA. See Figure~\ref{fig:forecast-djia-vol-gwtest} for the reading of the heatmap.}
\label{fig:forecast-djia-var-es-gwtest-5p}
\end{figure}

\begin{figure}[tbp]
\centering
\includegraphics[width = .9\textwidth]{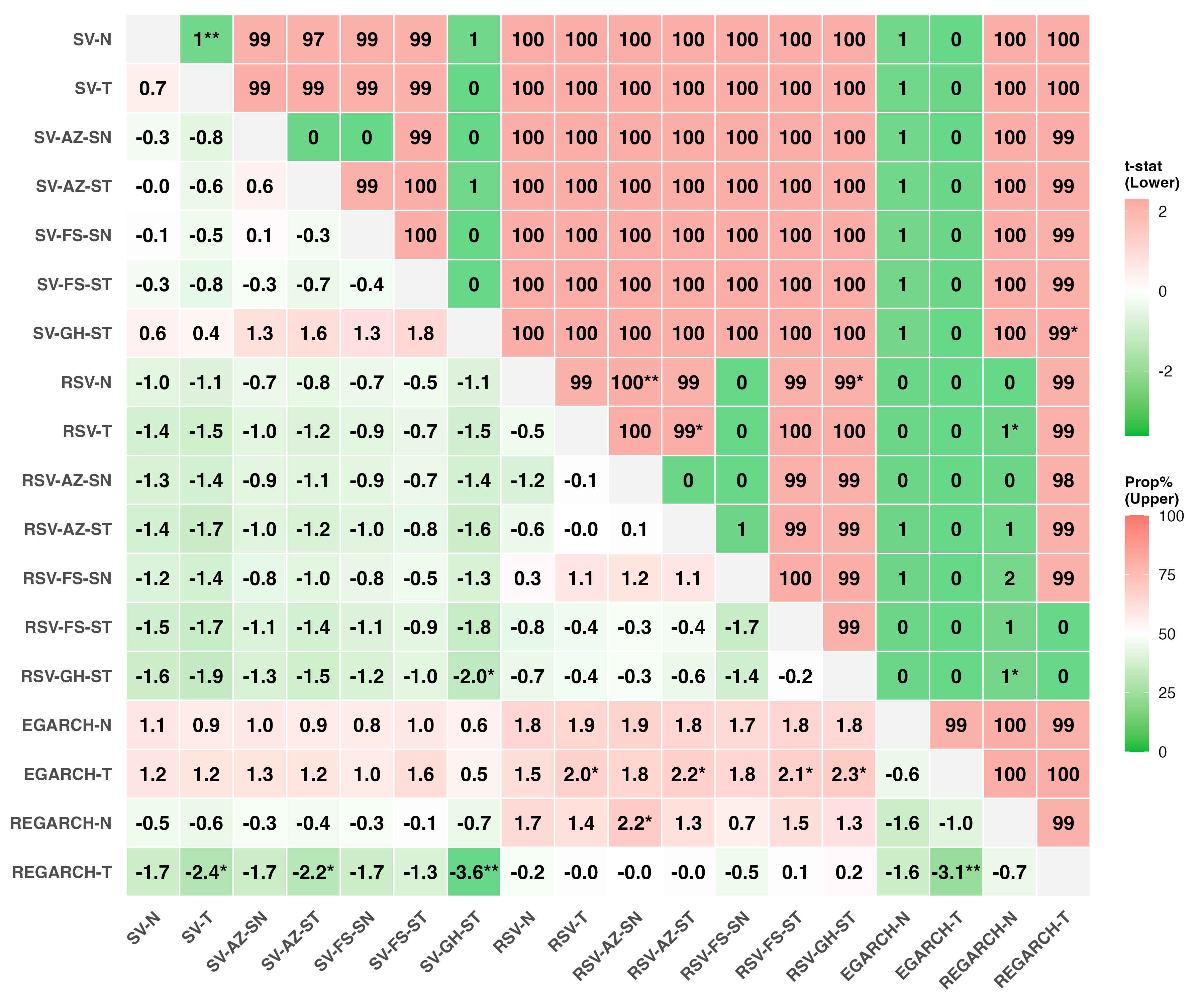}
\caption{Pairwise comparison of GW tests on FZ0 loss of VaR and ES forecasts ($\alpha = 1\%$) for the N225. See Figure~\ref{fig:forecast-djia-vol-gwtest-suppl} for additional details. Main sample (2009--2019).}
\label{fig:forecast-n225-var-es-gwtest-1p}
\end{figure}

\begin{figure}[tbp]
\centering
\includegraphics[width = .9\textwidth]{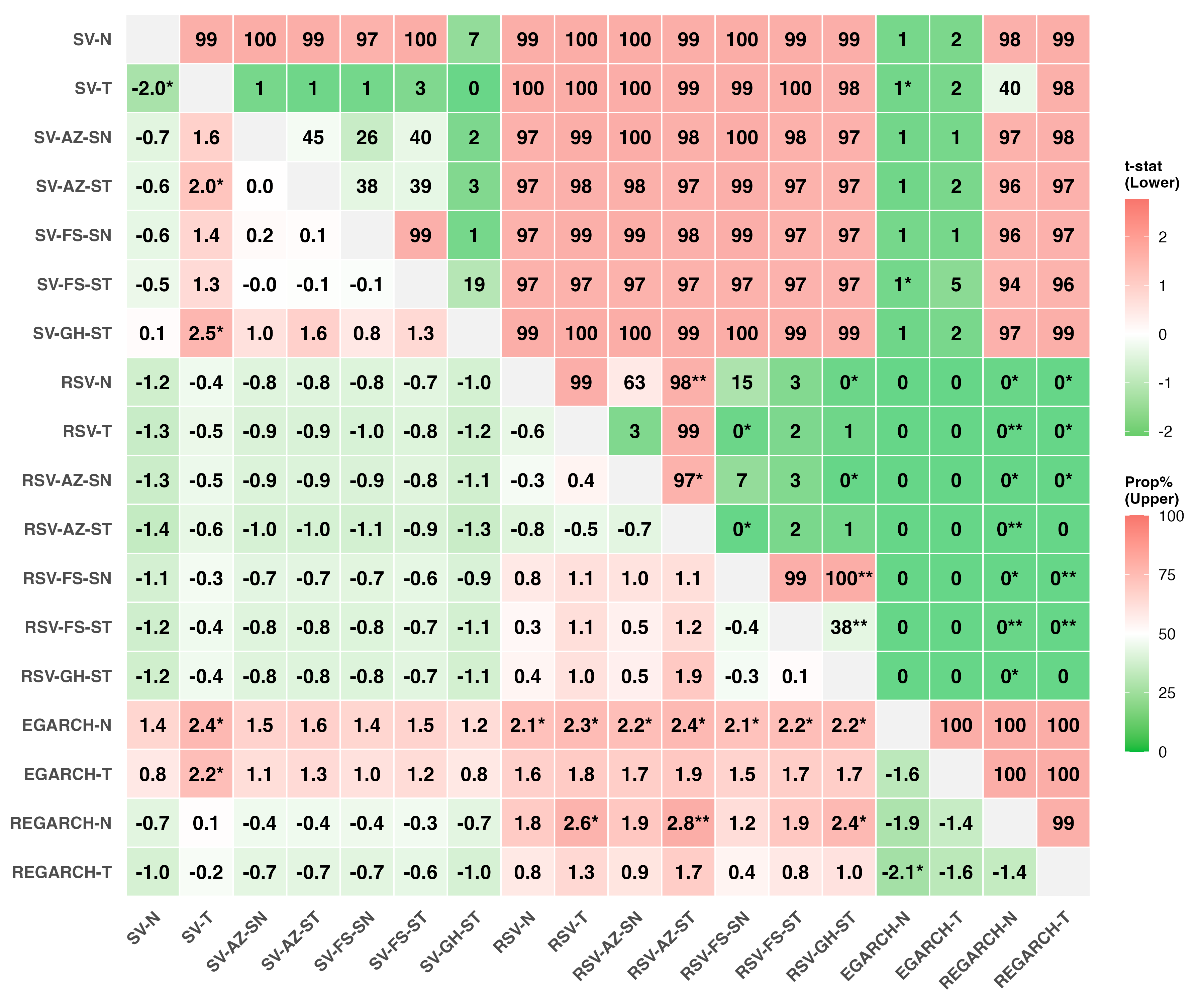}
\caption{Pairwise comparison of GW tests on FZ0 loss of VaR and ES forecasts ($\alpha = 5\%$) for the N225. See Figure~\ref{fig:forecast-djia-vol-gwtest-suppl} for additional details. Main sample (2009--2019).}
\label{fig:forecast-n225-var-es-gwtest-5p}
\end{figure}

Figures~\ref{fig:forecast-djia-var-es-gwtest-1p} to \ref{fig:forecast-n225-var-es-gwtest-5p} report the pairwise Giacomini--White tests on the FZ0 loss in the heatmap format of Section~\ref{sec:suppl-gw}.
For the seven matched pairs the unconditional test rejects equal predictive ability at the $5\%$ level only for the Azzalini skew-$t$ and GH skew-$t$ pairs for the DJIA at $\alpha = 5\%$, in favour of the RSV model (Table~3 of the main text); the remaining entries compare models across classes and across distributions.

Table~\ref{tab:stat-return-forecast} reports the mean and standard deviation of the skewness, the kurtosis, and several lower percentiles ($0.1\%$, $1\%$, $5\%$, and $10\%$) of the one-day-ahead predictive return distributions of the RSV models across the forecast days.
For both indices the RSV-FS-SN and RSV-GH-ST models generate the most negatively skewed and heavy-tailed predictive distributions, RSV-GH-ST with the most negative skewness and RSV-FS-SN with the highest kurtosis; for the N225 the RSV-FS-SN model produces the most conservative forecasts of the $0.1\%$ and $1\%$ percentiles while its $10\%$ percentile is higher than those of the other models.
These differences are what distinguishes the RSV specifications in the VaR and ES comparisons of Tables~4 and 5 of the main text.

\clearpage
\section{Sensitivity Analysis: Freely Estimated $\psi$}

As discussed in the main text, both the RSV and REGARCH measurement equations fix the slope coefficient $\psi$ at unity for parsimony and comparability.
This section reports the posterior distribution of $\psi$ when it is instead estimated freely, using a single in-sample MCMC estimation over the main (full) sample for each index.
The prior is $\psi \sim \mathcal{N}(1,1)$ for both the RSV and the REGARCH specifications.

Table~\ref{tab:psi-fullsample} reports the posterior mean, standard deviation, and 95\% credible interval of $\psi$ for each RSV distribution and for REGARCH, separately for the DJIA and the N225.
RSV-GH-ST is excluded from this table: the free-$\psi$ MCMC chain for this specification repeatedly diverges to a degenerate posterior mode (an issue documented further in the COVID-19 sensitivity analysis below), so its $\psi$ estimate is not reliable.

For every RSV specification and both indices, the 95\% credible interval for $\psi$ excludes unity and lies entirely below it, with posterior means ranging from $0.83$ to $0.93$.
For REGARCH, by contrast, the posterior mean of $\psi$ exceeds unity for both indices; the credible interval excludes unity for the DJIA (both REGARCH-N and REGARCH-T) but includes it for the N225 (both specifications, with REGARCH-N bordering on exclusion).
Fixing $\psi=1$ is therefore not uniformly supported as ``data-consistent'' across model classes: RSV estimates place $\psi$ below unity, while REGARCH estimates place it at or above unity.
Sections~\ref{sec:psi-covid-djia} and \ref{sec:psi-covid-n225} below report complementary sensitivity analyses based on out-of-sample forecast accuracy during the COVID-19 period.
They show that the forecasting consequences of this departure from unity depend on both the index and the model class: for the DJIA a freely estimated $\psi$ improves the RSV forecasts of volatility and of tail risk alike, for the N225 the RSV comparison is mixed, and for REGARCH a free $\psi$ is no better than $\psi=1$ on either index.

\section{COVID-19 Period Analysis: DJIA}
\label{sec:covid-djia}

To assess the robustness of our empirical findings during periods of extreme market stress, we conduct an independent analysis using data spanning from January 2012 to December 2020.
This sample includes the COVID-19 crisis and is analyzed separately from the main sample to avoid overlap and contamination of inference across subsamples.
The forecasts are evaluated over the 2020 calendar year against the same four volatility proxies used for the main sample.
Section~\ref{sec:covid-n225} repeats the exercise for the N225.
The RSV-GH-ST specification is excluded from this sample, for a reason specific to how that distribution is standardized; the SV-GH-ST specification is retained.\footnote{The GH skew-$t$ innovation is standardized by $\sqrt{\beta^2\sigma_\lambda^2+\mu_\lambda}$ with $\sigma_\lambda^2=\mathrm{Var}[\lambda_t]=2\nu^2/((\nu-2)^2(\nu-4))$, which exists only for $\nu>4$. On this sample the posterior of $\nu$ of the RSV-GH-ST model concentrates at that bound in most estimation windows, so $\mathrm{Var}[\lambda_t]$ diverges and the predictive mean of volatility ceases to be a usable point forecast. Reseeding every window three times under a prior that vanishes at the bound raises the number of windows with a well-identified $\nu$ from $90$ to $169$ of $253$, but the remaining failures concentrate after the March 2020 crash enters the estimation sample: $70$ of the first $75$ windows converge, against $99$ of the remaining $178$. The Azzalini specifications are unaffected on the same data, because their standardization never invokes $\sigma_\lambda^2$. We therefore treat this as an estimation problem of the RSV-GH-ST specification on this sample, under the standardization and the prior used here, and not as evidence about the RSV framework; whether it reflects a more general limit of the GH skew-$t$ distribution under extreme sample kurtosis is not something this sample can establish. The SV-GH-ST model, whose estimates pass the checks of Section~\ref{sec:checks} in every window of this sample, is retained.}

Table~\ref{tab:stats-djia-covid} reports descriptive statistics of daily returns and the logarithm of 5-minute RV for the DJIA over this sample.
Daily returns display pronounced negative skewness and excess kurtosis, and the Jarque--Bera test strongly rejects normality.
The Ljung--Box test also rejects the absence of linear autocorrelation in DJIA returns at conventional levels, which is not the case for the N225 (Section~\ref{sec:covid-n225}).
The log-RV series again exhibits strong serial dependence, together with positive skewness and leptokurtosis.

\begin{figure}[tbp]
\centering
\begin{tabular}{cc}
Return & Log-RV \\
\includegraphics[width=.45\textwidth]{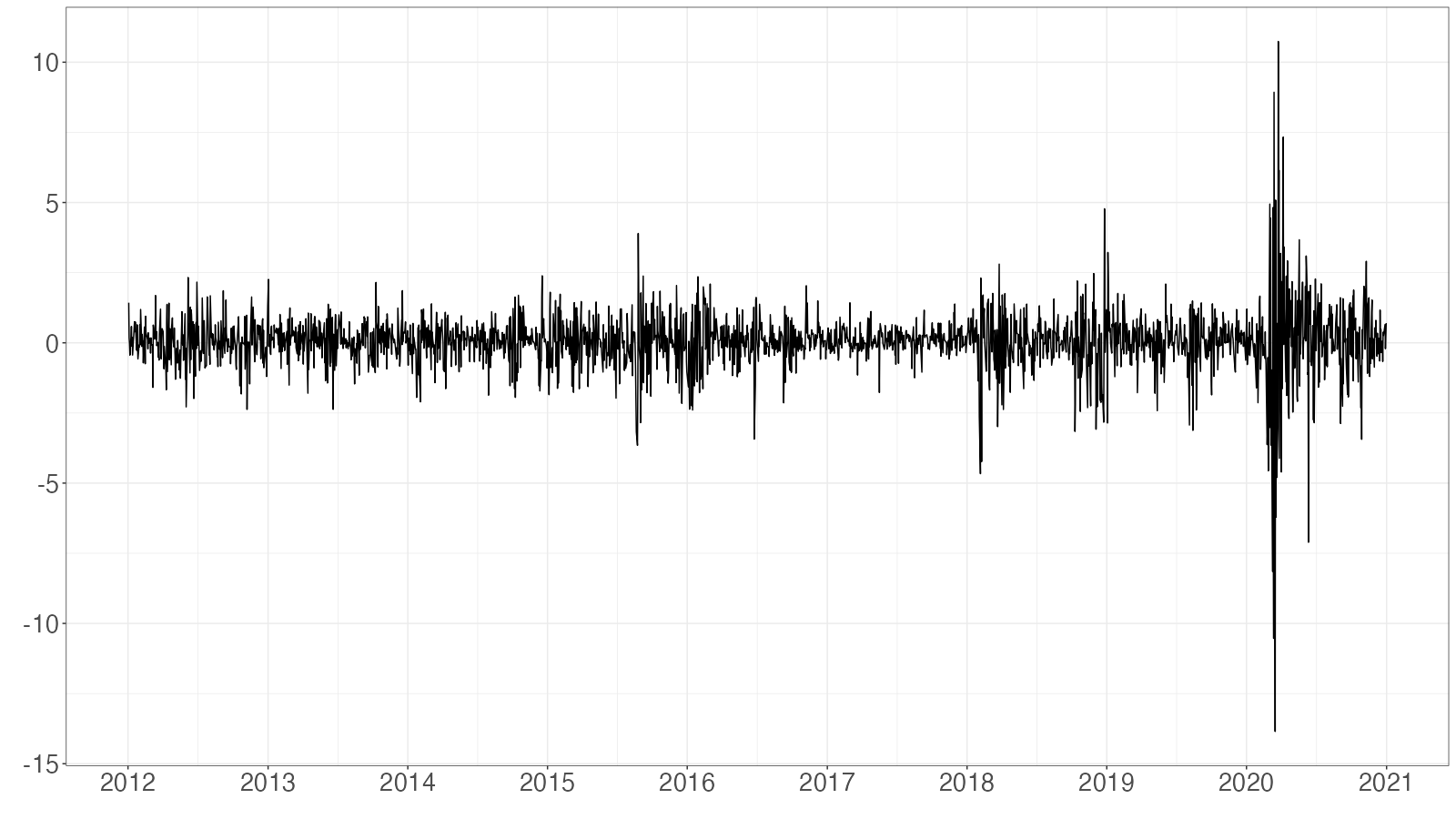} & \includegraphics[width=.45\textwidth]{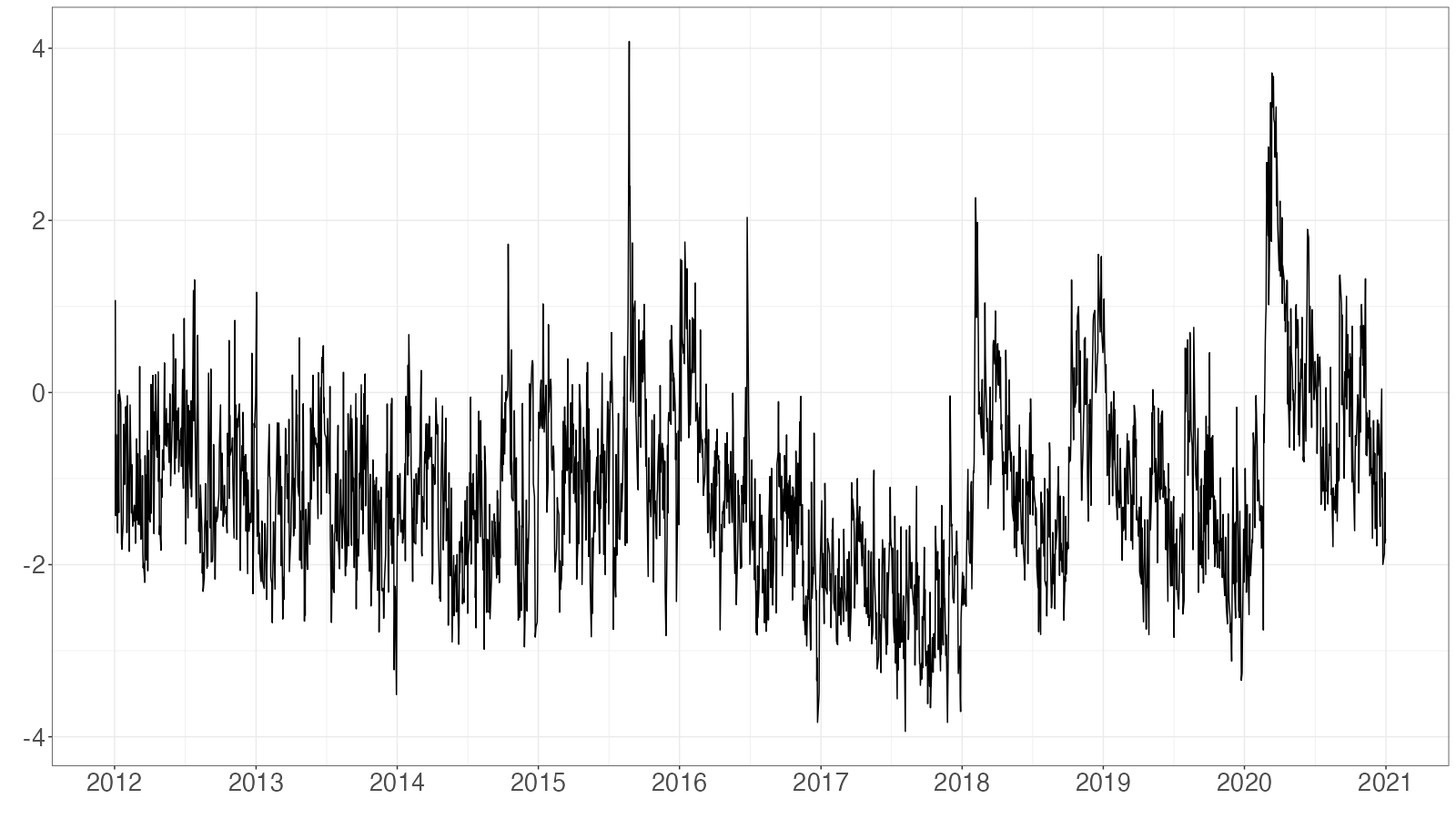} \\
\includegraphics[width=.45\textwidth]{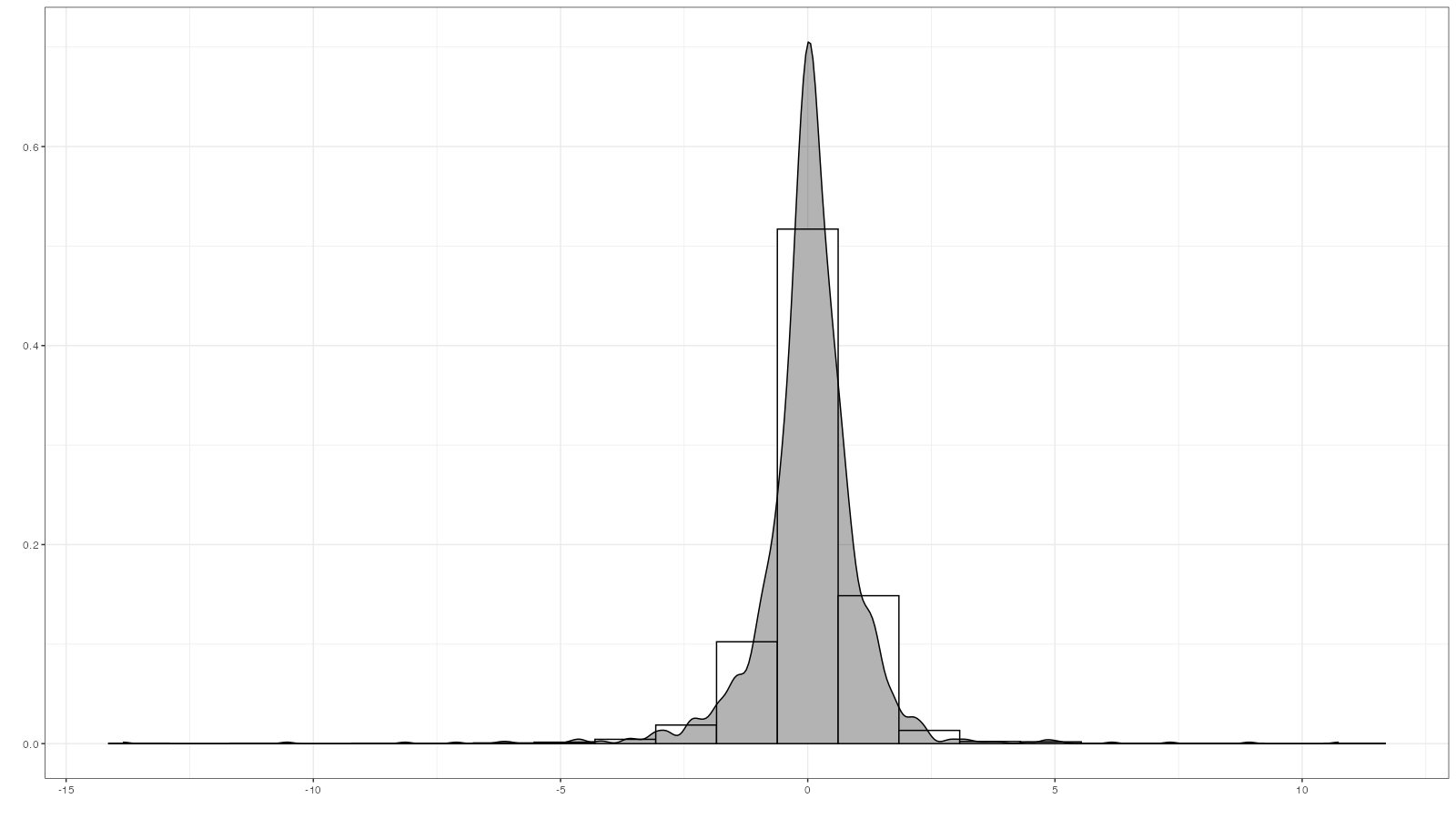} & \includegraphics[width=.45\textwidth]{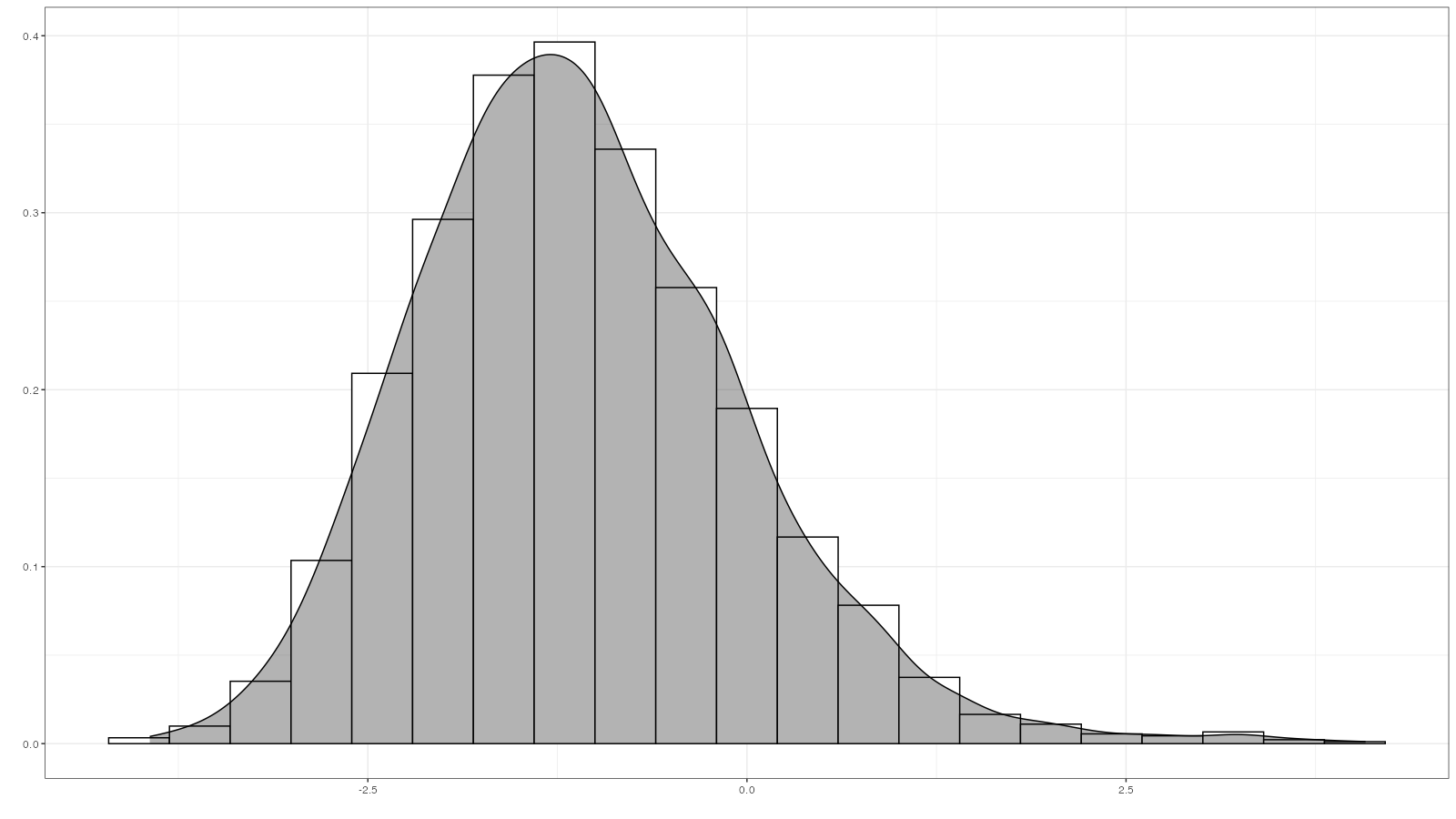}
\end{tabular}
\caption{Time series plots (top) and histograms (bottom) of daily returns (in percentage points) and the logarithms of 5-minute RVs for the DJIA (2012--2020).}
\label{fig:return-logrv-djia-covid}
\end{figure}

Figure~\ref{fig:return-logrv-djia-covid} plots the two series together with their histograms.
The COVID-19 episode stands out clearly.
The average log-RV rises from $-1.31$ over the first three weeks of February 2020 to $2.72$ over 9--31 March, an increase of roughly four log points, and the largest daily move of the entire sample in absolute value, $-13.85\%$, occurs on 16 March.
The elevation is also persistent: $72$ of the $253$ trading days in 2020 lie above the $95$th percentile of the 2012--2019 log-RV distribution, and every month from March to December has an average above the 2012--2019 mean of $-1.23$.
This combination of an abrupt level shift and a sustained aftermath provides a stringent environment for evaluating volatility and risk forecasting models.

\subsection{Volatility forecasts}

Table~\ref{tab:forecast-vol-qlike-djia-covid} reports the QLIKE scores across the four volatility proxies for the seventeen models.
The advantage of realized volatility observed over the main sample disappears on this sample: under RV5 the SV specifications attain scores of $0.258$ (SV-N) to $0.276$ (SV-FS-ST) against $0.264$ (RSV-AZ-ST) to $0.289$ (RSV-FS-SN) for the RSV specifications, and the same ordering holds under RK, BV, and Med.
REGARCH-T attains the lowest score under every proxy ($0.246$ under RV5) and the best average rank ($1.2$), followed by REGARCH-N ($0.257$) and SV-N ($0.258$); the EGARCH specifications are last ($0.362$ and $0.375$).
Within the RSV class the Azzalini skew-$t$ and Student's $t$ variants ($0.264$ and $0.266$) are ahead of the normal, Azzalini skew-normal, and Fern\'{a}ndez--Steel variants ($0.272$ to $0.289$); within the SV class the normal, GH skew-$t$, and Azzalini variants ($0.258$ to $0.264$) are ahead of the Fern\'{a}ndez--Steel variants ($0.274$ and $0.276$).
Under RV5 the $75\%$ MCS contains thirteen models, every SV specification, both REGARCH specifications, and the RSV-N, RSV-T, RSV-AZ-SN, and RSV-AZ-ST specifications; the two Fern\'{a}ndez--Steel RSV specifications and EGARCH-T belong only to the $90\%$ MCS, and EGARCH-N to neither.
For the N225 the RSV specifications remain ahead of the SV specifications on the same period (Section~\ref{sec:covid-n225}), so the loss of the RSV advantage is specific to the DJIA sample.

\subsection{VaR and ES forecasts}

\begin{figure}[tbp]
\centering
\begin{tabular}{cc}
RSV-AZ-ST (1\%) & RSV-AZ-ST (5\%) \\
\includegraphics[width = .45\textwidth]{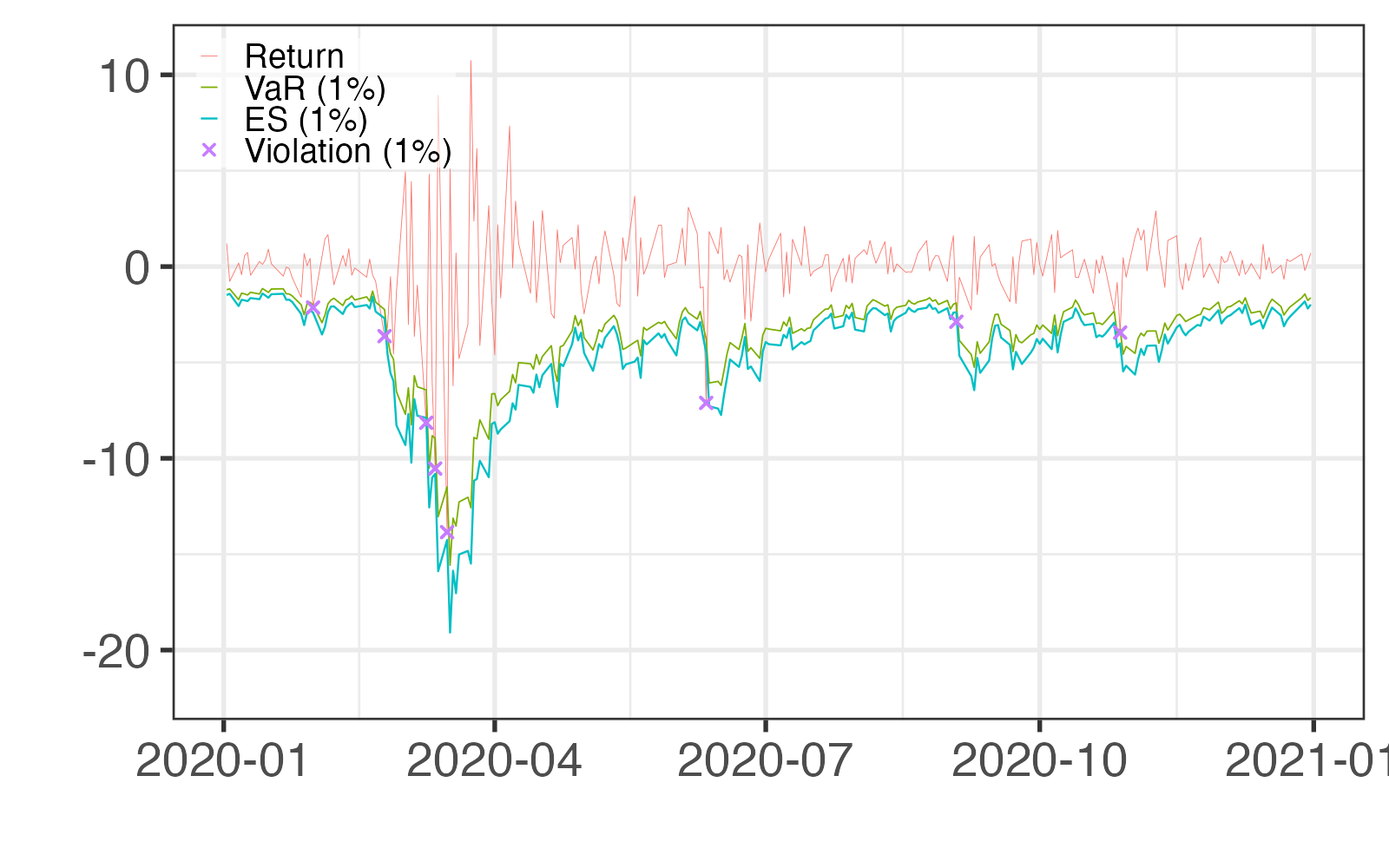} &
\includegraphics[width = .45\textwidth]{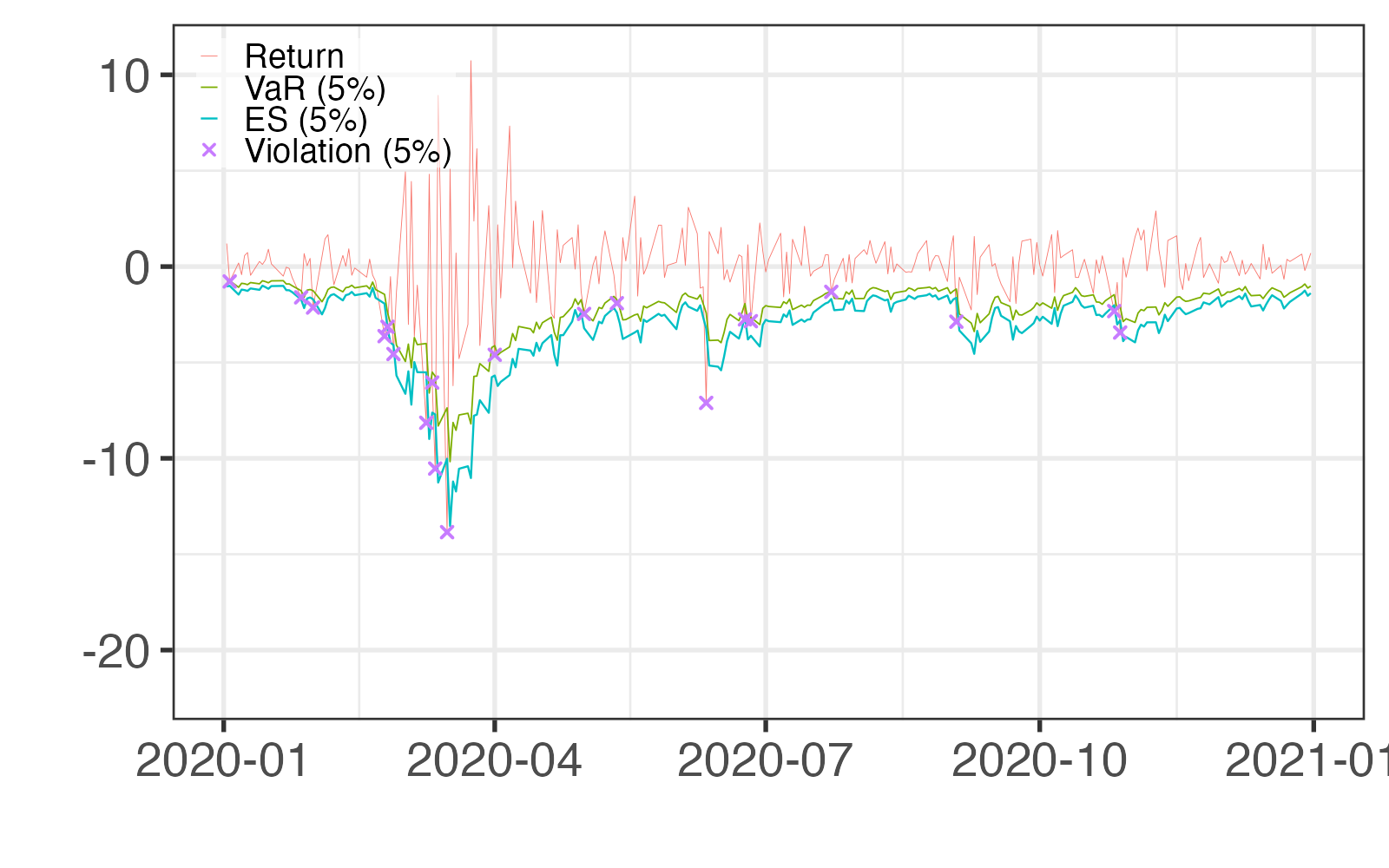} \\
REGARCH-T (1\%) & REGARCH-T (5\%) \\
\includegraphics[width = .45\textwidth]{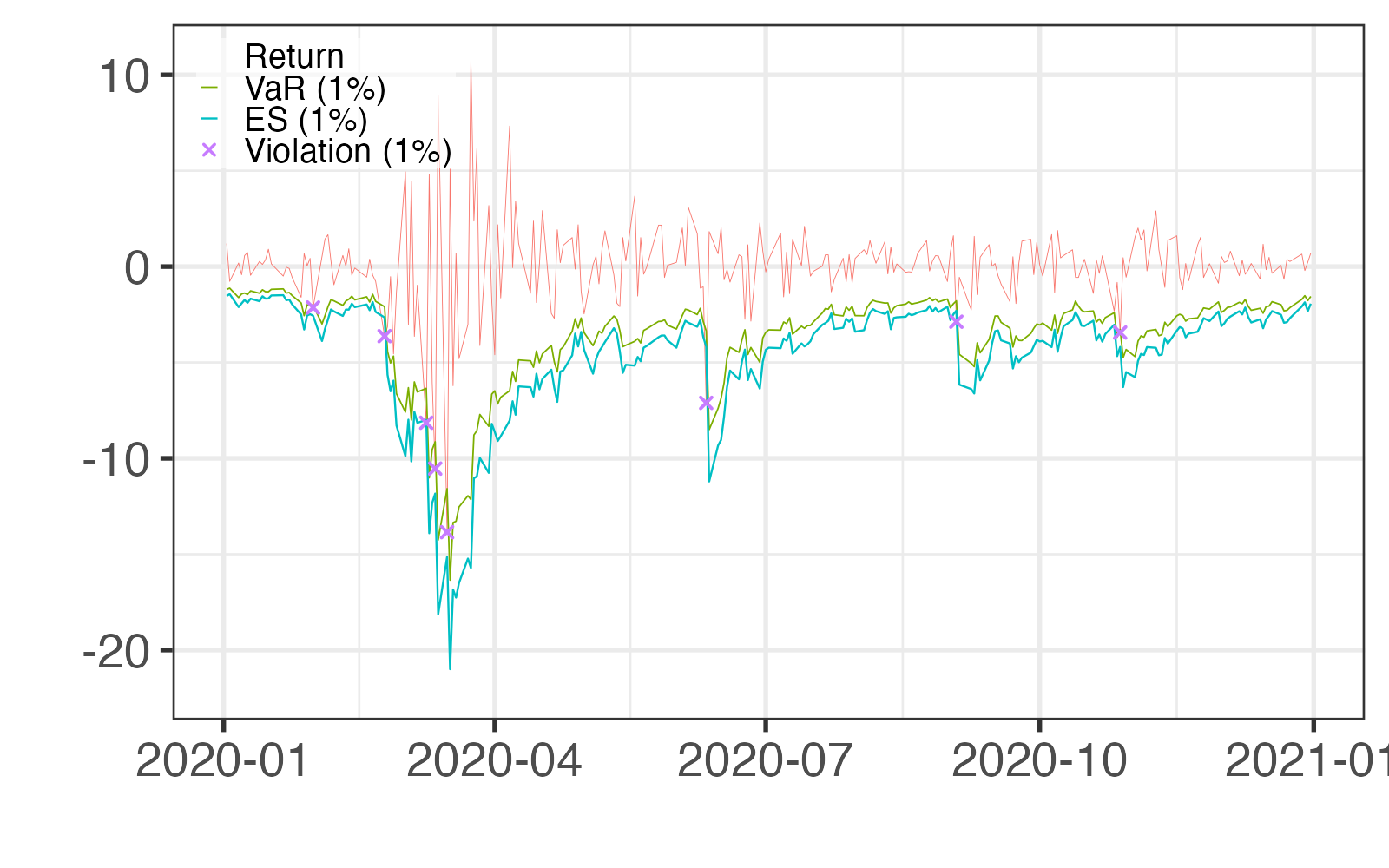} &
\includegraphics[width = .45\textwidth]{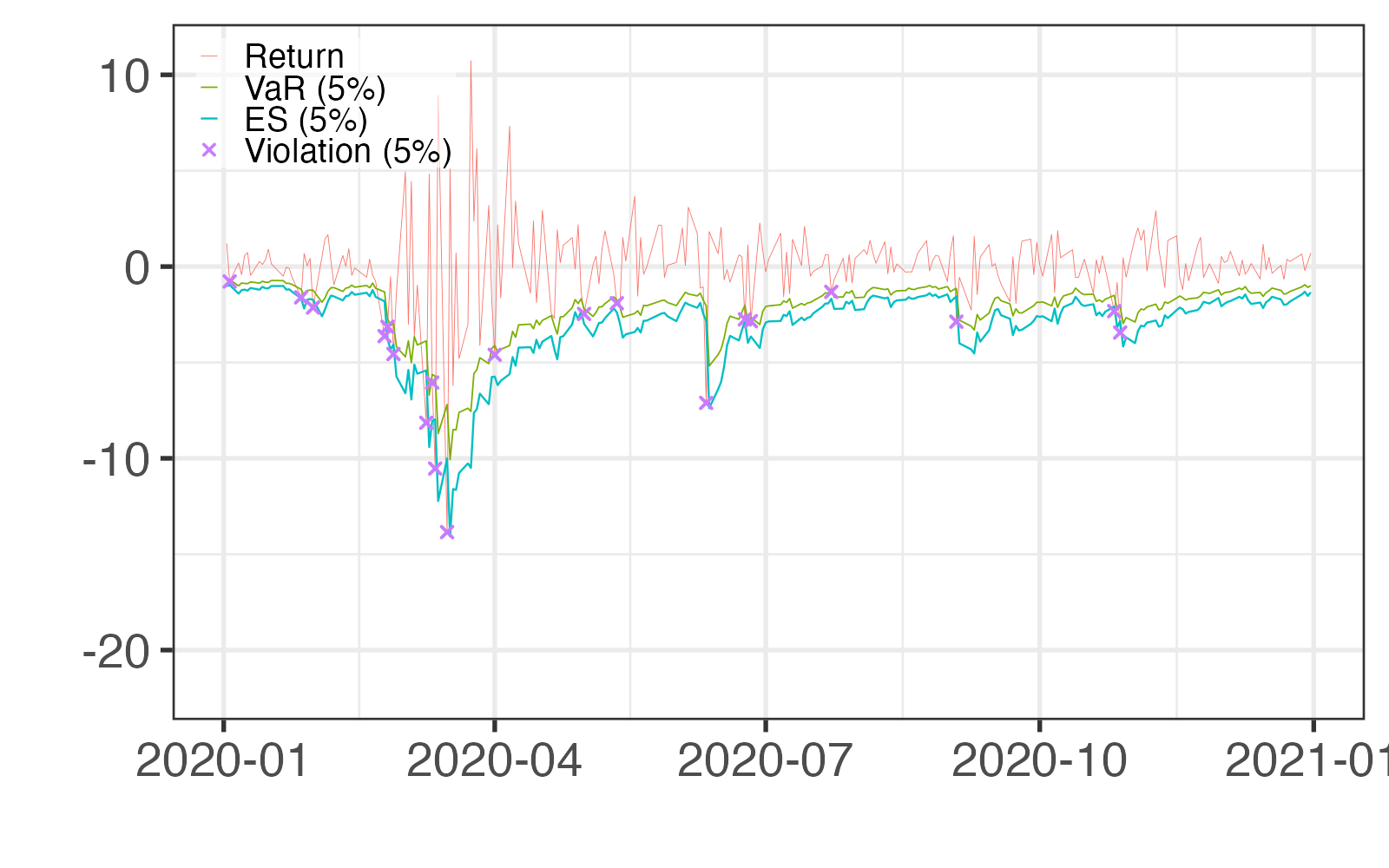}
\end{tabular}
\caption{VaR and ES forecasts of the RSV-AZ-ST and REGARCH-T models for the DJIA over the COVID-19 period (2020).}
\label{fig:pred-var-es-djia-covid}
\end{figure}

Figure~\ref{fig:pred-var-es-djia-covid} illustrates the VaR and ES forecasts generated by the RSV-AZ-ST and REGARCH-T models.
The two models violate the VaR on exactly the same dates, eight at the $1\%$ level and twenty at the $5\%$ level, three of the $1\%$ violations falling in March 2020.
For clarity, the forecasts from other models are omitted.

Table~\ref{tab:forecast-var-es-loss-djia-covid} reports violation rates, the $p$-values of the unconditional coverage (UC) test of \cite{kupiec_techniques_1995} and of the dynamic quantile (DQ) test of \cite{engle_caviar_2004}, as defined in Section~4.3.2 of the main text, FZ0 losses, and MCS $p$-values.
At the $1\%$ level the FZ0 losses range from $1.92$ (SV-GH-ST) to $2.81$ (EGARCH-N), and the comparison of matched pairs is uniformly unfavourable to the RSV models: all six RSV specifications have a higher loss than their SV counterparts.
The violation rates tell the same story, with the RSV models between $3.16\%$ and $4.35\%$ against $1.98\%$ to $3.16\%$ for the SV models at a nominal $1\%$, so incorporating realized volatility makes the models react more strongly than the data warrant at this quantile.
All seventeen models belong to the $90\%$ MCS, and all but EGARCH-N and REGARCH-N to the $75\%$ MCS, so the procedure does not separate them sharply.
The coverage tests, by contrast, reject for almost every model: at the $1\%$ level the UC test rejects at the $5\%$ significance level for 14 of the 17 models, the exceptions being SV-N, SV-FS-SN, and SV-FS-ST, and the DQ test rejects at the $1\%$ significance level for all but SV-FS-SN, SV-FS-ST, and EGARCH-T; at the $5\%$ level, where the violation rates lie between $6.72\%$ and $9.09\%$, the DQ test rejects for every model and the UC test for all but SV-FS-SN, SV-FS-ST, and SV-GH-ST.
At the $5\%$ level the evidence is more balanced: SV-FS-SN attains the lowest loss ($1.36$) with full MCS support, SV-AZ-SN and RSV-AZ-ST follow within $0.02$, four of the six matched pairs still favour the SV specification, and every model belongs to the $75\%$ MCS.

\begin{figure}[tbp]
\centering
\includegraphics[width = .9\textwidth]{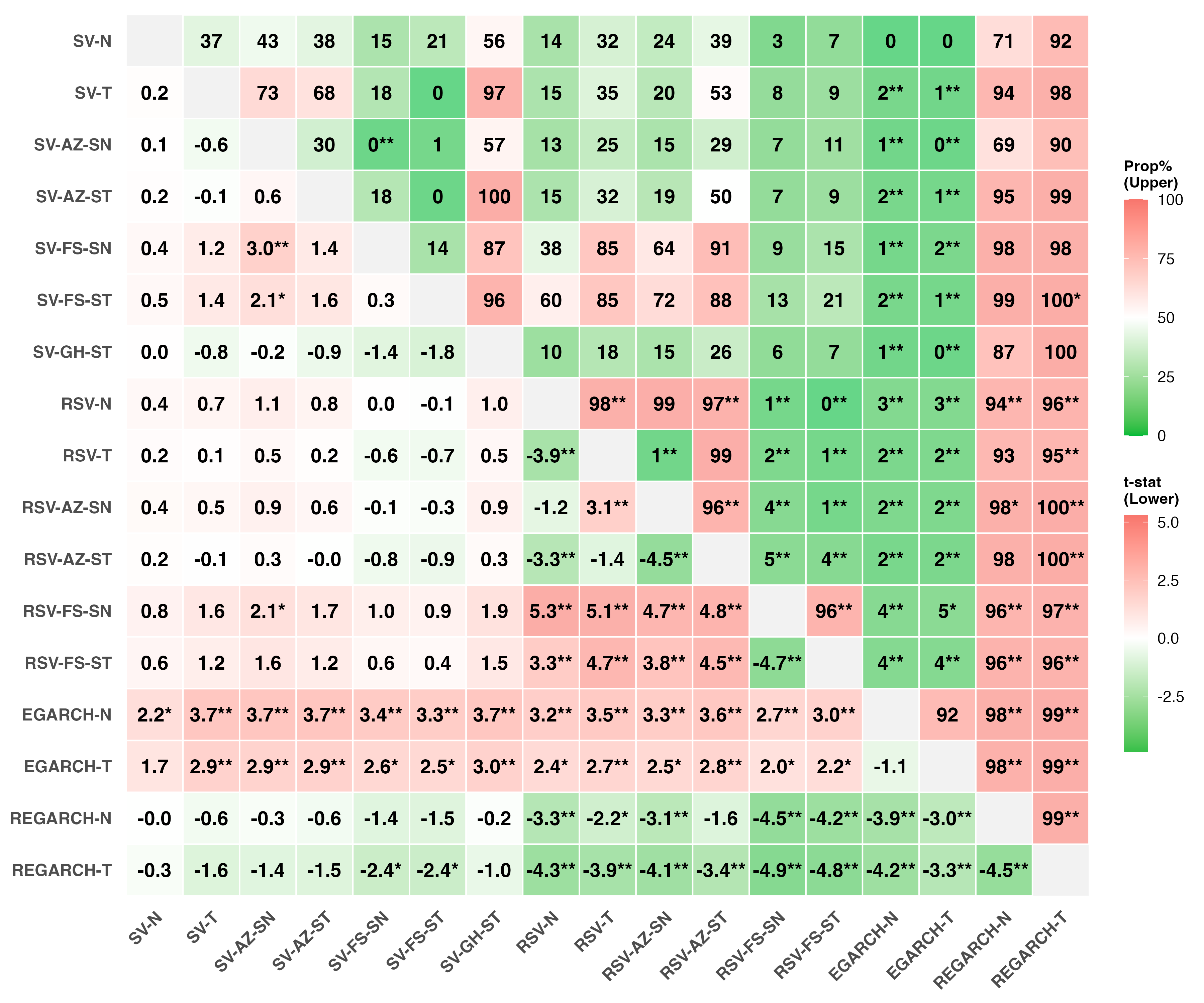}
\caption{Pairwise comparison of GW tests on QLIKE of volatility forecasts, based on RV5 as the volatility proxy, for the DJIA over the COVID-19 period (2020).
See Figure~\ref{fig:forecast-djia-vol-gwtest-suppl} for additional details.}
\label{fig:forecast-djia-vol-gwtest-covid}
\end{figure}

\begin{figure}[tbp]
\centering
\includegraphics[width = .9\textwidth]{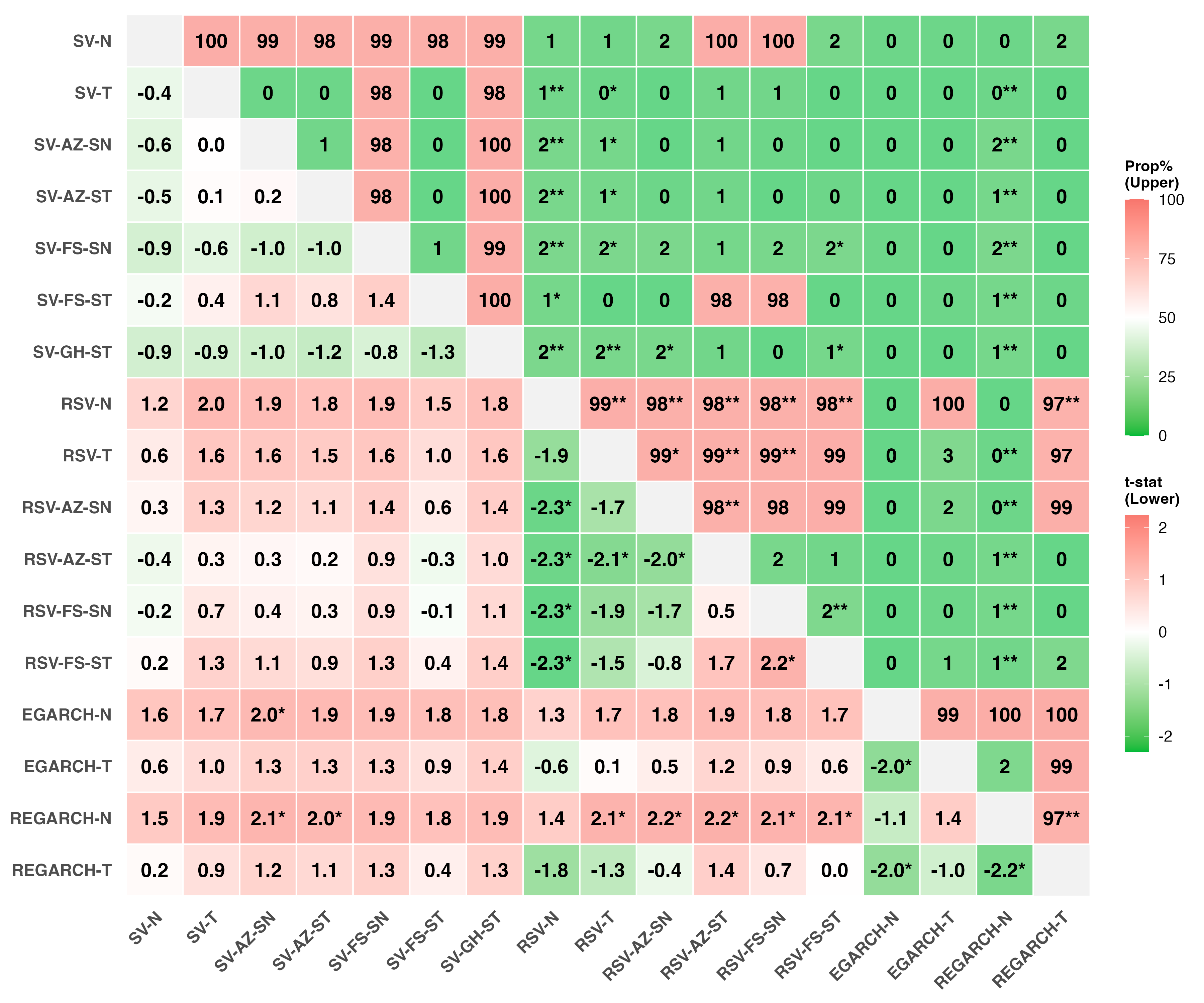}
\caption{Pairwise comparison of GW tests on FZ0 loss of VaR and ES forecasts ($\alpha = 1\%$) for the DJIA over the COVID-19 period (2020).
See Figure~\ref{fig:forecast-djia-vol-gwtest-suppl} for additional details.}
\label{fig:forecast-djia-var-es-gwtest-1p-covid}
\end{figure}

\begin{figure}[tbp]
\centering
\includegraphics[width = .9\textwidth]{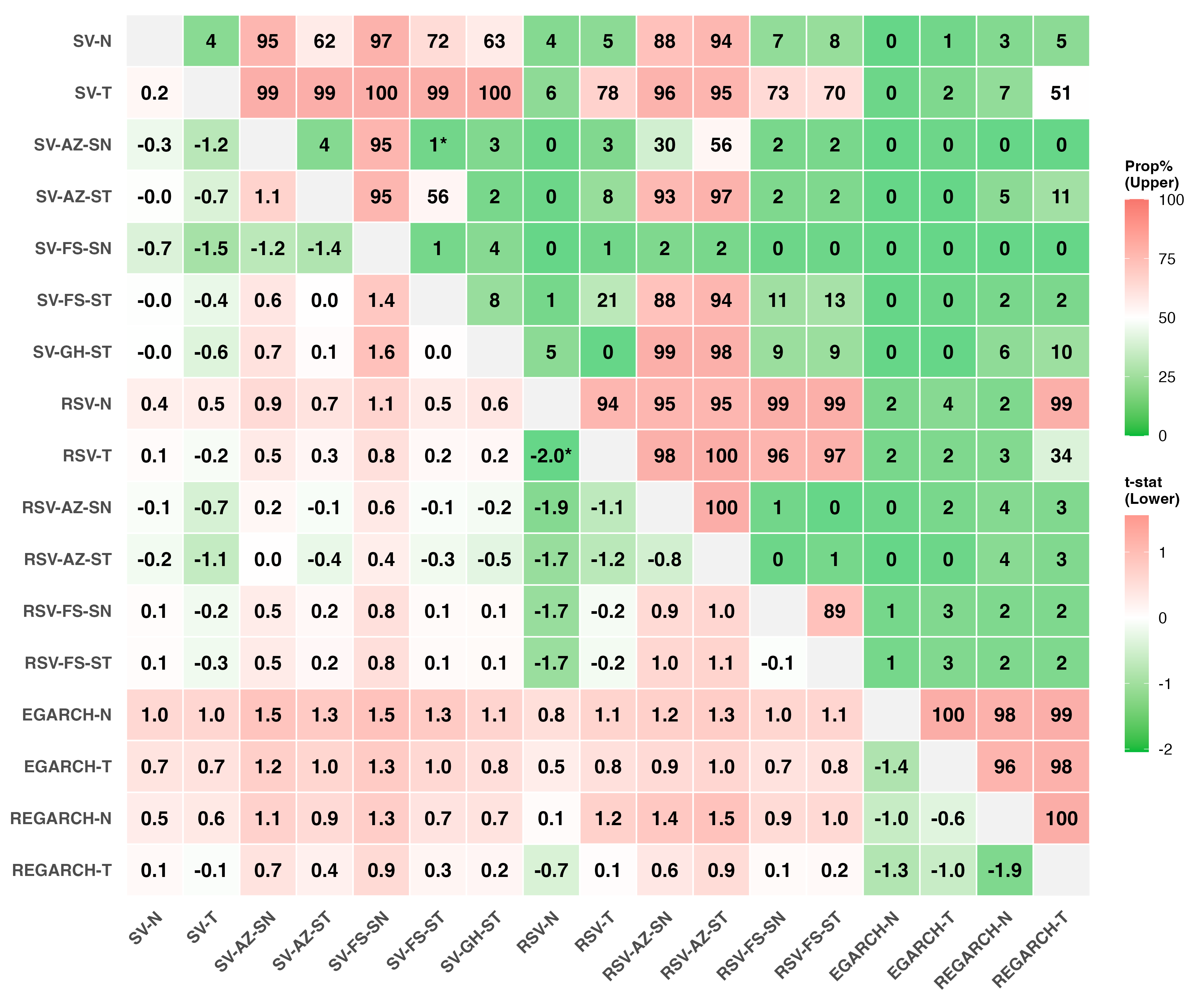}
\caption{Pairwise comparison of GW tests on FZ0 loss of VaR and ES forecasts ($\alpha = 5\%$) for the DJIA over the COVID-19 period (2020).
See Figure~\ref{fig:forecast-djia-vol-gwtest-suppl} for additional details.}
\label{fig:forecast-djia-var-es-gwtest-5p-covid}
\end{figure}

Figure~\ref{fig:forecast-djia-vol-gwtest-covid} reports the pairwise GW tests on the QLIKE loss under RV5 for this period, and Figures~\ref{fig:forecast-djia-var-es-gwtest-1p-covid} and \ref{fig:forecast-djia-var-es-gwtest-5p-covid} report them on the FZ0 loss.
For the six matched pairs the unconditional test on the QLIKE loss does not reject equal accuracy under any proxy, and on the FZ0 loss it rejects only for the Student's $t$ pair at the $1\%$ level, in favour of the SV model (Section~4.4 of the main text); the remaining entries compare models across classes and across distributions.

\subsection{Sensitivity analysis: freely estimated $\psi$} \label{sec:psi-covid-djia}

As discussed in the main text, the RSV and REGARCH measurement equations fix $\psi = 1$ for parsimony and comparability.
This subsection repeats the volatility and tail-risk evaluation with $\psi$ estimated freely instead, for the six RSV distributions and for REGARCH; Section~\ref{sec:psi-covid-n225} does the same for the N225.
RSV-GH-ST is excluded for the same reason as in the main comparison above: the posterior of $\nu$ concentrates at the lower bound required by the standardization of that distribution, so the predictive mean is not a usable point forecast.

Table~\ref{tab:forecast-vol-qlike-djia-covid-psi} reports QLIKE scores under $\psi=1$ and a freely estimated $\psi$.
For every RSV distribution, a freely estimated $\psi$ lowers the QLIKE score under all four proxies, by between $14\%$ and $17\%$ on average.
RSV-AZ-SN improves the most in relative terms, its average QLIKE falling from $0.281$ to $0.233$, RSV-AZ-ST attains the lowest free-$\psi$ score ($0.230$), and RSV-FS-SN falls from $0.300$ to $0.257$.
Under RV5 and RK the free-$\psi$ variants of all six distributions enter the $75\%$ MCS and none of the $\psi=1$ variants does; under BV and Med the free-$\psi$ variants of the Fern\'{a}ndez--Steel distributions receive weaker support.
For REGARCH, the effect goes the other way and is an order of magnitude smaller: a free $\psi$ raises the average QLIKE of REGARCH-N from $0.261$ to $0.267$ and of REGARCH-T from $0.249$ to $0.250$, and both variants of REGARCH-T remain in the $75\%$ MCS under every proxy.

Table~\ref{tab:forecast-var-es-loss-djia-covid-psi} reports the corresponding violation rates, UC and DQ test $p$-values, and FZ0 losses for VaR and ES forecasts.
The RSV results point the same way as for volatility: at both the $1\%$ and $5\%$ levels a freely estimated $\psi$ lowers the FZ0 loss for all six distributions, and the violation rates move markedly closer to the nominal level, falling from between $3.16\%$ and $4.35\%$ to between $1.98\%$ and $3.56\%$ at the $1\%$ level.
Under $\psi = 1$ both coverage tests reject for every RSV specification at both levels; under a free $\psi$ neither test rejects at the $5\%$ significance level for the normal, Student's $t$, and Azzalini specifications at either risk level, whereas for the Fern\'{a}ndez--Steel specifications the UC test still rejects at $\alpha = 1\%$ and the DQ test at both levels.
RSV-AZ-ST with a freely estimated $\psi$ attains the lowest FZ0 loss at both levels and retains full MCS support.
For REGARCH, the differences are again small: a free $\psi$ raises the FZ0 loss for REGARCH-N at both levels, lowers it slightly for REGARCH-T, and leaves the violation rates unchanged at the $1\%$ level.

Comparing these figures with the SV and EGARCH results of Table~\ref{tab:forecast-var-es-loss-djia-covid} sharpens the point, because $\psi$ is what decides the matched-pair comparison reported in the main text.
Under $\psi=1$ every RSV specification carries a higher FZ0 loss than its SV counterpart at the $1\%$ level, which is the finding on which that comparison rests.
Under a freely estimated $\psi$ every RSV specification carries a \emph{lower} FZ0 loss than its SV counterpart, at both the $1\%$ and the $5\%$ level, and the best free-$\psi$ RSV specification attains an FZ0 loss of $1.78$ at the $1\%$ level and $1.30$ at the $5\%$ level, against $1.92$ and $1.36$ for the best SV specification and $2.26$ and $1.47$ for the best EGARCH specification.
The apparent tail-risk disadvantage of the RSV models on this sample is therefore an artefact of fixing $\psi$ at unity rather than a property of the RSV framework.

For the DJIA, then, estimating $\psi$ freely improves the RSV forecasts on both dimensions and for every distribution, while for REGARCH it delivers no gain.
Fixing $\psi=1$ remains defensible as a device for keeping the RSV and REGARCH measurement equations comparable, but it is not what a free estimate would select for the RSV models on this sample.
Section~\ref{sec:psi-covid-n225} shows that the N225 does not reproduce this pattern.

Overall, the DJIA COVID-19 evidence departs from the main-sample findings on both dimensions under $\psi = 1$: incorporating realized volatility does not lower the QLIKE score under any of the four proxies, and the tail-risk forecasts of the RSV models fall behind their SV counterparts.
Once the free-$\psi$ RSV models are included in the comparison, however, the RSV specifications improve on \emph{both} dimensions.
They attain a lower QLIKE than any $\psi=1$ model under all four proxies, and a lower FZ0 loss than their SV counterparts in all six matched pairs at both risk levels, so that neither shortfall recorded under $\psi=1$ survives relaxing that restriction.
RSV-AZ-ST is the best RSV specification for volatility under $\psi=1$ and, once $\psi$ is estimated freely, the best specification for both volatility and tail risk.
Section~\ref{sec:covid-n225} reports the corresponding N225 results, for which the RSV specifications remain ahead of the SV specifications in volatility forecasting but a free $\psi$ brings no improvement.
\section{COVID-19 Period Analysis: N225}
\label{sec:covid-n225}

We repeat the exercise of Section~\ref{sec:covid-djia} for the N225, using the same January 2012--December 2020 sample and the same four volatility proxies.
The model set differs in one respect: here we exclude only the SV-GH-ST model, whose volatility forecasts become highly unstable and exhibit extreme spikes during this period, whereas for the DJIA both GH skew-$t$ specifications had to be dropped for the reason set out above.

The analysis focuses on daily close-to-close returns and the logarithms of 5-minute RVs for the N225. Table~\ref{tab:stats-n225-covid} reports the corresponding descriptive statistics.
As in the full sample, daily returns exhibit pronounced non-normality, with negative skewness and excess kurtosis.
The Jarque--Bera test strongly rejects normality, while the Ljung--Box statistic does not indicate significant linear autocorrelation in returns.
The log-RV series displays strong serial dependence, as evidenced by the Ljung--Box test, together with positive skewness and leptokurtosis.

\begin{figure}[tbp]
\centering
\begin{tabular}{cc}
Return & Log-RV \\
\includegraphics[width=.45\textwidth]{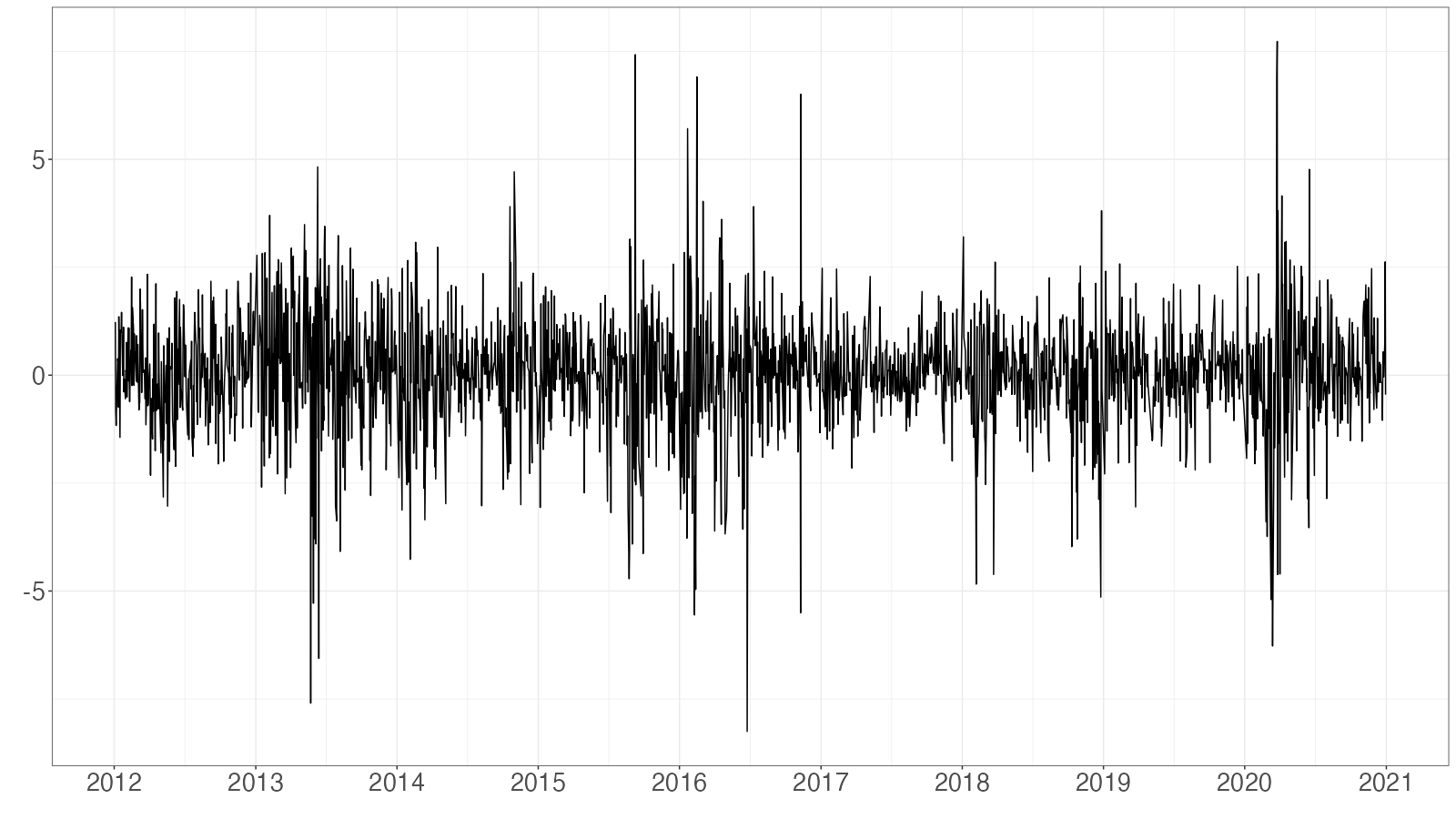} & \includegraphics[width=.45\textwidth]{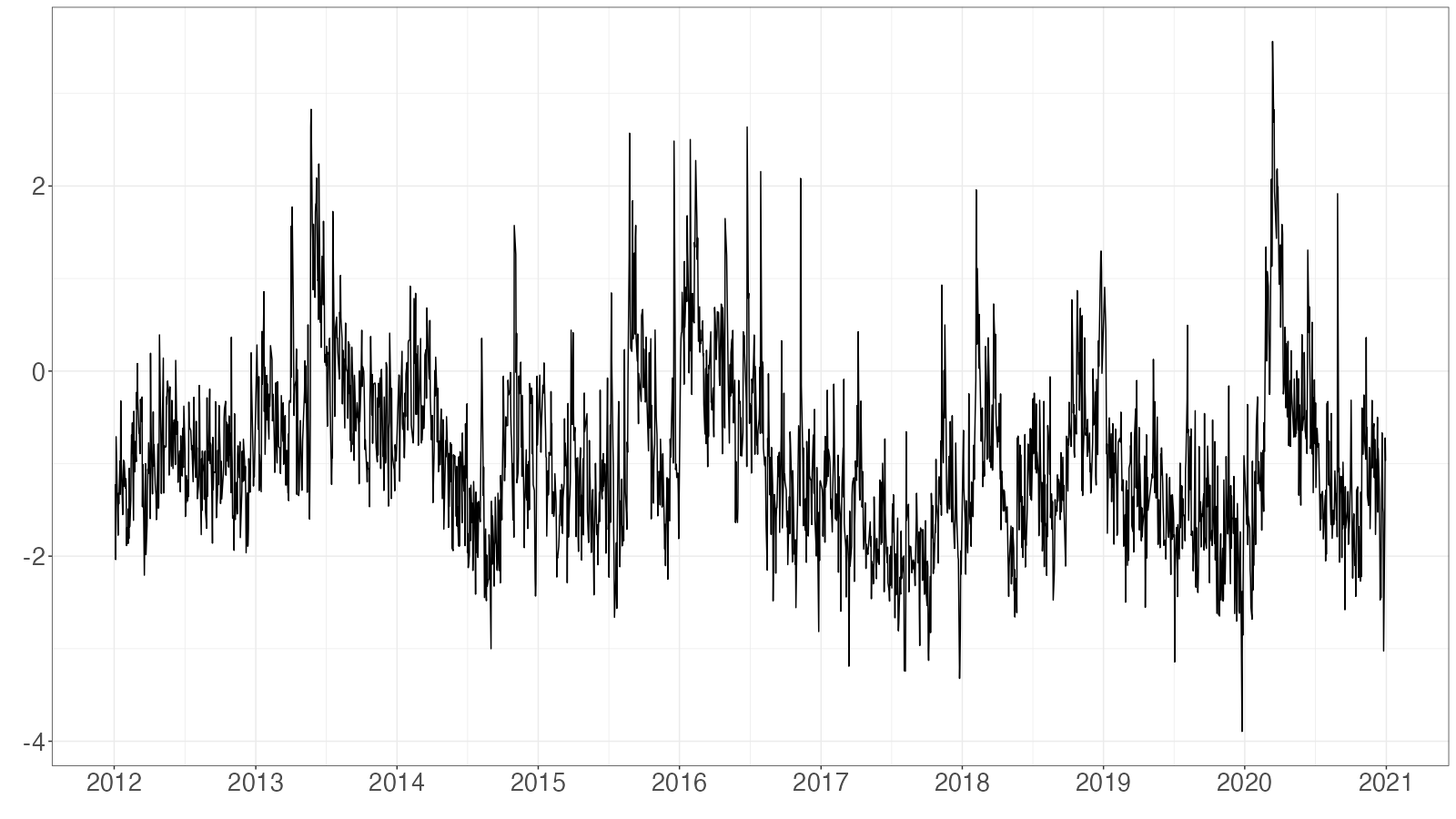} \\
\includegraphics[width=.45\textwidth]{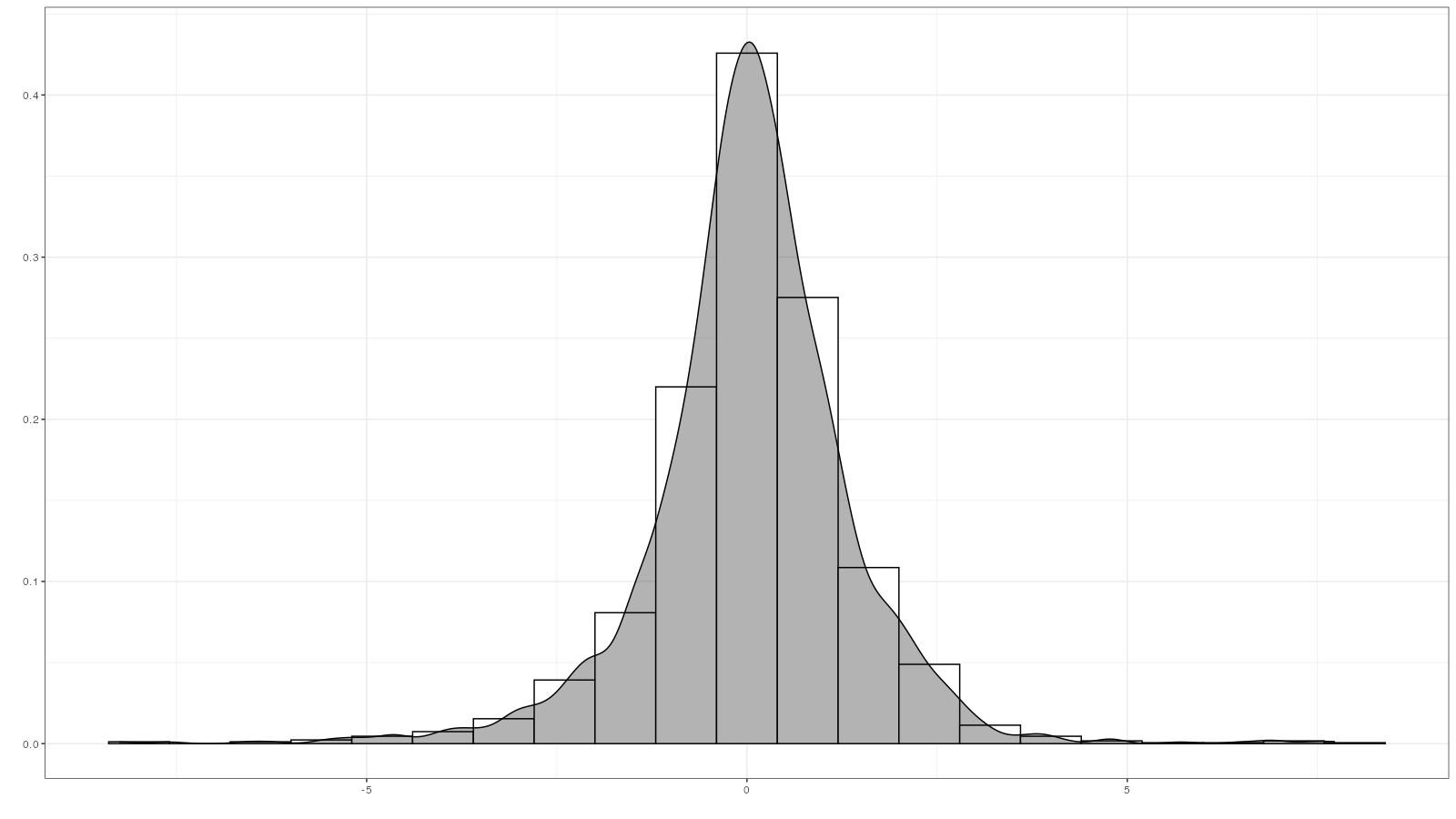} & \includegraphics[width=.45\textwidth]{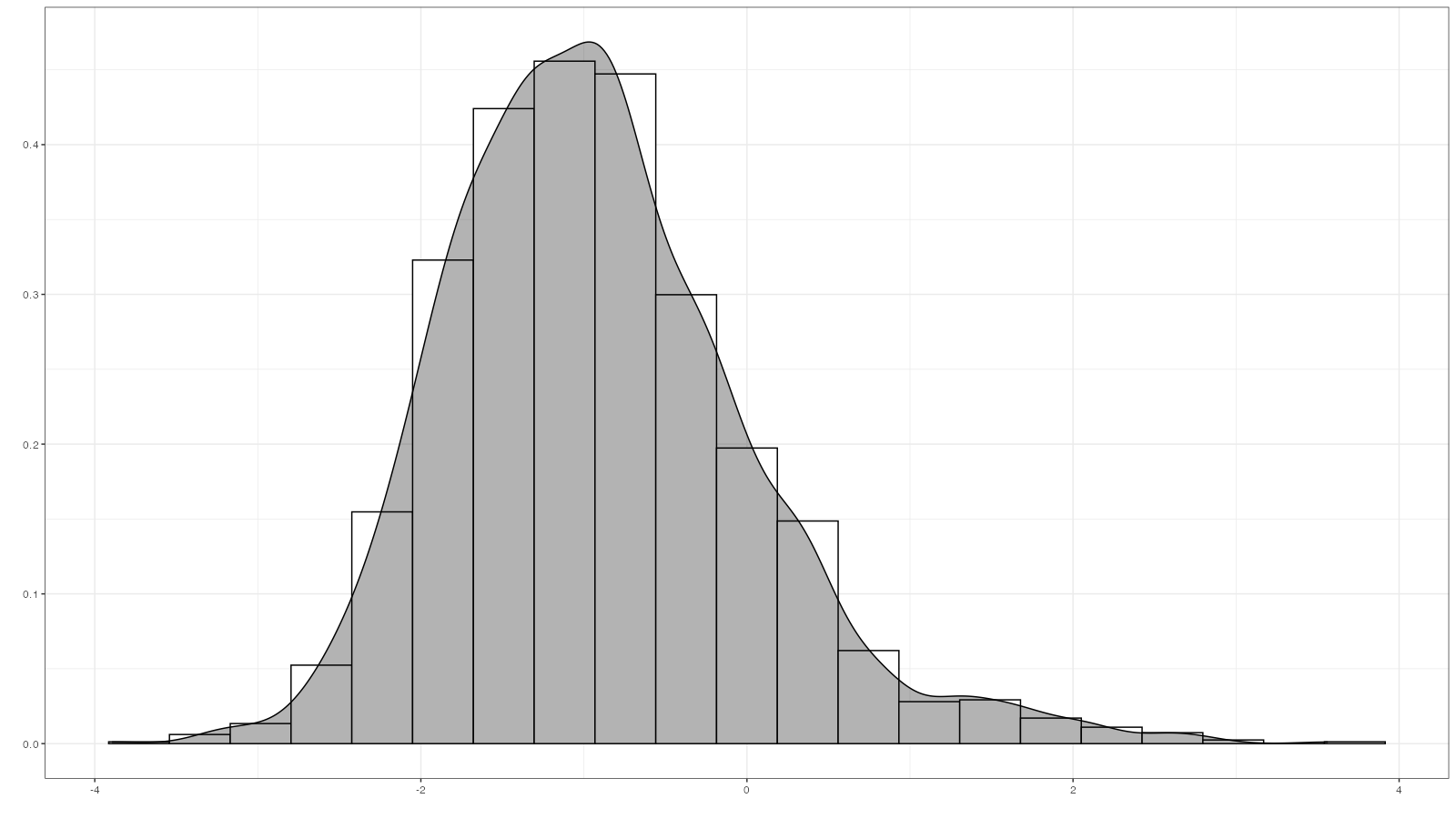}
\end{tabular}
\caption{Time series plots (top) and histograms (bottom) of daily returns (in percentage points) and the logarithms of 5-minute RVs for the N225 (2012--2020).}
\label{fig:return-logrv-n225-covid}
\end{figure}

The time series plots in Figure~\ref{fig:return-logrv-n225-covid} show an abrupt increase in RV from late February 2020 onward, coinciding with the global spread of COVID-19.
The average log-RV rises from $-1.10$ over the first three weeks of February to $1.94$ over 9--31 March, and the highest value of the sample, $3.56$, is reached on 13 March.
The episode is nonetheless short-lived, and in level terms it is less exceptional than for the DJIA.
Only $27$ of the $242$ trading days in 2020 lie above the $95$th percentile of the 2012--2019 log-RV distribution, against $72$ of $253$ for the DJIA.
The monthly average falls from $1.52$ in March to $0.31$ in April, already below the $0.71$ and $0.75$ recorded in January and February 2016, and the mean for 2020 as a whole, $-0.74$, is close to the 2012--2019 mean of $-0.94$.
The elevated level of March 2020 is thus comparable in magnitude with earlier episodes such as mid-2013 and early 2016, the monthly averages for June 2013 and February 2016 being $1.14$ and $0.75$ against $1.52$ for March 2020.
What distinguishes the period is therefore the speed of the transition rather than a persistently higher level of volatility, which still provides a demanding environment for evaluating volatility and risk forecasting models.

\subsection{Volatility forecasts}

\begin{figure}[tbp]
\centering
\includegraphics[width = \textwidth]{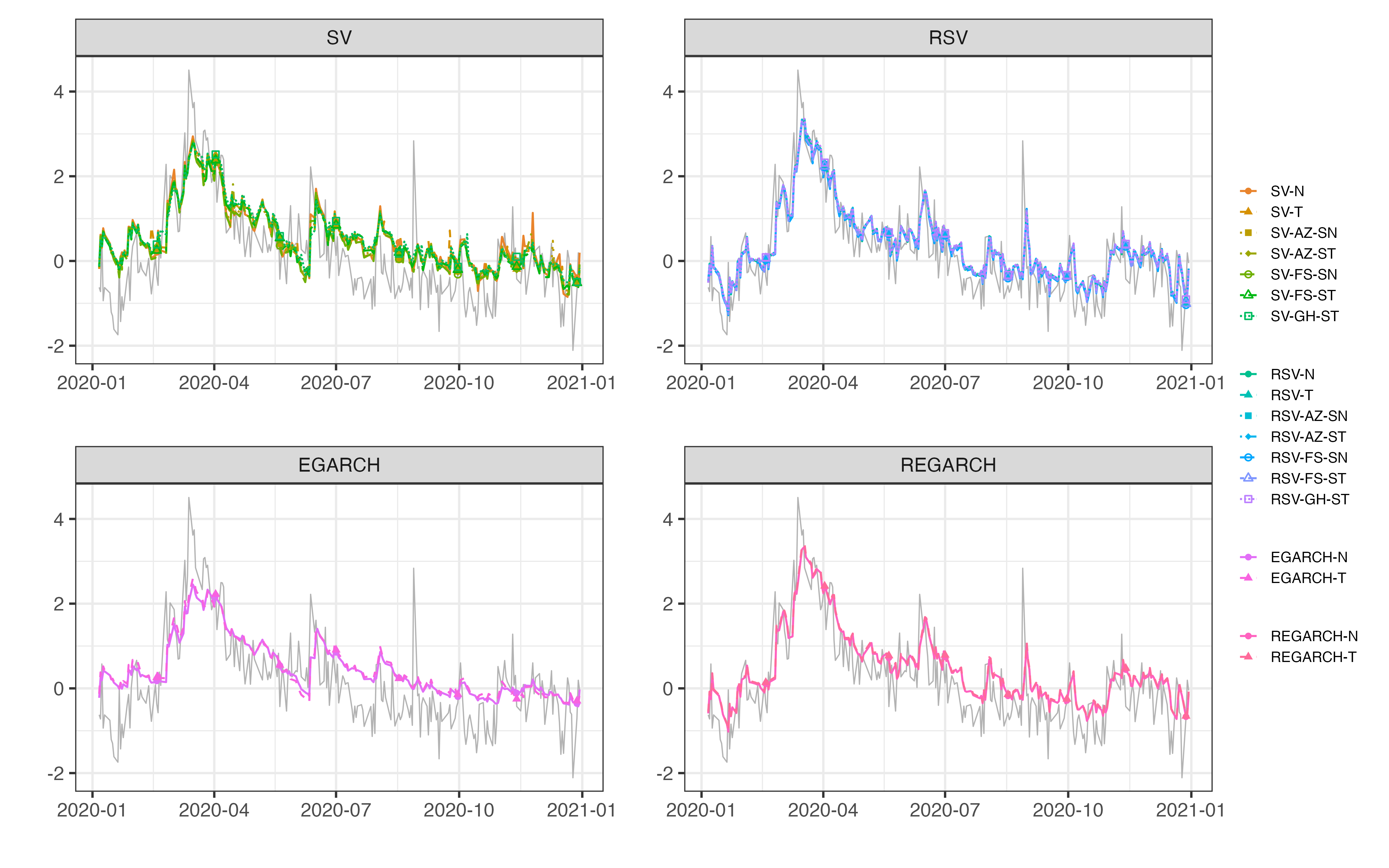}
\caption{Log-scale volatility forecasts together with the adjusted RV5 for the N225 over the COVID-19 period (2020).}
\label{fig:pred-vol-n225-covid}
\end{figure}

Figure~\ref{fig:pred-vol-n225-covid} displays the log-scale volatility forecasts alongside the adjusted RV5 series.
Forecast patterns are similar across models, while the RSV and REGARCH models follow the volatility spikes more closely.
These observations are consistent with the QLIKE scores in Table~\ref{tab:forecast-vol-qlike-n225-covid} for the eighteen models: under RV5 the RSV specifications range from $0.289$ (RSV-AZ-ST and RSV-GH-ST) to $0.300$ (RSV-FS-SN) and the SV specifications from $0.326$ (SV-GH-ST) to $0.350$ (SV-T), with REGARCH at $0.284$ and EGARCH at $0.340$.
Every RSV specification attains a lower score than its SV counterpart under all four proxies, but the margins are smaller than over the main sample, and the two REGARCH specifications attain the lowest scores under every proxy and the best average ranks ($1.2$ and $1.8$), followed by RSV-AZ-ST ($3.2$) and RSV-GH-ST ($3.8$).
Every RSV and REGARCH specification, SV-GH-ST, and EGARCH-N belong to the $75\%$ MCS under all four proxies, SV-FS-ST and EGARCH-T under all but BV, and the two Azzalini SV specifications to the $90\%$ MCS under all but BV; SV-N, SV-T, and SV-FS-SN enter neither confidence set, except SV-T under Med.
Within the SV class the average scores range from $0.316$ (SV-GH-ST) to $0.339$ (SV-T).

\begin{figure}[tbp]
\centering
\includegraphics[width = \textwidth]{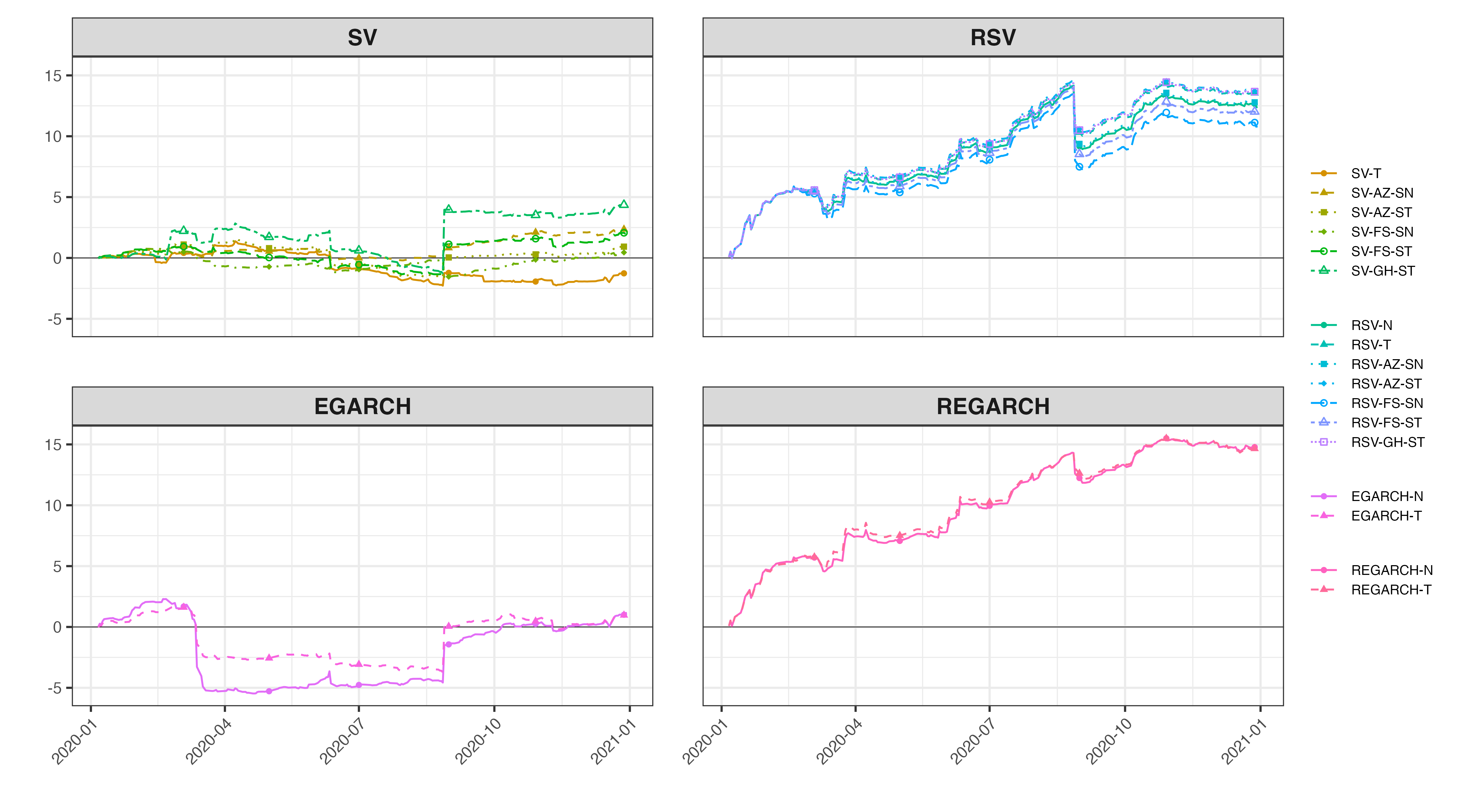}
\caption{Cumulative QLIKE loss differences relative to the SV-N model, based on RV5 as the volatility proxy, for the N225 over the COVID-19 period (2020).}
\label{fig:pred-vol-cld-n225-covid}
\end{figure}

\begin{figure}[tbp]
\centering
\includegraphics[width = .9\textwidth]{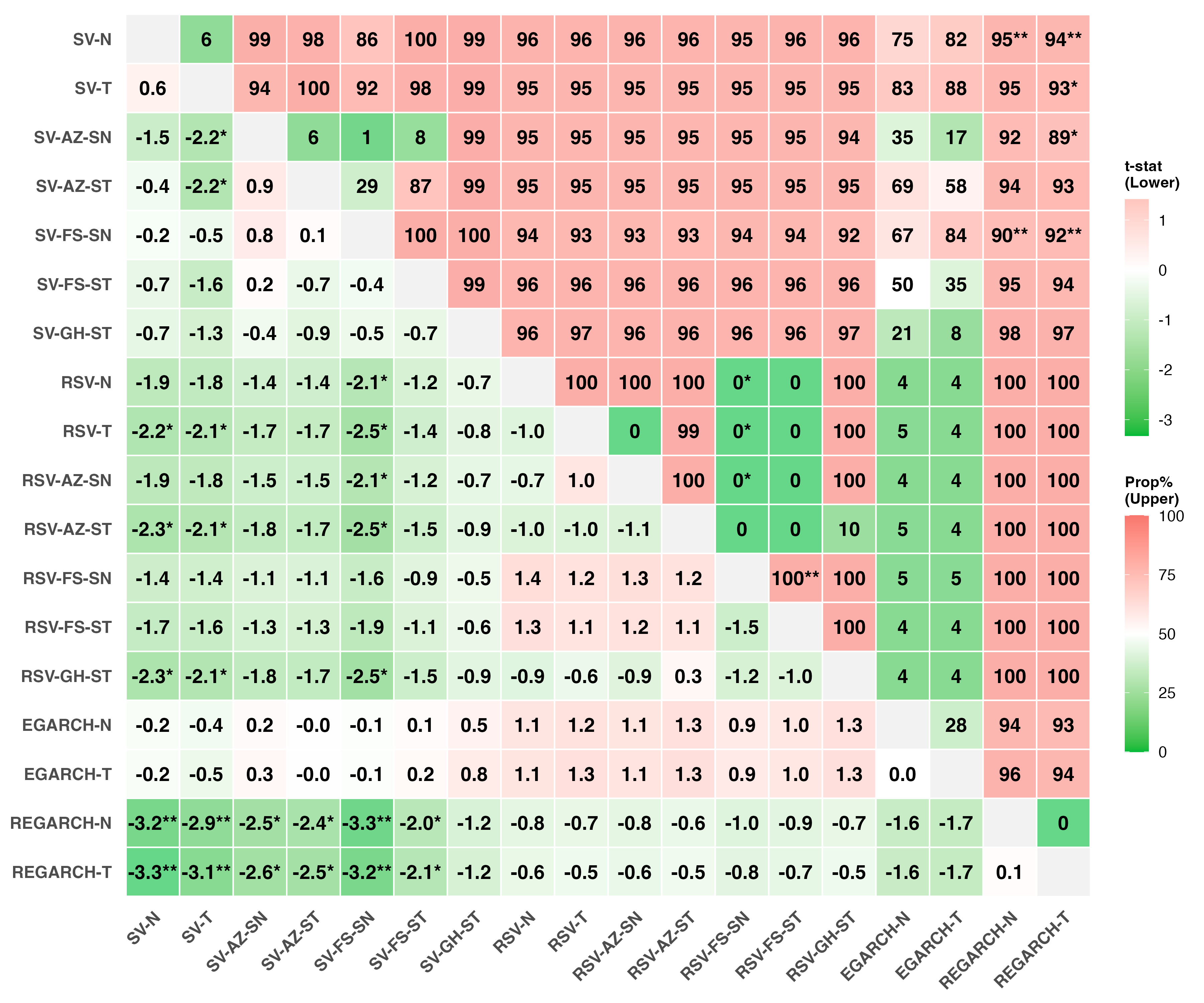}
\caption{Pairwise comparison of GW tests on QLIKE of volatility forecasts, based on RV5 as the volatility proxy, for the N225 over the COVID-19 period (2020).
See Figure~\ref{fig:forecast-djia-vol-gwtest-suppl} for additional details.}
\label{fig:forecast-n225-vol-gwtest-covid}
\end{figure}

The CLDs relative to the SV-N model under RV5 in Figure~\ref{fig:pred-vol-cld-n225-covid} show every specification except SV-T above the benchmark, and Figure~\ref{fig:forecast-n225-vol-gwtest-covid} reports the pairwise GW tests; for the seven matched pairs the unconditional test rejects equal accuracy at the $5\%$ level only for the Student's $t$ pair under RV5, RK, and Med and for four pairs under BV (Section~4.4 of the main text).
Within the RSV class the Azzalini skew-$t$, GH skew-$t$, and Student's $t$ distributions attain the lowest scores and the Fern\'{a}ndez--Steel distributions the highest, the reverse of the ordering over the main sample.
In summary, incorporating RV improves volatility forecast accuracy under extreme market stress for the N225, in line with the main-sample findings, with smaller margins and with the REGARCH benchmarks marginally ahead.

\subsection{VaR and ES forecasts}

\begin{figure}[tbp]
\centering
\begin{tabular}{cc}
RSV-AZ-ST (1\%) & RSV-AZ-ST (5\%) \\
\includegraphics[width = .45\textwidth]{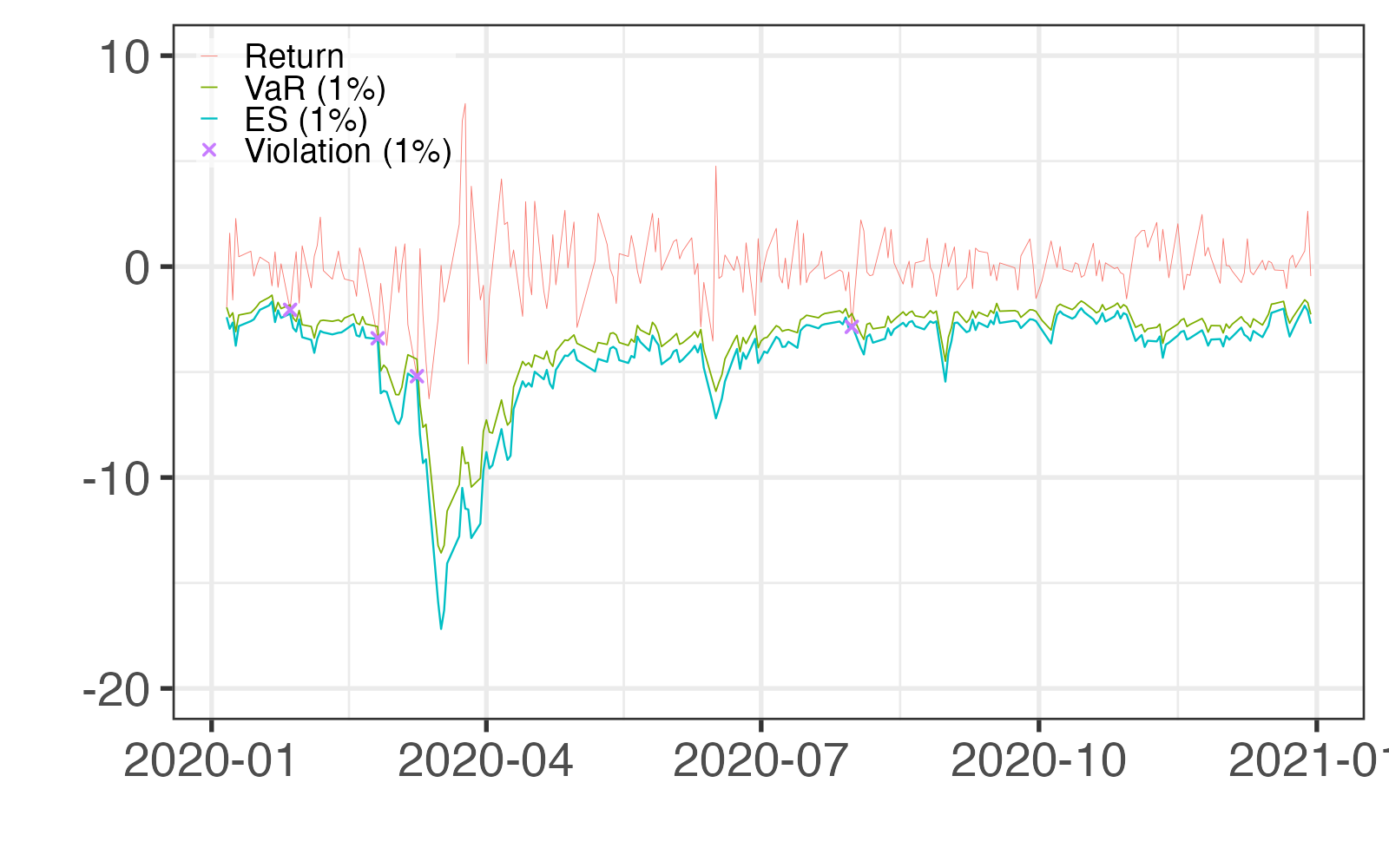} &
\includegraphics[width = .45\textwidth]{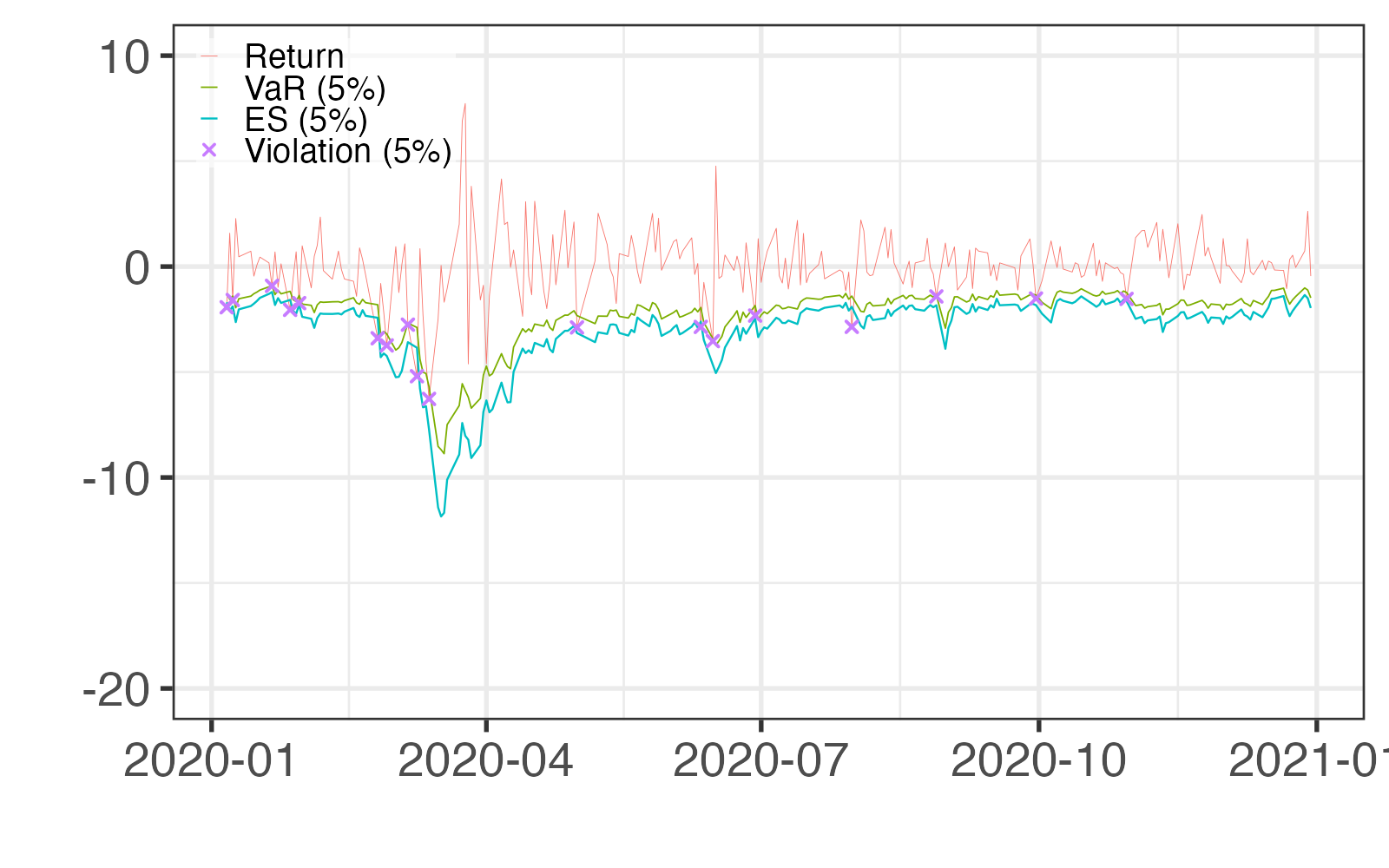} \\
REGARCH-T (1\%) & REGARCH-T (5\%) \\
\includegraphics[width = .45\textwidth]{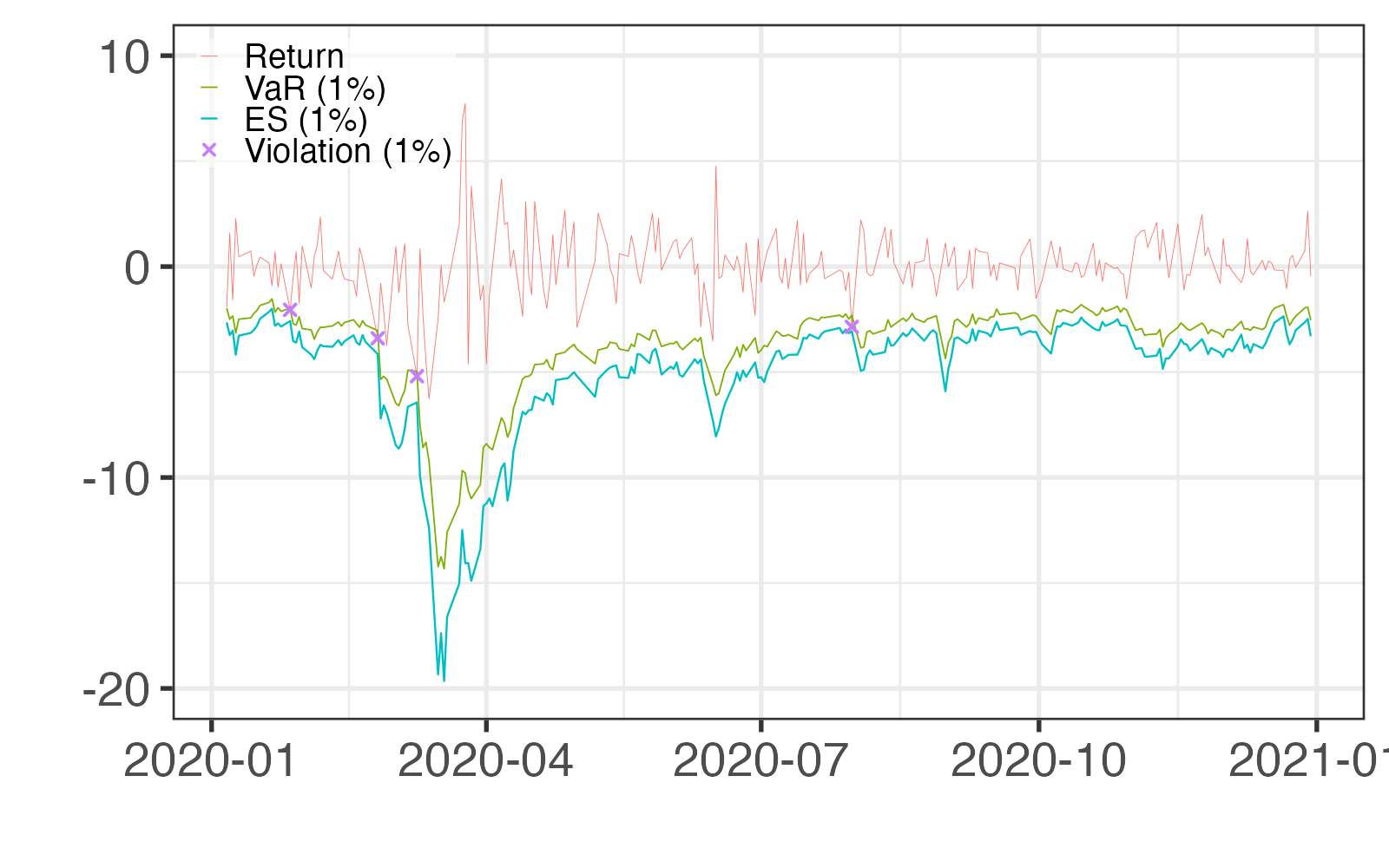} &
\includegraphics[width = .45\textwidth]{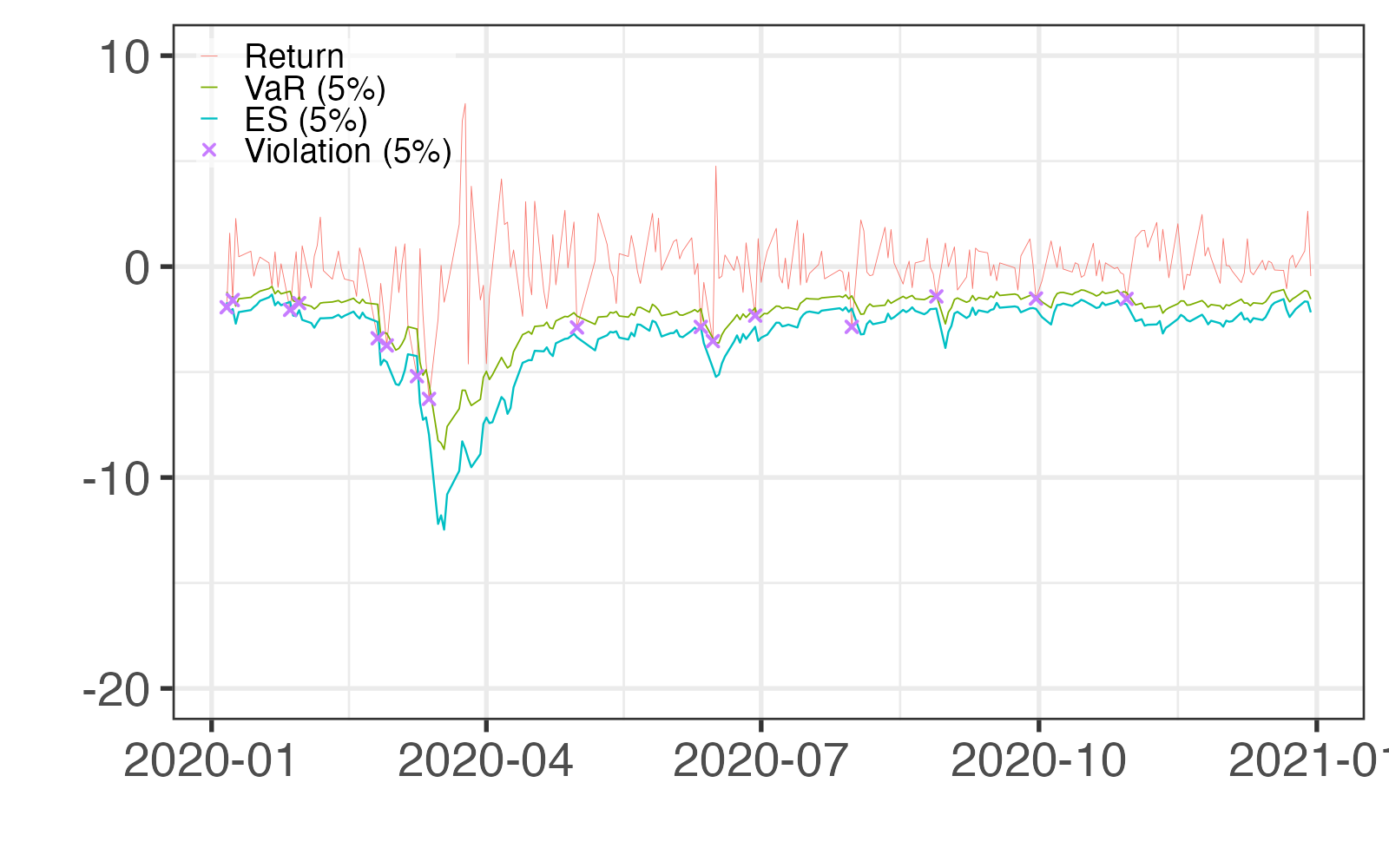}
\end{tabular}
\caption{VaR and ES forecasts of the RSV-AZ-ST and REGARCH-T models for the N225 over the COVID-19 period (2020).}
\label{fig:pred-var-es-covid}
\end{figure}

Figure~\ref{fig:pred-var-es-covid} illustrates the VaR and ES forecasts generated by the RSV-AZ-ST and REGARCH-T models.
VaR violations are observed around similar dates over the forecasting period.
For clarity, the forecasts from other models are omitted.

Table~\ref{tab:forecast-var-es-loss-n225-covid} presents the empirical violation rates ($\hat{\alpha}$), the average FZ0 losses, and the corresponding $p$-values of the UC and DQ tests and of the MCS procedure.
At the $1\%$ level the violation rates of the SV models, between $0.41\%$ and $0.83\%$, lie below the nominal level, whereas those of the RSV models, between $1.65\%$ and $2.48\%$, lie above it.
At the $5\%$ level all models except SV-AZ-ST and SV-GH-ST have violation rates above the nominal level, the RSV models between $7.44\%$ and $7.85\%$.
In terms of the FZ0 loss, REGARCH-T attains the lowest loss at the $1\%$ level ($1.374$) and SV-FS-SN at the $5\%$ level ($1.048$).
The matched pairs favour the SV specification in six of the seven pairs at the $1\%$ level, the exception being the GH skew-$t$ pair, and in all seven pairs at the $5\%$ level.
All models are included in both the $75\%$ and the $90\%$ MCS at both levels, indicating broadly comparable performance; the UC test does not reject for any model at either level, and the DQ test rejects only for EGARCH-N at the $1\%$ level.

\begin{figure}[tbp]
\centering
\includegraphics[width = \textwidth]{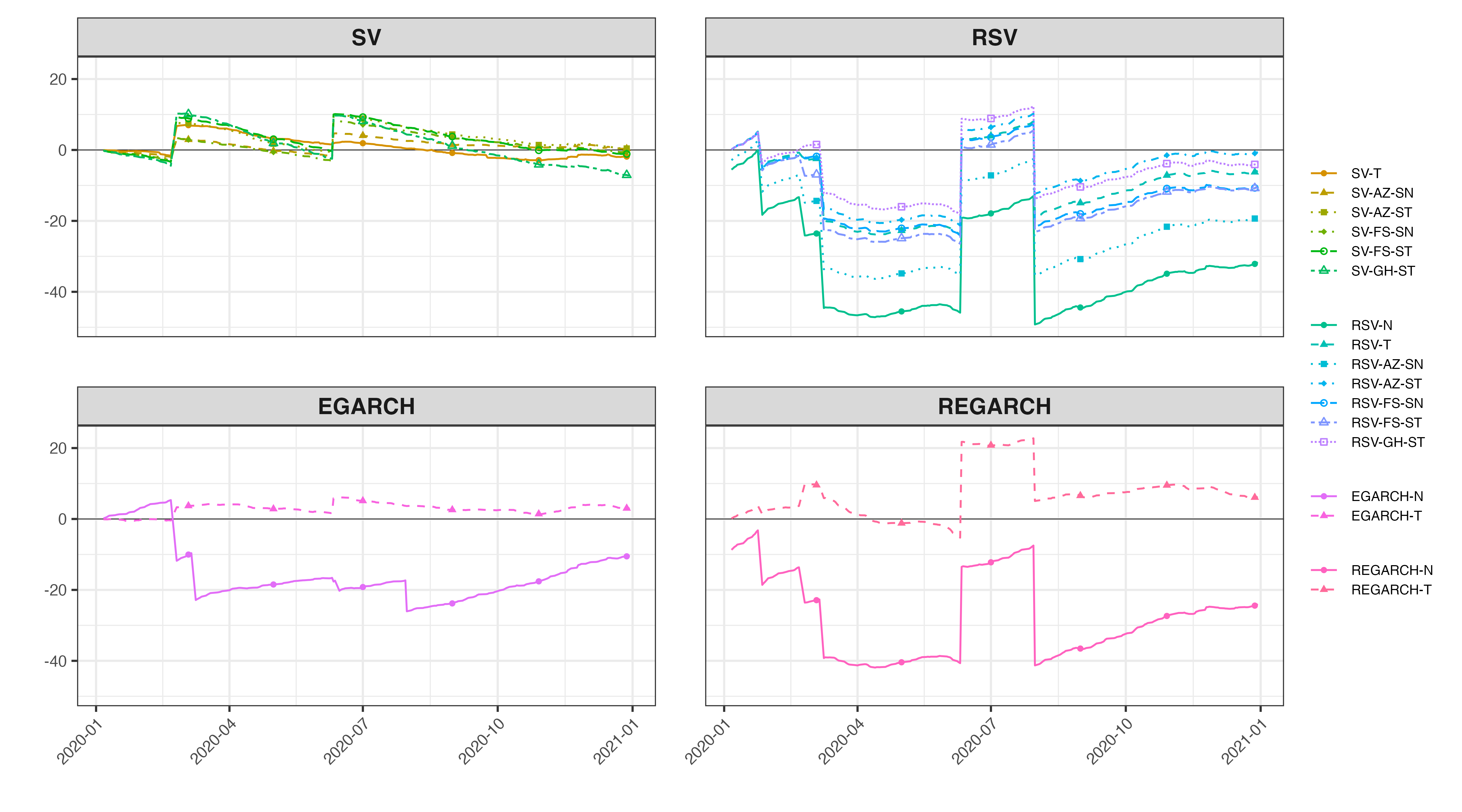}
\caption{Cumulative loss differences (FZ0) relative to the SV-N model ($\alpha = 1\%$) for the N225 over the COVID-19 period (2020).}
\label{fig:pred-fz0-1p-cld-n225-covid}
\end{figure}

\begin{figure}[tbp]
\centering
\includegraphics[width = \textwidth]{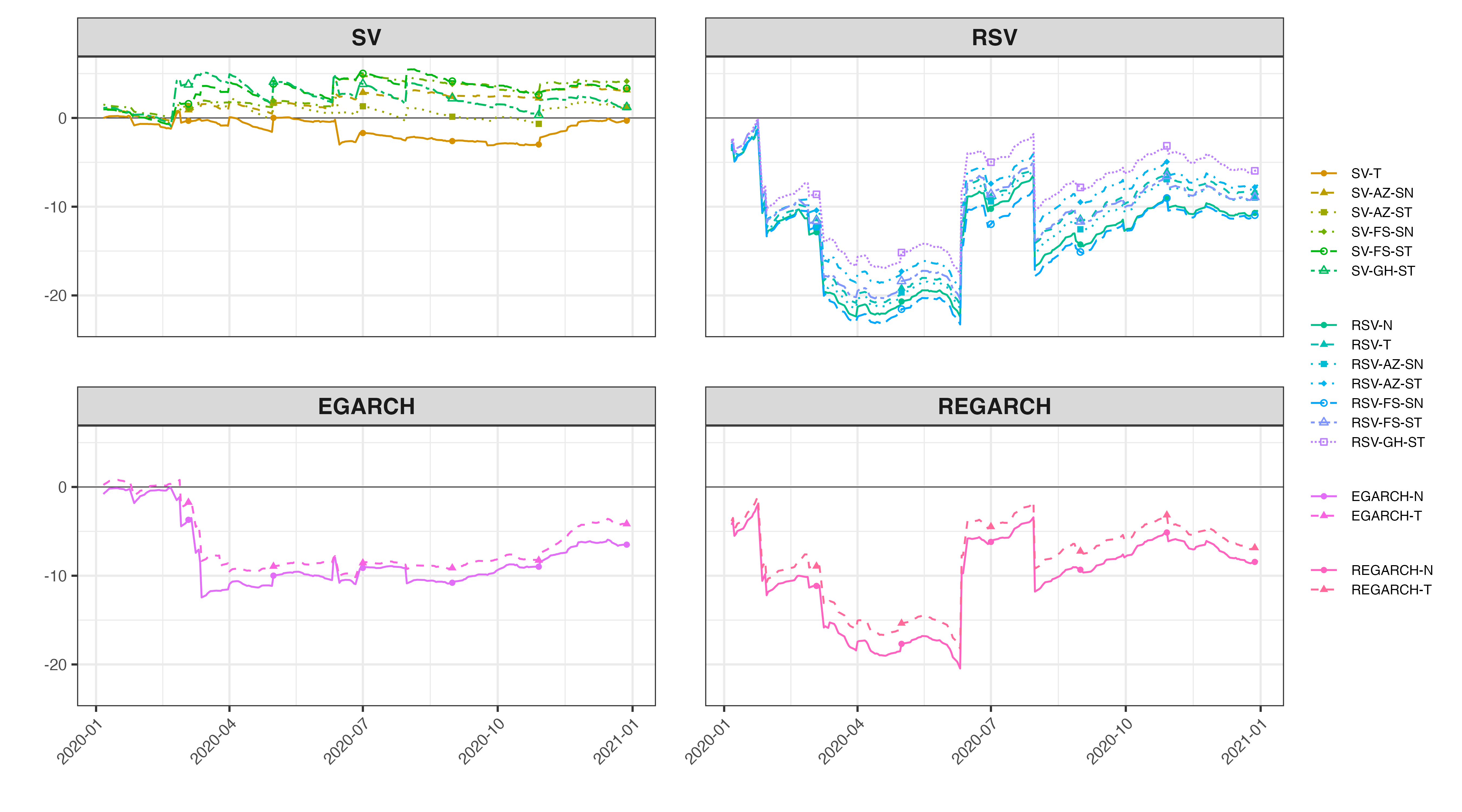}
\caption{Cumulative loss differences (FZ0) relative to the SV-N model ($\alpha = 5\%$) for the N225 over the COVID-19 period (2020).}
\label{fig:pred-fz0-5p-cld-n225-covid}
\end{figure}

Figures~\ref{fig:pred-fz0-1p-cld-n225-covid} and \ref{fig:pred-fz0-5p-cld-n225-covid} show the CLDs relative to the SV-N benchmark.
At the $1\%$ level only SV-AZ-SN, SV-AZ-ST, EGARCH-T, and REGARCH-T end above the benchmark, and at the $5\%$ level only the SV specifications with skewed innovations do.
Contrary to the results for the main sample, incorporating RV does not improve the VaR and ES forecasts in terms of the FZ0 loss during this period.

\begin{figure}[tbp]
\centering
\includegraphics[width = .9\textwidth]{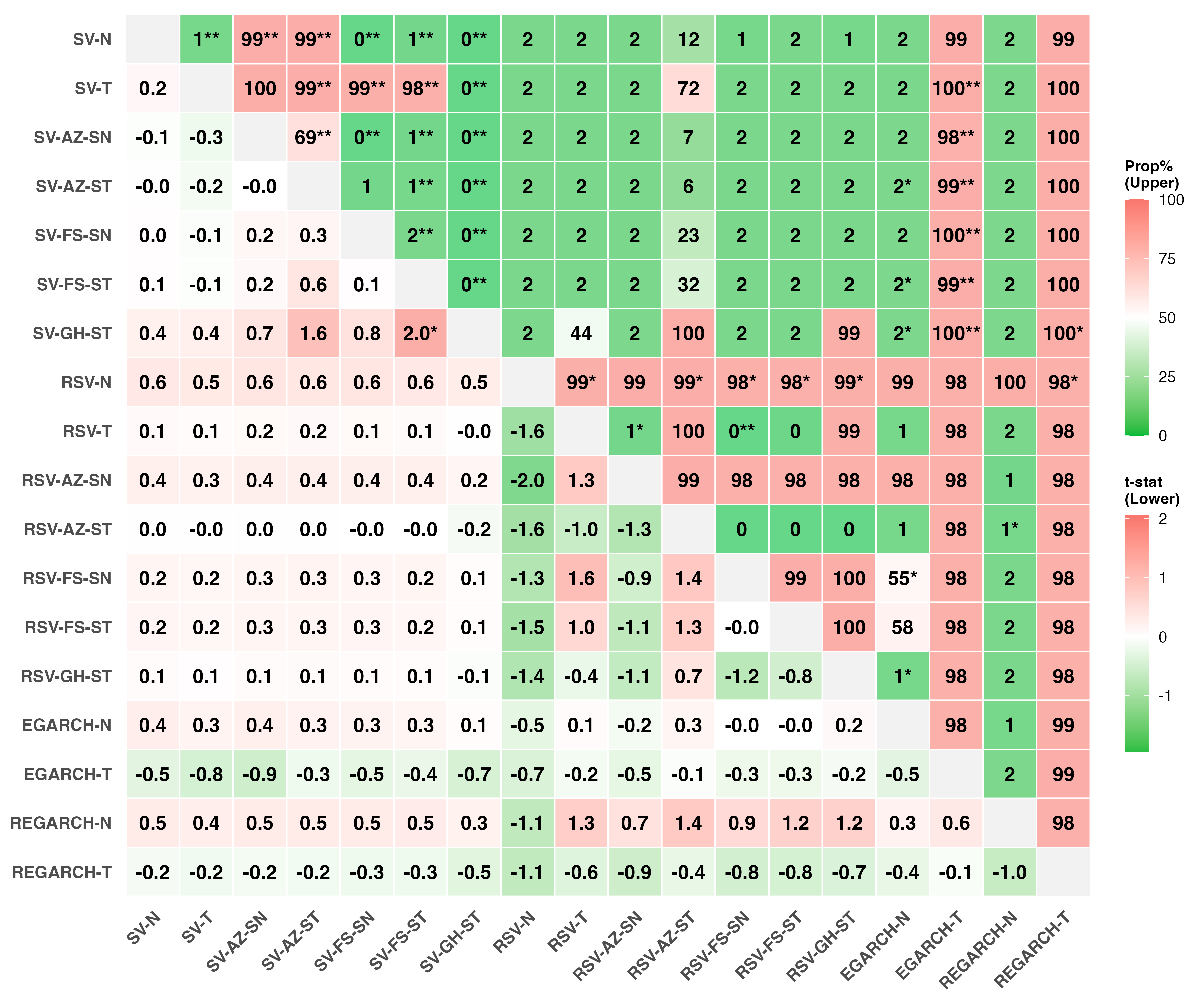}
\caption{Pairwise comparison of GW tests on FZ0 loss of VaR and ES forecasts ($\alpha = 1\%$) for the N225 over the COVID-19 period (2020).
See Figure~\ref{fig:forecast-djia-vol-gwtest-suppl} for additional details.}
\label{fig:forecast-n225-var-es-gwtest-1p-covid}
\end{figure}

\begin{figure}[tbp]
\centering
\includegraphics[width = .9\textwidth]{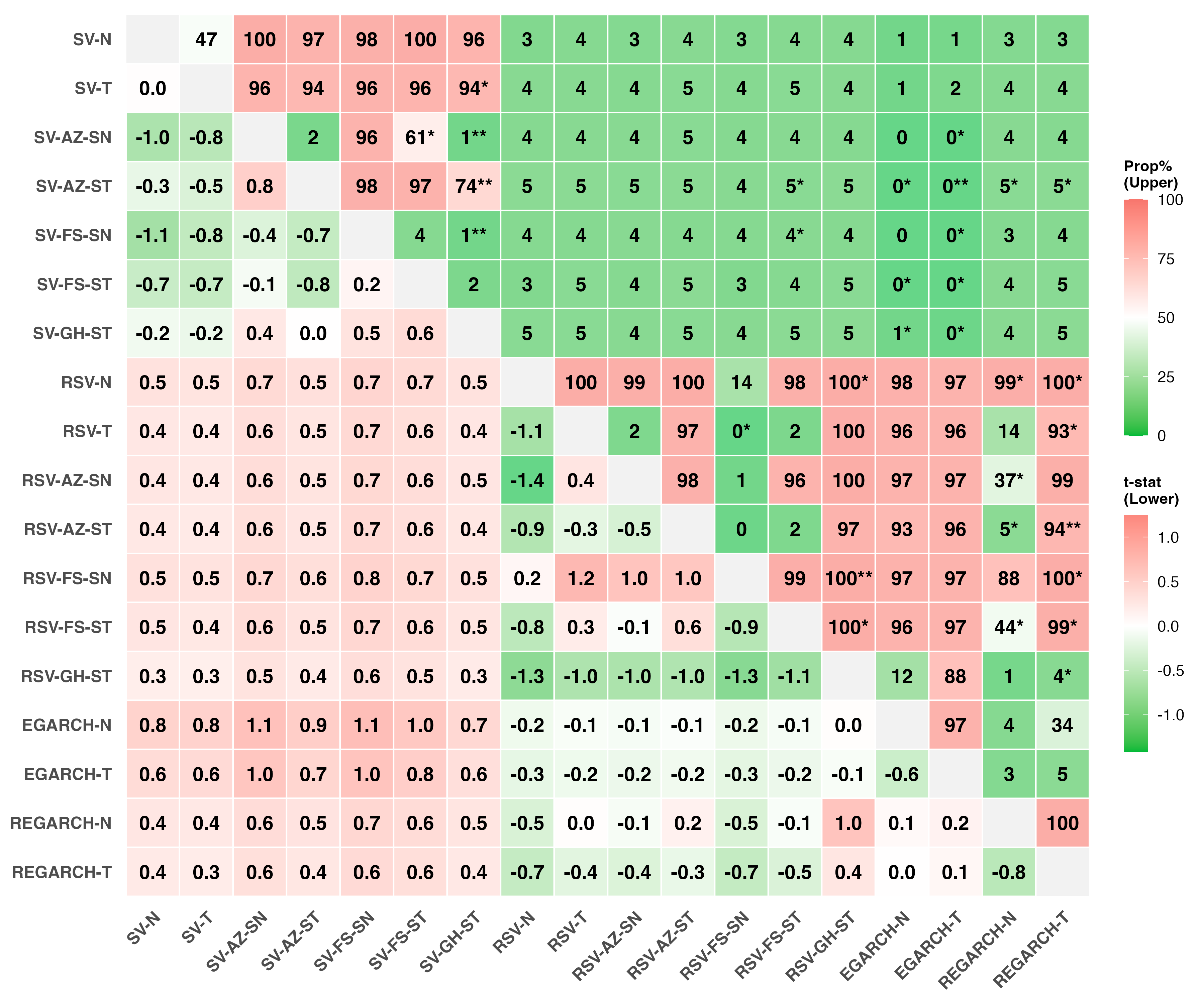}
\caption{Pairwise comparison of GW tests on FZ0 loss of VaR and ES forecasts ($\alpha = 5\%$) for the N225 over the COVID-19 period (2020).
See Figure~\ref{fig:forecast-djia-vol-gwtest-suppl} for additional details.}
\label{fig:forecast-n225-var-es-gwtest-5p-covid}
\end{figure}

Figures~\ref{fig:forecast-n225-var-es-gwtest-1p-covid} and \ref{fig:forecast-n225-var-es-gwtest-5p-covid} visualize the GW test results based on the FZ0 loss.
For the seven matched pairs the unconditional test does not reject equal predictive ability at the $1\%$ level and is borderline at the $5\%$ level (Section~4.4 of the main text); the remaining entries compare models across classes and across distributions.

\subsection{Sensitivity analysis: freely estimated $\psi$}
\label{sec:psi-covid-n225}

This subsection is the N225 counterpart of Section~\ref{sec:psi-covid-djia}, reporting a sensitivity analysis in which $\psi$ is estimated freely rather than fixed at unity, for the RSV and REGARCH models, over the N225 COVID-19 sample.
RSV-GH-ST with a free $\psi$ diverges to a degenerate posterior mode, as it does over the full sample, and is therefore excluded from this comparison.

Table~\ref{tab:forecast-vol-qlike-n225-covid-psi} reports QLIKE scores under $\psi = 1$ and under a freely estimated $\psi$.
A free $\psi$ raises the average QLIKE of every RSV specification, by between $0.5\%$ (RSV-FS-SN) and $8\%$ (RSV-T), and under every proxy except BV and Med for RSV-FS-SN.
For REGARCH-N it raises the score under all four proxies, and for REGARCH-T it lowers it marginally, from $0.268$ to $0.267$ on average.
The MCS does not separate the variants: every row of the table belongs to the $75\%$ MCS under every proxy.

Table~\ref{tab:forecast-var-es-loss-n225-covid-psi} reports the corresponding violation rates, UC and DQ test $p$-values, and FZ0 losses for VaR and ES forecasts.
The pattern runs the same way: a freely estimated $\psi$ lowers the FZ0 loss for only one RSV specification, RSV-N at the $1\%$ level, and raises it for every RSV specification at the $5\%$ level and for the other five at the $1\%$ level, although it brings the violation rates closer to the nominal levels, from between $7.44\%$ and $7.85\%$ to between $6.20\%$ and $7.02\%$ at the $5\%$ level.
Neither coverage test rejects at the $5\%$ significance level for any row of the table; the smallest UC $p$-values are $0.05$, for RSV-N and RSV-AZ-SN under $\psi = 1$ at $\alpha = 1\%$.
For REGARCH, the differences between $\psi=1$ and a freely estimated $\psi$ are small at both risk levels, the only reduction being REGARCH-N at the $5\%$ level, from $1.0997$ to $1.0982$.

Set against the SV and EGARCH results of Table~\ref{tab:forecast-var-es-loss-n225-covid}, this leaves the N225 conclusion unchanged.
Under $\psi=1$ every RSV specification carries a higher FZ0 loss than its SV counterpart at the $5\%$ level, and all but RSV-GH-ST do at the $1\%$ level, and a freely estimated $\psi$ does not reverse a single one of those comparisons.
The best free-$\psi$ RSV specification attains an FZ0 loss of $1.44$ at the $1\%$ level and $1.11$ at the $5\%$ level, against $1.40$ and $1.05$ for the best SV specification and $1.39$ and $1.08$ for EGARCH-T.
The contrast with the DJIA of Section~\ref{sec:psi-covid-djia}, where freeing $\psi$ reversed every matched-pair comparison at both levels, is therefore sharp: relaxing $\psi$ recovers the RSV tail-risk performance on one index but not on the other.

\clearpage

\begin{table}[tbp]
\centering
\begin{threeparttable}
\caption{Simulation results for the RSV-N, -T-, and GH-ST models. Main sample (2009--2019).}
\label{tab:sim-study-1}
\begin{tabular}{lcrrrrrrr}
\toprule
Model & Parameter & \multicolumn{1}{c}{True} & \multicolumn{1}{c}{Mean} & \multicolumn{1}{c}{SD} & \multicolumn{1}{c}{95\%L} & \multicolumn{1}{c}{95\%U} & \multicolumn{1}{c}{CD} & \multicolumn{1}{c}{IF} \\
\midrule
RSV-N
      & $\mu$         & 0.00     & 0.0391 & 0.0855 & $-0.1298$ & 0.2073 & 0.184 & 9.76 \\
      & $\phi$        & 0.95  & 0.9452 & 0.0084 & 0.9282 & 0.9609 & 0.164 & 10.14 \\
      & $\sigma_\eta$ & 0.20  & 0.1934 & 0.0096 & 0.1754 & 0.2131 & 0.088 & 29.36 \\
      & $\rho$        & $-0.30$ & $-0.2729$ & 0.0375 & $-0.3458$ & $-0.1983$ & 0.537 & 6.31 \\
      & $\xi$         & $-0.80$ & $-0.8480$ & 0.0364 & $-0.9198$ & $-0.7781$ & 0.255 & 47.38 \\
      & $\sigma_u$    & 0.30  & 0.3058 & 0.0078 & 0.2908 & 0.3213 & 0.144 & 12.72 \\
\midrule
RSV-T
      & $\mu$         & 0.00  & 0.0796 & 0.0995 & $-0.1176$ & 0.2763 & 0.726 & 21.65 \\
      & $\phi$        & 0.95  & 0.9529 & 0.0077 & 0.9376 & 0.9676 & 0.984 & 7.73 \\
      & $\sigma_\eta$      & 0.20  & 0.1918 & 0.0087 & 0.1754 & 0.2095 & 0.730 & 23.13 \\
      & $\rho$        & $-0.30$ & $-0.3524$ & 0.0374 & $-0.4254$ & $-0.2791$ & 0.870 & 8.17 \\
      & $\nu$         & 10.00 & 9.0892 & 1.7902 & 6.2375 & 13.3261 & 0.404 & 113.82 \\
      & $\xi$         & $-0.80$ & $-0.8844$ & 0.0465 & $-0.9800$ & $-0.7985$ & 0.765 & 98.70 \\
      & $\sigma_u$    & 0.30  & 0.2949 & 0.0075 & 0.2804 & 0.3098 & 0.912 & 8.91 \\
\midrule
RSV-GH-ST
      & $\mu$         & 0.00  & 0.0484 & 0.0918 & $-0.1310$ & 0.2290 & 0.294 & 58.56 \\
      & $\phi$        & 0.95  & 0.9460 & 0.0084 & 0.9289 & 0.9619 & 0.204 & 8.08 \\
      & $\sigma_\eta$ & 0.20  & 0.1925 & 0.0095 & 0.1744 & 0.2118 & 0.327 & 22.75 \\
      & $\rho$        & $-0.30$ & $-0.2675$ & 0.0426 & $-0.3500$ & $-0.1833$ & 0.888 & 12.28 \\
      & $\beta$       & $-0.70$ & $-0.7808$ & 0.2373 & $-1.3339$ & $-0.4186$ & 0.063 & 356.67 \\
      & $\nu$         & 10.00 & 10.6874 & 2.6314 & 6.8163 & 16.8797 & 0.060 & 447.17 \\
      & $\xi$         & $-0.80$ & $-0.8647$ & 0.0537 & $-0.9773$ & $-0.7642$ & 0.457 & 197.59 \\
      & $\sigma_u$    & 0.30  & 0.3066 & 0.0079 & 0.2913 & 0.3223 & 0.619 & 10.69 \\
\bottomrule
\end{tabular}
\begin{tablenotes}
\footnotesize
\item \textit{Notes:} The true parameter values correspond to the data-generating process used in the simulation study with sample size $T=2{,}000$.
The estimation is based on 50{,}000 MCMC draws after discarding the first 10{,}000 as burn-in.
95\%L and 95\%U denote the lower and upper limits of the 95\% credible interval.
CD denotes the $p$-value of the convergence diagnostic of \citet{geweke_evaluating_1992}, and IF denotes the inefficiency factor.
\end{tablenotes}
\end{threeparttable}
\end{table}

\begin{table}[tbp]
\centering
\begin{threeparttable}
\caption{Simulation results for the RSV-AZ-SN, AZ-ST-, FS-SN, and FS-ST models. Main sample (2009--2019).}
\label{tab:sim-study-2}
\begin{tabular}{lcrrrrrrr}
\toprule
Model & Parameter & \multicolumn{1}{c}{True} & \multicolumn{1}{c}{Mean} & \multicolumn{1}{c}{SD} & \multicolumn{1}{c}{95\%L} & \multicolumn{1}{c}{95\%U} & \multicolumn{1}{c}{CD} & \multicolumn{1}{c}{IF} \\
\midrule
RSV-AZ-SN
      & $\mu$         & $-0.50$ & $-0.3960$ & 0.0744 & $-0.5425$ & $-0.2498$ & 0.865 & 21.37 \\
      & $\phi$        & 0.90  & 0.9003 & 0.0108 & 0.8787 & 0.9213 & 0.889 & 9.63 \\
      & $\sigma_\eta$      & 0.30  & 0.3096 & 0.0102 & 0.2899 & 0.3299 & 0.943 & 23.90 \\
      & $\rho$        & $-0.40$ & $-0.3829$ & 0.0431 & $-0.4680$ & $-0.2995$ & 0.447 & 27.49 \\
      & $\delta$      & $-0.90$ & $-0.8814$ & 0.0206 & $-0.9157$ & $-0.8350$ & 0.716 & 46.90 \\
      & $\xi$         & $-0.50$ & $-0.5284$ & 0.0374 & $-0.6015$ & $-0.4573$ & 0.740 & 123.31 \\
      & $\sigma_u$    & 0.20  & 0.1958 & 0.0108 & 0.1743 & 0.2167 & 0.868 & 30.28 \\
\midrule
RSV-AZ-ST
      & $\mu$         & $-0.50$ & $-0.3462$ & 0.0800 & $-0.5038$ & $-0.1888$ & 0.877 & 43.18 \\
      & $\phi$        & 0.90  & 0.9063 & 0.0105 & 0.8851 & 0.9264 & 0.691 & 4.69 \\
      & $\sigma_\eta$      & 0.30  & 0.2967 & 0.0100 & 0.2775 & 0.3167 & 0.617 & 14.32 \\
      & $\rho$        & $-0.40$ & $-0.4031$ & 0.0501 & $-0.5027$ & $-0.3076$ & 0.218 & 48.62 \\
      & $\delta$      & $-0.90$ & $-0.9006$ & 0.0195 & $-0.9341$ & $-0.8570$ & 0.309 & 103.90 \\
      & $\nu$         & 10.00 & 9.6449 & 1.9002 & 6.7854 & 14.1460 & 0.018 & 265.25 \\
      & $\xi$         & $-0.50$ & $-0.5888$ & 0.0447 & $-0.6781$ & $-0.5000$ & 0.977 & 175.57 \\
      & $\sigma_u$    & 0.20  & 0.2133 & 0.0100 & 0.1933 & 0.2327 & 0.238 & 18.48 \\
\midrule
RSV-FS-SN
      & $\mu$         & $-0.50$ & $-0.5628$ & 0.0693 & $-0.6991$ & $-0.4258$ & 0.035 & 23.20 \\
      & $\phi$        & 0.90  & 0.9015 & 0.0102 & 0.8813 & 0.9211 & 0.371 & 1.94 \\
      & $\sigma_\eta$ & 0.30  & 0.3029 & 0.0092 & 0.2852 & 0.3212 & 0.207 & 7.83 \\
      & $\rho$        & $-0.40$ & $-0.4024$ & 0.0248 & $-0.4506$ & $-0.3534$ & 0.014 & 6.92 \\
      & $\gamma$      & 0.50  & 0.4942 & 0.0226 & 0.4513 & 0.5393 & 0.030 & 36.54 \\
      & $\xi$         & $-0.50$ & $-0.5316$ & 0.0279 & $-0.5855$ & $-0.4774$ & 0.025 & 149.58 \\
      & $\sigma_u$    & 0.20  & 0.1932 & 0.0089 & 0.1754 & 0.2103 & 0.142 & 12.59 \\
\midrule
RSV-FS-ST
      & $\mu$         & $-0.50$ & $-0.5665$ & 0.0729 & $-0.7100$ & $-0.4240$ & 0.540 & 81.70 \\
      & $\phi$        & 0.90  & 0.9023 & 0.0096 & 0.8830 & 0.9209 & 0.445 & 4.79 \\
      & $\sigma_\eta$ & 0.30  & 0.3025 & 0.0088 & 0.2857 & 0.3202 & 0.779 & 22.58 \\
      & $\rho$        & $-0.40$ & $-0.4098$ & 0.0247 & $-0.4579$ & $-0.3612$ & 0.508 & 27.46 \\
      & $\gamma$      & 0.50  & 0.5059 & 0.0224 & 0.4629 & 0.5506 & 0.638 & 33.37 \\
      & $\xi$         & $-0.50$ & $-0.4946$ & 0.0362 & $-0.5663$ & $-0.4235$ & 0.545 & 333.69 \\
      & $\sigma_u$    & 0.20  & 0.1890 & 0.0086 & 0.1717 & 0.2054 & 0.713 & 23.84 \\
      & $\nu$         & 10.00 & 11.8080 & 2.8157 & 7.7866 & 18.7360 & 0.547 & 86.69 \\
\bottomrule
\end{tabular}
\begin{tablenotes}
\footnotesize
\item \textit{Notes:} The true parameter values correspond to the data-generating process used in the simulation study with sample size $T=2{,}000$.
The estimation is based on 50{,}000 MCMC draws after discarding the first 10{,}000 as burn-in.
95\%L and 95\%U denote the lower and upper limits of the 95\% credible interval.
CD denotes the $p$-value of the convergence diagnostic of \citet{geweke_evaluating_1992}, and IF denotes the inefficiency factor.
\end{tablenotes}
\end{threeparttable}
\end{table}

\begin{table}[tbp]
\centering
\begin{threeparttable}
\caption{MCMC diagnostics for the RSV-AZ-ST model applied to the DJIA over representative estimation windows. Main sample (2009--2019).}
\label{tab:mcmc-rsvaz-rep}
\begin{tabular}{lcrrrrrrrr}
\toprule
Window & Parameter
& \multicolumn{2}{c}{Prior} 
& \multicolumn{6}{c}{Posterior} 
\\
\cmidrule(lr){3-4} \cmidrule(lr){5-10}
& & Mean & SD & Mean & SD & 95\%L & 95\%U & CD & IF   \\
\midrule
20090609--20170503
 & $\mu$ & 0.00 & 10.00 & $-0.4819$ & 0.0884 & $-0.6503$ & $-0.3038$ & 0.404 & 10.58 \\
 & $\phi$ & 0.86 & 0.11 & 0.9316 & 0.0092 & 0.9122 & 0.9483 & 0.119 & 49.57 \\
 & $\sigma_\eta$ & 0.119 & 0.049 & 0.3056 & 0.0165 & 0.2759 & 0.3407 & 0.178 & 75.66 \\
 & $\rho$ & $-0.33$ & 0.47 & $-0.7512$ & 0.0556 & $-0.8558$ & $-0.6342$ & 0.013 & 128.41 \\
 & $\delta$ & $-0.33$ & 0.47 & $-0.8122$ & 0.0361 & $-0.8658$ & $-0.7237$ & 0.024 & 106.60 \\
 & $\nu$ & 10.00 & 4.47 & 21.9987 & 5.0949 & 13.9802 & 33.7558 & 0.003 & 217.51 \\
 & $\xi$ & 0.00 & 1.00 & $-0.2237$ & 0.0403 & $-0.3041$ & $-0.1469$ & 0.700 & 37.56 \\
 & $\sigma_u$ & 0.238 & 0.104 & 0.5328 & 0.0121 & 0.5094 & 0.5569 & 0.178 & 24.42 \\
\midrule
20100414--20180314
 & $\mu$ & 0.00 & 10.00 & $-0.5561$ & 0.0979 & $-0.7417$ & $-0.3607$ & 0.054 & 6.74 \\
 & $\phi$ & 0.86 & 0.11 & 0.9273 & 0.0090 & 0.9089 & 0.9440 & 0.250 & 28.58 \\
 & $\sigma_\eta$ & 0.119 & 0.049 & 0.3308 & 0.0175 & 0.2978 & 0.3651 & 0.580 & 57.13 \\
 & $\rho$ & $-0.33$ & 0.47 & $-0.6631$ & 0.0530 & $-0.7651$ & $-0.5544$ & 0.303 & 76.82 \\
 & $\delta$ & $-0.33$ & 0.47 & $-0.7881$ & 0.0478 & $-0.8563$ & $-0.6703$ & 0.446 & 90.48 \\
 & $\nu$ & 10.00 & 4.47 & 20.2736 & 4.8926 & 11.9686 & 31.4003 & 0.004 & 149.34 \\
 & $\xi$ & 0.00 & 1.00 & $-0.2883$ & 0.0451 & $-0.3811$ & $-0.2039$ & 0.000 & 30.45 \\
 & $\sigma_u$ & 0.238 & 0.104 & 0.5257 & 0.0128 & 0.5003 & 0.5510 & 0.331 & 24.51 \\
\midrule
20110328--20190301
 & $\mu$ & 0.00 & 10.00 & $-0.5699$ & 0.0984 & $-0.7601$ & $-0.3722$ & 0.141 & 7.87 \\
 & $\phi$ & 0.86 & 0.11 & 0.9280 & 0.0087 & 0.9101 & 0.9443 & 0.759 & 23.00 \\
 & $\sigma_\eta$ & 0.119 & 0.049 & 0.3249 & 0.0159 & 0.2941 & 0.3571 & 0.891 & 42.96 \\
 & $\rho$ & $-0.33$ & 0.47 & $-0.6179$ & 0.0534 & $-0.7204$ & $-0.5093$ & 0.675 & 96.39 \\
 & $\delta$ & $-0.33$ & 0.47 & $-0.7484$ & 0.0682 & $-0.8413$ & $-0.5667$ & 0.216 & 141.42 \\
 & $\nu$ & 10.00 & 4.47 & 20.7675 & 5.0344 & 12.7083 & 32.2019 & 0.148 & 169.91 \\
 & $\xi$ & 0.00 & 1.00 & $-0.3233$ & 0.0434 & $-0.4098$ & $-0.2375$ & 0.937 & 38.82 \\
 & $\sigma_u$ & 0.238 & 0.104 & 0.5058 & 0.0126 & 0.4815 & 0.5307 & 0.292 & 16.91 \\
\bottomrule
\end{tabular}
\begin{tablenotes}
\footnotesize
\item \textit{Notes:} The table reports prior moments, posterior summaries, and MCMC diagnostics for representative rolling estimation windows of the RSV-AZ-ST model.
For variance parameters, priors are specified on $\sigma_\eta^2$ and $\sigma_u^2$; reported prior moments for $\sigma_\eta$ and $\sigma_u$ are computed from the implied distributions.
CD denotes the $p$-value of the convergence diagnostic of \citet{geweke_evaluating_1992}, and IF denotes the inefficiency factor.
\end{tablenotes}
\end{threeparttable}
\end{table}

\begin{table}[tbp]
\centering
\begin{threeparttable}
\caption{Windows of the Azzalini series re-estimated with three independent chains (confirmation pass).}
\label{tab:az-confirm}
\tiny
\setlength{\tabcolsep}{1.5pt}
\begin{tabular}{lllllllrp{3.6cm}}
\toprule
Sample & Model & Forecast date & Stage $\times$ draws & Outcome & $\max \hat{R}$ & $\min$ ESS$_b$ & MCSE (\%) & Criteria not met \\
\midrule
DJIA, main & RSV-AZ-SN & 2017-08-08 & 2 $\times$ 45,000 & accepted & $1.006$ ($\rho$) & $835$ ($\rho$) & $0.17$ &  \\
 & RSV-AZ-ST & 2017-08-15 & 2 $\times$ 74,000 & accepted & $1.004$ ($\rho$) & $472$ ($\nu$) & $0.17$ &  \\
 & SV-AZ-SN & 2017-08-04 & 2 $\times$ 60,000 & not accepted & $1.010$ ($\rho$) & $496$ ($\rho$) & $0.51$ & MCSE$_{\exp h}$ \\
 & SV-AZ-SN & 2017-08-08 & 2 $\times$ 90,000 & not completed & \multicolumn{3}{l}{third chain of stage 2 not completed} &  \\
 & SV-AZ-SN & 2017-10-12 & 2 $\times$ 150,000 & accepted & $1.003$ ($\rho$) & $1321$ ($\rho$) & $0.42$ &  \\
 & SV-AZ-SN & 2017-11-08 & 2 $\times$ 90,000 & not completed & \multicolumn{3}{l}{third chain of stage 2 not completed} &  \\
 & SV-AZ-ST & 2017-06-13 & 2 $\times$ 150,000 & not completed & \multicolumn{3}{l}{third chain of stage 2 not completed} &  \\
 & SV-AZ-ST & 2017-06-16 & 2 $\times$ 147,000 & not accepted & $1.010$ ($\nu$) & $281$ ($\nu$) & $0.32$ & $\hat R$($\nu$), ESS$_b$($\nu$) \\
 & SV-AZ-ST & 2017-06-20 & 2 $\times$ 120,000 & not accepted & $1.019$ ($\nu$) & $149$ ($\nu$) & $0.36$ & $\hat R$($\nu$), ESS$_b$($\nu$) \\
 & SV-AZ-ST & 2017-08-14 & 2 $\times$ 406,000 & not accepted & $1.004$ ($\nu$) & $580$ ($\nu$) & $0.54$ & MCSE$_{\exp h}$ \\
 & SV-AZ-ST & 2017-10-03 & 2 $\times$ 178,000 & not accepted & $1.020$ ($\nu$) & $264$ ($\nu$) & $0.27$ & Geweke, $\hat R$($\nu$), ESS$_b$($\nu$) \\
 & SV-AZ-ST & 2017-10-13 & 2 $\times$ 150,000 & not completed & \multicolumn{3}{l}{third chain of stage 2 not completed} &  \\
 & SV-AZ-ST & 2017-10-16 & 2 $\times$ 194,000 & not accepted & $1.010$ ($\nu$) & $178$ ($\nu$) & $9.13$ & ESS$_b$($\nu$), ESS$_t$($\nu$), MCSE$_{\exp h}$, MCSE$_{\mathrm{VaR}1}$, MCSE$_{\mathrm{VaR}5}$, MCSE$_{\mathrm{ES}1}$, MCSE$_{\mathrm{ES}5}$ \\
 & SV-AZ-ST & 2017-10-17 & 2 $\times$ 143,000 & not accepted & $1.026$ ($\nu$) & $137$ ($\nu$) & $0.51$ & $\hat R$($\delta$,$\nu$), ESS$_b$($\nu$), MCSE$_{\exp h}$ \\
 & SV-AZ-ST & 2017-11-08 & 2 $\times$ 150,000 & not completed & \multicolumn{3}{l}{third chain of stage 2 not completed} &  \\
 & SV-AZ-ST & 2019-02-14 & 2 $\times$ 397,000 & accepted & $1.002$ ($\nu$) & $1024$ ($\nu$) & $0.27$ &  \\
N225, main & SV-AZ-SN & 2017-05-31 & 2 $\times$ 45,000 & accepted & $1.006$ ($\sigma_\eta$) & $526$ ($\sigma_\eta$) & $0.41$ &  \\
 & SV-AZ-SN & 2017-10-04 & 2 $\times$ 45,000 & not accepted & $1.003$ ($\rho$) & $520$ ($\rho$) & $1.96$ & MCSE$_{\exp h}$, MCSE$_{\mathrm{VaR}1}$, MCSE$_{\mathrm{ES}1}$, MCSE$_{\mathrm{ES}5}$ \\
 & SV-AZ-SN & 2017-10-19 & 2 $\times$ 93,000 & not accepted & $1.002$ ($\delta$) & $862$ ($\rho$) & $2.03$ & MCSE$_{\exp h}$, MCSE$_{\mathrm{VaR}1}$, MCSE$_{\mathrm{ES}1}$, MCSE$_{\mathrm{ES}5}$ \\
 & SV-AZ-SN & 2017-10-30 & 2 $\times$ 45,000 & not accepted & $1.004$ ($\sigma_\eta$) & $534$ ($\rho$) & $0.77$ & MCSE$_{\exp h}$ \\
 & SV-AZ-SN & 2018-01-25 & 2 $\times$ 45,000 & not accepted & $1.006$ ($\rho$) & $576$ ($\sigma_\eta$) & $0.50$ & MCSE$_{\exp h}$ \\
 & SV-AZ-SN & 2018-04-13 & 2 $\times$ 45,000 & not accepted & $1.008$ ($\rho$) & $430$ ($\rho$) & $3.16$ & MCSE$_{\exp h}$, MCSE$_{\mathrm{VaR}1}$, MCSE$_{\mathrm{ES}1}$, MCSE$_{\mathrm{ES}5}$ \\
 & SV-AZ-SN & 2018-05-18 & 2 $\times$ 59,000 & not accepted & $1.005$ ($\rho$) & $778$ ($\sigma_\eta$) & $0.73$ & MCSE$_{\exp h}$ \\
 & SV-AZ-ST & 2017-11-06 & 2 $\times$ 86,000 & not accepted & $1.004$ ($\nu$) & $423$ ($\nu$) & $2.10$ & MCSE$_{\exp h}$, MCSE$_{\mathrm{VaR}1}$, MCSE$_{\mathrm{ES}1}$, MCSE$_{\mathrm{ES}5}$ \\
 & SV-AZ-ST & 2019-03-26 & 2 $\times$ 64,000 & not accepted & $1.007$ ($\sigma_\eta$) & $310$ ($\nu$) & $0.37$ & ESS$_b$($\nu$) \\
DJIA, 2020 & RSV-AZ-SN (free $\psi$) & 2020-04-07 & 2 $\times$ 45,000 & accepted & $1.003$ ($\delta$) & $903$ ($\delta$) & $0.26$ &  \\
 & RSV-AZ-ST (free $\psi$) & 2020-03-30 & 2 $\times$ 45,000 & accepted & $1.006$ ($\delta$) & $684$ ($\nu$) & $0.30$ &  \\
N225, 2020 & SV-AZ-SN & 2020-12-10 & 2 $\times$ 45,000 & not accepted & $1.006$ ($\rho$) & $601$ ($\rho$) & $0.56$ & MCSE$_{\exp h}$, MCSE$_{\mathrm{ES}1}$ \\
 & SV-AZ-ST & 2020-04-15 & 2 $\times$ 145,000 & not accepted & $1.007$ ($\nu$) & $344$ ($\nu$) & $0.38$ & ESS$_b$($\nu$) \\
 & SV-AZ-ST & 2020-11-24 & 2 $\times$ 110,000 & not accepted & $1.007$ ($\nu$) & $454$ ($\nu$) & $0.87$ & MCSE$_{\exp h}$, MCSE$_{\mathrm{ES}1}$ \\
\bottomrule
\end{tabular}
\begin{tablenotes}
\footnotesize
\item \textit{Notes:} Stage 1: three chains with $15{,}000$ retained draws each; stage 2: the same chains re-run with the number of retained draws per chain shown. $\max \hat{R}$ and $\min$ ESS$_b$ are the largest rank-normalized split-$\hat{R}$ and the smallest bulk effective sample size over $h_n$, the parameters, $\exp(h_{n+1})$, and $y_{n+1}$, with the quantity attaining them in parentheses. MCSE is the Monte Carlo standard error of the predictive mean of volatility in percent of its value. ESS$_t$ denotes the tail effective sample size and MCSE$_{\mathrm{VaR}}$, MCSE$_{\mathrm{ES}}$ the Monte Carlo standard errors of the VaR and ES forecasts at the indicated level. For accepted windows the pooled chains replace the single-chain values; the other windows keep the single-chain values (Section~\ref{sec:suppl-az-recomputation}).
\end{tablenotes}
\end{threeparttable}
\end{table}

\begin{table}[tbp]
\centering
\begin{threeparttable}
\caption{Mean characteristics of one-day-ahead return forecasts under RSV models}
\label{tab:stat-return-forecast}
{\small
\begin{tabular}{lrrrrrr}
DJIA \\
\toprule
& \multicolumn{1}{c}{Skewness} & \multicolumn{1}{c}{Kurtosis} & \multicolumn{1}{c}{$0.1\%$} & \multicolumn{1}{c}{$1\%$} & \multicolumn{1}{c}{$5\%$} & \multicolumn{1}{c}{$10\%$} \\
\midrule
RSV-N & $0.000$ $(0.03)$ & $3.588$ $(0.36)$ & $-3.473$ $(0.18)$ & $-2.436$ $(0.05)$ & $-1.633$ $(0.02)$ & $-1.245$ $(0.02)$ \\ 
RSV-T & $0.002$ $(0.04)$ & $4.050$ $(0.20)$ & $-3.716$ $(0.15)$ & $-2.496$ $(0.03)$ & $-1.624$ $(0.01)$ & $-1.222$ $(0.01)$ \\ 
RSV-AZ-SN & $-0.240$ $(0.05)$ & $3.684$ $(0.11)$ & $-3.780$ $(0.13)$ & $-2.597$ $(0.04)$ & $-1.696$ $(0.02)$ & $-1.265$ $(0.01)$ \\
RSV-AZ-ST & $-0.258$ $(0.06)$ & $4.176$ $(0.19)$ & $-4.022$ $(0.15)$ & $-2.657$ $(0.04)$ & $-1.684$ $(0.02)$ & $-1.242$ $(0.01)$ \\
RSV-FS-SN & $-0.244$ $(0.07)$ & $4.764$ $(0.35)$ & $-4.225$ $(0.19)$ & $-2.690$ $(0.05)$ & $-1.659$ $(0.02)$ & $-1.219$ $(0.01)$ \\ 
RSV-FS-ST & $-0.216$ $(0.05)$ & $4.004$ $(0.22)$ & $-3.894$ $(0.17)$ & $-2.613$ $(0.04)$ & $-1.679$ $(0.02)$ & $-1.249$ $(0.01)$ \\ 
RSV-GH-ST & $-0.388$ $(0.07)$ & $4.390$ $(0.32)$ & $-4.250$ $(0.20)$ & $-2.716$ $(0.05)$ & $-1.694$ $(0.02)$ & $-1.241$ $(0.01)$ \\ 
\bottomrule
\\
N225 \\
\toprule
& \multicolumn{1}{c}{Skewness} & \multicolumn{1}{c}{Kurtosis} & \multicolumn{1}{c}{$0.1\%$} & \multicolumn{1}{c}{$1\%$} & \multicolumn{1}{c}{$5\%$} & \multicolumn{1}{c}{$10\%$} \\
\midrule
RSV-N & $-0.001$ $(0.03)$ & $3.487$ $(0.08)$ & $-3.420$ $(0.10)$ & $-2.425$ $(0.03)$ & $-1.636$ $(0.01)$ & $-1.249$ $(0.01)$ \\ 
RSV-T & $0.001$ $(0.04)$ & $3.892$ $(0.14)$ & $-3.639$ $(0.12)$ & $-2.475$ $(0.03)$ & $-1.627$ $(0.01)$ & $-1.229$ $(0.01)$ \\ 
RSV-AZ-SN & $-0.017$ $(0.03)$ & $3.491$ $(0.08)$ & $-3.447$ $(0.10)$ & $-2.435$ $(0.03)$ & $-1.639$ $(0.01)$ & $-1.252$ $(0.01)$ \\
RSV-AZ-ST & $-0.017$ $(0.04)$ & $3.895$ $(0.12)$ & $-3.670$ $(0.12)$ & $-2.488$ $(0.04)$ & $-1.632$ $(0.01)$ & $-1.231$ $(0.01)$ \\
RSV-FS-SN & $-0.064$ $(0.07)$ & $4.661$ $(0.28)$ & $-4.041$ $(0.20)$ & $-2.577$ $(0.05)$ & $-1.624$ $(0.02)$ & $-1.204$ $(0.01)$ \\ 
RSV-FS-ST & $-0.066$ $(0.05)$ & $3.850$ $(0.17)$ & $-3.670$ $(0.16)$ & $-2.510$ $(0.04)$ & $-1.645$ $(0.02)$ & $-1.238$ $(0.01)$ \\ 
RSV-GH-ST & $-0.081$ $(0.04)$ & $3.903$ $(0.15)$ & $-3.749$ $(0.14)$ & $-2.524$ $(0.04)$ & $-1.644$ $(0.01)$ & $-1.235$ $(0.01)$ \\ 
\bottomrule
\end{tabular}
}
\begin{tablenotes}
\footnotesize
\item \textit{Notes:} The table reports the means of skewness, kurtosis, and lower percentiles ($0.1\%$, $1\%$, $5\%$, and $10\%$) computed from the posterior predictive distributions of one-day-ahead return forecasts.
Standard deviations are shown in parentheses.
These statistics are calculated across 603 and 590 forecast days for the DJIA and N225, respectively.
\end{tablenotes}
\end{threeparttable}
\end{table}

\begin{table}[tbp]
\centering
\begin{threeparttable}
\caption{Posterior mean, standard deviation, and 95\% credible interval of the freely estimated slope coefficient $\psi$, main sample.}
\label{tab:psi-fullsample}
\begin{tabular}{lccc ccc}
\toprule
& \multicolumn{3}{c}{DJIA} & \multicolumn{3}{c}{N225} \\
\cmidrule(lr){2-4} \cmidrule(lr){5-7}
& Mean & SD & 95\% CI & Mean & SD & 95\% CI \\
\midrule
RSV-N & $0.862$ & $0.035$ & $[0.793,\ 0.932]$ & $0.836$ & $0.041$ & $[0.758,\ 0.915]$ \\
RSV-T & $0.859$ & $0.037$ & $[0.787,\ 0.931]$ & $0.830$ & $0.043$ & $[0.747,\ 0.913]$ \\
RSV-AZ-SN & $0.888$ & $0.032$ & $[0.826,\ 0.952]$ & $0.840$ & $0.040$ & $[0.758,\ 0.918]$ \\
RSV-AZ-ST & $0.882$ & $0.033$ & $[0.818,\ 0.947]$ & $0.833$ & $0.043$ & $[0.746,\ 0.916]$ \\
RSV-FS-SN & $0.925$ & $0.030$ & $[0.869,\ 0.986]$ & $0.899$ & $0.035$ & $[0.834,\ 0.970]$ \\
RSV-FS-ST & $0.926$ & $0.031$ & $[0.869,\ 0.989]$ & $0.898$ & $0.037$ & $[0.829,\ 0.972]$ \\
\cmidrule(lr){1-7}
REGARCH-N & $1.117$ & $0.033$ & $[1.056,\ 1.184]$ & $1.080$ & $0.045$ & $[0.997,\ 1.172]$ \\
REGARCH-T & $1.102$ & $0.035$ & $[1.036,\ 1.173]$ & $1.048$ & $0.047$ & $[0.960,\ 1.146]$ \\
\bottomrule
\end{tabular}
\begin{tablenotes}
\footnotesize
\item \textit{Notes:} Estimates are based on a single MCMC run over the main sample (2009--2019) for each index, with $\psi$ estimated freely rather than fixed at unity.
RSV-GH-ST is excluded because its posterior for $\psi$ is unstable (see text).
The prior is $\psi \sim \mathcal{N}(1,1)$ for all specifications.
\end{tablenotes}
\end{threeparttable}
\end{table}

\clearpage

\begin{table}[tbp]
\centering
\begin{threeparttable}
\caption{Descriptive statistics of daily returns and the logarithms of 5-minute RVs for the DJIA (2012--2020).}
\label{tab:stats-djia-covid}
\begin{tabular}{lcccccccc}
\toprule
& Mean & SD & Skew & Kurt & Min & Max & JB & LB \\
\midrule
Return & $0.041$ & $1.078$ & $-1.121$ & $31.049$ & $-13.850$ & $10.741$ & $0.00$ & $0.00$ \\
& $(0.023)$ & & $(0.051)$ & $(0.103)$ \\
Log-RV & $-1.101$ & $1.079$ & $0.646$ & $4.123$ & $-3.938$ & $4.080$ & $0.00$ & $0.00$ \\
& $(0.023)$ & & $(0.051)$ & $(0.103)$ \\
\bottomrule
\end{tabular}
\begin{tablenotes}
\footnotesize
\item \textit{Notes:} Standard errors are shown in parentheses.
JB refers to the $p$-value of the Jarque--Bera test.
LB denotes the $p$-value of the \citet{ljung_measure_1978} statistic, adjusted as in \citet{diebold_empirical_1988}.
\end{tablenotes}
\end{threeparttable}
\end{table}

\begin{table}[tbp]
\centering
\begin{threeparttable}
\caption{QLIKE scores of volatility forecasts for the DJIA over the COVID-19 period (2020).}
\label{tab:forecast-vol-qlike-djia-covid}
\begin{tabular}{llllllrr}
\toprule
& \multicolumn{4}{l}{Proxy} & & \multicolumn{2}{l}{Average} \\
\cmidrule(lr){2-5} \cmidrule(lr){7-8}
& RV5 & RK & BV & Med & & Score & Rank \\
\midrule
SV-N & $0.258^{**}$ & $0.253^{**}$ & $0.260^{**}$ & $0.231^{**}$ & & $0.250$ & $2.5$ \\
SV-T & $0.264^{**}$ & $0.255^{**}$ & $0.288^{**}$ & $0.242^{**}$ & & $0.262$ & $6.5$ \\
SV-AZ-SN & $0.260^{**}$ & $0.250^{**}$ & $0.285^{**}$ & $0.239^{**}$ & & $0.259$ & $3.5$ \\
SV-AZ-ST & $0.264^{**}$ & $0.253^{**}$ & $0.291^{**}$ & $0.244^{**}$ & & $0.263$ & $6.2$ \\
SV-FS-SN & $0.274^{**}$ & $0.265^{**}$ & $0.301^{**}$ & $0.253^{**}$ & & $0.273$ & $9.5$ \\
SV-FS-ST & $0.276^{**}$ & $0.267^{**}$ & $0.305^{**}$ & $0.259^{**}$ & & $0.277$ & $11.2$ \\
SV-GH-ST & $0.259^{**}$ & $0.251^{**}$ & $0.287^{**}$ & $0.242^{**}$ & & $0.260$ & $4.0$ \\
\cmidrule(lr){1-8}
RSV-N & $0.274^{**}$ & $0.274^{**}$ & $0.323^{*}$ & $0.260^{**}$ & & $0.283$ & $12.8$ \\
RSV-T & $0.266^{**}$ & $0.266^{**}$ & $0.312^{*}$ & $0.251^{**}$ & & $0.274$ & $9.8$ \\
RSV-AZ-SN & $0.272^{**}$ & $0.272^{**}$ & $0.321^{*}$ & $0.258^{**}$ & & $0.281$ & $11.2$ \\
RSV-AZ-ST & $0.264^{**}$ & $0.264^{**}$ & $0.309^{*}$ & $0.249^{**}$ & & $0.272$ & $8.0$ \\
RSV-FS-SN & $0.289^{*}$ & $0.290^{*}$ & $0.343$ & $0.278^{*}$ & & $0.300$ & $15.0$ \\
RSV-FS-ST & $0.281^{*}$ & $0.282^{**}$ & $0.333$ & $0.269^{**}$ & & $0.291$ & $14.0$ \\
\cmidrule(lr){1-8}
EGARCH-N & $0.375$ & $0.361$ & $0.404$ & $0.355$ & & $0.374$ & $17.0$ \\
EGARCH-T & $0.362^{*}$ & $0.348^{**}$ & $0.384^{*}$ & $0.337^{**}$ & & $0.358$ & $16.0$ \\
\cmidrule(lr){1-8}
REGARCH-N & $0.257^{**}$ & $0.255^{**}$ & $0.295^{**}$ & $0.238^{**}$ & & $0.261$ & $4.5$ \\
REGARCH-T & $0.246^{**}$ & $0.244^{**}$ & $0.279^{**}$ & $0.225^{**}$ & & $0.249$ & $1.2$ \\
\bottomrule
\end{tabular}
\begin{tablenotes}
\footnotesize
\item \textit{Notes:} A double asterisk ($^{**}$) and a single asterisk ($^{*}$) indicate that the model belongs to the 75\% and the 90\% model confidence set (MCS), respectively; the 75\% MCS is nested within the 90\% MCS.
The average columns report the mean QLIKE scores and ranks across the four volatility proxies.
RSV-GH-ST is excluded (see text).
\end{tablenotes}
\end{threeparttable}
\end{table}

\clearpage

\begin{table}[tbp]
\centering
\begin{threeparttable}
\caption{Violation rates and FZ0 losses of VaR and ES forecasts for the DJIA over the COVID-19 period (2020).}
\label{tab:forecast-var-es-loss-djia-covid}
\begin{tabular}{lrrrrlrrrrrl}
\toprule
& \multicolumn{5}{l}{$\alpha = 1\%$} & & \multicolumn{5}{l}{$\alpha = 5\%$} \\
\cline{2-6} \cline{8-12}
& \multicolumn{1}{c}{$\hat{\alpha}$} & \multicolumn{1}{c}{$p_{UC}$} & \multicolumn{1}{c}{$p_{DQ}$} & \multicolumn{1}{c}{FZ0} & \multicolumn{1}{c}{$p_{MCS}$} & & \multicolumn{1}{c}{$\hat{\alpha}$} & \multicolumn{1}{c}{$p_{UC}$} & \multicolumn{1}{c}{$p_{DQ}$} & \multicolumn{1}{c}{FZ0} & \multicolumn{1}{c}{$p_{MCS}$} \\
\midrule
SV-N & $2.37$ & $0.06$ & $0.00$ & $2.1068$ & $0.95^{**}$ &  & $8.70$ & $0.01$ & $0.00$ & $1.3950$ & $0.99^{**}$ \\
SV-T & $2.77$ & $0.02$ & $0.00$ & $2.0141$ & $0.95^{**}$ &  & $8.70$ & $0.01$ & $0.00$ & $1.4082$ & $0.88^{**}$ \\
SV-AZ-SN & $3.16$ & $0.01$ & $0.00$ & $2.0181$ & $0.95^{**}$ &  & $7.91$ & $0.05$ & $0.00$ & $1.3783$ & $0.99^{**}$ \\
SV-AZ-ST & $2.77$ & $0.02$ & $0.00$ & $2.0239$ & $0.95^{**}$ &  & $7.91$ & $0.05$ & $0.00$ & $1.3924$ & $0.99^{**}$ \\
SV-FS-SN & $1.98$ & $0.17$ & $0.15$ & $1.9575$ & $0.95^{**}$ &  & $7.51$ & $0.09$ & $0.00$ & $1.3595$ & $1.00^{**}$ \\
SV-FS-ST & $1.98$ & $0.17$ & $0.26$ & $2.0788$ & $0.84^{**}$ &  & $6.72$ & $0.23$ & $0.00$ & $1.3929$ & $0.99^{**}$ \\
SV-GH-ST & $2.77$ & $0.02$ & $0.00$ & $1.9222$ & $1.00^{**}$ &  & $7.11$ & $0.15$ & $0.00$ & $1.3943$ & $0.99^{**}$ \\
\cmidrule(lr){1-12}
RSV-N & $4.35$ & $0.00$ & $0.00$ & $2.4074$ & $0.26^{**}$ &  & $8.30$ & $0.03$ & $0.00$ & $1.4272$ & $0.87^{**}$ \\
RSV-T & $3.95$ & $0.00$ & $0.00$ & $2.2427$ & $0.39^{**}$ &  & $8.30$ & $0.03$ & $0.00$ & $1.4031$ & $0.98^{**}$ \\
RSV-AZ-SN & $3.56$ & $0.00$ & $0.00$ & $2.1697$ & $0.75^{**}$ &  & $8.30$ & $0.03$ & $0.00$ & $1.3869$ & $0.99^{**}$ \\
RSV-AZ-ST & $3.16$ & $0.01$ & $0.00$ & $2.0386$ & $0.95^{**}$ &  & $7.91$ & $0.05$ & $0.00$ & $1.3788$ & $0.99^{**}$ \\
RSV-FS-SN & $3.56$ & $0.00$ & $0.00$ & $2.0603$ & $0.95^{**}$ &  & $8.30$ & $0.03$ & $0.00$ & $1.4004$ & $0.99^{**}$ \\
RSV-FS-ST & $3.95$ & $0.00$ & $0.00$ & $2.1353$ & $0.75^{**}$ &  & $8.30$ & $0.03$ & $0.00$ & $1.4001$ & $0.99^{**}$ \\
\cmidrule(lr){1-12}
EGARCH-N & $3.95$ & $0.00$ & $0.00$ & $2.8115$ & $0.19^{*}$ &  & $8.30$ & $0.03$ & $0.01$ & $1.5021$ & $0.64^{**}$ \\
EGARCH-T & $2.77$ & $0.02$ & $0.05$ & $2.2640$ & $0.75^{**}$ &  & $9.09$ & $0.01$ & $0.00$ & $1.4661$ & $0.87^{**}$ \\
\cmidrule(lr){1-12}
REGARCH-N & $3.95$ & $0.00$ & $0.00$ & $2.5639$ & $0.15^{*}$ &  & $7.91$ & $0.05$ & $0.00$ & $1.4301$ & $0.73^{**}$ \\
REGARCH-T & $3.16$ & $0.01$ & $0.00$ & $2.1367$ & $0.75^{**}$ &  & $7.91$ & $0.05$ & $0.00$ & $1.4050$ & $0.97^{**}$ \\
\bottomrule
\end{tabular}
\begin{tablenotes}
\footnotesize
\item \textit{Notes:} $\hat{\alpha}$ represents the empirical violation rate (\%).
$p_{UC}$ indicates the $p$-value of the unconditional coverage test of \cite{kupiec_techniques_1995}.
$p_{DQ}$ indicates the $p$-value of the dynamic quantile test of \cite{engle_caviar_2004}, with four lags of the violation indicator and the VaR forecast as regressors.
FZ0 denotes the average FZ0 loss.
$p_{MCS}$ indicates the MCS $p$-value.
A double asterisk ($^{**}$) and a single asterisk ($^{*}$) on $p_{MCS}$ indicate that the model belongs to the 75\% and the 90\% MCS, respectively (equivalently, $p_{MCS}>0.25$ and $p_{MCS}>0.10$); the 75\% MCS is nested within the 90\% MCS.
RSV-GH-ST is excluded (see text).
\end{tablenotes}
\end{threeparttable}
\end{table}

\begin{table}[tbp]
\centering
\begin{threeparttable}
\caption{QLIKE scores of volatility forecasts under $\psi=1$ and freely estimated $\psi$, for the DJIA over the COVID-19 period (2020).}
\label{tab:forecast-vol-qlike-djia-covid-psi}
\begin{tabular}{lllllll}
\toprule
& & \multicolumn{4}{c}{Proxy} & \\
\cmidrule(lr){3-6}
& $\psi$ & RV5 & RK & BV & Med & Average \\
\midrule
RSV-N & $1$ & $0.2741$ & $0.2739$ & $0.3225$ & $0.2595$ & $0.2825$ \\
 & free & $0.2431^{**}$ & $0.2411^{**}$ & $0.2607^{**}$ & $0.2102^{**}$ & $0.2388$ \\
RSV-T & $1$ & $0.2659$ & $0.2657$ & $0.3119$ & $0.2506$ & $0.2735$ \\
 & free & $0.2395^{**}$ & $0.2399^{**}$ & $0.2544^{**}$ & $0.2055^{**}$ & $0.2349$ \\
RSV-AZ-SN & $1$ & $0.2715$ & $0.2721$ & $0.3206$ & $0.2584$ & $0.2807$ \\
 & free & $0.2349^{**}$ & $0.2355^{**}$ & $0.2577^{**}$ & $0.2053^{**}$ & $0.2333$ \\
RSV-AZ-ST & $1$ & $0.2635$ & $0.2642^{*}$ & $0.3094$ & $0.2491$ & $0.2715$ \\
 & free & $0.2336^{**}$ & $0.2341^{**}$ & $0.2522^{**}$ & $0.2019^{**}$ & $0.2304$ \\
RSV-FS-SN & $1$ & $0.2885$ & $0.2897$ & $0.3433$ & $0.2781$ & $0.2999$ \\
 & free & $0.2525^{**}$ & $0.2533^{**}$ & $0.2908$ & $0.2305$ & $0.2568$ \\
RSV-FS-ST & $1$ & $0.2811$ & $0.2822$ & $0.3331$ & $0.2695$ & $0.2915$ \\
 & free & $0.2464^{**}$ & $0.2468^{**}$ & $0.2814$ & $0.2230^{*}$ & $0.2494$ \\
\cmidrule(lr){1-7}
REGARCH-N & $1$ & $0.2568^{**}$ & $0.2549^{**}$ & $0.2951$ & $0.2383^{*}$ & $0.2613$ \\
 & free & $0.2630^{**}$ & $0.2605^{**}$ & $0.3017$ & $0.2445^{*}$ & $0.2674$ \\
REGARCH-T & $1$ & $0.2461^{**}$ & $0.2439^{**}$ & $0.2793^{**}$ & $0.2250^{**}$ & $0.2486$ \\
 & free & $0.2485^{**}$ & $0.2460^{**}$ & $0.2806^{**}$ & $0.2266^{**}$ & $0.2504$ \\
\bottomrule
\end{tabular}
\begin{tablenotes}
\footnotesize
\item \textit{Notes:} A double asterisk ($^{**}$) and a single asterisk ($^{*}$) indicate that the model belongs to the 75\% and the 90\% model confidence set (MCS), respectively, among the models shown, computed separately for each proxy; the 75\% MCS is nested within the 90\% MCS.
The $\psi$ column indicates whether the slope coefficient of the measurement equation is fixed at unity ($1$) or estimated freely (free).
RSV-GH-ST is excluded (see text).
The average column reports the mean QLIKE score across the four proxies.
\end{tablenotes}
\end{threeparttable}
\end{table}

\begin{table}[tbp]
\centering
\begin{threeparttable}
\caption{Violation rates and FZ0 losses of VaR and ES forecasts under $\psi=1$ and freely estimated $\psi$, for the DJIA over the COVID-19 period (2020).}
\label{tab:forecast-var-es-loss-djia-covid-psi}
\begin{tabular}{llllll lllll}
\toprule
& & \multicolumn{4}{l}{$\alpha = 1\%$} & & \multicolumn{4}{l}{$\alpha = 5\%$} \\
\cline{3-6} \cline{8-11}
& \multicolumn{1}{c}{$\psi$} & \multicolumn{1}{c}{$\hat{\alpha}$} & \multicolumn{1}{c}{$p_{UC}$} & \multicolumn{1}{c}{$p_{DQ}$} & \multicolumn{1}{c}{FZ0} & & \multicolumn{1}{c}{$\hat{\alpha}$} & \multicolumn{1}{c}{$p_{UC}$} & \multicolumn{1}{c}{$p_{DQ}$} & \multicolumn{1}{c}{FZ0} \\
\midrule
RSV-N & $1$ & $4.35$ & $0.00$ & $0.00$ & $2.4074^{*}$ & & $8.30$ & $0.03$ & $0.00$ & $1.4272^{**}$ \\
 & free & $2.37$ & $0.06$ & $0.32$ & $1.9372^{**}$ & & $5.93$ & $0.51$ & $0.12$ & $1.3260^{**}$ \\
RSV-T & $1$ & $3.95$ & $0.00$ & $0.00$ & $2.2427^{*}$ & & $8.30$ & $0.03$ & $0.00$ & $1.4031^{**}$ \\
 & free & $1.98$ & $0.17$ & $0.73$ & $1.8382^{**}$ & & $5.93$ & $0.51$ & $0.13$ & $1.3079^{**}$ \\
RSV-AZ-SN & $1$ & $3.56$ & $0.00$ & $0.00$ & $2.1697^{*}$ & & $8.30$ & $0.03$ & $0.00$ & $1.3869^{**}$ \\
 & free & $1.98$ & $0.17$ & $0.74$ & $1.8126^{**}$ & & $5.93$ & $0.51$ & $0.13$ & $1.3043^{**}$ \\
RSV-AZ-ST & $1$ & $3.16$ & $0.01$ & $0.00$ & $2.0386^{**}$ & & $7.91$ & $0.05$ & $0.00$ & $1.3788^{**}$ \\
 & free & $1.98$ & $0.17$ & $0.74$ & $1.7770^{**}$ & & $5.93$ & $0.51$ & $0.13$ & $1.3017^{**}$ \\
RSV-FS-SN & $1$ & $3.56$ & $0.00$ & $0.00$ & $2.0603^{**}$ & & $8.30$ & $0.03$ & $0.00$ & $1.4004^{**}$ \\
 & free & $2.77$ & $0.02$ & $0.00$ & $1.8431^{**}$ & & $7.51$ & $0.09$ & $0.00$ & $1.3295^{**}$ \\
RSV-FS-ST & $1$ & $3.95$ & $0.00$ & $0.00$ & $2.1353^{**}$ & & $8.30$ & $0.03$ & $0.00$ & $1.4001^{**}$ \\
 & free & $3.56$ & $0.00$ & $0.00$ & $1.9316^{**}$ & & $7.11$ & $0.15$ & $0.03$ & $1.3300^{**}$ \\
\cmidrule(lr){1-11}
REGARCH-N & $1$ & $3.95$ & $0.00$ & $0.00$ & $2.5639^{*}$ & & $7.91$ & $0.05$ & $0.00$ & $1.4301^{*}$ \\
 & free & $3.95$ & $0.00$ & $0.00$ & $2.6763^{*}$ & & $8.30$ & $0.03$ & $0.00$ & $1.4620^{*}$ \\
REGARCH-T & $1$ & $3.16$ & $0.01$ & $0.00$ & $2.1367^{*}$ & & $7.91$ & $0.05$ & $0.00$ & $1.4050^{**}$ \\
 & free & $3.16$ & $0.01$ & $0.00$ & $2.0969^{**}$ & & $8.30$ & $0.03$ & $0.00$ & $1.4034^{**}$ \\
\bottomrule
\end{tabular}
\begin{tablenotes}
\footnotesize
\item \textit{Notes:} $\hat{\alpha}$ represents the empirical violation rate (\%).
$p_{UC}$ indicates the $p$-value of the unconditional coverage test of \cite{kupiec_techniques_1995}.
$p_{DQ}$ indicates the $p$-value of the dynamic quantile test of \cite{engle_caviar_2004}, with four lags of the violation indicator and the VaR forecast as regressors.
FZ0 denotes the average FZ0 loss.
A double asterisk ($^{**}$) and a single asterisk ($^{*}$) on FZ0 indicate that the model belongs to the 75\% and the 90\% MCS, respectively, among the models shown; the 75\% MCS is nested within the 90\% MCS.
The $\psi$ column indicates whether the slope coefficient of the measurement equation is fixed at unity ($1$) or estimated freely (free).
RSV-GH-ST is excluded (see text).
\end{tablenotes}
\end{threeparttable}
\end{table}

\begin{table}[tbp]
\centering
\begin{threeparttable}
\caption{Descriptive statistics of daily returns and the logarithms of 5-minute RVs for the N225 (2012--2020).}
\label{tab:stats-n225-covid}
\begin{tabular}{lcccccccc}
\toprule
& Mean & SD & Skew & Kurt & Min & Max & JB & LB \\
\midrule
Return & $0.053$ & $1.323$ & $-0.274$ & $7.617$ & $-8.253$ & $7.731$ & $0.00$ & $0.64$ \\ 
& $(0.028)$ & & $(0.052)$ & $(0.104)$ \\
Log-RV & $-0.921$ & $0.925$ & $0.735$ & $4.230$ & $-3.894$ & $3.562$ & $0.00$ & $0.00$ \\ 
& $(0.020)$ & & $(0.052)$ & $(0.104)$ \\
\bottomrule
\end{tabular}
\begin{tablenotes}
\footnotesize
\item \textit{Notes:} Standard errors are shown in parentheses.
JB refers to the $p$-value of the Jarque–Bera test.
LB denotes the $p$-value of the \citet{ljung_measure_1978} statistic, adjusted as in \citet{diebold_empirical_1988}.
\end{tablenotes}
\end{threeparttable}
\end{table}

\begin{table}[tbp]
\centering
\begin{threeparttable}
\caption{QLIKE scores of volatility forecasts for the N225 over the COVID-19 period (2020).}
\label{tab:forecast-vol-qlike-n225-covid}
\begin{tabular}{llllllrr}
\toprule
& \multicolumn{4}{l}{Proxy} & & \multicolumn{2}{l}{Average} \\
\cmidrule(lr){2-5} \cmidrule(lr){7-8}
& RV5 & RK & BV & Med & & Score & Rank \\
\midrule
SV-N & $0.345$ & $0.365$ & $0.328$ & $0.296$ & & $0.333$ & $16.0$ \\
SV-T & $0.350$ & $0.370$ & $0.335$ & $0.299^{*}$ & & $0.339$ & $17.8$ \\
SV-AZ-SN & $0.335^{*}$ & $0.355^{*}$ & $0.321$ & $0.290^{*}$ & & $0.325$ & $11.0$ \\
SV-AZ-ST & $0.341^{*}$ & $0.361^{*}$ & $0.326$ & $0.292^{*}$ & & $0.330$ & $13.8$ \\
SV-FS-SN & $0.343$ & $0.363$ & $0.326$ & $0.301$ & & $0.333$ & $15.8$ \\
SV-FS-ST & $0.336^{**}$ & $0.357^{**}$ & $0.323^{*}$ & $0.290^{**}$ & & $0.326$ & $12.0$ \\
SV-GH-ST & $0.326^{**}$ & $0.343^{**}$ & $0.318^{**}$ & $0.275^{**}$ & & $0.316$ & $10.0$ \\
\cmidrule(lr){1-8}
RSV-N & $0.293^{**}$ & $0.311^{**}$ & $0.267^{**}$ & $0.245^{**}$ & & $0.279$ & $7.0$ \\
RSV-T & $0.290^{**}$ & $0.308^{**}$ & $0.264^{**}$ & $0.238^{**}$ & & $0.275$ & $5.0$ \\
RSV-AZ-SN & $0.293^{**}$ & $0.311^{**}$ & $0.267^{**}$ & $0.244^{**}$ & & $0.279$ & $6.0$ \\
RSV-AZ-ST & $0.289^{**}$ & $0.307^{**}$ & $0.264^{**}$ & $0.238^{**}$ & & $0.274$ & $3.2$ \\
RSV-FS-SN & $0.300^{**}$ & $0.317^{**}$ & $0.273^{**}$ & $0.255^{**}$ & & $0.286$ & $9.0$ \\
RSV-FS-ST & $0.296^{**}$ & $0.313^{**}$ & $0.270^{**}$ & $0.250^{**}$ & & $0.282$ & $8.0$ \\
RSV-GH-ST & $0.289^{**}$ & $0.308^{**}$ & $0.264^{**}$ & $0.238^{**}$ & & $0.275$ & $3.8$ \\
\cmidrule(lr){1-8}
EGARCH-N & $0.340^{**}$ & $0.362^{**}$ & $0.327^{**}$ & $0.297^{**}$ & & $0.332$ & $14.5$ \\
EGARCH-T & $0.340^{**}$ & $0.361^{**}$ & $0.329^{*}$ & $0.297^{**}$ & & $0.332$ & $15.2$ \\
\cmidrule(lr){1-8}
REGARCH-N & $0.284^{**}$ & $0.305^{**}$ & $0.259^{**}$ & $0.227^{**}$ & & $0.268$ & $1.2$ \\
REGARCH-T & $0.284^{**}$ & $0.305^{**}$ & $0.260^{**}$ & $0.224^{**}$ & & $0.268$ & $1.8$ \\
\bottomrule
\end{tabular}
\begin{tablenotes}
\footnotesize
\item \textit{Notes:} A double asterisk ($^{**}$) and a single asterisk ($^{*}$) indicate that the model belongs to the 75\% and the 90\% model confidence set (MCS), respectively; the 75\% MCS is nested within the 90\% MCS.
The average columns report the mean QLIKE scores and ranks across the four volatility proxies.
\end{tablenotes}
\end{threeparttable}
\end{table}

\begin{table}[tbp]%
\centering
\begin{threeparttable}
\caption{Violation rates and FZ0 losses of VaR and ES forecasts for the N225 over the COVID-19 period (2020).}
\label{tab:forecast-var-es-loss-n225-covid}
\begin{tabular}{lrrrrlrrrrrl}
\toprule
& \multicolumn{5}{l}{$\alpha = 1\%$} & & \multicolumn{5}{l}{$\alpha = 5\%$} \\
\cline{2-6} \cline{8-12}
& \multicolumn{1}{c}{$\hat{\alpha}$} & \multicolumn{1}{c}{$p_{UC}$} & \multicolumn{1}{c}{$p_{DQ}$} & \multicolumn{1}{c}{FZ0} & \multicolumn{1}{c}{$p_{MCS}$} & & \multicolumn{1}{c}{$\hat{\alpha}$} & \multicolumn{1}{c}{$p_{UC}$} & \multicolumn{1}{c}{$p_{DQ}$} & \multicolumn{1}{c}{FZ0} & \multicolumn{1}{c}{$p_{MCS}$} \\
\midrule
SV-N & $0.83$ & $0.78$ & $0.99$ & $1.3999$ & $1.00^{**}$ &  & $6.20$ & $0.41$ & $0.23$ & $1.0655$ & $0.98^{**}$ \\
SV-T & $0.83$ & $0.78$ & $0.99$ & $1.4068$ & $1.00^{**}$ &  & $6.20$ & $0.41$ & $0.31$ & $1.0660$ & $0.98^{**}$ \\
SV-AZ-SN & $0.83$ & $0.78$ & $0.99$ & $1.3980$ & $1.00^{**}$ &  & $5.79$ & $0.58$ & $0.79$ & $1.0523$ & $0.98^{**}$ \\
SV-AZ-ST & $0.83$ & $0.78$ & $0.99$ & $1.3977$ & $1.00^{**}$ &  & $4.96$ & $0.98$ & $0.89$ & $1.0600$ & $0.98^{**}$ \\
SV-FS-SN & $0.83$ & $0.78$ & $0.99$ & $1.4024$ & $1.00^{**}$ &  & $5.79$ & $0.58$ & $0.75$ & $1.0483$ & $1.00^{**}$ \\
SV-FS-ST & $0.83$ & $0.78$ & $0.99$ & $1.4038$ & $1.00^{**}$ &  & $5.79$ & $0.58$ & $0.76$ & $1.0509$ & $0.98^{**}$ \\
SV-GH-ST & $0.41$ & $0.30$ & $0.98$ & $1.4291$ & $0.98^{**}$ &  & $4.55$ & $0.74$ & $0.80$ & $1.0603$ & $0.98^{**}$ \\
\cmidrule(lr){1-12}
RSV-N & $2.48$ & $0.05$ & $0.58$ & $1.5306$ & $0.75^{**}$ &  & $7.44$ & $0.10$ & $0.73$ & $1.1085$ & $0.83^{**}$ \\
RSV-T & $2.07$ & $0.14$ & $0.90$ & $1.4240$ & $0.99^{**}$ &  & $7.85$ & $0.06$ & $0.72$ & $1.0990$ & $0.92^{**}$ \\
RSV-AZ-SN & $2.48$ & $0.05$ & $0.59$ & $1.4777$ & $0.91^{**}$ &  & $7.44$ & $0.10$ & $0.55$ & $1.1015$ & $0.92^{**}$ \\
RSV-AZ-ST & $1.65$ & $0.35$ & $0.91$ & $1.4021$ & $1.00^{**}$ &  & $7.44$ & $0.10$ & $0.73$ & $1.0966$ & $0.93^{**}$ \\
RSV-FS-SN & $1.65$ & $0.35$ & $0.90$ & $1.4428$ & $0.98^{**}$ &  & $7.85$ & $0.06$ & $0.72$ & $1.1097$ & $0.80^{**}$ \\
RSV-FS-ST & $1.65$ & $0.35$ & $0.90$ & $1.4426$ & $0.98^{**}$ &  & $7.85$ & $0.06$ & $0.72$ & $1.1011$ & $0.90^{**}$ \\
RSV-GH-ST & $1.65$ & $0.35$ & $0.90$ & $1.4152$ & $1.00^{**}$ &  & $7.44$ & $0.10$ & $0.73$ & $1.0892$ & $0.98^{**}$ \\
\cmidrule(lr){1-12}
EGARCH-N & $2.07$ & $0.14$ & $0.00$ & $1.4420$ & $0.98^{**}$ &  & $5.79$ & $0.58$ & $0.63$ & $1.0922$ & $0.92^{**}$ \\
EGARCH-T & $0.83$ & $0.78$ & $0.99$ & $1.3870$ & $1.00^{**}$ &  & $6.61$ & $0.27$ & $0.19$ & $1.0821$ & $0.95^{**}$ \\
\cmidrule(lr){1-12}
REGARCH-N & $2.07$ & $0.14$ & $0.91$ & $1.4989$ & $0.84^{**}$ &  & $6.20$ & $0.41$ & $0.71$ & $1.0997$ & $0.91^{**}$ \\
REGARCH-T & $1.65$ & $0.35$ & $0.91$ & $1.3740$ & $1.00^{**}$ &  & $6.61$ & $0.27$ & $0.75$ & $1.0932$ & $0.95^{**}$ \\
\bottomrule
\end{tabular}
\begin{tablenotes}
\footnotesize
\item \textit{Notes:} $\hat{\alpha}$ represents the empirical violation rate (\%).
$p_{UC}$ indicates the $p$-value of the unconditional coverage test of \cite{kupiec_techniques_1995}.
$p_{DQ}$ indicates the $p$-value of the dynamic quantile test of \cite{engle_caviar_2004}, with four lags of the violation indicator and the VaR forecast as regressors.
FZ0 denotes the average FZ0 loss.
$p_{MCS}$ indicates the MCS $p$-value.
A double asterisk ($^{**}$) and a single asterisk ($^{*}$) on $p_{MCS}$ indicate that the model belongs to the 75\% and the 90\% MCS, respectively (equivalently, $p_{MCS}>0.25$ and $p_{MCS}>0.10$); the 75\% MCS is nested within the 90\% MCS.
\end{tablenotes}
\end{threeparttable}
\end{table}

\clearpage

\begin{table}[tbp]
\centering
\begin{threeparttable}
\caption{QLIKE scores of volatility forecasts under $\psi=1$ and freely estimated $\psi$, for the N225 over the COVID-19 period (2020).}
\label{tab:forecast-vol-qlike-n225-covid-psi}
\begin{tabular}{lllllll}
\toprule
& & \multicolumn{4}{c}{Proxy} & \\
\cmidrule(lr){3-6}
& $\psi$ & RV5 & RK & BV & Med & Average \\
\midrule
RSV-N & $1$ & $0.2933^{**}$ & $0.3109^{**}$ & $0.2671^{**}$ & $0.2447^{**}$ & $0.2790$ \\
 & free & $0.3119^{**}$ & $0.3311^{**}$ & $0.2815^{**}$ & $0.2549^{**}$ & $0.2949$ \\
RSV-T & $1$ & $0.2895^{**}$ & $0.3080^{**}$ & $0.2642^{**}$ & $0.2384^{**}$ & $0.2750$ \\
 & free & $0.3142^{**}$ & $0.3343^{**}$ & $0.2846^{**}$ & $0.2535^{**}$ & $0.2967$ \\
RSV-AZ-SN & $1$ & $0.2928^{**}$ & $0.3107^{**}$ & $0.2668^{**}$ & $0.2440^{**}$ & $0.2786$ \\
 & free & $0.3099^{**}$ & $0.3290^{**}$ & $0.2799^{**}$ & $0.2530^{**}$ & $0.2930$ \\
RSV-AZ-ST & $1$ & $0.2888^{**}$ & $0.3072^{**}$ & $0.2639^{**}$ & $0.2378^{**}$ & $0.2744$ \\
 & free & $0.3109^{**}$ & $0.3310^{**}$ & $0.2819^{**}$ & $0.2507^{**}$ & $0.2936$ \\
RSV-FS-SN & $1$ & $0.2998^{**}$ & $0.3165^{**}$ & $0.2730^{**}$ & $0.2550^{**}$ & $0.2861$ \\
 & free & $0.3027^{**}$ & $0.3196^{**}$ & $0.2729^{**}$ & $0.2547^{**}$ & $0.2875$ \\
RSV-FS-ST & $1$ & $0.2961^{**}$ & $0.3134^{**}$ & $0.2698^{**}$ & $0.2495^{**}$ & $0.2822$ \\
 & free & $0.3007^{**}$ & $0.3184^{**}$ & $0.2715^{**}$ & $0.2507^{**}$ & $0.2854$ \\
RSV-GH-ST & $1$ & $0.2890^{**}$ & $0.3077^{**}$ & $0.2640^{**}$ & $0.2375^{**}$ & $0.2746$ \\
\cmidrule(lr){1-7}
REGARCH-N & $1$ & $0.2836^{**}$ & $0.3046^{**}$ & $0.2590^{**}$ & $0.2266^{**}$ & $0.2685$ \\
 & free & $0.2859^{**}$ & $0.3064^{**}$ & $0.2639^{**}$ & $0.2309^{**}$ & $0.2718$ \\
REGARCH-T & $1$ & $0.2840^{**}$ & $0.3055^{**}$ & $0.2598^{**}$ & $0.2240^{**}$ & $0.2683$ \\
 & free & $0.2822^{**}$ & $0.3034^{**}$ & $0.2593^{**}$ & $0.2231^{**}$ & $0.2670$ \\
\bottomrule
\end{tabular}
\begin{tablenotes}
\footnotesize
\item \textit{Notes:} A double asterisk ($^{**}$) and a single asterisk ($^{*}$) indicate that the model belongs to the 75\% and the 90\% model confidence set (MCS), respectively, among the models shown, computed separately for each proxy; the 75\% MCS is nested within the 90\% MCS.
The $\psi$ column indicates whether the slope coefficient of the measurement equation is fixed at unity ($1$) or estimated freely (free).
RSV-GH-ST with a freely estimated $\psi$ is excluded (see text).
The average column reports the mean QLIKE score across the four proxies.
\end{tablenotes}
\end{threeparttable}
\end{table}

\begin{table}[tbp]
\centering
\begin{threeparttable}
\caption{Violation rates and FZ0 losses of VaR and ES forecasts under $\psi=1$ and freely estimated $\psi$, for the N225 over the COVID-19 period (2020).}
\label{tab:forecast-var-es-loss-n225-covid-psi}
\begin{tabular}{llllll lllll}
\toprule
& & \multicolumn{4}{l}{$\alpha = 1\%$} & & \multicolumn{4}{l}{$\alpha = 5\%$} \\
\cline{3-6} \cline{8-11}
& \multicolumn{1}{c}{$\psi$} & \multicolumn{1}{c}{$\hat{\alpha}$} & \multicolumn{1}{c}{$p_{UC}$} & \multicolumn{1}{c}{$p_{DQ}$} & \multicolumn{1}{c}{FZ0} & & \multicolumn{1}{c}{$\hat{\alpha}$} & \multicolumn{1}{c}{$p_{UC}$} & \multicolumn{1}{c}{$p_{DQ}$} & \multicolumn{1}{c}{FZ0} \\
\midrule
RSV-N & $1$ & $2.48$ & $0.05$ & $0.58$ & $1.5306^{**}$ & & $7.44$ & $0.10$ & $0.73$ & $1.1085^{**}$ \\
 & free & $2.07$ & $0.14$ & $0.90$ & $1.5071^{**}$ & & $6.20$ & $0.41$ & $0.43$ & $1.1371^{**}$ \\
RSV-T & $1$ & $2.07$ & $0.14$ & $0.90$ & $1.4240^{**}$ & & $7.85$ & $0.06$ & $0.72$ & $1.0990^{**}$ \\
 & free & $2.07$ & $0.14$ & $0.91$ & $1.4423^{**}$ & & $6.20$ & $0.41$ & $0.43$ & $1.1325^{**}$ \\
RSV-AZ-SN & $1$ & $2.48$ & $0.05$ & $0.59$ & $1.4777^{**}$ & & $7.44$ & $0.10$ & $0.55$ & $1.1015^{**}$ \\
 & free & $2.07$ & $0.14$ & $0.91$ & $1.4954^{**}$ & & $6.20$ & $0.41$ & $0.43$ & $1.1286^{**}$ \\
RSV-AZ-ST & $1$ & $1.65$ & $0.35$ & $0.91$ & $1.4021^{**}$ & & $7.44$ & $0.10$ & $0.73$ & $1.0966^{**}$ \\
 & free & $1.65$ & $0.35$ & $0.96$ & $1.4635^{**}$ & & $6.20$ & $0.41$ & $0.43$ & $1.1291^{**}$ \\
RSV-FS-SN & $1$ & $1.65$ & $0.35$ & $0.90$ & $1.4428^{**}$ & & $7.85$ & $0.06$ & $0.72$ & $1.1097^{**}$ \\
 & free & $1.65$ & $0.35$ & $0.90$ & $1.4562^{**}$ & & $7.02$ & $0.17$ & $0.51$ & $1.1188^{**}$ \\
RSV-FS-ST & $1$ & $1.65$ & $0.35$ & $0.90$ & $1.4426^{**}$ & & $7.85$ & $0.06$ & $0.72$ & $1.1011^{**}$ \\
 & free & $2.07$ & $0.14$ & $0.90$ & $1.4789^{**}$ & & $6.61$ & $0.27$ & $0.53$ & $1.1079^{**}$ \\
RSV-GH-ST & $1$ & $1.65$ & $0.35$ & $0.90$ & $1.4152^{**}$ & & $7.44$ & $0.10$ & $0.73$ & $1.0892^{**}$ \\
\cmidrule(lr){1-11}
REGARCH-N & $1$ & $2.07$ & $0.14$ & $0.91$ & $1.4989^{**}$ & & $6.20$ & $0.41$ & $0.71$ & $1.0997^{**}$ \\
 & free & $2.07$ & $0.14$ & $0.90$ & $1.5024^{**}$ & & $6.61$ & $0.27$ & $0.70$ & $1.0982^{**}$ \\
REGARCH-T & $1$ & $1.65$ & $0.35$ & $0.91$ & $1.3740^{**}$ & & $6.61$ & $0.27$ & $0.75$ & $1.0932^{**}$ \\
 & free & $1.65$ & $0.35$ & $1.00$ & $1.3855^{**}$ & & $7.02$ & $0.17$ & $0.70$ & $1.0975^{**}$ \\
\bottomrule
\end{tabular}
\begin{tablenotes}
\footnotesize
\item \textit{Notes:} $\hat{\alpha}$ represents the empirical violation rate (\%).
$p_{UC}$ indicates the $p$-value of the unconditional coverage test of \cite{kupiec_techniques_1995}.
$p_{DQ}$ indicates the $p$-value of the dynamic quantile test of \cite{engle_caviar_2004}, with four lags of the violation indicator and the VaR forecast as regressors.
FZ0 denotes the average FZ0 loss.
A double asterisk ($^{**}$) and a single asterisk ($^{*}$) on FZ0 indicate that the model belongs to the 75\% and the 90\% MCS, respectively, among the models shown; the 75\% MCS is nested within the 90\% MCS.
The $\psi$ column indicates whether the slope coefficient of the measurement equation is fixed at unity ($1$) or estimated freely (free).
RSV-GH-ST with a freely estimated $\psi$ is excluded (see text).
\end{tablenotes}
\end{threeparttable}
\end{table}

\clearpage

\clearpage

\clearpage

\clearpage

\clearpage

\clearpage

\clearpage

\bibliography{rsvst202608}
\bibliographystyle{agsm}